\documentclass[PAPER, cernpreprint, atlasdraft=false, texlive=2023, UKenglish, orcidlogo, texmf, orcidlogo]{atlasdoc}
\usepackage[full]{atlaspackage}
\usepackage{atlasbiblatex}

\usepackage[process=true]{atlasphysics}
\graphicspath{{logos/}{figures/}}

\usepackage{ANA-STDM-2018-17-PAPER-defs}
\usepackage{wasysym}
\usepackage{bm,nicefrac}

\AtlasTitle{Precise measurements of $W$- and $Z$-boson transverse momentum
spectra with the ATLAS detector using $pp$ collisions at $\sqrt{s} = 5.02$ TeV and $13$ TeV}

\AtlasAbstract{%
This paper describes measurements of the transverse momentum spectra
of $W$ and $Z$ bosons produced in proton--proton collisions at
centre-of-mass energies of $\sqrt{s}=5.02$ TeV and $\sqrt{s}=13$ TeV
with the ATLAS experiment at the Large Hadron Collider. Measurements
are performed in the electron and muon channels, $W\to\ell\nu$ and
$Z\to\ell\ell$ ($\ell=e$ or $\mu$), and for $W$ events further
separated by charge. The data were collected in 2017 and 2018, in
dedicated runs with reduced instantaneous luminosity, and correspond
to 255 pb$^{-1}$ and 338 pb$^{-1}$ at $\sqrt{s}=5.02$ TeV and 13
TeV, respectively.  These conditions optimise the reconstruction of
the $W$-boson transverse momentum. The distributions observed in the
electron and muon channels are unfolded, combined, and compared to QCD
calculations based on parton shower Monte Carlo event generators and
analytical resummation.  The description of the transverse momentum
distributions by Monte Carlo event generators is imperfect and shows
significant differences largely common to $W^-$, $W^+$ and $Z$
production. The agreement is better at $\sqrt{s}=5.02$ TeV, especially
for predictions that were tuned to $Z$ production data at $\sqrt{s}=7$
TeV. Higher-order, resummed predictions based on \textsc{DYTurbo}
generally match the data best across the spectra.  Distribution ratios
are also presented and test the understanding of differences between
the production processes.
}
\author{The ATLAS Collaboration}

\AtlasJournalRef{Eur. Phys. J. C 84 (2024) 1126}
\AtlasDOI{10.1140/epjc/s10052-024-13414-0}

\PreprintIdNumber{CERN-EP-2024-080}

\hypersetup{pdftitle={ATLAS document},pdfauthor={The ATLAS Collaboration}}

\begin{document}

\maketitle

\tableofcontents

%
\section{Introduction}
\label{sec:intro}

Measurements of $W$- and $Z$-boson production in proton--proton
($pp$) collisions at the Large Hadron Collider (LHC) constitute a sensitive test of Quantum
Chromodynamics (QCD).  The transverse momentum (\ptv) of the
gauge bosons $V$ (\Wboson, \Zboson) with respect to the initial
proton beam arises from higher order
corrections to the leading order Drell--Yan processes, and from
non-perturbative effects such as the primordial $k_\mathrm{T}$
of the incoming partons. Precise
measurements and predictions of the spectra in the region $\ptv\lesssim 30\,\GeV$ are of
particular interest for the measurement of the \Wboson-boson mass
at hadron colliders~\cite{ATLAS:2024erm,CDF:2022hxs,D0:2013jba,LHCb:2021bjt}.

Over the last years, theoretical predictions have been calculated
including increasingly higher order corrections. These comprise the
fixed-order corrections, known up to third order in the strong
coupling $\alpha_S$~\cite{Boughezal:2015ded,
Gehrmann-DeRidder:2015wbt, Boughezal:2016dtm}, and the resummation
of logarithmic terms from soft and collinear emissions using complex
formalisms to obtain reliable predictions in the low \ptv
region. Analytic approaches based \textit{e.g.} on transverse momentum
resummation~\cite{Bozzi:2005wk} have been extended up to fourth
order~\cite{Bizon:2019zgf,Neumann:2022lft,Camarda:2023dqn}. Parton
shower and hadronisation models, such as those implemented in
\PYTHIA~\cite{Sjostrand:2007gs}, \HERWIG~\cite{Bellm:2015jjp} or
\SHERPA~\cite{Bothmann:2019yzt} provide an alternative approach and
predict the complete event topology.

Experimentally, a measurement of the \Zboson-boson \pt\ spectrum (\ptz) is
relatively straightforward through the transverse momentum of the
pair of charged decay leptons, \ptdilep. It was measured
previously in $pp$ collisions at the LHC by
the ATLAS Collaboration at centre-of-mass energies of $\sqrt{s} = 7$,
$8$, and $13\,\TeV$~\cite{STDM-2012-23,STDM-2014-12,STDM-2018-14}. Similar measurements were performed by CMS~\cite{CMS-EWK-10-010, CMS-SMP-13-013, CMS-SMP-20-003} and LHCb~\cite{LHCb:2015okr,LHCb:2015mad,LHCb:2021huf}.

Even though precise data on \ptz\ spectra exist, these can only be employed in a measurement of the \Wboson-boson mass ($m_W$) through a theoretical modelling of the differences between the $W$ and $Z$ production processes, \textit{e.g.} those induced by different energy scales, couplings, and flavour and momentum fractions of the initial-state quarks. A direct measurement of the \Wboson-boson \pt\ distribution (\ptw) is therefore of significant interest to test this modelling and reduce the related uncertainties in a measurement of $m_W$.
While samples of \Wboson-boson events are easily selected using the leptonic decays \Wln, a measurement of the \ptw spectrum is a much more significant challenge, as the decay neutrino escapes direct detection. The measurement requires the reconstruction of the hadronic system the boson recoils against, the \textit{hadronic recoil}. Additional $pp$ interactions
occurring in the same bunch crossing as the signal (pile-up), degrade the resolution and constitute
the main limitation in this measurement. Consequently, previous measurements at the LHC used relatively small, low pile-up datasets, at $\sqrt{s} = 7\,\TeV$ by ATLAS~\cite{STDM-2011-15} and at $\sqrt{s} = 8\,\TeV$ by CMS~\cite{CMS-SMP-14-012}, with integrated luminosities of $31~\ipb$ and $18~\ipb$, respectively.

This paper
presents measurements of the $W$- and $Z$-boson \ptv distributions in $pp$ collisions, using data collected in dedicated LHC runs with low instantaneous luminosity at centre-of-mass energies of $\sqrt{s} = \fivetev$ and
$\sqrt{s} = 13 \,\TeV$. These datasets respectively correspond to about 255 and 338~\ipb; two inelastic $pp$ collisions take place on average in the same bunch
crossing. The clean experimental conditions optimise the resolution on the hadronic recoil while maintaining high yields of signal events, allowing a
precise measurement of the \ptw distribution. The required calibrations for the detector response to electrons, muons and the hadronic recoil are derived or cross-checked with \Zll\ events in the same datasets.

The analysis is performed in leptonic final states, where lepton refers to electrons and muons ($\ell = e$, $\mu$). The \Wboson-boson measurement is performed separately for $W^-$ and $W^+$ production, and \ptw is
unfolded from the hadronic recoil, $\vecut$, reconstructed in the plane transverse to the beam using charged-particle tracks and energy deposits in the calorimeter.
For \Zll events, the \ptz\ spectrum is primarily measured through the dilepton
system, \ptdilep. An alternative measurement, based on $\vecut$, is compared to the \ptdilep result and used to validate the reconstruction and
measurement techniques. The measurements in the electron and muon
channels are combined accounting for correlations of statistical and systematic
uncertainties.


%
\newcommand{\AtlasCoordFootnote}{%
ATLAS uses a right-handed coordinate system with its origin at the nominal interaction point (IP)
in the centre of the detector and the \(z\)-axis along the beam pipe.
The \(x\)-axis points from the IP to the centre of the LHC ring,
and the \(y\)-axis points upwards.
Polar coordinates \((r,\phi)\) are used in the transverse plane,
\(\phi\) being the azimuthal angle around the \(z\)-axis.
The pseudorapidity is defined in terms of the polar angle \(\theta\) as \(\eta = -\ln \tan(\theta/2)\) and is equal to the rapidity
$ y = \frac{1}{2} \ln \left( \frac{E + p_z}{E - p_z} \right) $ in the relativistic limit.
Angular distance is measured in units of \(\Delta R \equiv \sqrt{(\Delta y)^{2} + (\Delta\phi)^{2}}\).}

\section{The ATLAS detector and datasets}
\label{sec:atlas}

\subsection{ATLAS detector}
\label{sec:detector}
The ATLAS detector~\cite{PERF-2007-01} at the LHC covers nearly the entire solid angle around the collision point.\footnote{\AtlasCoordFootnote}
It consists of an inner tracking detector surrounded by a thin superconducting solenoid, electromagnetic and hadronic calorimeters,
and a muon spectrometer incorporating three large superconducting air-core toroidal magnets.

The inner-detector system (ID) is immersed in a \qty{2}{\tesla} axial magnetic field
and provides charged-particle tracking in the range \(|\eta| < 2.5\).
The high-granularity silicon pixel detector covers the vertex region and typically provides four measurements per track,
the first hit generally being in the insertable B-layer (IBL) installed before Run~2~\cite{ATLAS-TDR-19,PIX-2018-001}.
It is followed by the SemiConductor Tracker (SCT), which usually provides eight measurements per track.
These silicon detectors are complemented by the transition radiation tracker (TRT),
which enables radially extended track reconstruction up to \(|\eta| = 2.0\).
The TRT also provides electron identification information
based on the fraction of hits (typically 30 in total) above a higher energy-deposit threshold corresponding to transition radiation.

The calorimeter system covers the pseudorapidity range \(|\eta| < 4.9\).
Within the region \(|\eta|< 3.2\), electromagnetic calorimetry is provided by barrel and
endcap high-granularity lead/liquid-argon (LAr) calorimeters,
with an additional thin LAr presampler covering \(|\eta| < 1.8\)
to correct for energy loss in material upstream of the calorimeters.
Hadronic calorimetry is provided by the steel/scintillator-tile calorimeter,
segmented into three barrel structures within \(|\eta| < 1.7\), and two copper/LAr hadronic endcap calorimeters.
The solid angle coverage is completed with forward copper/LAr and tungsten/LAr calorimeter modules
optimised for electromagnetic and hadronic energy measurements respectively.

The muon spectrometer (MS) comprises separate trigger and
high-precision tracking chambers measuring the deflection of muons in a magnetic field generated by the superconducting air-core toroidal magnets.
The field integral of the toroids ranges between \num{2.0} and \qty{6.0}{\tesla\metre}
across most of the detector.
Three layers of precision chambers, each consisting of layers of monitored drift tubes, cover the region \(|\eta| < 2.7\),
complemented by cathode-strip chambers in the forward region, where the background is highest.
The muon trigger system covers the range \(|\eta| < 2.4\) with resistive-plate chambers in the barrel, and thin-gap chambers in the endcap regions.

The luminosity is measured mainly by the LUCID--2~\cite{LUCID2} detector that records Cherenkov light produced in the quartz windows of photomultipliers located close to the beampipe.

Events are selected by the first-level trigger system implemented in custom hardware,
followed by selections made by algorithms implemented in software in the high-level trigger~\cite{TRIG-2016-01}.
The first-level trigger accepts events from the \qty{40}{\MHz} bunch crossings at a rate below \qty{100}{\kHz},
which the high-level trigger further reduces in order to record complete events to disk at about \qty{1}{\kHz}.

A software suite~\cite{ATL-SOFT-PUB-2021-001} is used in data simulation, in the reconstruction
and analysis of real and simulated data, in detector operations, and in the trigger and data acquisition
systems of the experiment.

\subsection{Data sets}
\label{sec:datasets}

The analysis is conducted using datasets corresponding to integrated luminosities of $254.9 \pm 2.6\,\ipb$  at
$\sqrt{s} = \fivetev$ and $338.1 \pm 3.1\,\ipb$ at $\sqrt{s} = 13\,\tev$.
The luminosity determination for the \fivetev\ data sample was carried out in a similar way to that described for the low-instantaneous-luminosity 13~TeV dataset in Ref.~\cite{DAPR-2021-01}, and has similar uncertainties.
They were collected in 2017 and 2018 during dedicated
LHC low pile-up runs with an average number of $pp$
interactions, $\avg{\mu}$, of about two, as compared to $\avg{\mu}\sim 34$ for
the nominal LHC Run 2 operation between 2015 and 2018.
To fully exploit these conditions, the thresholds applied to suppress noise in the reconstruction of clusters of energy in the calorimeters were lowered, optimising the reconstruction of the hadronic recoil as described in Section~\ref{sec:recoil}. The data was collected with
triggers that require at least one electron or muon with transverse
momentum thresholds of $\pt^e > 15\,\gev$ and $\pt^\mu > 14\,\gev$,
respectively~\cite{TRIG-2016-01, TRIG-2018-05,TRIG-2018-01}. Loose
identification criteria are applied at the trigger level for electron or muon
candidates.

\subsection{Signal and background simulation}
\label{sec:mc}

Samples of Monte Carlo (MC) simulated events are used to model the
signal and background processes. All MC samples were processed through
the full ATLAS detector simulation~\cite{SOFT-2010-01} based on
$\GEANT$~\cite{Agostinelli:2002hh} using settings specific
to the low pile-up run conditions. The effects of pile-up collisions
in the same or neighbouring bunch crossings were included in the MC
simulation by overlaying inelastic $pp$ interactions produced using
\PYTHIAV{8} with the NNPDF2.3LO set of parton distribution functions (PDFs)~\cite{Ball:2012cx} and the
A3 set of tuned parameters~\cite{ATL-PHYS-PUB-2016-017}.

The main signal event samples for $W$ and $Z$ production were generated
using the \Powheg\ event
generator at next-to-leading order in QCD (NLO)~\cite{Nason:2004rx,Frixione:2007vw,Alioli:2008gx,Alioli:2010xd}
using the CT10 PDF~\cite{Lai:2010vv}, interfaced to \PYTHIAV{8}~\cite{Sjostrand:2007gs} using the
AZNLO tune~\cite{STDM-2012-23}. These \Powheg+\PYTHIAV{8} samples were interfaced
to \Photos~\cite{Golonka:2005pn} to simulate the effect of final-state
QED radiation. The effective sample size is typically $10$--$20$ times larger than
the data to minimise the impact of MC statistical uncertainties.
Alternative signal samples were prepared with
\Sherpa\ 2.2.1 ($\sqrt{s}=13\,\tev$) and 2.2.5
($\sqrt{s}= \fivetev$)~\cite{Bothmann:2019yzt} using the NNPDF3.0 NNLO PDFs~\cite{Ball:2014uwa}
and merging matrix element calculations from Comix~\cite{Gleisberg:2008fv} and \OPENLOOPS~\cite{Buccioni:2019sur,Cascioli:2011va,Denner:2016kdg} for $V+0,1,2$ partons at NLO accuracy with $V+3,4$ partons at leading order (LO) accuracy in the MEPS@NLO scheme~\cite{Schumann:2007mg,Hoeche:2011fd,Hoeche:2012yf,Catani:2001cc,Hoeche:2009rj}. These samples are used to evaluate the
uncertainty arising from the choice of the program used to simulate
the signal events. The $W$ and $Z$ samples are normalised to next-to-next-to-leading (NNLO)
calculations performed using the \textsc{DYTurbo}~\cite{Camarda:2021ict} program, an optimised version
of \textsc{DYNNLO}~\cite{Catani:2007vq,Catani:2009sm} using the MMHT2014 NNLO PDF
set~\cite{Harland-Lang:2014zoa}. A conservative total
uncertainty of $5\%$ is taken as normalisation uncertainty in
background predictions or control regions, covering the uncertainties in PDFs, $\alpha_S$ and missing higher order corrections.

Backgrounds from top-antitop-quark pair (\ttbar) production and single-top-quark production ($Wt$ associated production, $t$-channel,
$s$-channel) were generated with \Powheg+\PYTHIAV{8}~\cite{ATL-PHYS-PUB-2020-023} and normalised to the NNLO predictions with resummation at next-to-next-to-leading-logarithmic (NNLL) accuracy~\cite{Beneke:2011mq,Cacciari:2011hy,Baernreuther:2012ws,Czakon:2012zr,Czakon:2012pz,Czakon:2013goa,Czakon:2011xx}.
Diboson production $VV (V=W,Z)$ was generated with \Sherpa\ in all decay
channels with at least one real lepton in the final state and treated as background~\cite{ATL-PHYS-PUB-2017-005}.


%
\section{Lepton and event selection}
\label{sec:leptonreco}

This analysis relies on the reconstruction and selection of electrons and muons. The calibrations and efficiency measurements used for
the low pile-up datasets are briefly summarised in this section. The methodology closely follows that
established for the full ATLAS Run 2 dataset~\cite{EGAM-2018-01,MUON-2022-01}, referred to as the
high pile-up dataset. A full set of corrections are determined from the low pile-up datasets. However, where possible,
corrections determined from the large high pile-up dataset are used instead to benefit from their better statistical precision. Corrections derived from the high pile-up dataset are complemented
by additional corrections or uncertainties reflecting the extrapolation to the low pile-up regime.

The \Wboson- and \Zboson-boson event selections are also described in this section, as well as the determination of the primary vertex reconstruction efficiency.
The hadronic recoil, $\vec{u}_\mathrm{T}$, is a detector-level measure of \ptw, and is used to infer the transverse momentum of the decay neutrino. Its reconstruction and calibration are specific to this analysis and are described in Section~\ref{sec:recoil}.

\subsection{Electrons}
\label{sec:electrons}

Electron candidates are reconstructed from clusters of energy
deposited in the electromagnetic calorimeter and associated with at
least one track in the ID\@. Electrons are required to be within the
coverage of the ID and the precision region of the EM calorimeter,
$|\eta| < 2.47$. Electrons in the transition region between the
barrel and endcap calorimeters, $1.37<|\eta|<1.52$, are
excluded. Electron candidates are required to have a transverse
momentum of $\ptl > 25\,\gev$ and pass the \textit{Medium} likelihood
identification requirements~\cite{EGAM-2018-01}.  They are also
required to be isolated from nearby activity, as measured by tracks in
a cone of size $\Delta R < 0.2$ around the candidate. The
scalar sum of the \pt of these tracks is required to be less than $10\%$ of the electron
\ptl and may not exceed $5\,\GeV$ for electrons with $\ptl >
50\,\gev$, i.e. $\pt^\mathrm{cone20}/\min(\ptl, 50\,\gev) < 0.1$.

Corrections and uncertainties in the electron energy calibration are
estimated by using a combination of simulation-based and data-driven
procedures as described in Ref.~\cite{EGAM-2018-01}. The energy scale
and resolution differences between data and simulation are determined
in the low pile-up datasets using \Zee decays and the statistical
uncertainties are dominant. Smaller systematic uncertainties
on the simulation of the material and on the relative calibration
of the calorimeter layers are taken from the analysis of
the high pile-up dataset instead.

Corrections to the electron trigger, identification and isolation
efficiencies are determined separately in the $13\,\tev$ and
\fivetev low pile-up data from $Z$-boson events using tag-and-probe
methods~\cite{EGAM-2018-01}. The electron reconstruction efficiency
corrections are instead extrapolated from the high pile-up
dataset.

\subsection{Muons}
\label{sec:muons}

The muon reconstruction is performed independently in the ID and in
the MS, and a muon candidate is formed from the combination
of a muon spectrometer track with an ID track. The muon candidates are
required to have an absolute pseudorapidity of $|\eta| < 2.4$, a
transverse momentum of $\ptl > 25\,\gev$ and to satisfy the \textit{Medium}
identification criteria~\cite{MUON-2022-01}. Muons are required to be
isolated from nearby activity with the same criterion as
electrons.

Corrections are applied to the reconstructed muon momentum in
simulation to precisely match the data. The corrections to
the simulated momentum resolution and momentum scale are determined using both \Zmm\ and
$\Jpsi \to \mu\mu$ events in the high pile-up dataset as described in
Ref.~\cite{MUON-2022-01}. They are cross-checked with the low pile-up dataset
and good agreement is found. A dedicated
correction for charge-dependent momentum biases in data is determined in situ and
applied as a function of muon $\eta$ using a combination of different
methods~\cite{MUON-2022-01,STDM-2014-18}. These correlated systematic biases can be introduced either
by detector deformations to which the alignment procedure has little sensitivity or by
the procedure used to determine the alignment parameters~\cite{IDTR-2019-05}.

The efficiency of the selection criteria for muon candidates is
measured using tag-and-probe methods~\cite{MUON-2022-01} in \Zmm\ events
separately for the $13\,\tev$ and \fivetev data. Efficiency
correction factors for muon trigger and isolation are evaluated as a
function of muon $\eta$ and \pt in the low pile-up dataset with a
precision limited by the size of the \Zmm\ data samples. The muon
reconstruction efficiency correction is instead determined with the
high pile-up dataset and extrapolated to low pile-up.

\subsection{Primary vertex reconstruction}
\label{sec:pv}

Primary vertices are reconstructed requiring at least two
charged-particle tracks~\cite{ATL-PHYS-PUB-2015-026} and the hard-scatter
vertex is chosen as the one with the largest sum of squared track transverse
momenta. Lepton candidates are required to originate from this
vertex. The significance of the track's transverse impact parameter
calculated relative to the beam line, $|d_0/\sigma_{d_{0}}|$, must be
smaller than three for muons and smaller than five for
electrons. Furthermore, the longitudinal impact parameter, $z_0$ (the
difference between the $z$-coordinate of the point on the track at
which $d_0$ is defined and the longitudinal position of the primary
vertex), is required to satisfy $|z_0 \cdot \sin(\theta)| <$ 0.5 mm.

The efficiency of this requirement in \Zll events is generally above $99\%$
and included in the tag-and-probe procedures described
above. However, for \Wboson-boson events at low \pT, the
reconstruction of the correct primary vertex is more challenging as it
requires at least one additional charged-particle track from hadronic
activity in addition to the lepton. A dedicated measurement of this
efficiency is therefore derived. It is performed in
\Zll events as function of the number of reconstructed additional tracks, i.e. excluding the leptons, and the boson \pT. The data measurement can
then be applied to the \Wln simulation. The vertex reconstruction inefficiency in \Wboson-boson events with $\ptw < 5\,\GeV$ at
$\sqrt{s} = \fivetev\,(13\,\TeV)$ is found to be about 5\% (3\%) for
the electron channels and 3\% (2\%) for the muon channels. It decreases towards higher \ptw,  and asymptotically reaches about $1\%$ in the
electron channel and $0.5\%$ in the muon channel. The inefficiency is larger in the electron channels because bremsstrahlung complicates the track reconstruction and track-to-vertex association.
The modelling of the soft activity at low \Wboson\ boson \pt\ is the main
source of uncertainty on the vertex efficiency and its effect is estimated by observing
the closure between different generators, as discussed in section~\ref{sec:sysunc}.

\subsection{Selection of $W $- and $Z$-boson candidate events}
\label{sec:selection}

The \Wln{} event selection requires exactly one reconstructed
and isolated electron or muon satisfying the criteria described in
Sections~\ref{sec:electrons} or \ref{sec:muons}, respectively.
Events with additional leptons of the same flavour with transverse momentum $\ptl > 20\,\GeV$ are discarded to reduce the \Zboson\ background; the identification requirement for these veto leptons is loosened with respect to the nominal selection to make the veto more effective. There are no requirements on the number of leptons with flavour different than the channel under study.
The missing transverse momentum, $\vec{p}_\mathrm{T}^\mathrm{miss}$, and its magnitude \met are defined from $\vec{p}_\mathrm{T}^{\ell}$ and $\vecut$ following $\vec{p}_\mathrm{T}^\mathrm{miss} = - \left(\vec{p}_\mathrm{T}^{\ell} +\vec{u}_\mathrm{T}\right)$,
and represent a measure of the transverse momentum of the neutrino.
The background from QCD multijet events is reduced by the requirement $\MET > 25\,\gev$. Furthermore, the \Wboson-boson transverse mass
$\mt = \sqrt{2 \ptl \met (1-\cos\Delta\phi_{\ell\nu})}$ must exceed $50\,\gev$, where $\Delta\phi_{\ell\nu}$ is the azimuthal angle
between  $\vec{p}_\mathrm{T}^{\ell}$ and $\vec{p}_\mathrm{T}^\mathrm{miss}$.
After all selections, a total number of $7.1\times 10^{5}$
($2.2\times 10^{6}$) W-boson candidate events are selected in the
\Wen{} channel and $7.5\times 10^{5}$ ($2.2\times 10^{6}$ ) in the
\Wmn{} channel for the \fivetev ($13\,\TeV$) data.

\Zll events are selected by requiring exactly two same-flavour, opposite-charge leptons. The same reconstruction and isolation criteria as discussed above are applied,
and the lepton-pair invariant mass is restricted to the range $66<\mll<116 \,\gev$. A total of $5.2\times 10^{4}$ ($1.7\times 10^{5}$)
\Zboson-boson candidates are found in the electron channel and  $7.0\times 10^{4}$ ($2.1\times 10^{5}$) in the muon channel for the \fivetev ($13\,\TeV$) data.


%
\section{Hadronic recoil reconstruction and calibration}
\label{sec:recoil}

The final state of a \Wln or \Zll event naturally separates into the leptonic system and the associated QED radiation,
and the hadronic recoil, $\vecut$, defined in the plane transverse to the beam from the remaining reconstructed particles in the event.
Up to the detector resolution, the transverse momentum of the hadronic recoil is equal in magnitude to the vector boson transverse
momentum and points in the opposite direction, $\vecut=-\vec{p}_{\mathrm{T}}^{V}$.

In \Zll events, the $Z$-boson transverse momentum can be estimated from both the lepton pair \ptll and \ut. The well-measured dilepton system can thus be used to calibrate
the hadronic recoil response, and the unfolded \ptll distribution provides a cross-check of the \ptz spectrum measured from \ut.

\subsection{Hadronic recoil reconstruction}
\label{sec:recoilreco}

The hadronic recoil was used successfully in ATLAS analyses with
$\sqrt{s}=7\,\tev$ data~\cite{STDM-2011-15, STDM-2014-18}.  For these
measurements, the reconstruction was based on calorimetric energy deposits reconstructed using a topological clustering algorithm~\cite{PERF-2014-07}. In this analysis, the
hadronic recoil is reconstructed from \textit{Particle Flow Objects} (PFOs)~\cite{PERF-2015-09, JETM-2018-05} that combine information from ID
tracks and the calorimeters in an optimal way and avoid double counting. Compared to the calorimeter-based measurement, the particle flow technique improves the recoil resolution by $3\%$ at low \ptv, increasing to $15\%$ at $\ptv \approx 50\,\gev$.

The recoil vector $\vecut$ is computed from the vector sum of all charged and neutral PFO momenta. To reject pile-up contributions, charged PFOs are required to be associated with the same primary vertex as the $W$ lepton candidate.
PFOs within $\Delta R = 0.2$ around identified electrons and muons are not included in the recoil, to avoid energy deposits originating from charged leptons and accompanying photons.
This procedure also removes small underlying event and pile-up contributions. To compensate this bias, the PFO momentum measured in a region
at the same $\eta$ as the lepton, and randomly chosen $\phi$, with the constraint of being separated from the lepton and recoil directions by $\Delta \phi > 0.4$, is rotated to the lepton axis and added to the recoil.

The resolution of \ut primarily depends on the event activity variable \set, the scalar sum of the
transverse energies of all PFOs included in the recoil. The quantity $\setue = \set
- \ut$ is a modified event activity variable with reduced correlation to the vector boson dynamics, and that mostly represents the underlying event and pile-up.
Events are characterised by the vector boson transverse momentum and \setue.

In \Zll events, the projections of the hadronic recoil on the axes parallel and perpendicular to the momentum of the lepton pair
are denoted \upar and \uperp, respectively. These projections are used to calibrate the hadronic recoil scale and resolution. The
perpendicular component \uperp\ is expected to have a mean of zero and
a width given by the resolution. The sum of \upar and \ptll is also called \textit{bias},
\begin{equation}
b = \upar + \ptll \label{eq:bias}\,,
\end{equation}
and would ideally be close to zero. However, $b > 0$ is observed on average, due to particles escaping detection, energy losses in passive detector material and the non-compensating
nature of the calorimeters.

\subsection{Hadronic recoil calibration}
\label{sec:recoilcalib}

The hadronic recoil scale, resolution and direction need to be calibrated before the measurement is performed.
The \Zll final state is the primary reference, since the final-state kinematics are fully determined through the
precise reconstruction of the decay lepton pair. The calibration derived from the $Z$-boson sample at a given
value of dilepton transverse momentum \ptll, is applied to the \Wln simulation at the same value in the generated transverse momentum \pttruthv of the \Wboson-boson. This replacement is possible as
the lepton resolution is better than that of the hadronic recoil by about one order of
magnitude. The corrections are derived in four sequential steps:
\begin{itemize}
\item the modelling of the modified event activity \setue in the signal simulation is corrected, separately in $W$- and $Z$-boson events;
\item the simulated azimuthal distribution of the recoil is corrected to that observed in data;
\item the simulated recoil scale and core resolution are corrected to those observed in data, using the \uperp and \upar projections in $Z$-boson events;
\item residual corrections for non-Gaussian resolution tails are derived in data and applied to the simulation.
\end{itemize}

All corrections are obtained separately for $\sqrt{s} = \fivetev$
and $\sqrt{s} = 13\,\tev$ data. As the recoil response is independent of the decay lepton
flavour, the \Zee and \Zmm samples are merged to derive the corrections. The corrections are then applied to the \Wln MC samples, and the closure of the calibration is verified by application to \Zll samples.
Small differences in the hadronic activity between $W$ and $Z$ bosons, for example induced by the different initial-state quark flavours and
momentum fractions, are accounted for by dedicated corrections or systematic uncertainties, as explained at the end of this section.

\subsubsection*{Modelling of \setue}

Three event weights are obtained for the \Powheg+\PYTHIAV{8} MC samples to model the observed \setue distribution
and its correlation with the \ptv distribution. A first two-dimensional reweighting of the (\setue, \ptll) distribution is
derived from the ratio of data and simulation for the $Z$-boson sample. The weights are
applied to the \Wboson simulation as a function of \pttruthv, as explained above.
After this correction, the \setue\ distributions in $W$-boson data and simulation agree at the percent level.

The residual disagreement in the $W$-boson sample remaining after the first reweighting using the $Z$-boson data, is removed using a second event weight obtained from the $W$-boson sample itself. This correction is performed
separately for each charge, as a function of \setue\ and in bins of
\ut after a first calibration of \ut in the simulation
using the resolution and response corrections discussed below.

The application of
these weights to the \Zboson and
\Wboson samples modifies the generated transverse momentum spectrum of the bosons in the simulation.
The initial spectrum is restored using a third, one-dimensional weight as a function of \pttruthv.
The total weight applied to simulated \Wln events is the product of the three weights described above.
Only the first and third weight are applied to \Zboson\ events.
The whole procedure is equivalent to making a reweighting in slices of \ptv\ (and then \ut for \Wln events), which is why it is not
required to have the same distribution of \ptv\ in data and in simulation.
The total number of events is preserved, as only shapes are used.

A closure test is performed using simulated \SHERPA\ \Zboson and \Wboson events as pseudo-data, to verify that the correlation between \setue\ and
\pttruthv\ in \Powheg+\PYTHIA{8} is corrected to that present in \SHERPA. A good closure is observed and the small residual
non-closure is taken as a systematic uncertainty. This uncertainty also allows to check that in this procedure, there is no
dependence on how well the distribution of \ptv\ is modelled in \Powheg+\PYTHIA{8}, as the two generators differ significantly in their predicted \ptv\ spectra.

\subsubsection*{Azimuthal correction}
\label{subsec:calib_utphi}

The azimuthal direction of the recoil, $\phi(\vec{\ut})$, is not uniformly distributed due to effects
not fully modelled by the simulation, such as non-uniformity in the calorimeter response, the calorimeter mechanical deformation,
the beam displacement from the nominal beam line in $(x,y)$, or the beam-crossing angle.
An empirical correction is obtained from the differences between the mean values of the $x$ and $y$ components of the recoil in
\Zboson\ events in data and in the simulation ($\langle u_x\rangle$
and $\langle u_y\rangle$). A small dependence on \setue is observed and included in the correction. The components of the recoil
in the simulation are then corrected accordingly.

The impact of this stage of the calibration on the measured
\ut\ distribution in \Wboson\ simulated events is at most $0.5\%$,
giving confidence that the correlation between the direction and the
magnitude of the recoil is weak.

\subsubsection*{Resolution and response corrections}
\label{subsec:calib_resol_resp}

The data-to-MC resolution correction for \uperp\ is derived as a ratio of $\sigma(\uperp)$ values following
\begin{equation}
r(\setue,\ptll) = \frac{\sigma_{\uperp}(\setue,\ptll)^\mathrm{data}}{\sigma_{\uperp}(\setue,\ptll)^\mathrm{MC}},
\label{eq:r}
\end{equation}
where $\sigma_{\uperp}$ is derived from Gaussian fits to the $\uperp$ distributions in \Zll events, in bins of \setue and \ptll.
The distribution of \uperp in the $W$-boson simulation is then corrected following
\begin{equation}
\uperp^{\mathrm{MC,corr}} = \uperp^{\mathrm{MC}} \cdot
r(\setue,\pttruthv)\,.
\end{equation}

The data-to-MC corrections for \upar ($ r'\,$) use the standard deviation and mean of the
\textit{bias} defined in Eq.~\eqref{eq:bias} that relate to the
resolution and scale, respectively. The corrections are
extracted in \Zll events, $r'(\setue,\ptll)$, and transferred to the $W$-boson MC, $r'(\setue,\pttruthv)$, as:
\begin{equation}
\upar^\mathrm{MC, corr} = \avg{\upar^\mathrm{data}} +
(\avg{b^\mathrm{data}} - \avg{b^\mathrm{MC}})\cdot r' + (\upar^\mathrm{MC} -
\avg{\upar^\mathrm{data}})\cdot r'\,,
\label{eq:uparcor}
\end{equation}
where all quantities are defined in bins of $\setue$ and $\ptll$ or $\pttruthv$.

\subsubsection*{Tail correction}
\label{subsubsec:calib_tail}

The resolution corrections discussed above only correct the core Gaussian parts of the \upar and \uperp distributions. Non-Gaussian effects appear in the tails,
mostly driven by hadronic activity out of the detector acceptance. A final step in the recoil calibration therefore corrects the resolution
tails in simulation to that in data after applying all previous corrections.
The correction is derived in bins of \ptll from the cumulants of the  \upar and \uperp distributions in data and simulation with an inverse transform sampling (\textit{Smirnoff transform})~\cite{Devroye}. The full transformation is parameterised by empirical functions with two (four) parameters for \uperp (\upar). The transformation gives corrected values as of \upar and \uperp as a function of these same variables for each event. For \uperp this results in a symmetric stretching around zero and for \upar in an  asymmetric stretching around the median of the distribution.

\subsubsection*{Summary of recoil calibration}

\begin{figure}[htbp]
\centering
\subfloat[]{\includegraphics[width=.49\textwidth]{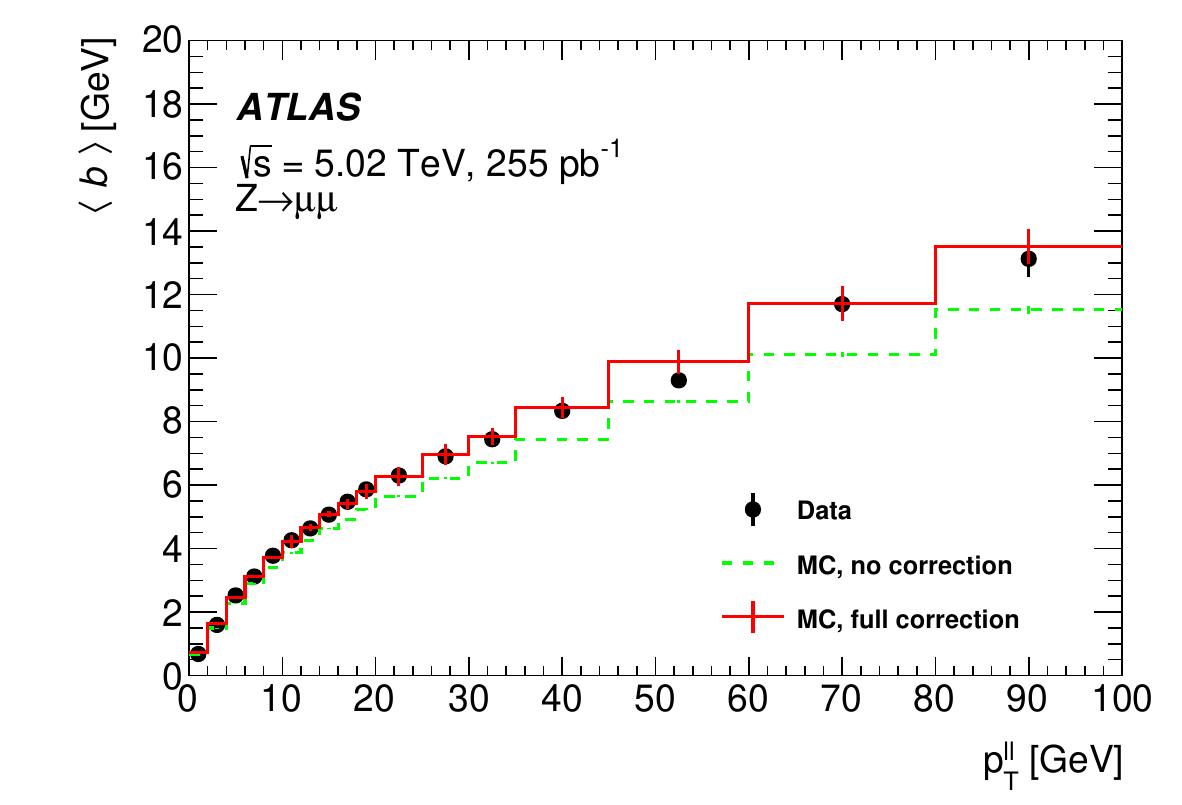}}
\subfloat[]{\includegraphics[width=.49\textwidth]{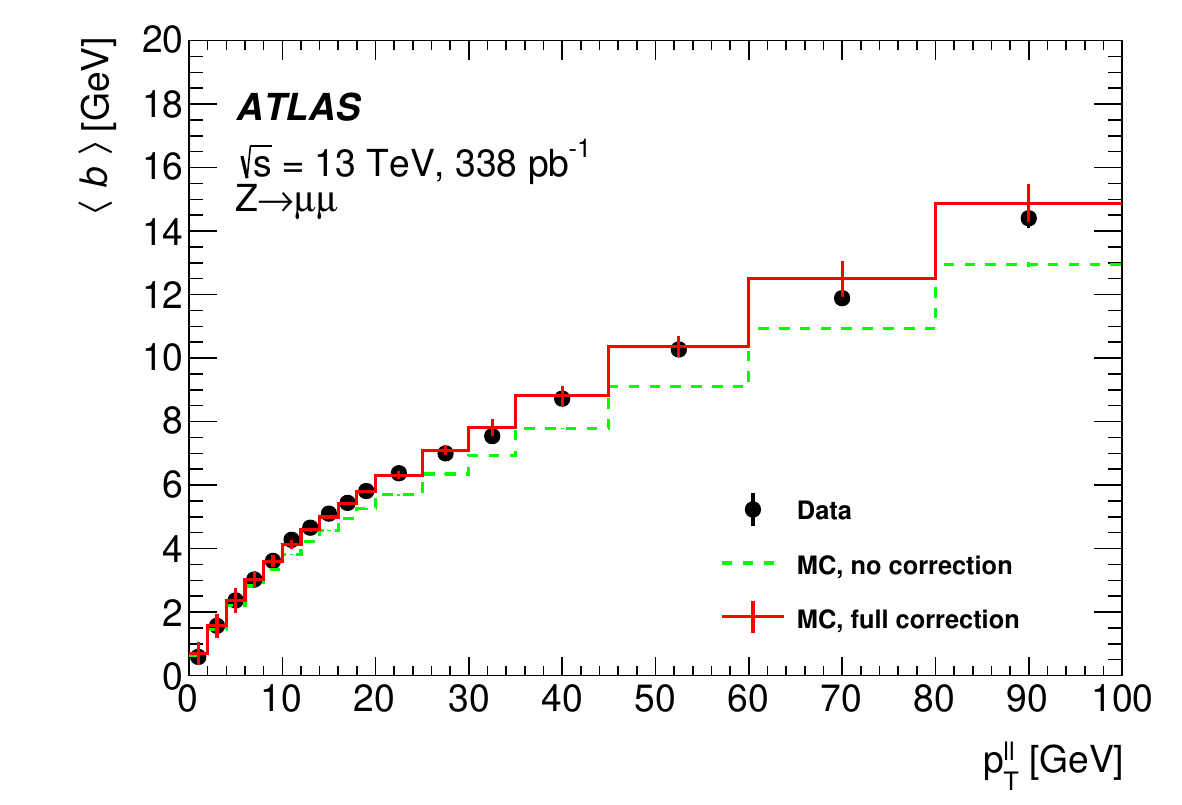}}

\subfloat[]{\includegraphics[width=.49\textwidth]{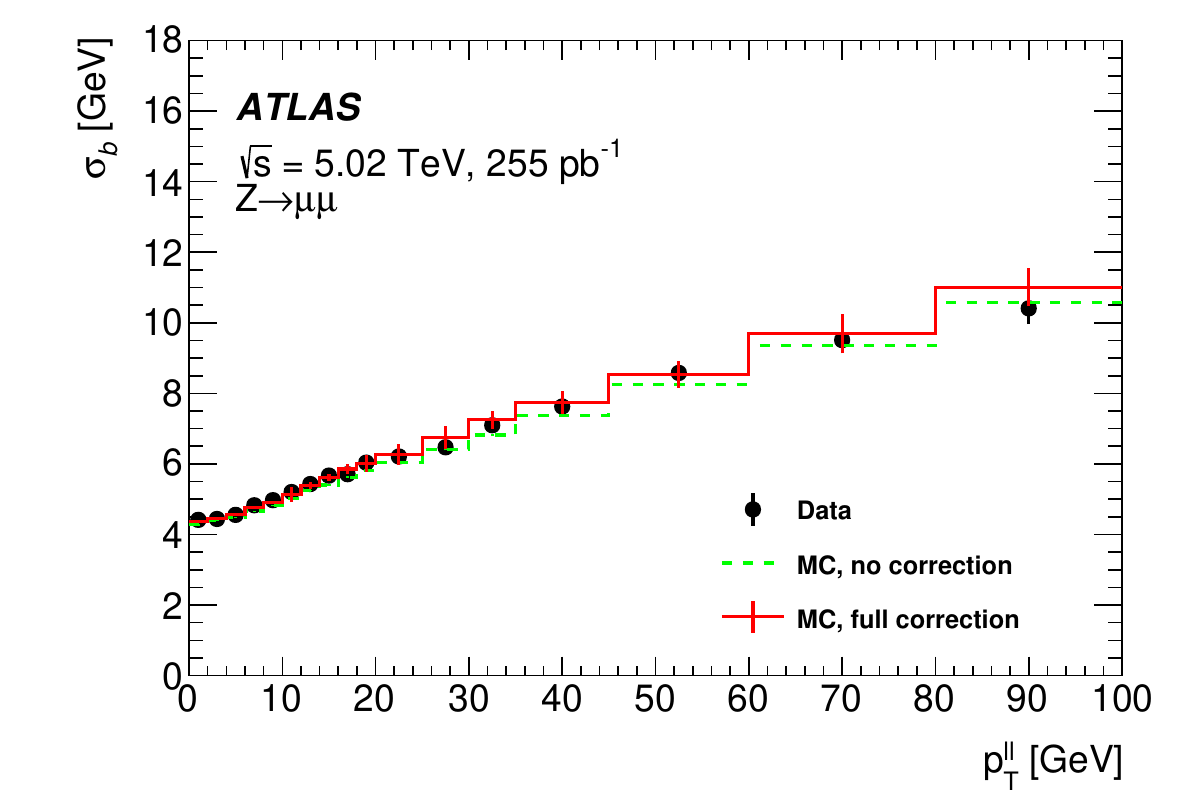}}
\subfloat[]{\includegraphics[width=.49\textwidth]{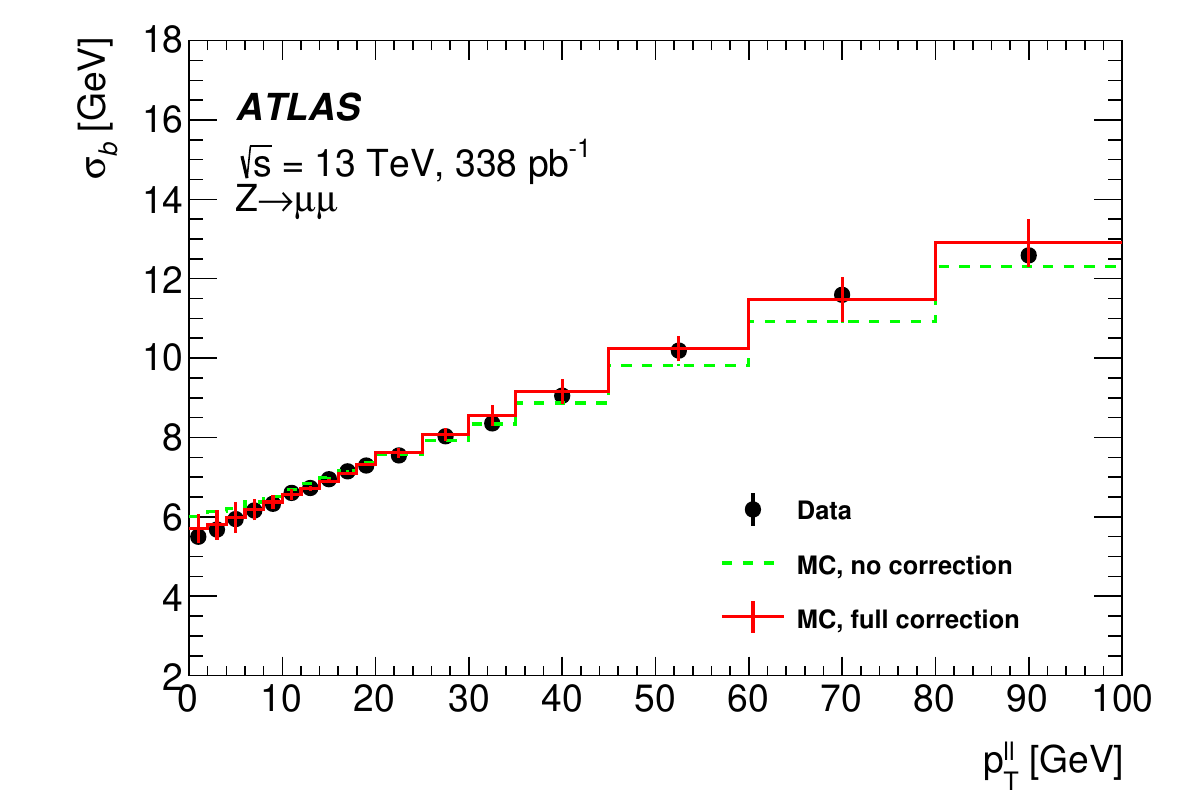}}

\subfloat[]{\includegraphics[width=.49\textwidth]{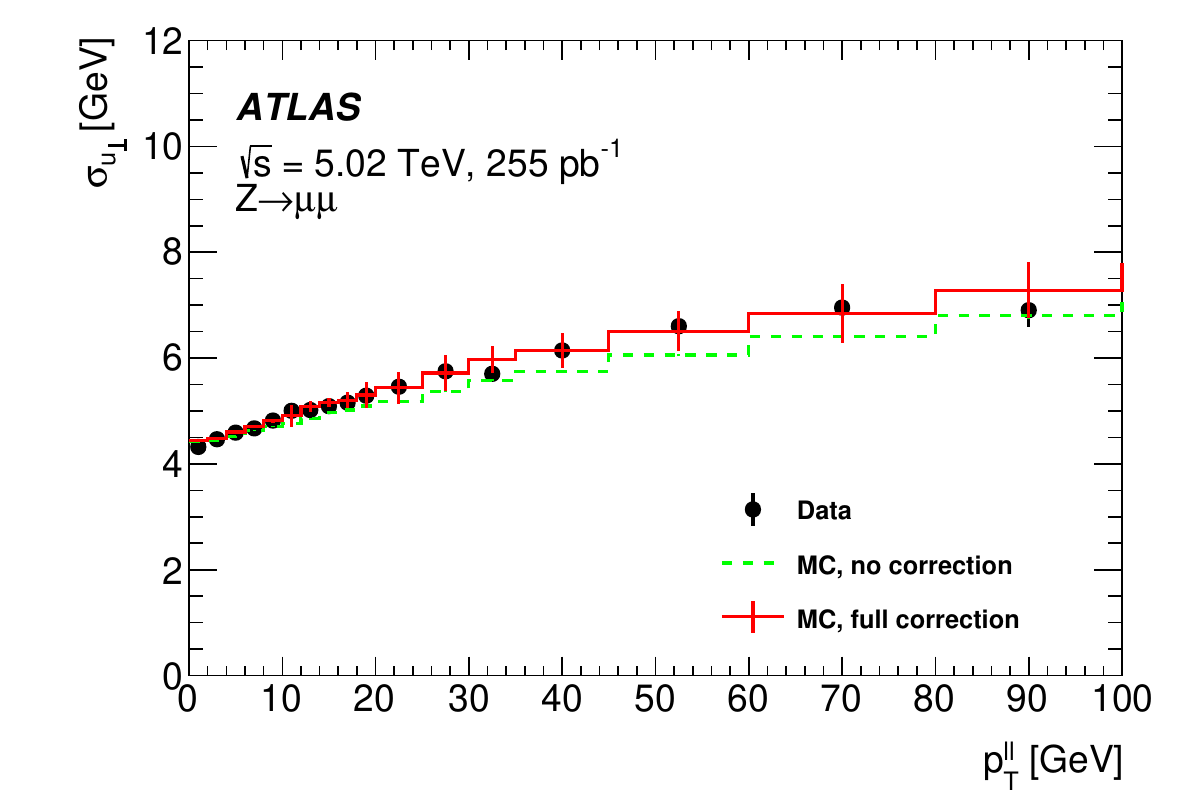}}
\subfloat[]{\includegraphics[width=.49\textwidth]{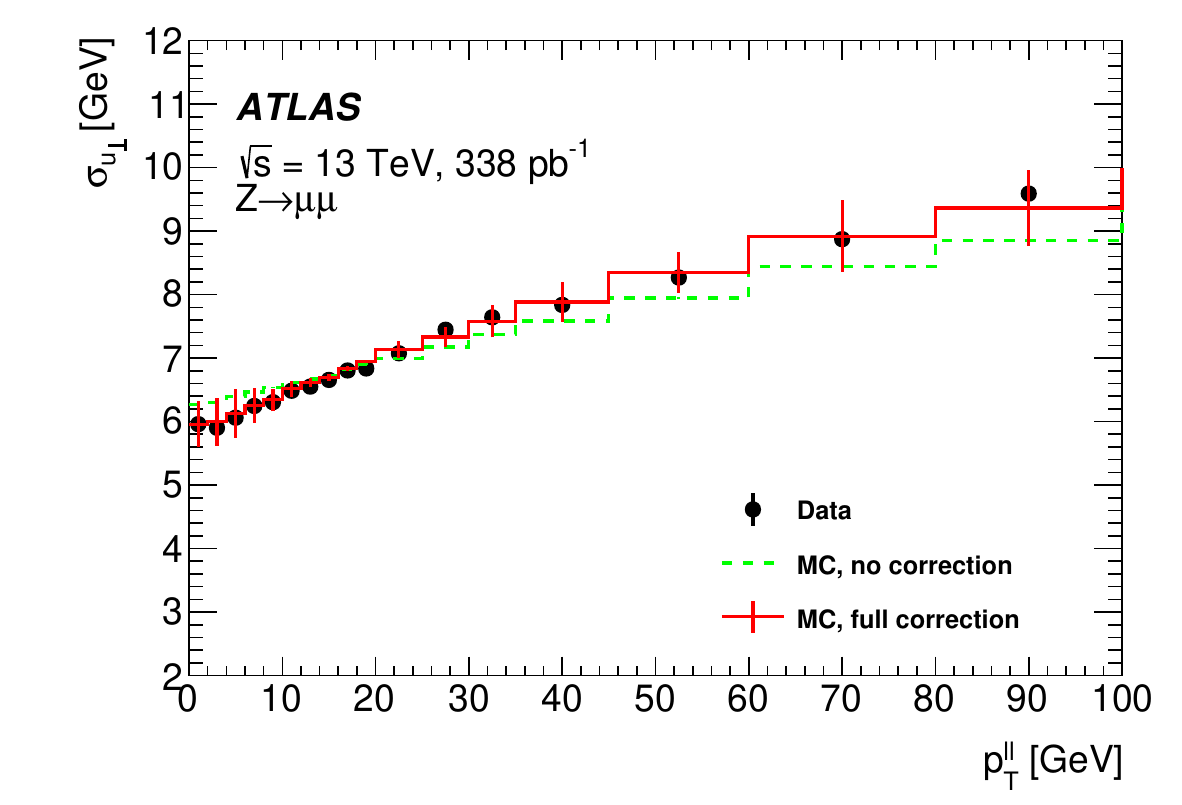}}

\caption{Mean value for the bias $b = \upar + \ptll$ at (a) $\sqrt{s} = \fivetev$ and (b) $\sqrt{s} = 13\,\TeV$,
standard deviation of the bias at (c) $\sqrt{s} = \fivetev$ and (d) $\sqrt{s} = 13\,\TeV$, and
standard deviation of $u_{\perp}$ at (e) $\sqrt{s} = \fivetev$ and (f) $\sqrt{s} = 13\,\TeV$ in
\Zmm\ events. Data (black dots) are compared to the simulation before (green dotted histogams) and after (red histograms) correction.
The values are shown as a function of \ptll. Error bars show the sum of statistical
uncertainties and systematic uncertainties from recoil calibration and from the choice of event generator (the difference
between \Sherpa\ and \Powheg+\PYTHIAV{8}). }
\label{fig:recoilCalib_illustration}
\end{figure}

Figure~\ref{fig:recoilCalib_illustration} shows the comparison of $\sigma_{\uperp}$ and $\avg{b}$ as a function of
\ptll in the \Zmm simulated and data samples.
Using PFOs instead of calorimeter-based inputs only improves the recoil resolution, $e.g$ at 13~\TeV\ $\sigma_{\uperp}$ decreases from
$\approx 12.1~\GeV$ to $\approx 10.5~\GeV$ around $\ptv~\approx~50~\GeV$, once it is calibrated and corrected for the scale.
Good closure of the calibration is observed. Similar results are obtained in the \Zee samples.
The uncertainty in the extrapolation from the \Zboson\ to the \Wboson\ is estimated from the differences between the standard deviation and mean of the \uperp and \upar distributions
in the simulation, at given values of $(\setue, \pttruthv)$, to derive alternative sets of resolution and response corrections $r$ and $r'$.

A final systematic uncertainty in the recoil calibration is
estimated by using \SHERPA signal events as pseudo data in the calibration procedure. The \Powheg+\PYTHIAV{8} sample is calibrated to \SHERPA, the full analysis described in Section~\ref{sec:measurement} is performed and
the difference between the measured \ptv distribution and the generated spectrum gives
the closure of the correction. To simplify the procedure and avoid
double-counting the extrapolation from \Zboson\ to \Wboson, the
calibration of each process uses only the corresponding signal sample
from both event generators. The \ptv spectrum of \SHERPA is also reweighted
to \Powheg+\PYTHIAV{8} to avoid double-counting the unfolding prior
uncertainty, i.e. the effects tested are residual differences in the simulation of hadronic jet production beyond the hardest emission, the underlying event, fragmentation and hadronisation. The difference between the two generators in observables relevant
to the description of the recoil is
larger than the one between data and \Powheg+\PYTHIAV{8} in \Zboson\
events, which gives confidence that this uncertainty component is not
accidentally underestimated.


\section{Background estimates and event yields}
\label{sec:selectionbkg}

\subsection{Background in the \Wln selection}
\label{sec:wbkg}

Background contributions from single-boson and diboson production (electroweak background)
and \ttbar and single-top-quark production (top-quark background)
are directly estimated from
the MC simulated samples described earlier. They are dominated by $W \to \tau\nu$
events with $\tau \to \ell\nu\nu$ at low \ut and top-quark pair
production at higher \ut. In the electron channels there is an
additional background contribution from \Wenu events where the reconstructed electron
charge is mismeasured.

Background from QCD multijet (MJ) production cannot be reliably
simulated and has to be derived from data. Depending on the lepton
flavour, the MJ background has significant contributions from
leptons produced in semileptonic decays of heavy quarks, pion and kaon decays,
or photon conversions, and is referred to as fake leptons below. The multijet yields in the electron and muon selections of the
\Wboson-boson analysis are estimated separately for positively and negatively
charged samples. As multijet production is concentrated at lower
values of \ptl, \met\ and \mt\ as compared to the signal, this
background is estimated with a template fit in these kinematic
distributions in a phase-space region with relaxed \met\ and \mt
selections, called the \textit{fit region}. The \ptl, \met\ and \mt\ templates in the fit region
for the \Wln signal and electroweak and top-quark
backgrounds are obtained using simulation. The multijet templates are
built from events passing the same kinematic selections as the nominal
selection except that the lepton isolation requirement is inverted.
In the final step, the multijet yields in the signal region are
determined by multiplying the fit region yields by a transfer factor
that corrects for the
acceptance of the \met and \mt selections and the impact of the inverted isolation
requirements needed to extract the multijet background templates.

The multijet yield is estimated three times using the \ptl, \met\ and
\mt\ distributions separately. The three results are found to be compatible, and the final yield estimate is derived
from their unweighted mean, neglecting the correlations.
As the shape of the multijet templates
depends on the anti-isolation criterion, mutually exclusive intervals
in the isolation variable are chosen to create statistically
independent multijet templates. These samples are designed to be
progressively closer to the signal-candidate selection and are used to
study the dependence of the multijet extraction on the considered isolation.

\begin{figure}[tbp]
\centering

\subfloat[]{\includegraphics[width=.49\textwidth]{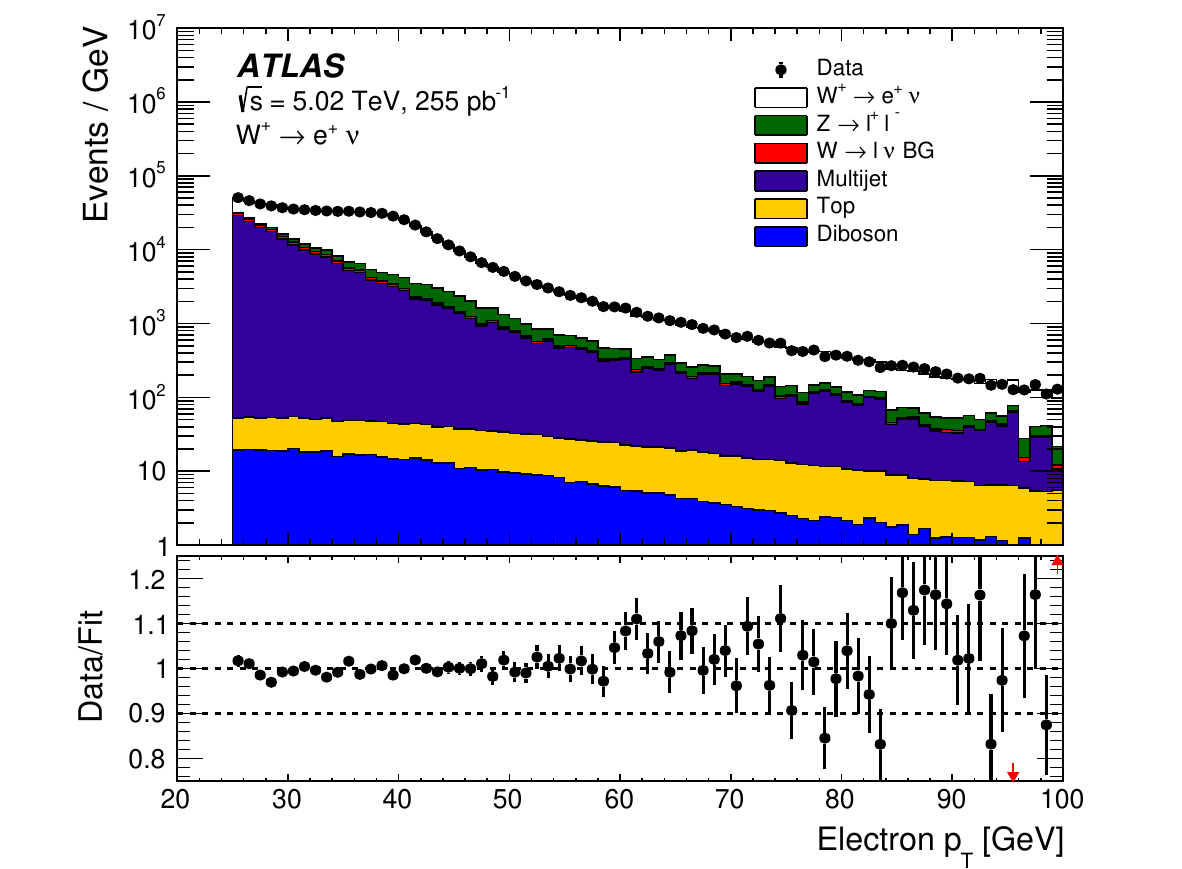}}
\subfloat[]{\includegraphics[width=.49\textwidth]{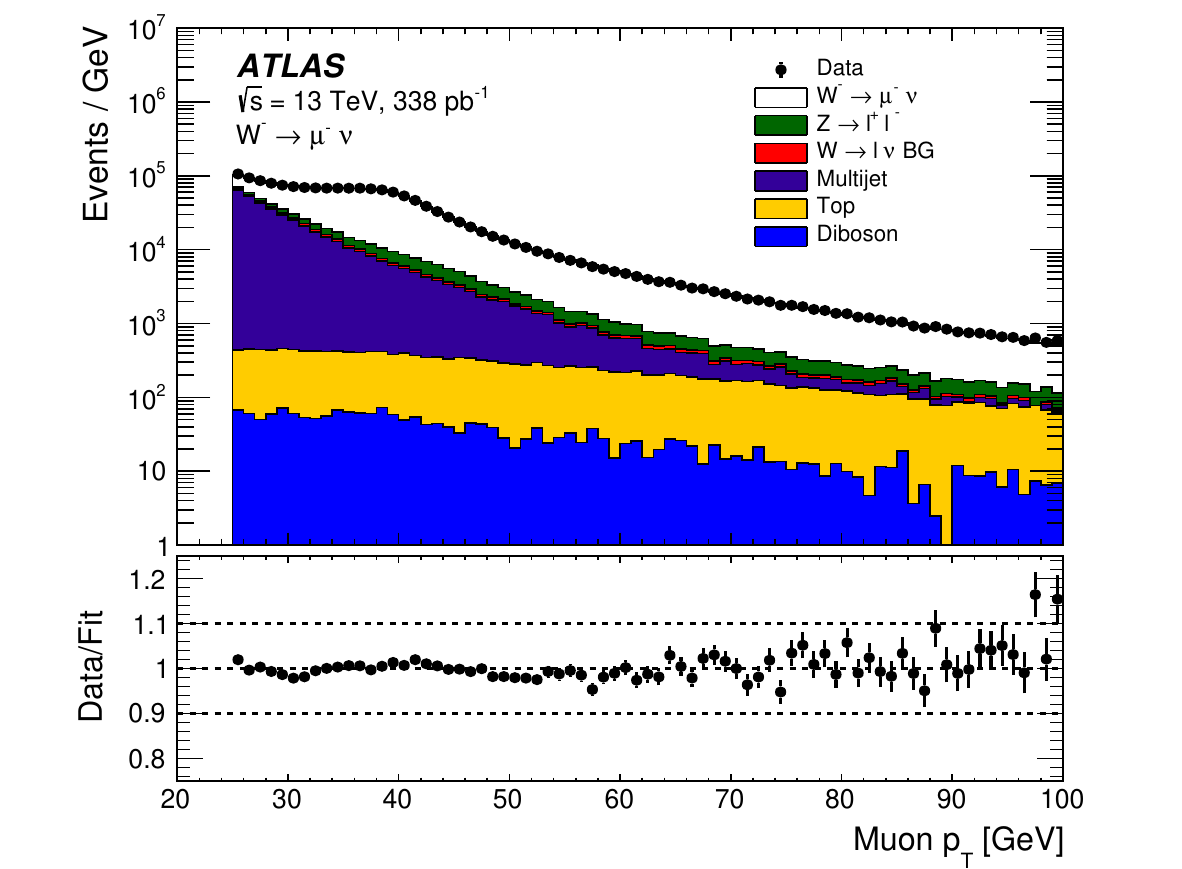}}

\subfloat[]{\includegraphics[width=.49\textwidth]{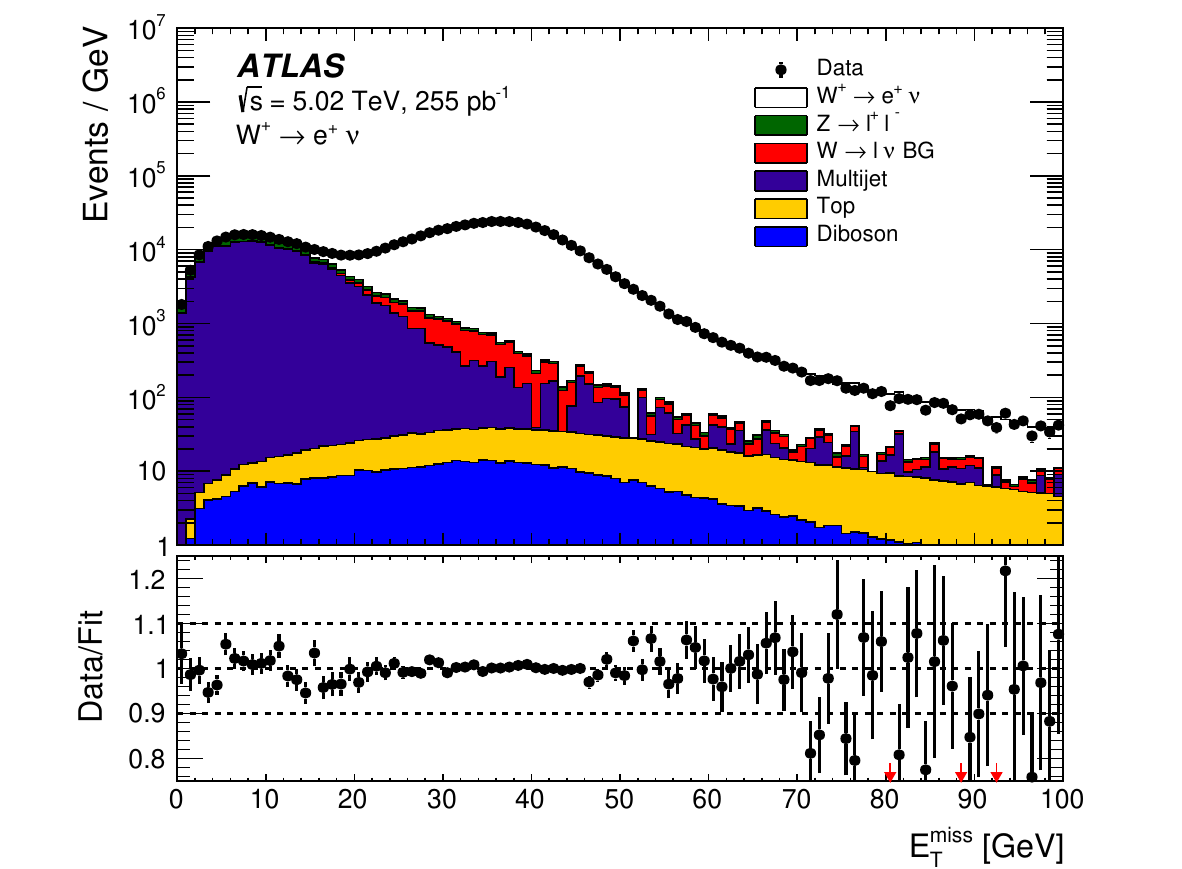}}
\subfloat[]{\includegraphics[width=.49\textwidth]{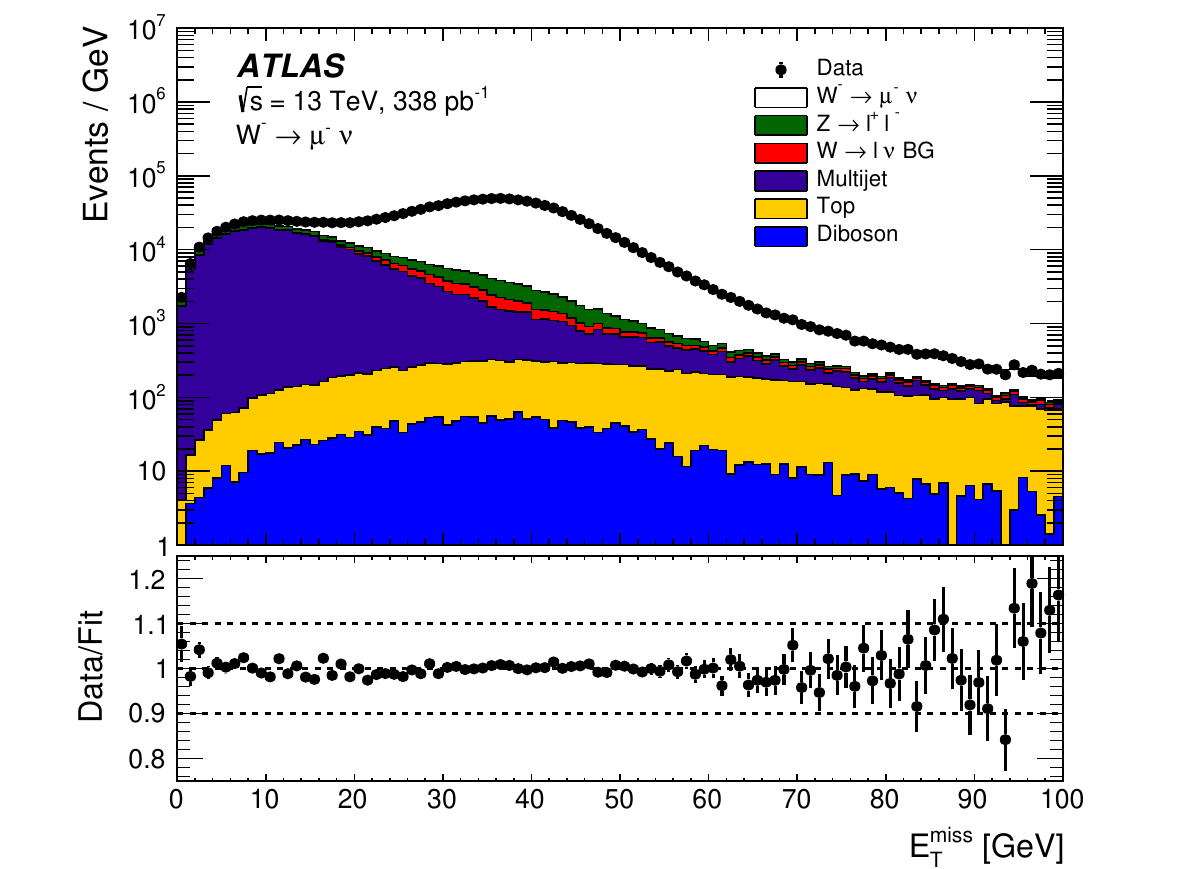}}

\subfloat[]{\includegraphics[width=.49\textwidth]{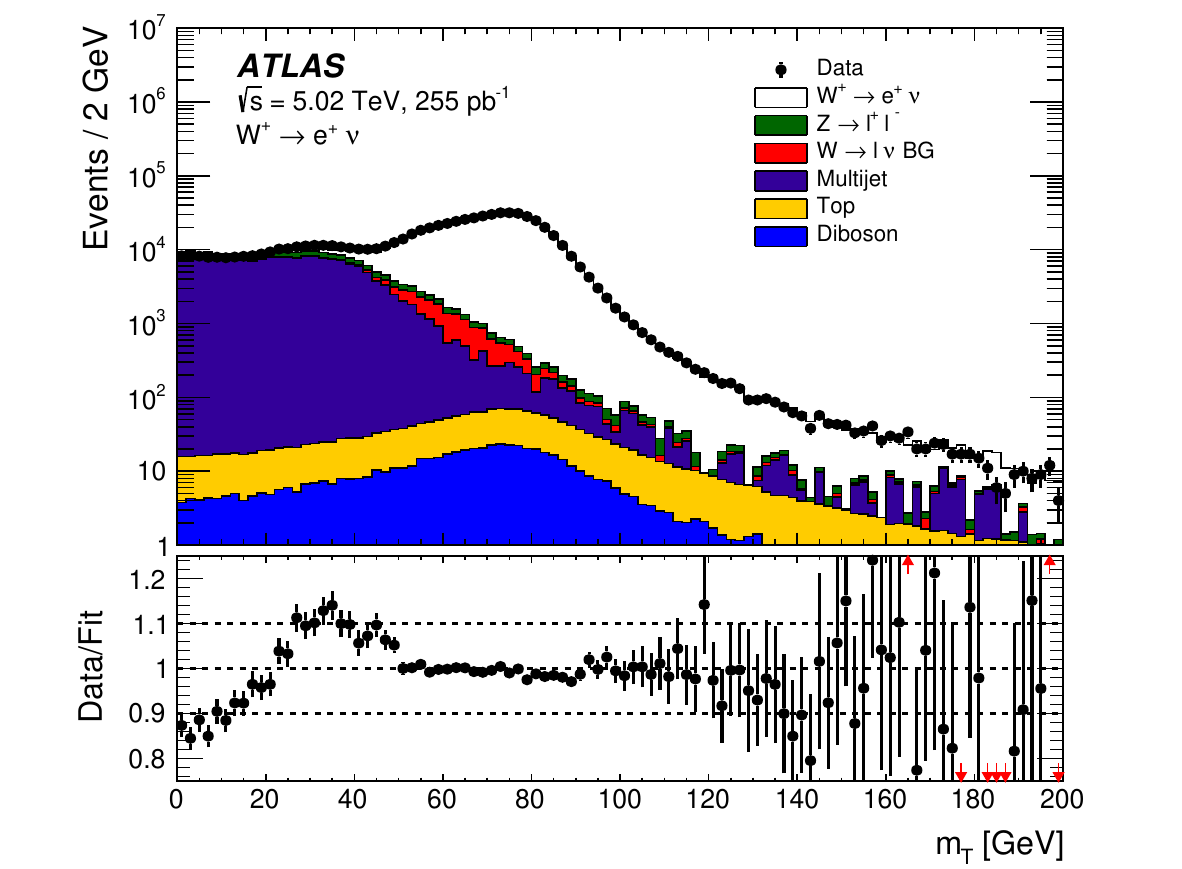}}
\subfloat[]{\includegraphics[width=.49\textwidth]{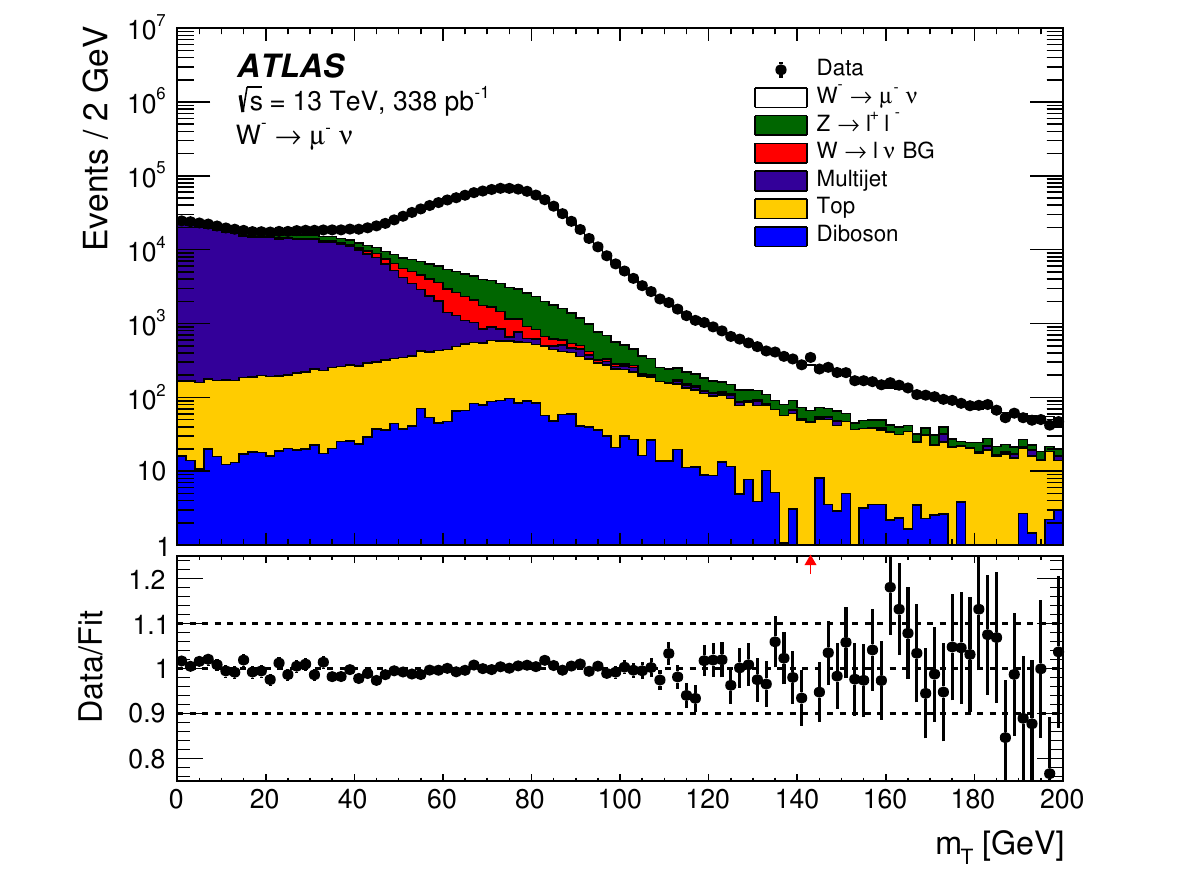}}

\caption{Distributions of (a,b) \ptl, (c,d) \met\ and (e,f) \mt\ in
a phase-space region without \met\ and \mt selections (\textit{fit
region}) used to extract multijet yields in template fits in the
\Wpenu in the (a,c,e) $\sqrt{s} = \fivetev$ dataset and \Wpmunu in
the (b,d,f) $\sqrt{s} = 13$~\TeV dataset . Multijet templates are
derived from the data combining the shape information from
different anti-isolation intervals. The data are compared to the
simulation including signal and background contributions.
The category $\Wln$
BG refers to background from $W\to\tau\nu$ and charge
misidentification in the electron channels.
}
\label{fig:MJtemplateFit}
\end{figure}

\begin{figure}[h]
\centering
\subfloat[]{\includegraphics[width=.490\textwidth]{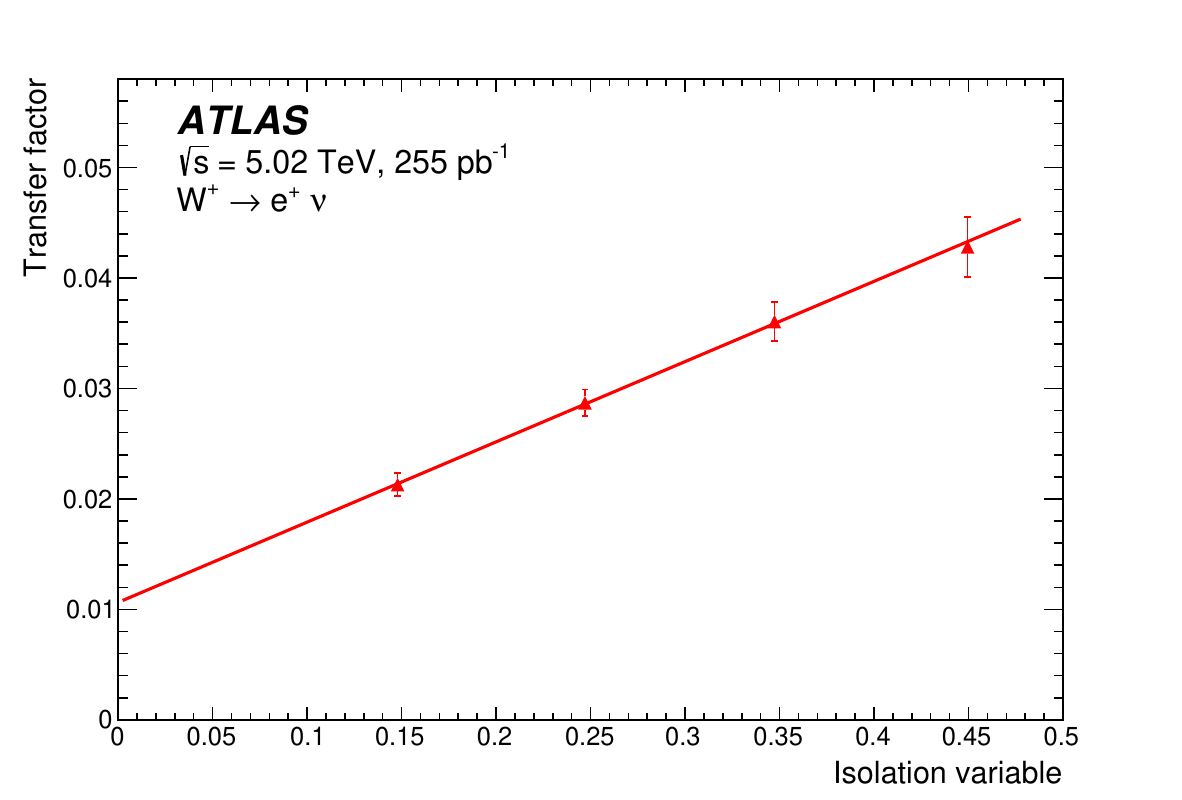}}
\subfloat[]{\includegraphics[width=.490\textwidth]{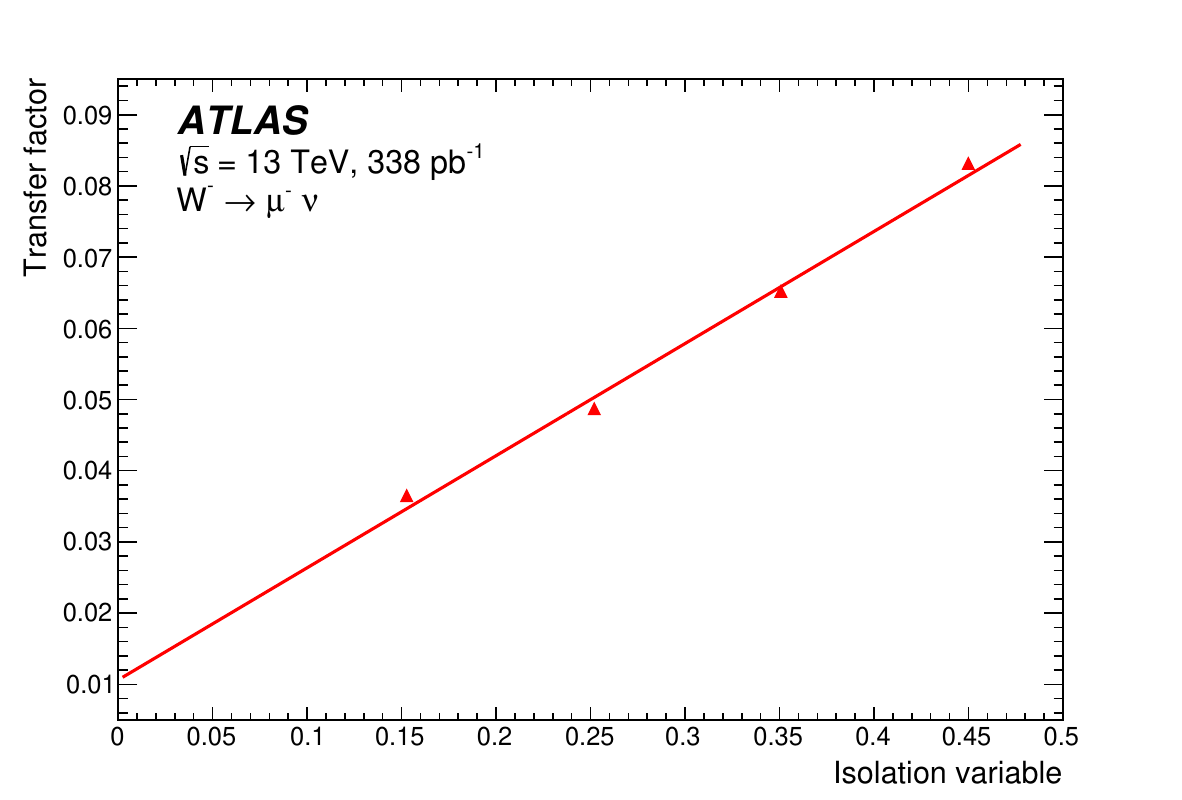}}

\caption{Illustration of the multijet yield transfer factors to account for
the different kinematic selections in \met and \mt and isolation requirements from the
\textit{fit region} to the signal region. The transfer factors are shown as a function of the lower bound of the range of the
isolation variable $\pt^\mathrm{cone20}/\min(\ptl, 50\,\gev)$ used to define the anti-isolation
intervals, for (a) the \Wpenu channel at $\sqrt{s} = \fivetev$ and (b)
the \Wmmunu channel at $\sqrt{s} = 13$~\TeV. The data
point are reasonably described by the linear fits
indicated by the solid lines. The uncertainties in the data points are statistical only.
}
\label{fig:MJextrapolation}
\end{figure}

Several improvements are included in this background estimate with
respect to a similar method described in Ref.~\cite{STDM-2014-18}. In
particular, the removal of all energy in the cone around lepton candidates
from the hadronic recoil calculation, as described in Section~\ref{sec:recoilreco},
is problematic for the case where leptons are required to fail the isolation selection.
The correct properties in the multijet-enriched sample are restored by adding back a fraction
of the hadronic activity around the lepton candidate and recalculating the kinematic variables
\ut, \met, and \mt accordingly.
Another significant improvement on the MJ yield estimate
is the use of MJ templates in the fits that have their shapes corrected. This correction is achieved
by combining the shapes from
different anti-isolation intervals to extrapolate into the fit region. Instead
of repeating the template fit in different anti-isolation intervals, it is carried out only once
for a given kinematic distribution, and the corresponding dependence of the measured
multijet yield is entirely located in the transfer factor.
Example results of template fits in the electron
and muon channels are shown in Figure~\ref{fig:MJtemplateFit}.
The disagreement observed in Figure~\ref{fig:MJtemplateFit} (e) is not a major concern,
since the estimation of the multijet yield using \mt\ is always still consistent with the ones using \ptl\ and \met.
In the final signal selection, the background-dominated region ($\mt<50~\GeV$) is removed.
In Figure~\ref{fig:MJextrapolation},
the transfer factors accounting for the efficiency of the kinematic cuts and the isolation
requirements to extrapolate from the \textit{fit region} to the signal region are shown as a function of anti-isolation intervals.
The extrapolation was also tested with alternative function forms, which produced estimations of yields consistent with the linear extrapolation within the yield uncertainty.

The final MJ background yield and total uncertainty
is presented in Table~\ref{tab:wyield}.
The dominant MJ yield uncertainties arise from: the statistical
uncertainty from the limited data and MC events in the fit region,
the linear extrapolation procedure to the average isolation value in the signal
region, and the possible mis-modelling of the jet activity in the
anti-isolated regions. The uncertainty arising from the latter effect is studied
by categorising events in broad bins of \ut.

Corrections to the shape of the multijet background contributions and
corresponding uncertainties in the distributions used in the final
measurement are estimated with a similar procedure. The MJ kinematic
distributions in the signal region are determined by linearly extrapolating the
shape of the distributions observed in the mutually exclusive
anti-isolation intervals into the isolated region. Uncertainties in
the extrapolated distributions are dominated by the statistical
uncertainty. The resulting multijet background distribution is
propagated to the final analysis. A smoothing is applied to the MJ \ut
spectra for $\ut>100\,\GeV$ to avoid large fluctuations due to low
event counts in the data samples.

The estimated multijet background fractions in the \fivetev dataset represent $0.8\%$ and $0.1\%$ of the total
number of observed \Wminus candidate events in the electron and muon
channels, respectively. In the 13~\TeV\ dataset the corresponding fractions are larger at $2.9\%$ and $0.6\%$ because the gluon-induced multijet production rises faster with $\sqrt{s}$ than the quark-induced signal.
For \Wplus candidate events the background yields are found to be similar to the \Wminus channels within uncertainties, and the relative fractions are
lower by the size of the $\Wplus/\Wminus$ cross-section ratio.
The different sources of the MJ
uncertainty components in yield and shape are treated as correlated
across the channels, while the statistical uncertainties are
uncorrelated.

The total impact of the uncertainties in the background estimation on the cross-section
measurements is shown in Figures~\ref{fig:unf_sys_ShapeUnc_5} and
\ref{fig:unf_sys_ShapeUnc_13} in Appendix~\ref{sec:uncbreakdown}, and is at the level of
$0.1\%$--$1\%$ in the low $\ptw < 63~\GeV$ region, depending on the channel.

\subsection{Background in the \Zll selection}
\label{sec:zbkg}

As in the \Wlnu analysis, the electroweak and
top-quark backgrounds are directly estimated from the MC simulated
samples described earlier. They are dominated by diboson and top-quark
pair production at higher \ut, but very small overall.

A data-driven two-dimensional sideband method is employed to estimate
MJ background. The signal region \textit{A} is complemented by three
orthogonal control regions \textit{B}, \textit{C}, and
\textit{D} enriched in events produced from background processes with
fake leptons.
For both channels, events in region \textit{B} have the nominal
requirements except that the subleading lepton is required to fail the
isolation selection. For events to be registered in region \textit{C}
in the electron channel, the primary requirement is to fail the
\textit{Medium} identification requirement, but pass the \textit{Loose}
criteria. In addition, only same-charge pairs are selected and
the \ptl requirement is loosened to $20\,\GeV$. For the muon channel,
region \textit{C} is defined to contain only same-charge muon pairs and
also here the \ptl requirement is loosened to $20\,\GeV$.  Finally,
for region \textit{D} events are selected with the modifications of
the \textit{B} and \textit{C} regions applied at the same time.

Multijet event yields $N_\mathrm{MJ}$ in each control region are
estimated from data by subtracting the contribution from prompt
leptons predicted by the simulation. Because of the very
high signal contamination in the background-enriched regions in the
\Zee channel, the events with $80<m_{\ell\ell}<100 \,\gev$ are
excluded. The total MJ yield in the signal region is then estimated
through $N_\mathrm{MJ}^{A} = N_\mathrm{MJ}^{B} \cdot N_\mathrm{MJ}^{C}/N_\mathrm{MJ}^{D}$.
In the electron channel the final MJ yield is also scaled back to the
full \mll range.

The final MJ event yield and uncertainty for each channel are
summarised in Table~\ref{tab:zyield}.  In all channels,
the MJ event estimate is below 0.1\% with respect to the expected
signal. In the \Zee channel at \fivetev the final
MJ event estimate was found to be consistent with zero. An upper
limit on the MJ yield in this channel is derived from the statistical
uncertainties of the yields in the background-enriched regions.

The shape of the MJ events is taken from the corresponding
distributions in the background-enriched regions. To suppress
statistical fluctuations for the \ptdilep\ and \ut\ distributions, the
templates are smoothed with a Landau function and normalised to
the inclusive number of MJ events for each channel.

The total impact of the background estimation on the unfolded \ptdilep\ and \ut\ distributions for the \fivetev and $13\,\TeV$ datasets
is provided in Figures~\ref{f:AlluncZ_5TeV_shape}~--~\ref{f:AlluncZ_13TeV_shape} in Appendix~\ref{sec:uncbreakdown} and is generally well below $0.1\%$.

%

\begin{table}[tb]
\caption{Observed and expected event yield comparison for the
\Wboson-boson selections in the $\sqrt{s} = \fivetev$ and $\sqrt{s} = 13\,\TeV$ datasets
for the electron and muon channels split by the charge of the
reconstructed lepton. The uncertainties in the
multijet background correspond to the total uncertainties, while for backgrounds
estimated from MC simulation only statistical uncertainties are given. The
category $\Wln$ BG refers to background from $W\to\tau\nu$ as well
as charge misidentification in the electron channels.}
\label{tab:wyield}
\setlength\tabcolsep{3pt} %
\resizebox{\textwidth}{!}{%
\begin{tabular}{l|crc|rcr|rcr|rcr|rcr|rcr|rcr}
\toprule
\toprule
\multicolumn{1}{l}{Channel} &  \multicolumn{3}{c}{Observed}  &  \multicolumn{3}{c}{Signal}  &  \multicolumn{3}{c}{$\Wln{}$ BG}  &  \multicolumn{3}{c}{$Z \rightarrow \ell\ell$}  &  \multicolumn{3}{c}{Top}  &  \multicolumn{3}{c}{Diboson}  &  \multicolumn{3}{c}{Multijet}  \\
\midrule
\multicolumn{22}{c}{$\sqrt{s} = \fivetev$ \; \Wln}\\
\midrule
$W^- \rightarrow e^-\nu$   & & 274375 & & 264510 & $\pm$ & 170 & 7303 & $\pm$ & 29 & 1031 & $\pm$ & 6 & 862.7 & $\pm$ & 2.1 & 340.5 & $\pm$ & 2.9 & 2200 & $\pm$ & 700 \\
$W^+ \rightarrow e^+\nu$  & & 430662 & &417090 & $\pm$ & 210 & 9064 & $\pm$ & 32 & 1102 & $\pm$ & 7 & 951.4 & $\pm$ & 2.9 & 376.2 & $\pm$ & 2.3 & 2300 & $\pm$ & 900 \\
$W^- \rightarrow \mu^-\nu$   & & 288026 & & 273900 & $\pm$ & 170 & 5160 & $\pm$ & 25 & 8517 & $\pm$ & 21 & 799.0 & $\pm$ & 2.0 & 339.1 & $\pm$ & 1.9 & 300 & $\pm$ & 340 \\
$W^+ \rightarrow \mu^+\nu$  & & 457223  & & 438020 & $\pm$ & 210 & 7854 & $\pm$ & 30 & 9773 & $\pm$ & 22 & 890.6 & $\pm$ & 2.8 & 384.6 & $\pm$ & 2.3 & 500 & $\pm$ & 400 \\
\midrule
\multicolumn{22}{c}{$\sqrt{s} = 13\,\TeV$ \; \Wln}\\
\midrule
$W^- \rightarrow e^-\nu$ & & 949297 & & 876490 & $\pm$ & 270 & 22360 & $\pm$ & 60 & 6546 & $\pm$ & 21 & 12780 & $\pm$ & 50 & 1720 & $\pm$ & 50 & 27000 & $\pm$ & 5000 \\
$W^+ \rightarrow e^+\nu$  & & 1207652 & & 1116800 & $\pm$ & 300 & 24680 & $\pm$ & 60 & 6883 & $\pm$ & 21 & 13260 & $\pm$ & 50 & 1720 & $\pm$ & 50 & 29000 & $\pm$ & 5000 \\
$W^- \rightarrow \mu^-\nu$ & & 964514 & & 897400 & $\pm$ & 270 & 14650 & $\pm$ & 50 & 33940 & $\pm$ & 40 & 11950 & $\pm$ & 50 & 1620 & $\pm$ & 40 & 6000 & $\pm$ & 2000 \\
$W^+ \rightarrow \mu^+\nu$  & & 1245755 & & 1153400 & $\pm$ & 300 & 18390 & $\pm$ & 50 & 37690 & $\pm$ & 40 & 12520 & $\pm$ & 50 & 1720 & $\pm$ & 50 & 6000 & $\pm$ & 2000 \\
\bottomrule
\bottomrule
\end{tabular}
}
\end{table}

\begin{table}[tb]
\setlength\tabcolsep{4pt} %
\caption{Observed and expected event yield comparison for \Zboson-boson selections in the $\sqrt{s} = \fivetev$ and $\sqrt{s} = 13\,\TeV$ datasets for the
electron and muon channels. The uncertainties in the
multijet background correspond to the total uncertainties, while for backgrounds
estimated from MC simulation only statistical uncertainties are given.}
\label{tab:zyield}
\centering
\begin{tabular}{l|crc|rcr|rcr|rcr|rcr|rcr}
\toprule
\toprule
\multicolumn{1}{l}{Channel} &  \multicolumn{3}{c}{Observed}  &  \multicolumn{3}{c}{Signal}  &  \multicolumn{3}{c}{$Z \rightarrow \tau\tau$}  &  \multicolumn{3}{c}{Top}  &  \multicolumn{3}{c}{Diboson}  &  \multicolumn{3}{c}{Multijet}  \\
\midrule
\multicolumn{19}{c}{$\sqrt{s} = \fivetev$ \;\; \Zll}\\
\midrule
\Zee & & 51772& & 51310 & $\pm$ & 40 & 23 & $\pm$ & 4 & 35.5 & $\pm$ & 0.4 & 109.1 & $\pm$ & 2.0 & 0 & $\pm$ & 100 \\
\Zmm & & 70447 & & 69820 & $\pm$ & 60 & 39 & $\pm$ & 5 & 44.0 & $\pm$ & 0.4 & 144.9 & $\pm$ & 2.1 & 16 & $\pm$ & 8 \\
\midrule
\multicolumn{19}{c}{$\sqrt{s} = 13\,\TeV$ \;\; \Zll}\\
\midrule
\Zee & & 165027 & & 158000 & $\pm$ & 100 & 80 & $\pm$ & 8 & 500 & $\pm$ & 10 & 324 & $\pm$ & 15 & 110 &$\pm$& 70 \\
\Zmm & & 214035 & & 207900 & $\pm$ & 100 & 80 & $\pm$ & 8 & 554 & $\pm$ & 9 & 414 & $\pm$ & 17 &  180 &$\pm$& 40 \\
\bottomrule
\bottomrule
\end{tabular}
\end{table}


\subsection{Detector-level event yields and distributions}
\label{sec:control}

The final observed and expected event yields for the full \Wboson and \Zboson selections obtained in the electron and muon channels for both datasets are summarised in Tables~\ref{tab:wyield} and \ref{tab:zyield}.
All calibrations and efficiency corrections discussed in Sections~\ref{sec:leptonreco} and \ref{sec:recoil} are applied.

\begin{figure}[h]
\centering
\subfloat[]{\includegraphics[width=.490\textwidth]{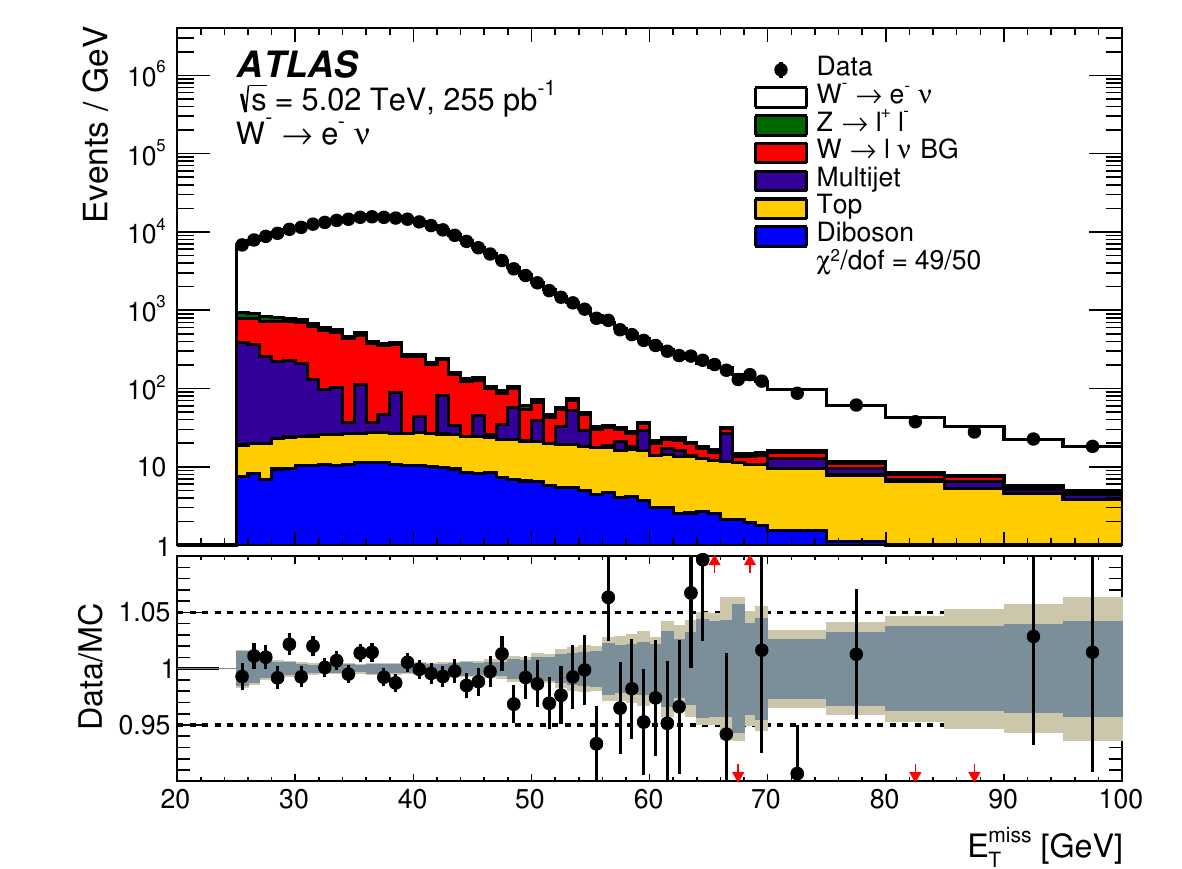}}
\subfloat[]{\includegraphics[width=.490\textwidth]{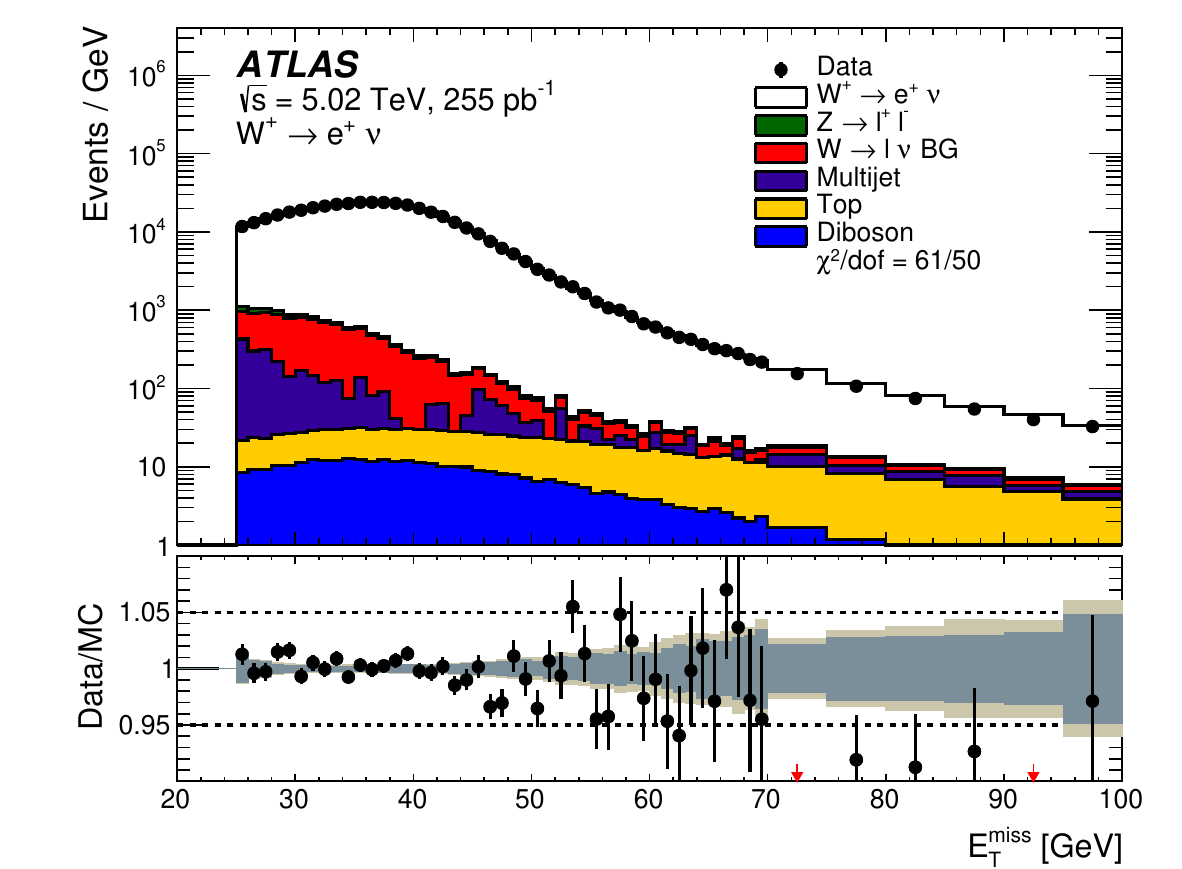}}

\subfloat[]{\includegraphics[width=.490\textwidth]{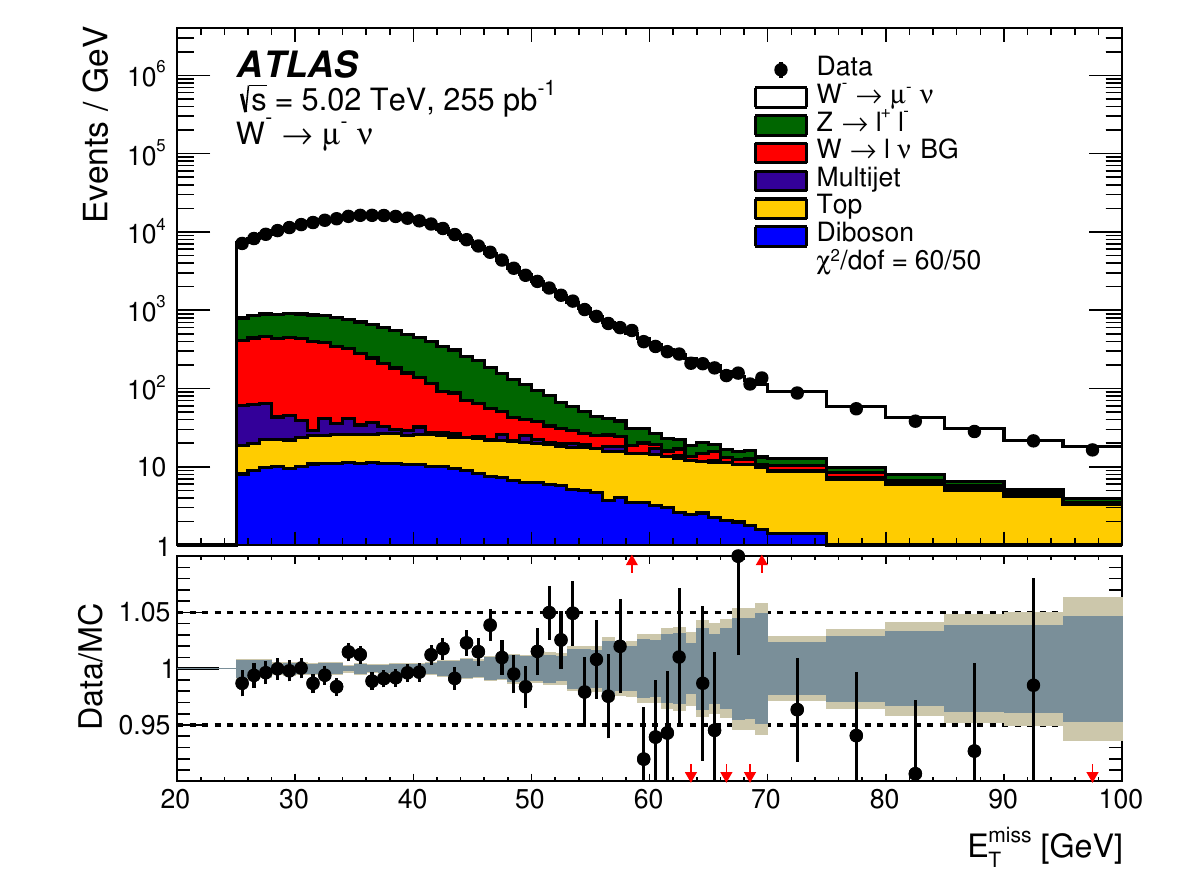}}
\subfloat[]{\includegraphics[width=.490\textwidth]{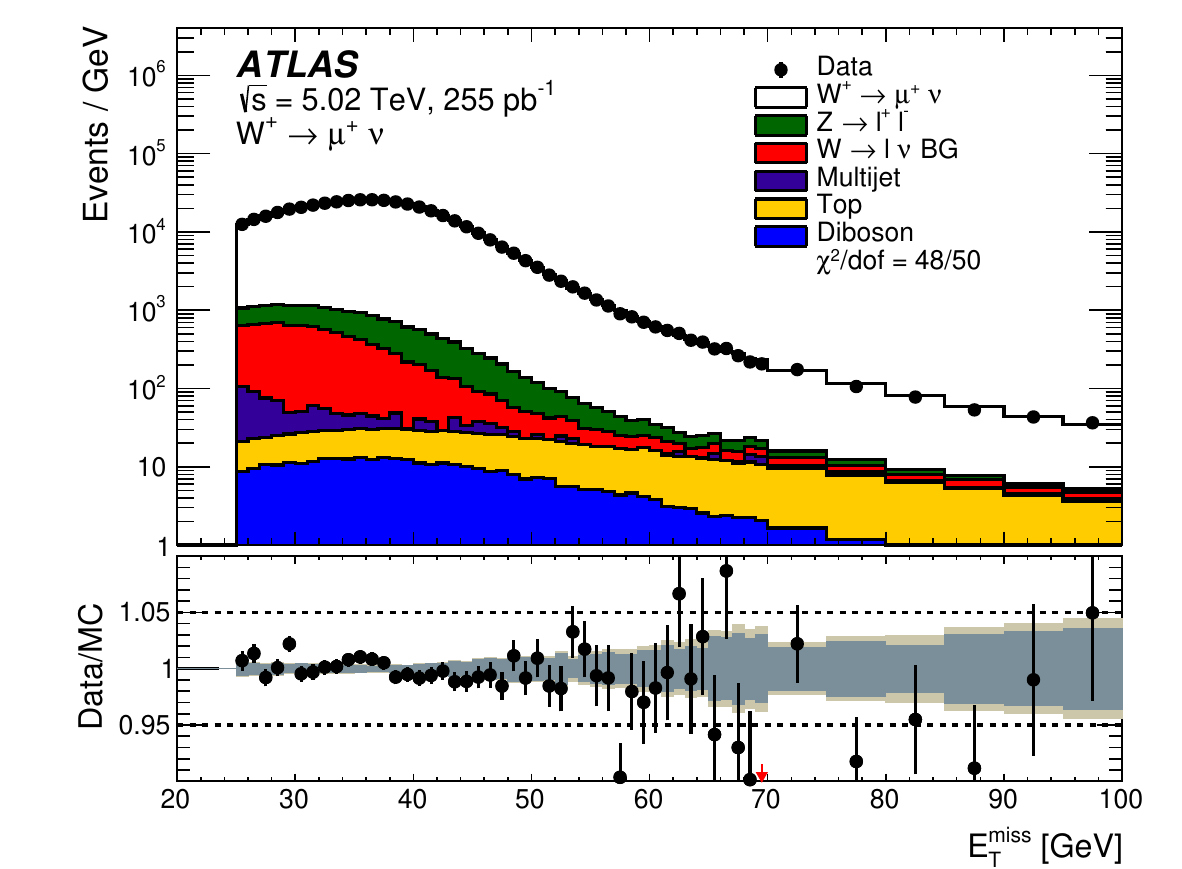}}

\caption{Missing transverse momentum \met distributions in the (a) \Wmenu, (b) \Wpenu, (c) \Wmmunu and (d) \Wpmunu channels for the $\sqrt{s} = \fivetev$ dataset. The data (black points) is compared to the sum
of all expected background and signal contributions, shown as a
solid line, normalised to the total data yield. The lower panel shows the
ratio of the data to the full prediction (black points) as well as the prediction uncertainties around $1$ excluding
(including) the MC simulation statistical component as dark (light) band.
The prediction uncertainties shown exclude those from the luminosity measurement
and the alternative signal modelling, see text. The
$\chi^2/\mathrm{dof}$ is computed using the full covariance matrix
of all shown statistical and systematic uncertainties. The category $\Wln$
BG refers to background from $W\to\tau\nu$ and charge
misidentification in the electron channels.}
\label{fig:met5_controlplot}
\end{figure}

\begin{figure}[h]
\centering
\subfloat[]{\includegraphics[width=.490\textwidth]{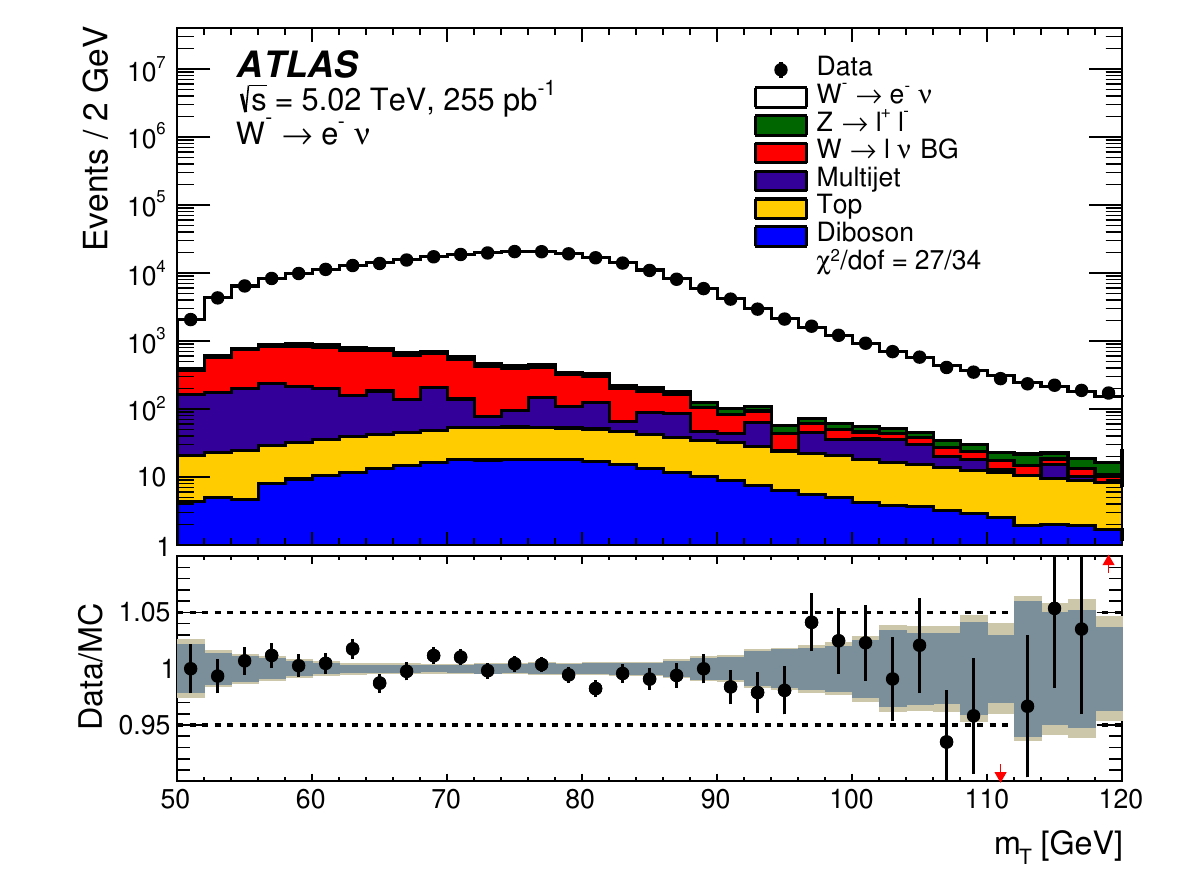}}
\subfloat[]{\includegraphics[width=.490\textwidth]{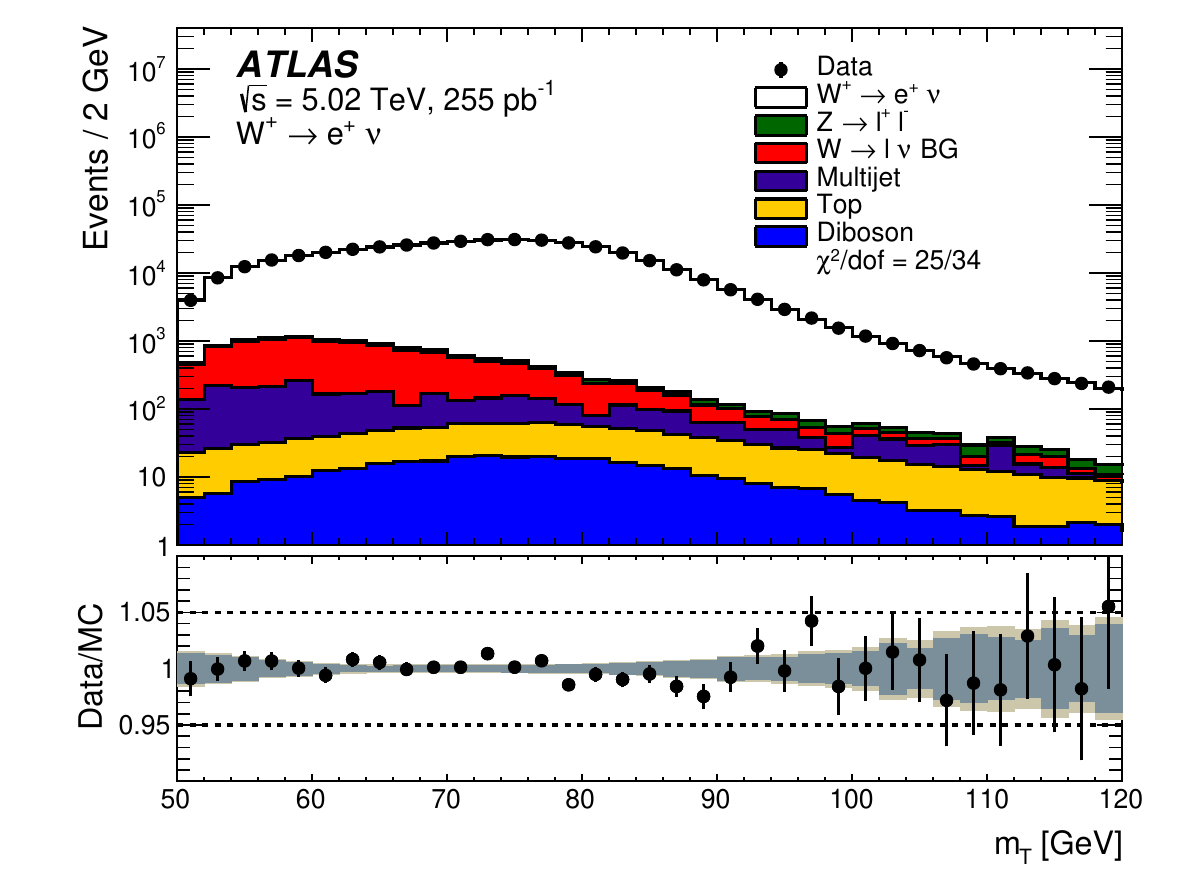}}

\subfloat[]{\includegraphics[width=.490\textwidth]{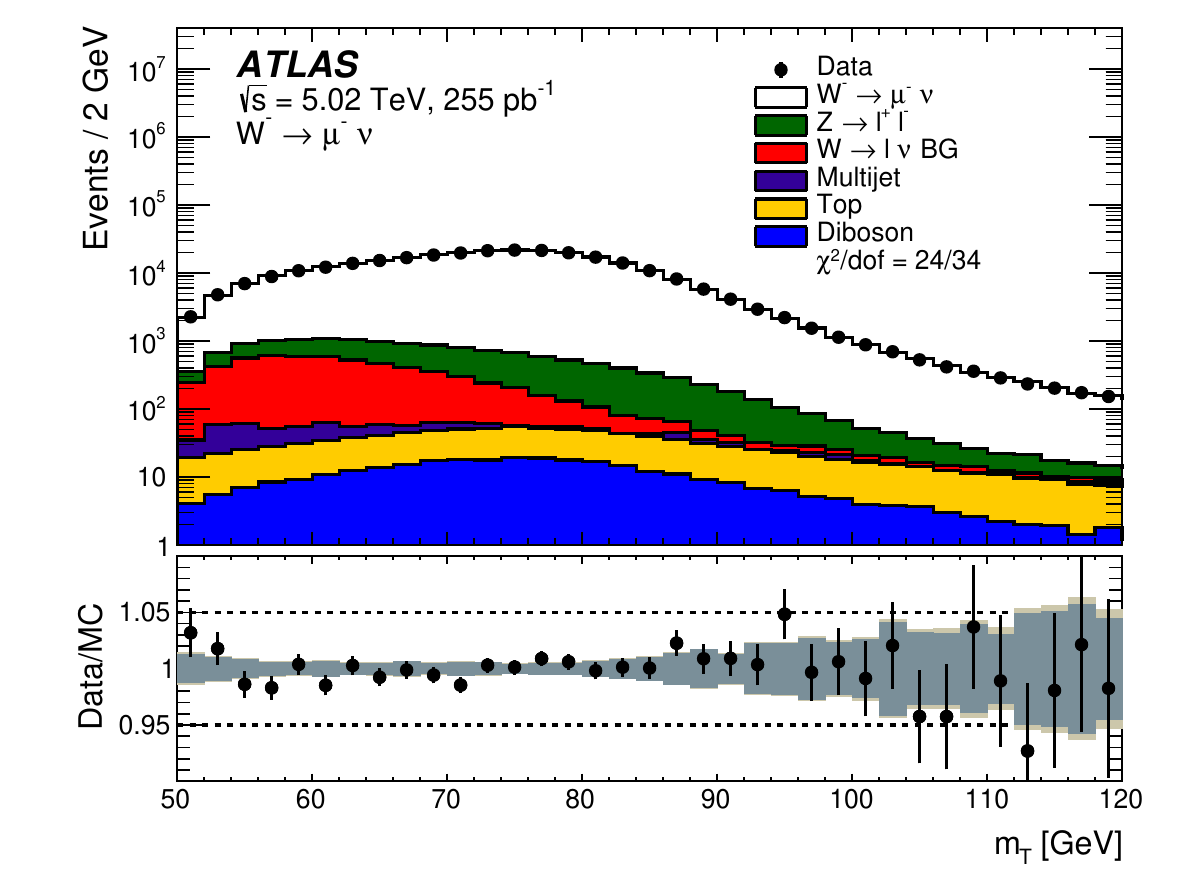}}
\subfloat[]{\includegraphics[width=.490\textwidth]{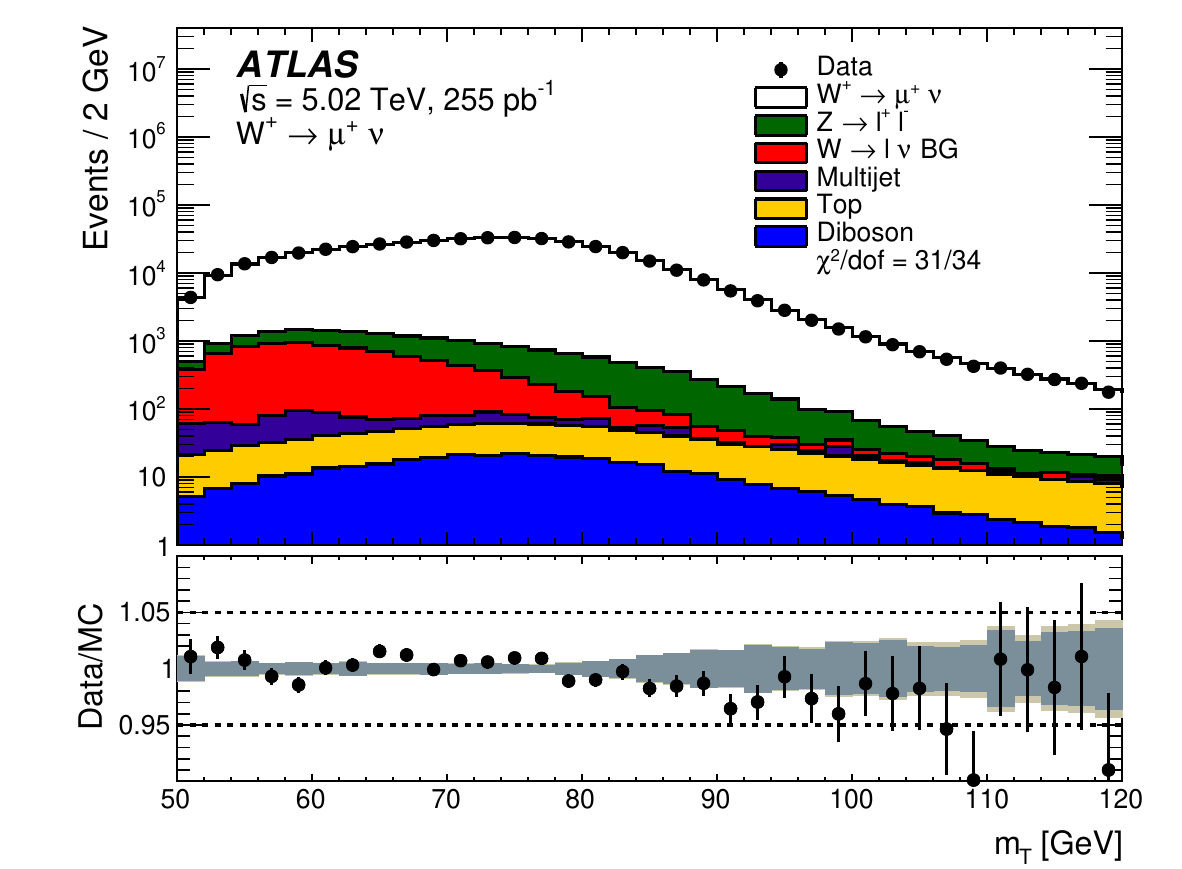}}
\caption{Transverse mass \mt distribution of the \Wboson\ boson in the (a) \Wmenu, (b) \Wpenu, (c) \Wmmunu and (d) \Wpmunu channels  for the $\sqrt{s} = \fivetev$ dataset. The data (black points) is compared to the sum
of all expected background and signal contributions, shown as a
solid line, normalised to the total data yield. The lower panel shows the
ratio of the data to the full prediction (black points) as well as the prediction uncertainties around $1$ excluding
(including) the MC simulation statistical component as dark (light) band.
The prediction uncertainties shown exclude those from the luminosity measurement
and the alternative signal modelling, see text. The
$\chi^2/\mathrm{dof}$ is computed using the full covariance matrix
of all shown statistical and systematic uncertainties. The category $\Wln$
BG refers to background from $W\to\tau\nu$ and charge
misidentification in the electron channels.}
\label{fig:mt5_controlplot}
\end{figure}

\begin{figure}[h]
\centering
\subfloat[]{\includegraphics[width=.490\textwidth]{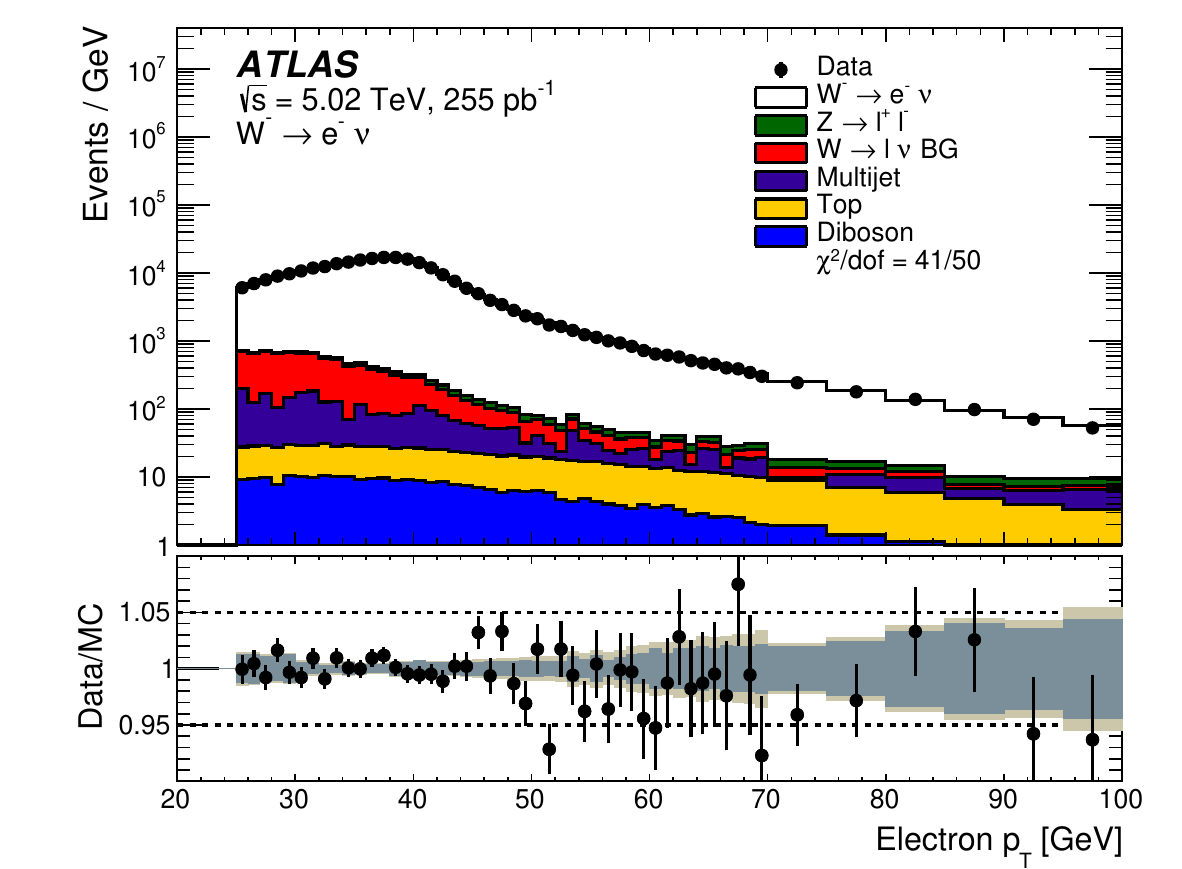}}
\subfloat[]{\includegraphics[width=.490\textwidth]{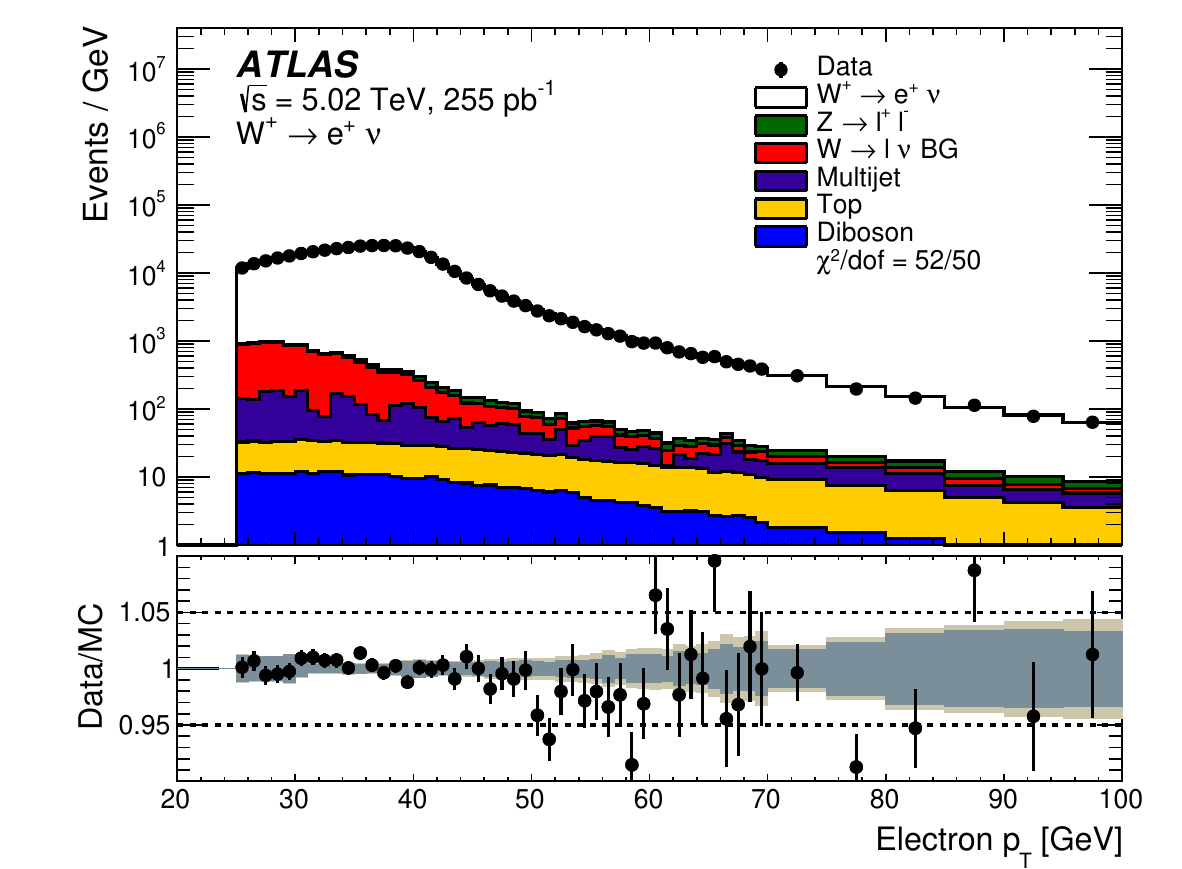}}

\subfloat[]{\includegraphics[width=.490\textwidth]{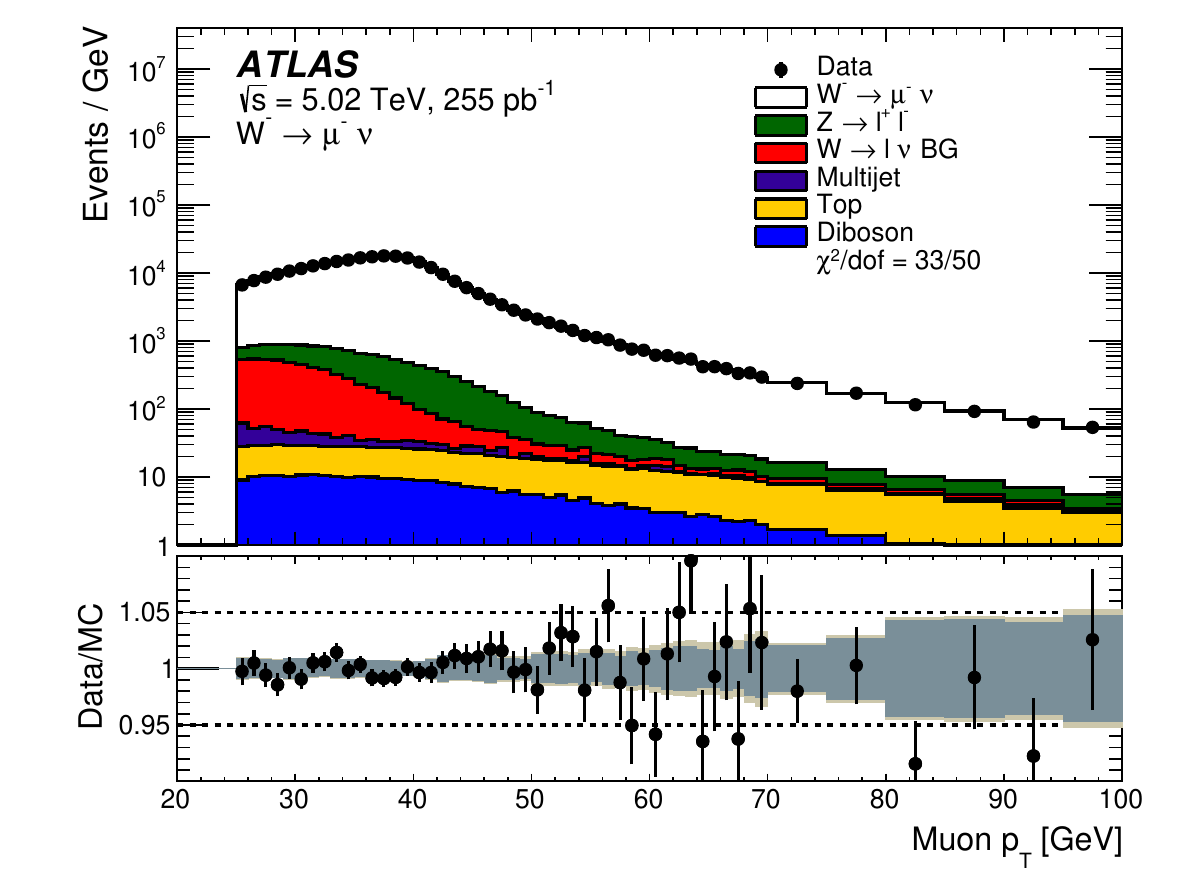}}
\subfloat[]{\includegraphics[width=.490\textwidth]{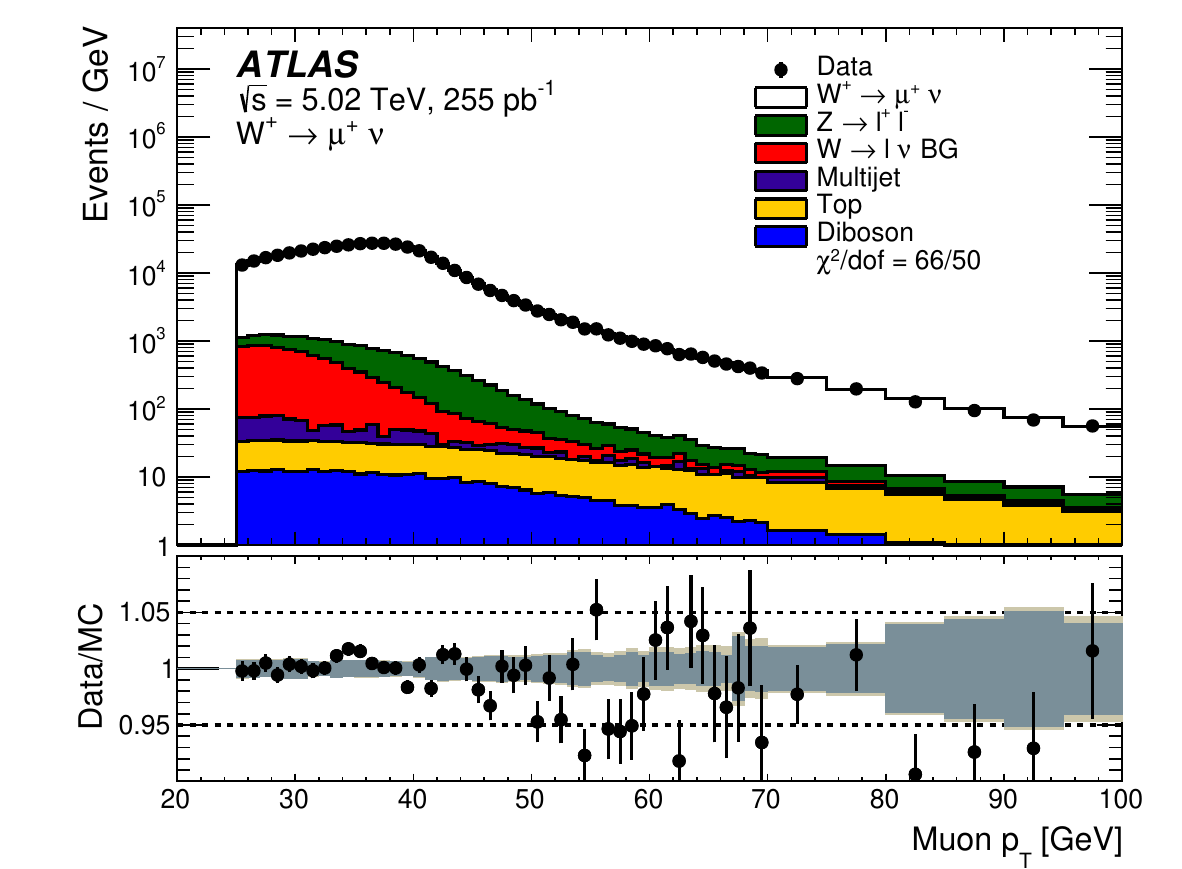}}
\caption{Lepton transverse momentum \pt distributions in the (a) \Wmenu, (b) \Wpenu, (c) \Wmmunu and (d) \Wpmunu channels  for the $\sqrt{s} = \fivetev$ dataset. The data (black points) is compared to the sum
of all expected background and signal contributions, shown as a
solid line, normalised to the total data yield. The lower panel shows the
ratio of the data to the full prediction (black points) as well as the prediction uncertainties around $1$ excluding
(including) the MC simulation statistical component as dark (light) band.
The prediction uncertainties shown exclude those from the luminosity measurement
and the alternative signal modelling, see text. The
$\chi^2/\mathrm{dof}$ is computed using the full covariance matrix
of all shown statistical and systematic uncertainties. The category $\Wln$
BG refers to background from $W\to\tau\nu$ and charge
misidentification in the electron channels.}
\label{fig:ptl5_controlplot}
\end{figure}

\begin{figure}[h]
\centering
\subfloat[]{\includegraphics[width=.490\textwidth]{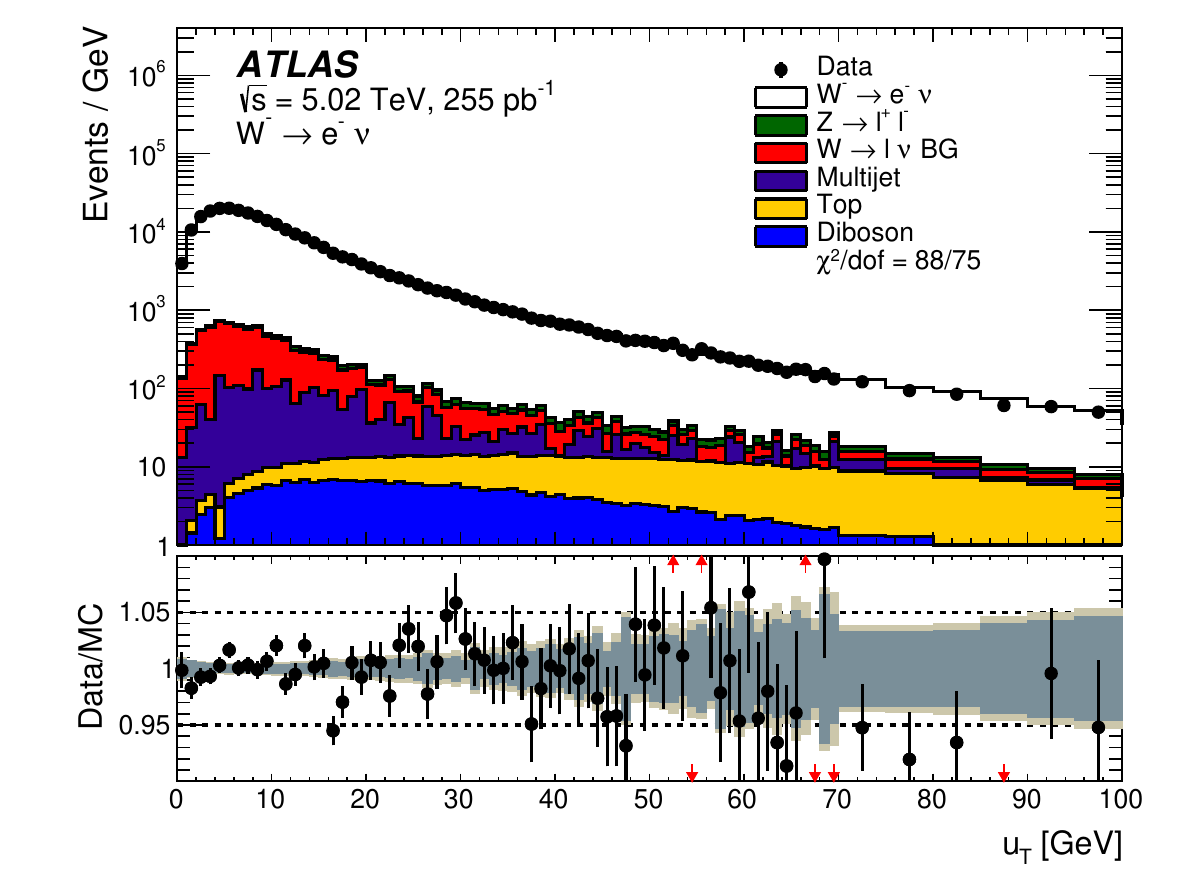}}
\subfloat[]{\includegraphics[width=.490\textwidth]{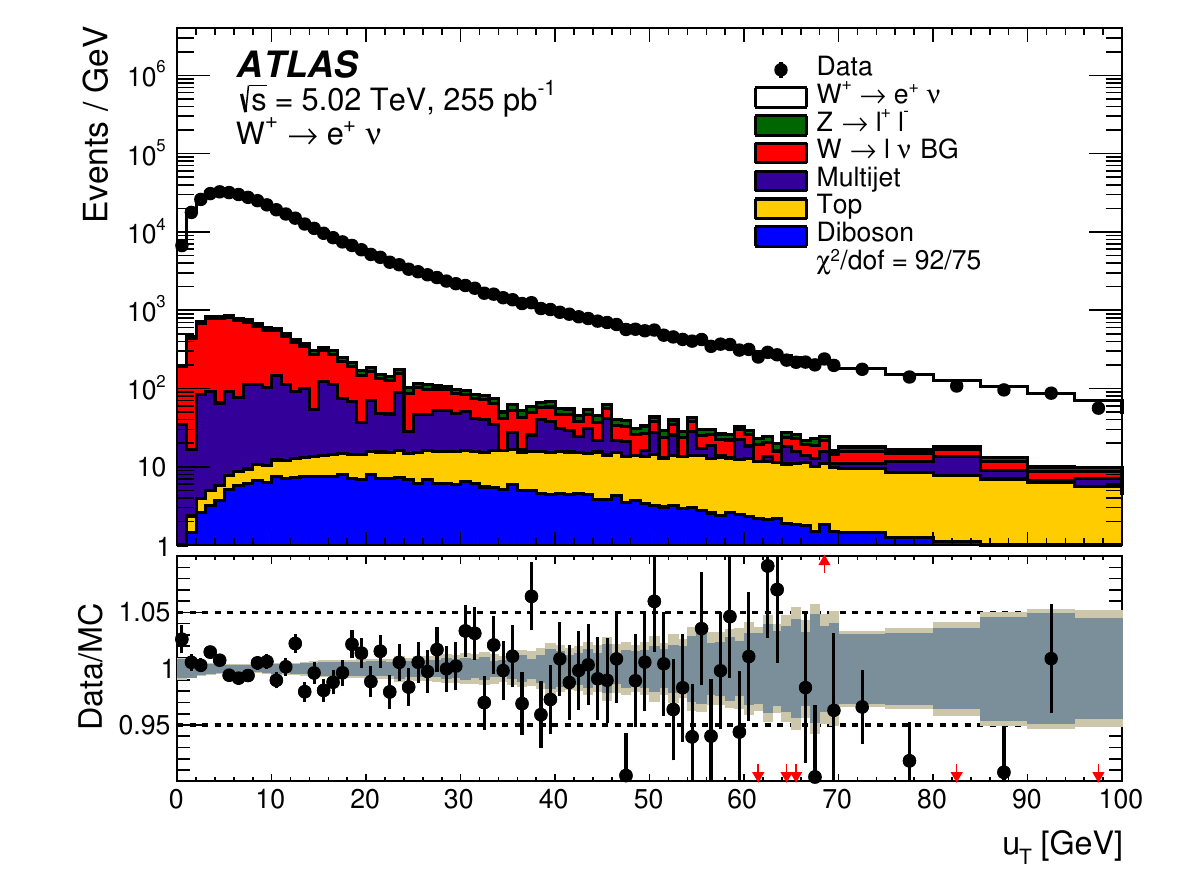}}

\subfloat[]{\includegraphics[width=.490\textwidth]{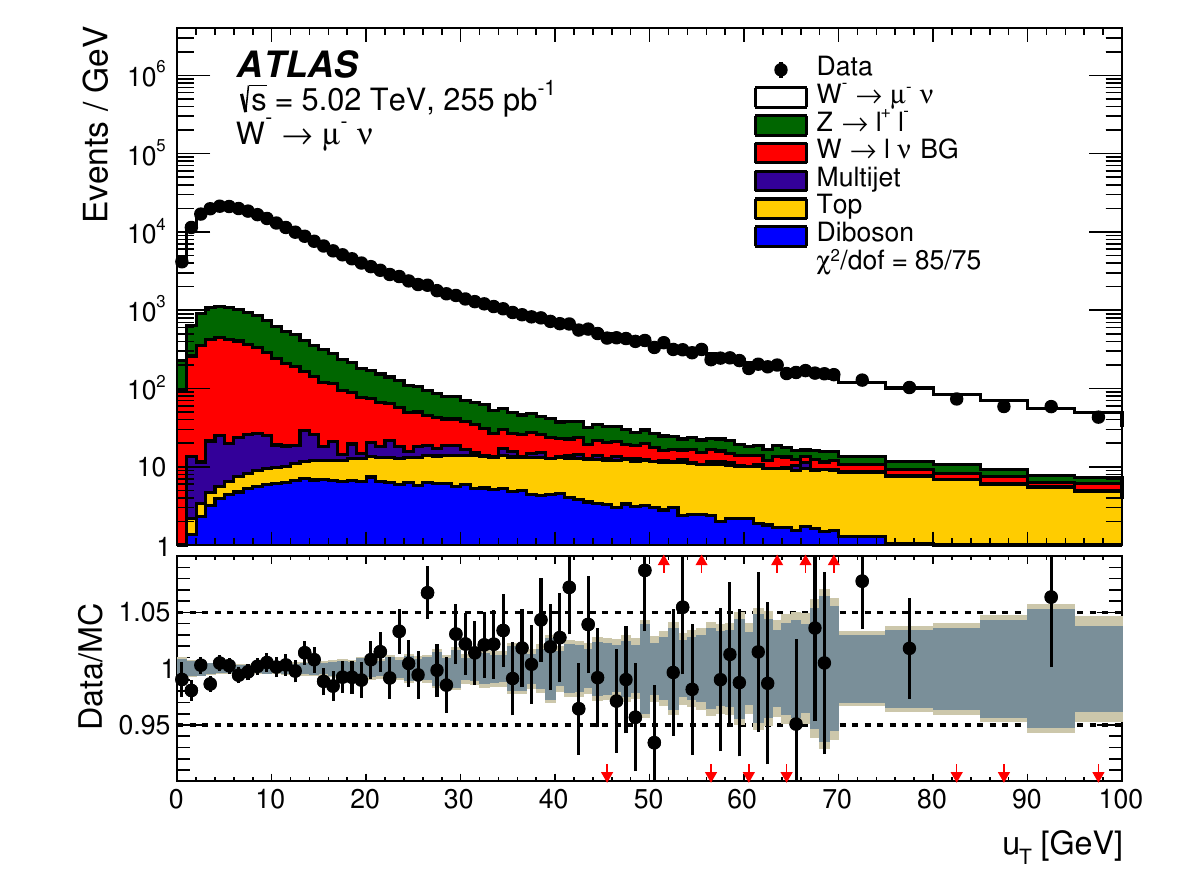}}
\subfloat[]{\includegraphics[width=.490\textwidth]{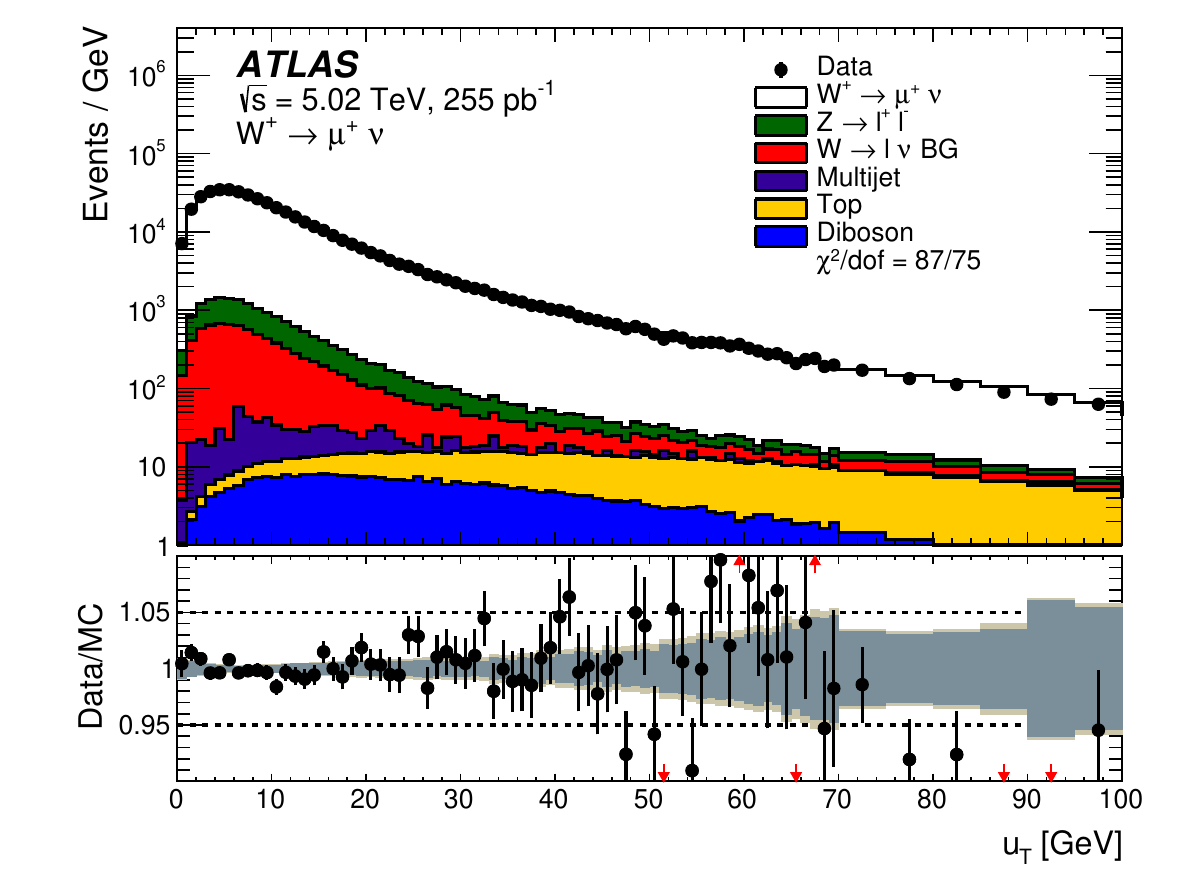}}
\caption{ Hadronic recoil \ut distributions (boson transverse momentum) in the (a) \Wmenu, (b) \Wpenu, (c) \Wmmunu and (d) \Wpmunu channels  for the $\sqrt{s} = \fivetev$ dataset. The data (black points) is compared to the sum
of all expected background and signal contributions, shown as a
solid line, normalised to the total data yield. The lower panel shows the
ratio of the data to the full prediction (black points) as well as the prediction uncertainties around $1$ excluding
(including) the MC simulation statistical component as dark (light) band.
The prediction uncertainties shown exclude those from the luminosity measurement
and the alternative signal modelling, see text. The
$\chi^2/\mathrm{dof}$ is computed using the full covariance matrix
of all shown statistical and systematic uncertainties. The category $\Wln$
BG refers to background from $W\to\tau\nu$ and charge
misidentification in the electron channels.}
\label{fig:ut5_controlplot}
\end{figure}

\begin{figure}[h]
\centering
\subfloat[]{\includegraphics[width=.49\textwidth]{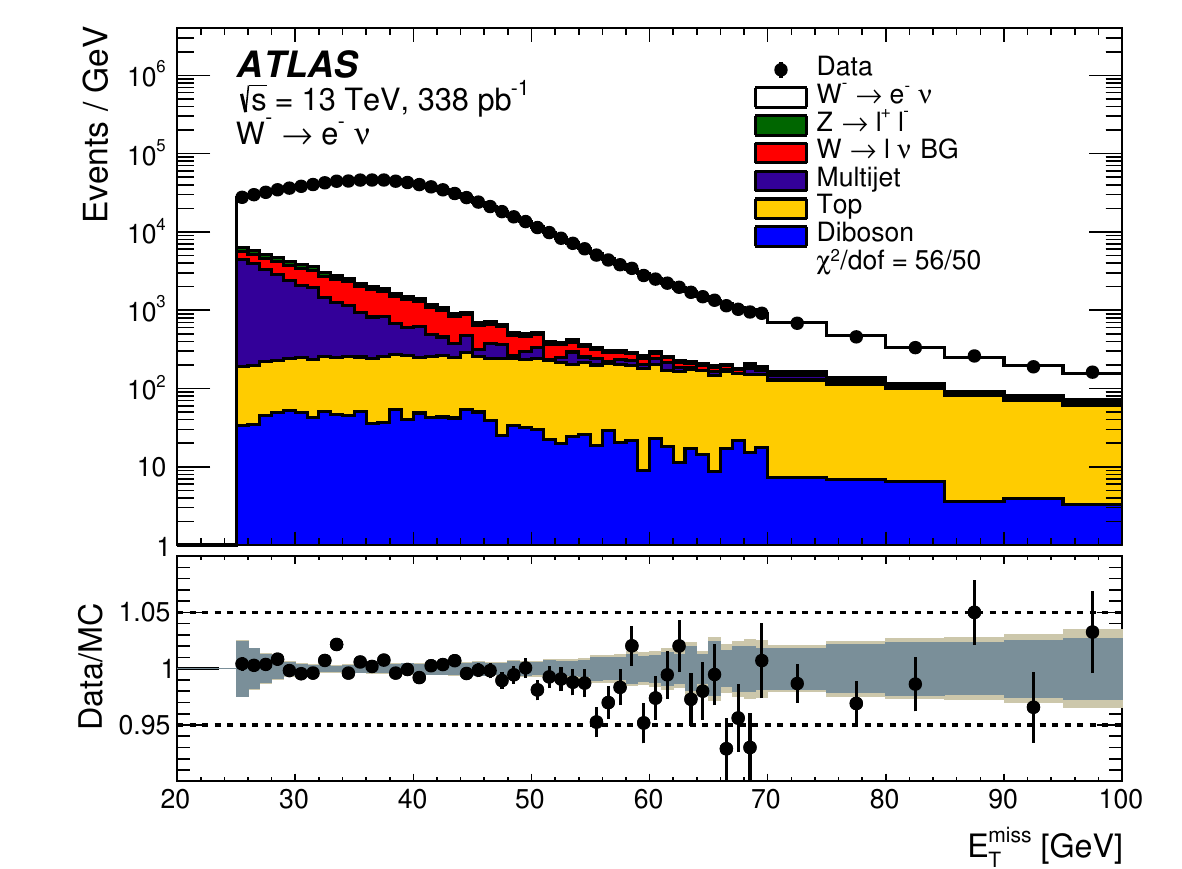}}
\subfloat[]{\includegraphics[width=.49\textwidth]{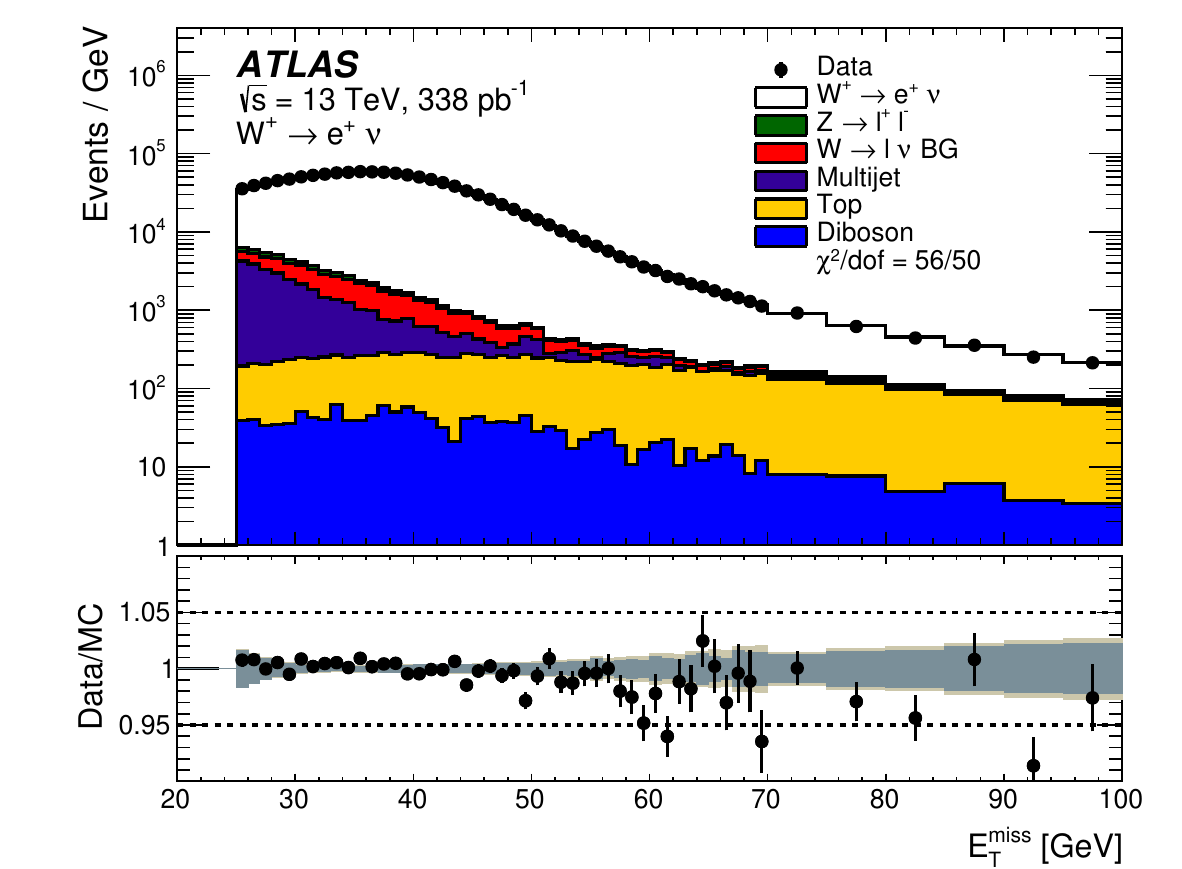}}

\subfloat[]{\includegraphics[width=.49\textwidth]{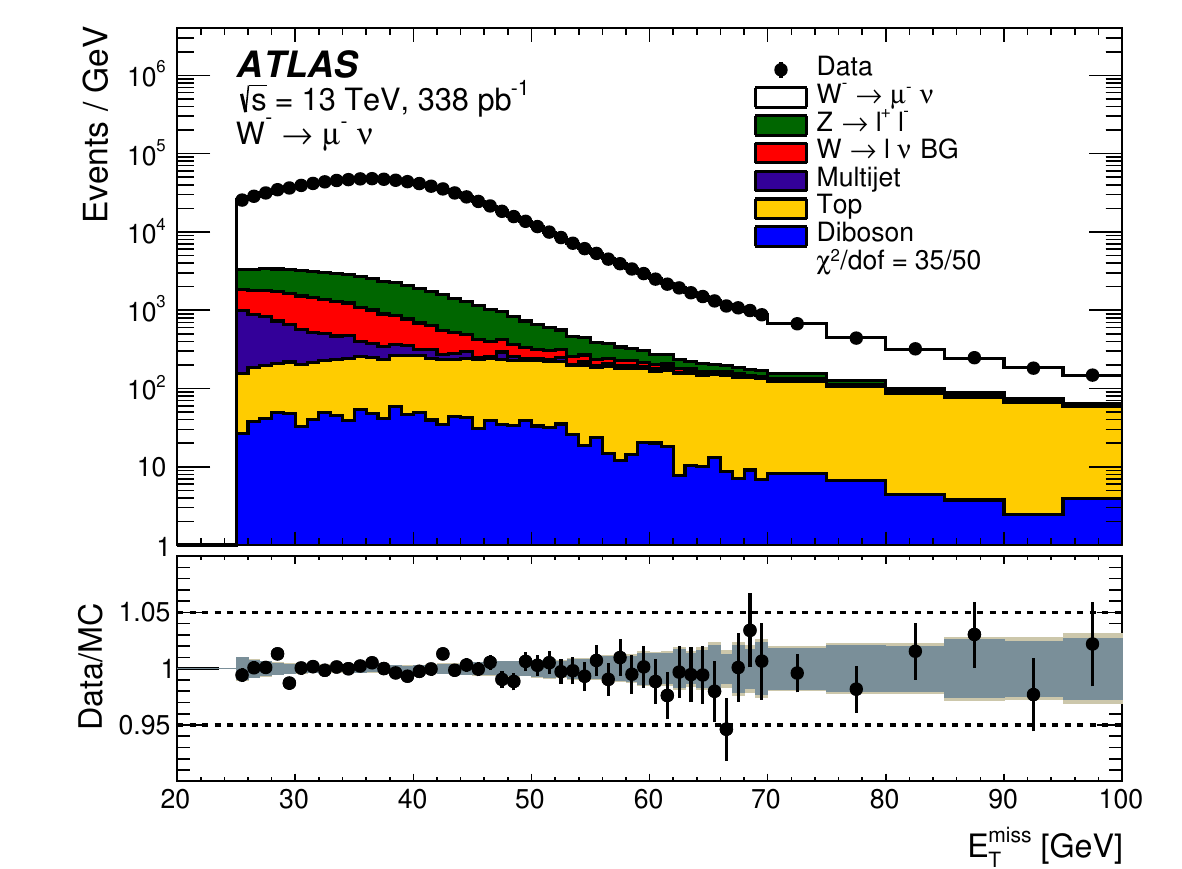}}
\subfloat[]{\includegraphics[width=.49\textwidth]{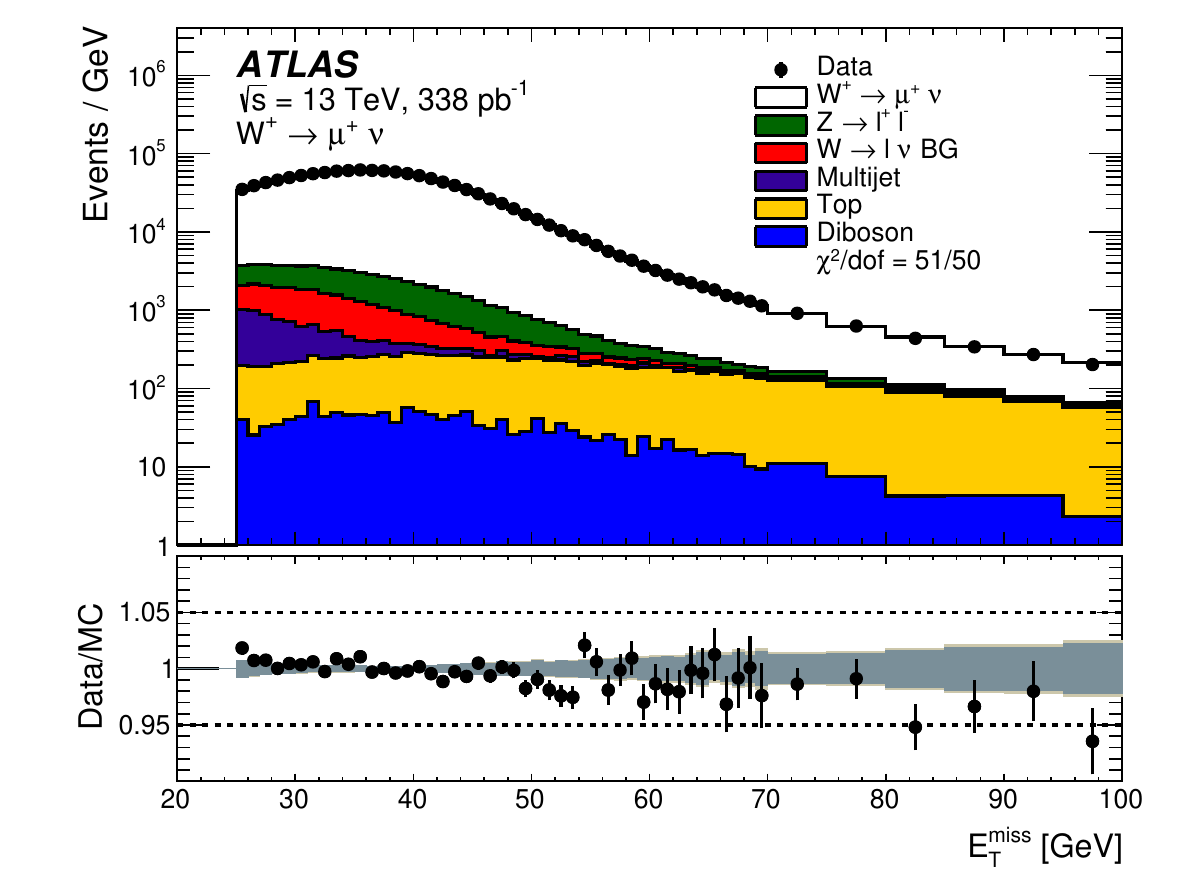}}

\caption{Missing transverse momentum \met distributions in the (a) \Wmenu, (b) \Wpenu, (c) \Wmmunu and (d) \Wpmunu channels  for the $\sqrt{s} = 13$~\TeV\ dataset. The data (black points) is compared to the sum
of all expected background and signal contributions, shown as a
solid line, normalised to the total data yield. The lower panel shows the
ratio of the data to the full prediction (black points) as well as the prediction uncertainties around $1$ excluding
(including) the MC simulation statistical component as dark (light) band.
The prediction uncertainties shown exclude those from the luminosity measurement
and the alternative signal modelling, see text. The
$\chi^2/\mathrm{dof}$ is computed using the full covariance matrix
of all shown statistical and systematic uncertainties. The category $\Wln$
BG refers to background from $W\to\tau\nu$ and charge
misidentification in the electron channels.}
\label{fig:met13_controlplot}
\end{figure}

\begin{figure}[h]
\centering
\subfloat[]{\includegraphics[width=.49\textwidth]{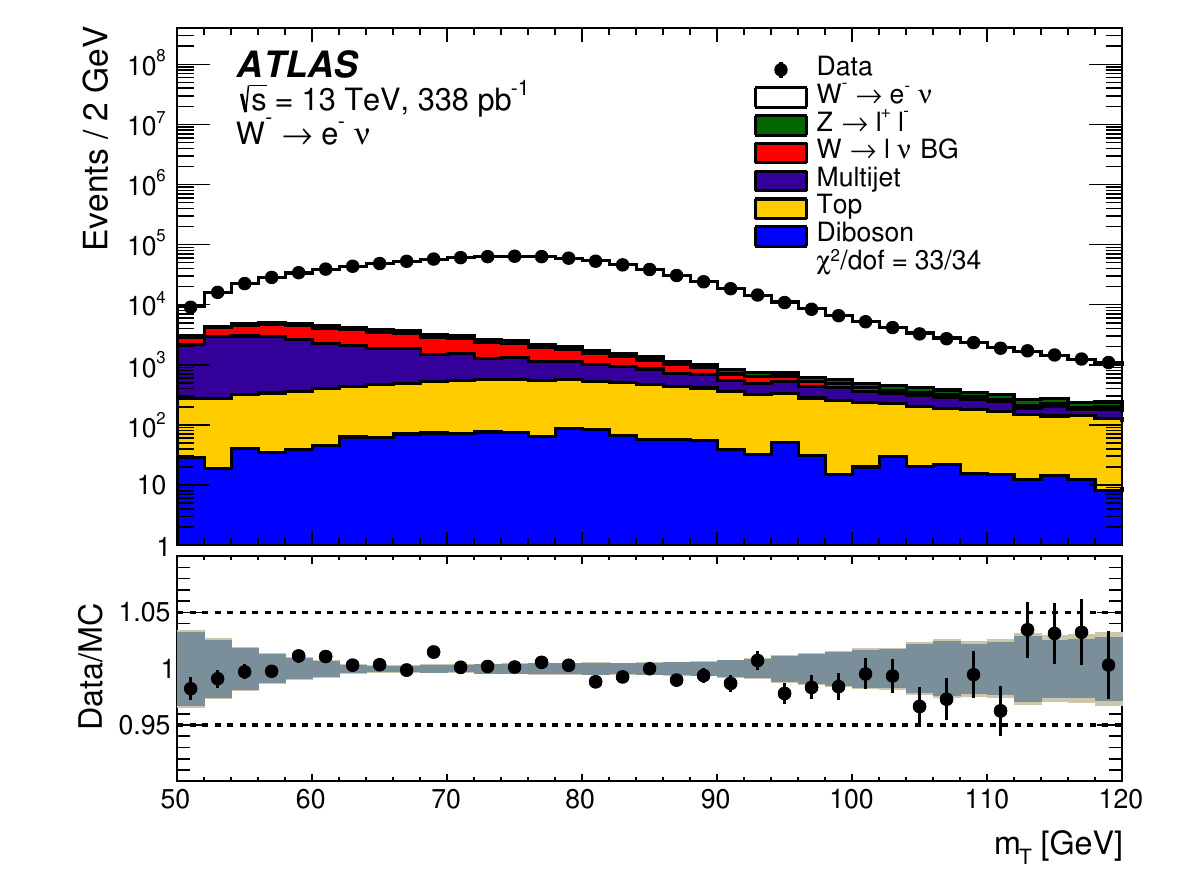}}
\subfloat[]{\includegraphics[width=.49\textwidth]{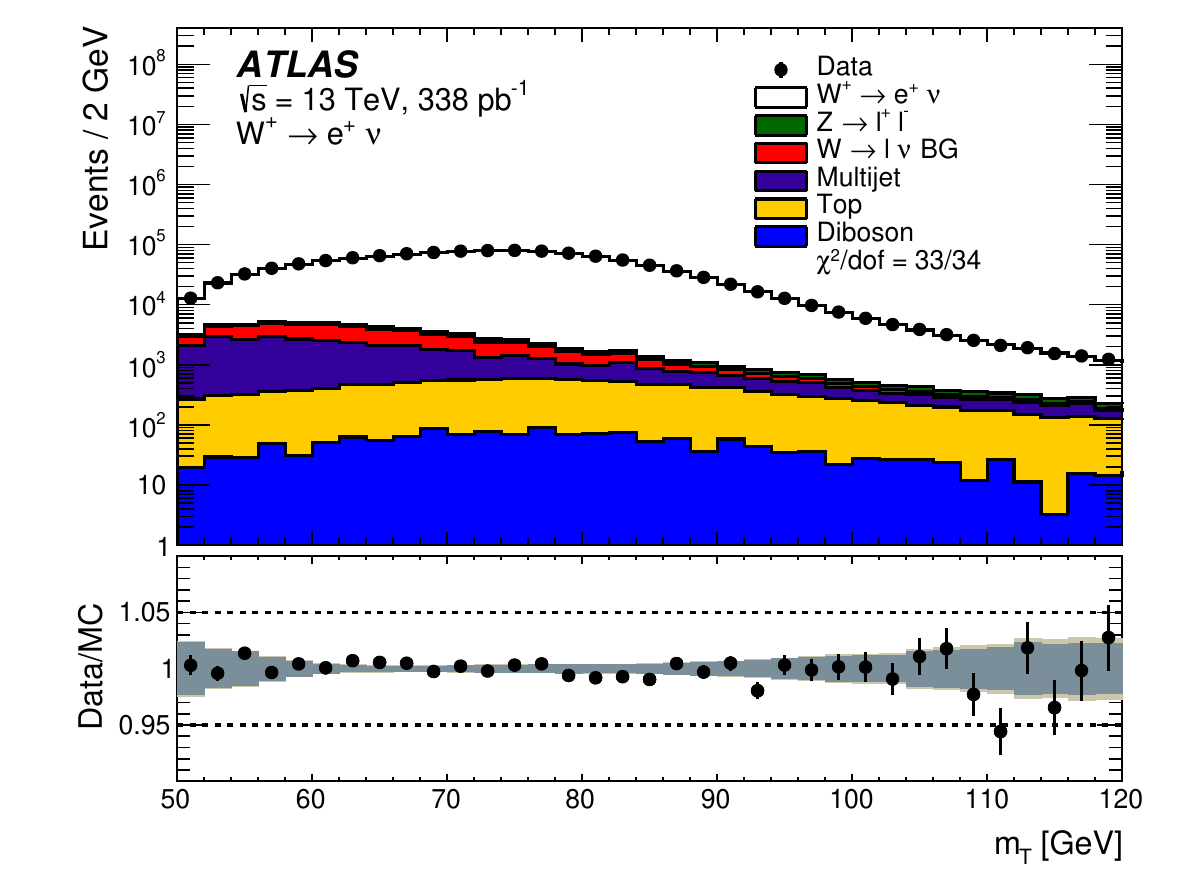}}

\subfloat[]{\includegraphics[width=.49\textwidth]{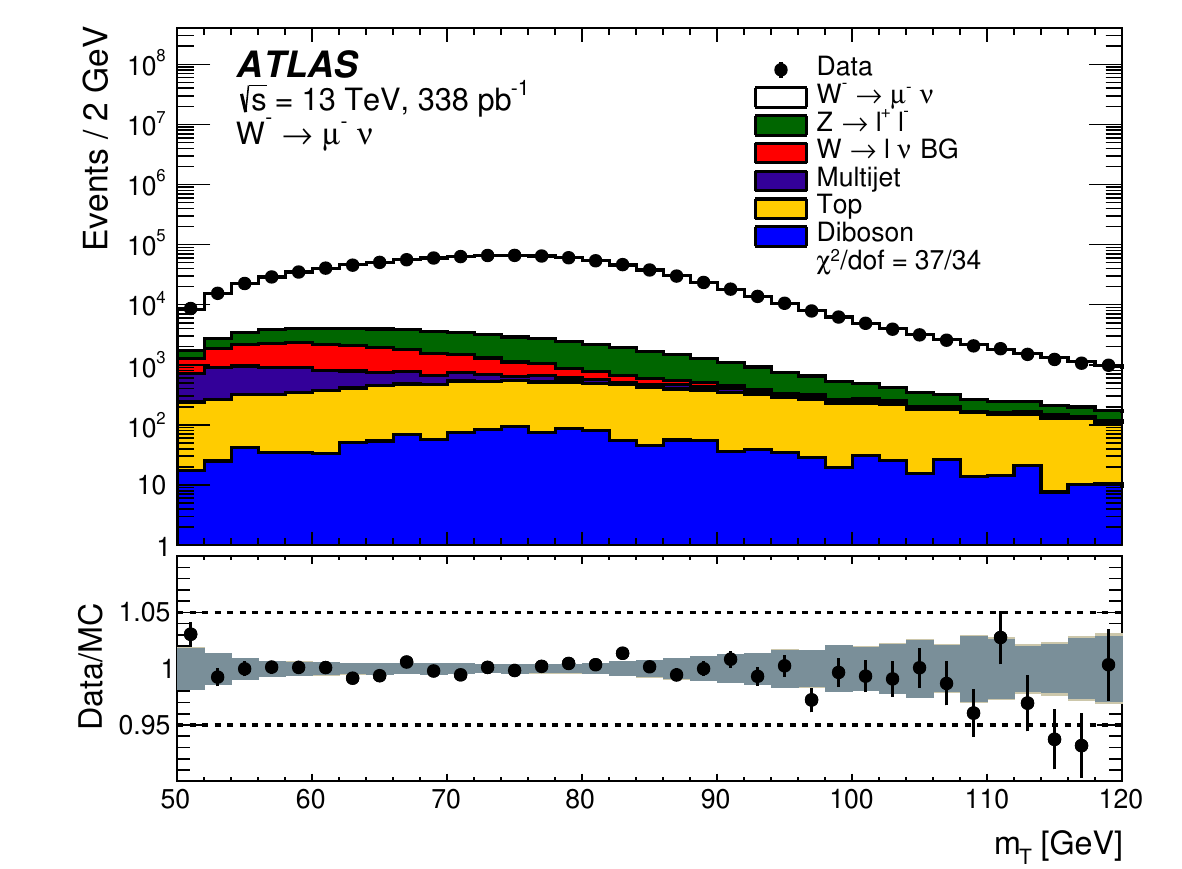}}
\subfloat[]{\includegraphics[width=.49\textwidth]{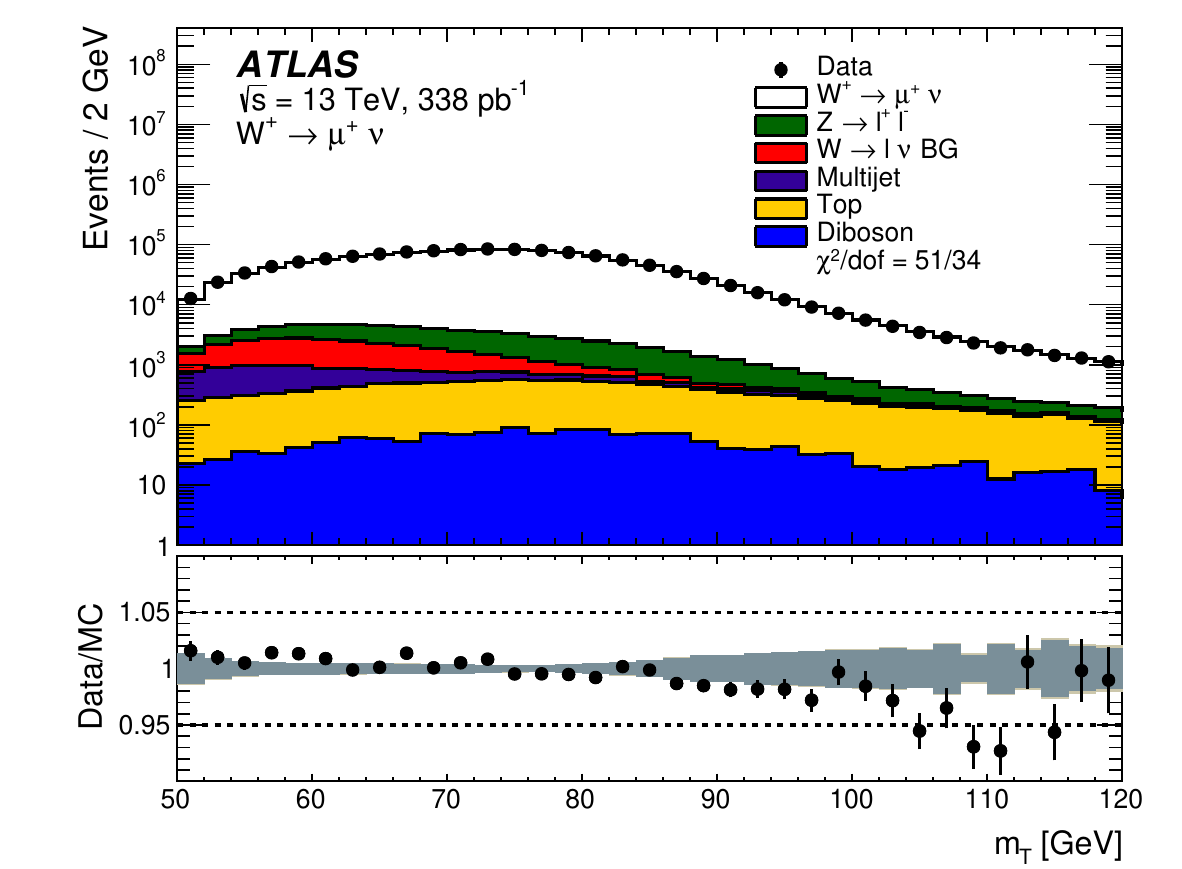}}
\caption{  Transverse mass \mt distributions of the \Wboson\ boson in the (a) \Wmenu, (b) \Wpenu, (c) \Wmmunu and (d) \Wpmunu channels  for the $\sqrt{s} = 13$~\TeV\ dataset. The data (black points) is compared to the sum
of all expected background and signal contributions, shown as a
solid line, normalised to the total data yield. The lower panel shows the
ratio of the data to the full prediction (black points) as well as the prediction uncertainties around $1$ excluding
(including) the MC simulation statistical component as dark (light) band.
The prediction uncertainties shown exclude those from the luminosity measurement
and the alternative signal modelling, see text. The
$\chi^2/\mathrm{dof}$ is computed using the full covariance matrix
of all shown statistical and systematic uncertainties. The category $\Wln$
BG refers to background from $W\to\tau\nu$ and charge
misidentification in the electron channels.}
\label{fig:mt13_controlplot}
\end{figure}

\begin{figure}[h]
\centering
\subfloat[]{\includegraphics[width=.490\textwidth]{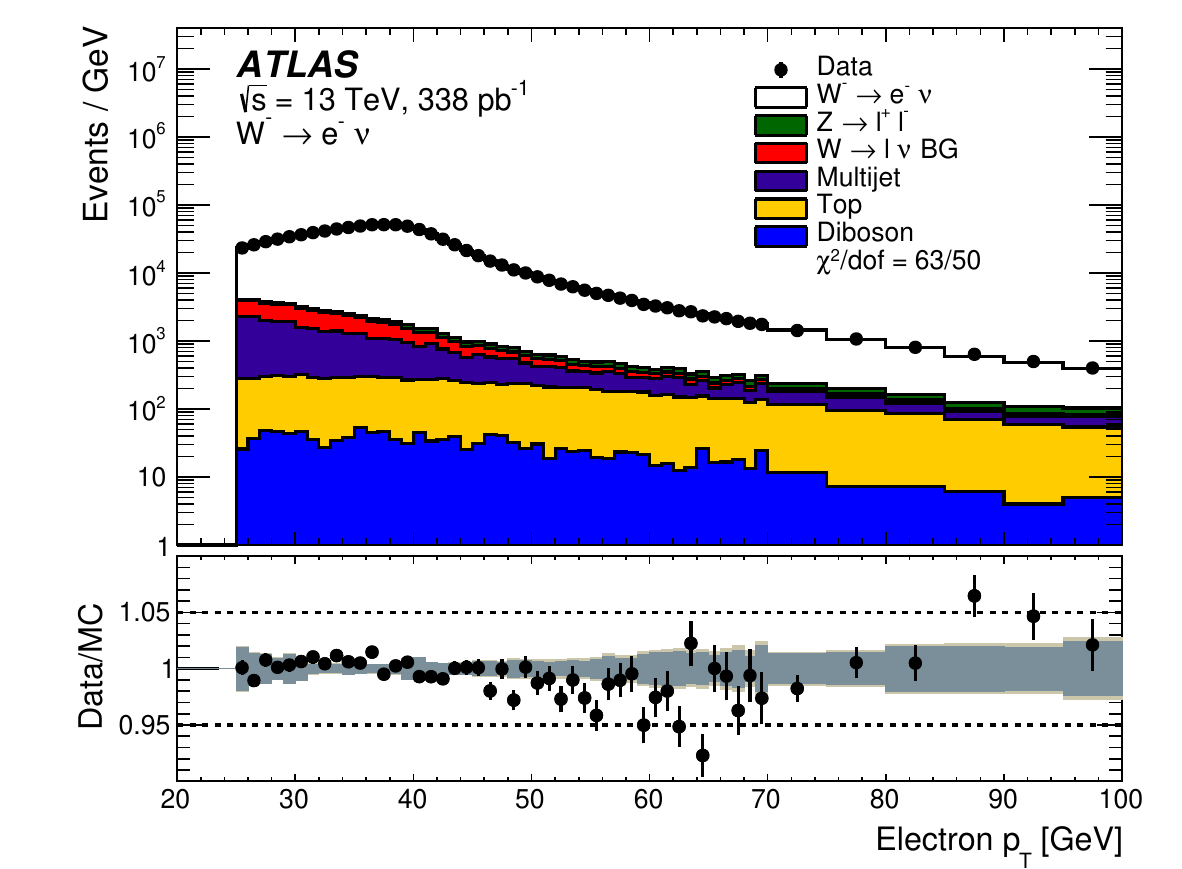}}
\subfloat[]{\includegraphics[width=.490\textwidth]{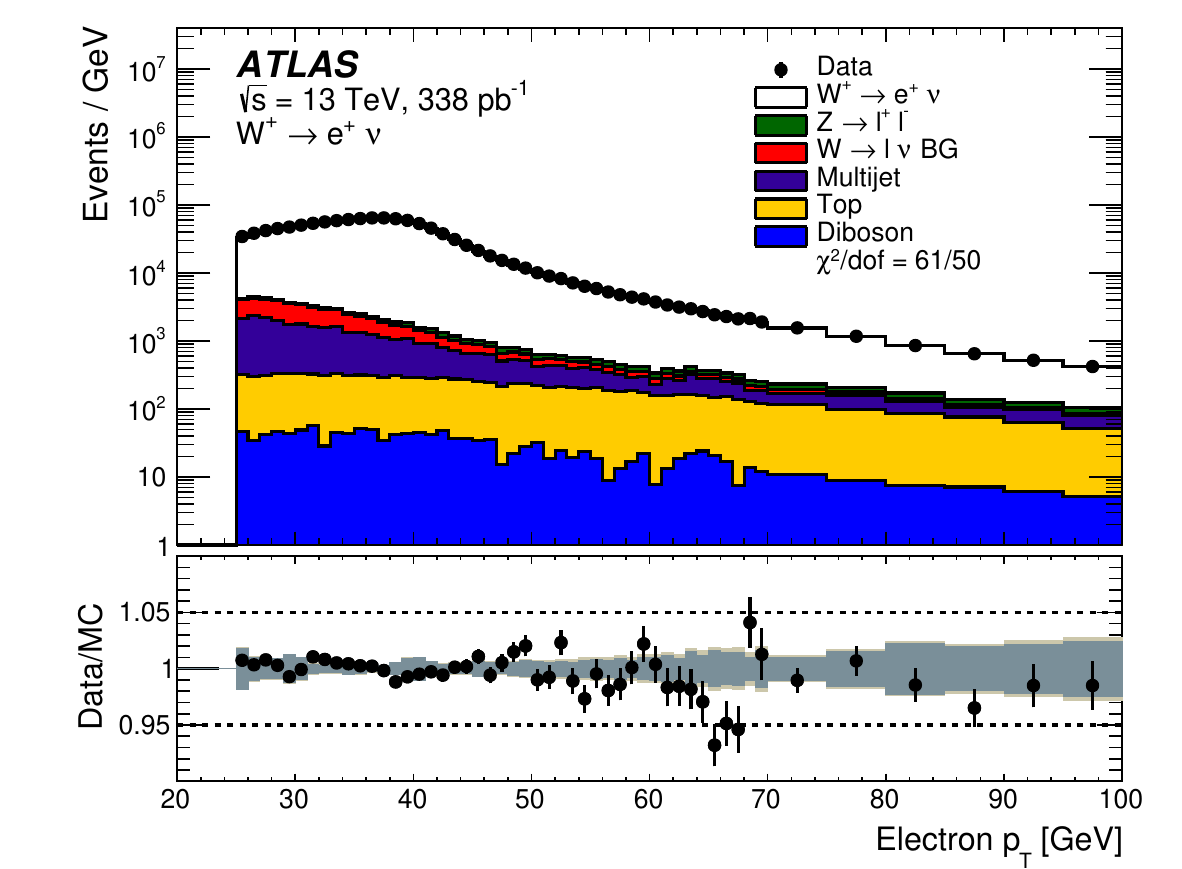}}

\subfloat[]{\includegraphics[width=.490\textwidth]{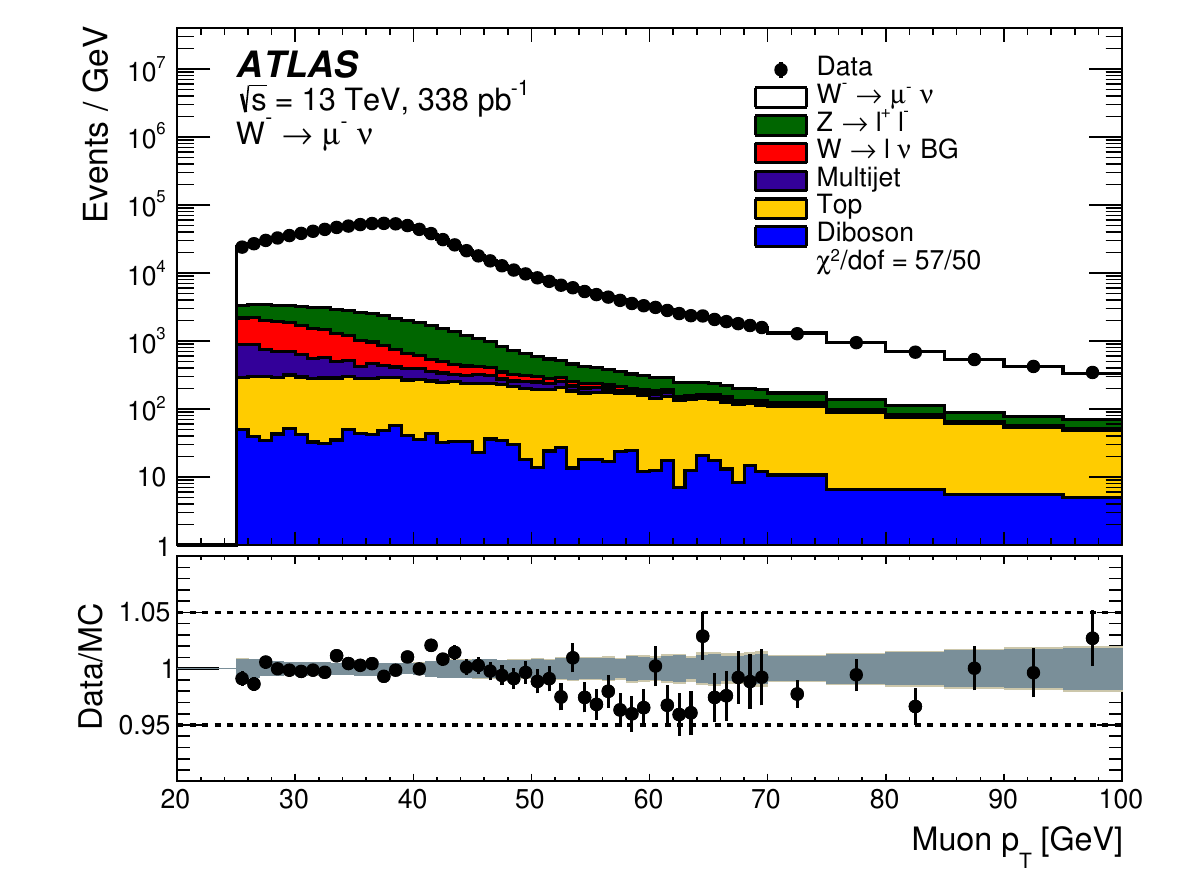}}
\subfloat[]{\includegraphics[width=.490\textwidth]{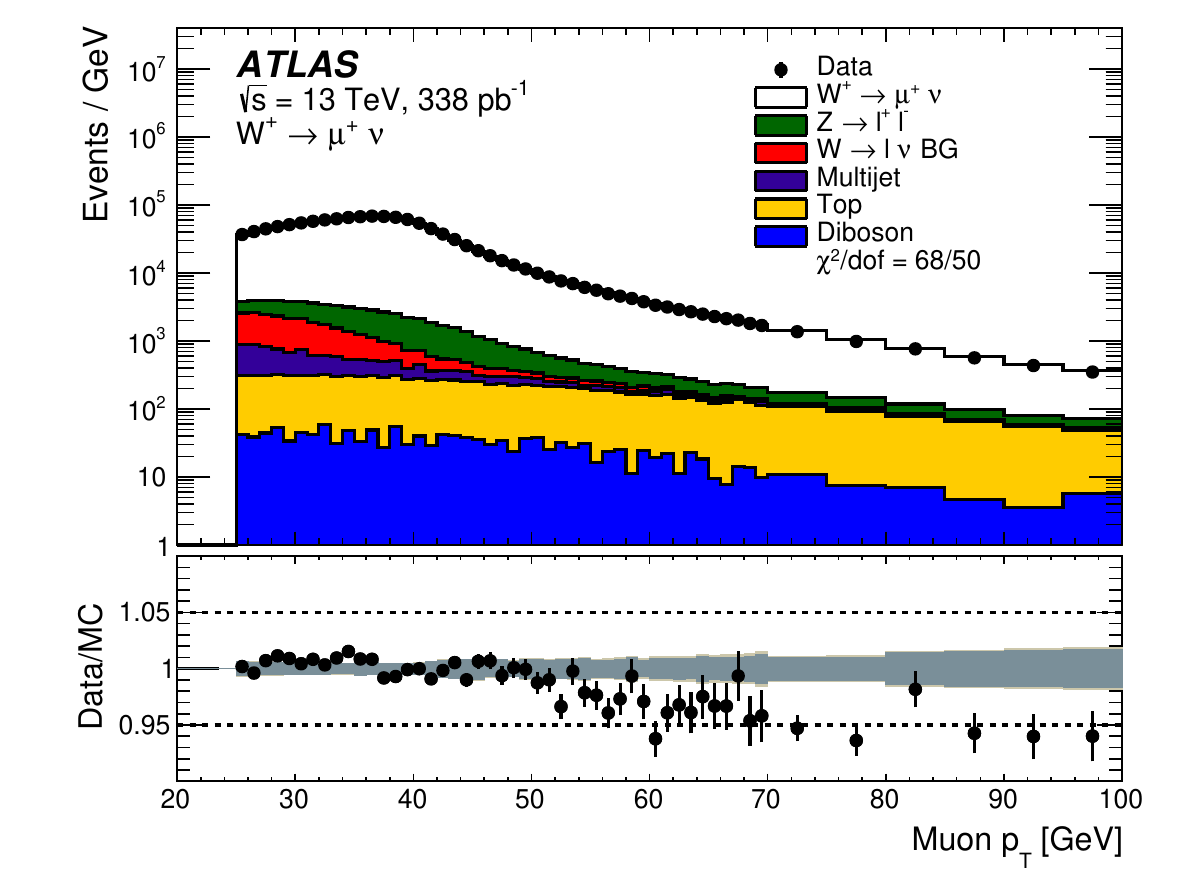}}
\caption{  Lepton transverse momentum \pt distributions in the (a) \Wmenu, (b) \Wpenu, (c) \Wmmunu and (d) \Wpmunu channels  for the $\sqrt{s} = 13$~\TeV\ dataset. The data (black points) is compared to the sum
of all expected background and signal contributions, shown as a
solid line, normalised to the total data yield. The lower panel shows the
ratio of the data to the full prediction (black points) as well as the prediction uncertainties around $1$ excluding
(including) the MC simulation statistical component as dark (light) band.
The prediction uncertainties shown exclude those from the luminosity measurement
and the alternative signal modelling, see text. The
$\chi^2/\mathrm{dof}$ is computed using the full covariance matrix
of all shown statistical and systematic uncertainties. The category $\Wln$
BG refers to background from $W\to\tau\nu$ and charge
misidentification in the electron channels.}
\label{fig:ptl13_controlplot}
\end{figure}

\begin{figure}[h]
\centering
\subfloat[]{\includegraphics[width=.490\textwidth]{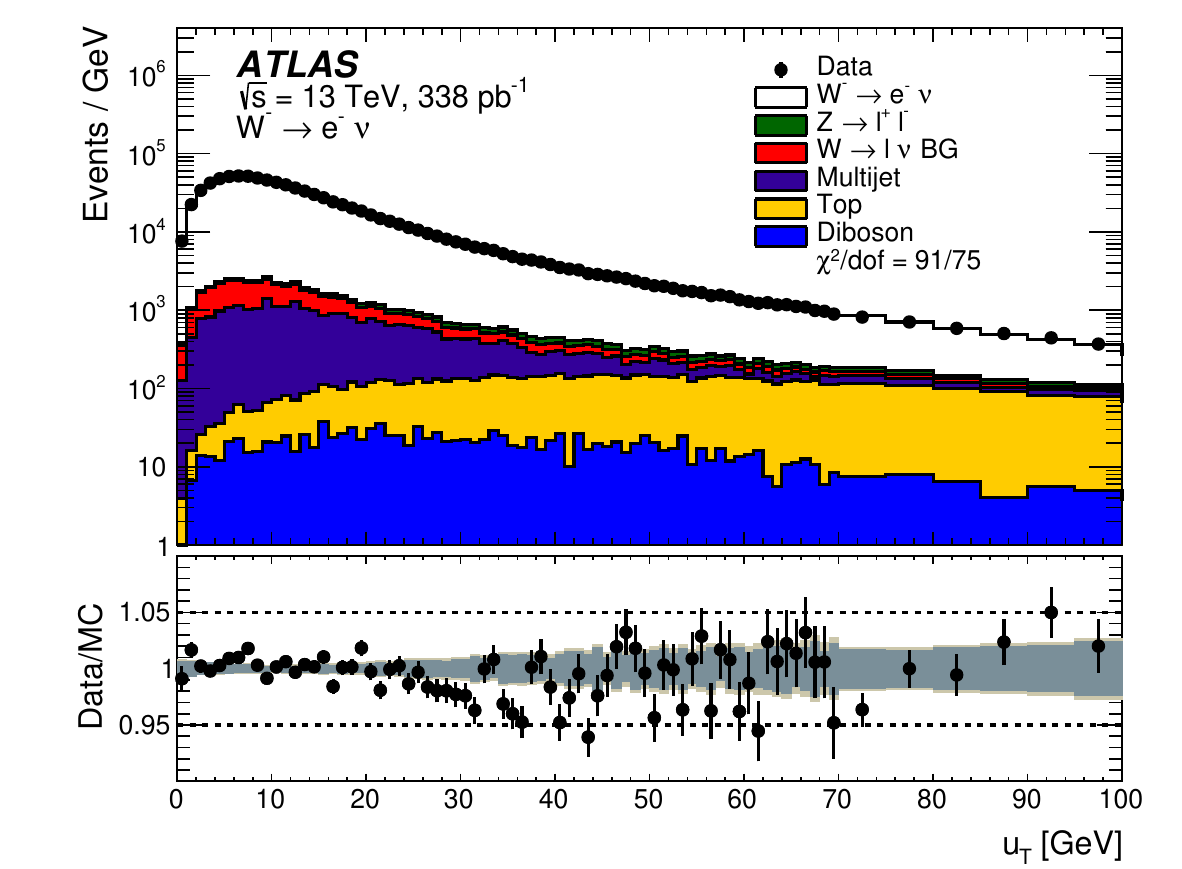}}
\subfloat[]{\includegraphics[width=.490\textwidth]{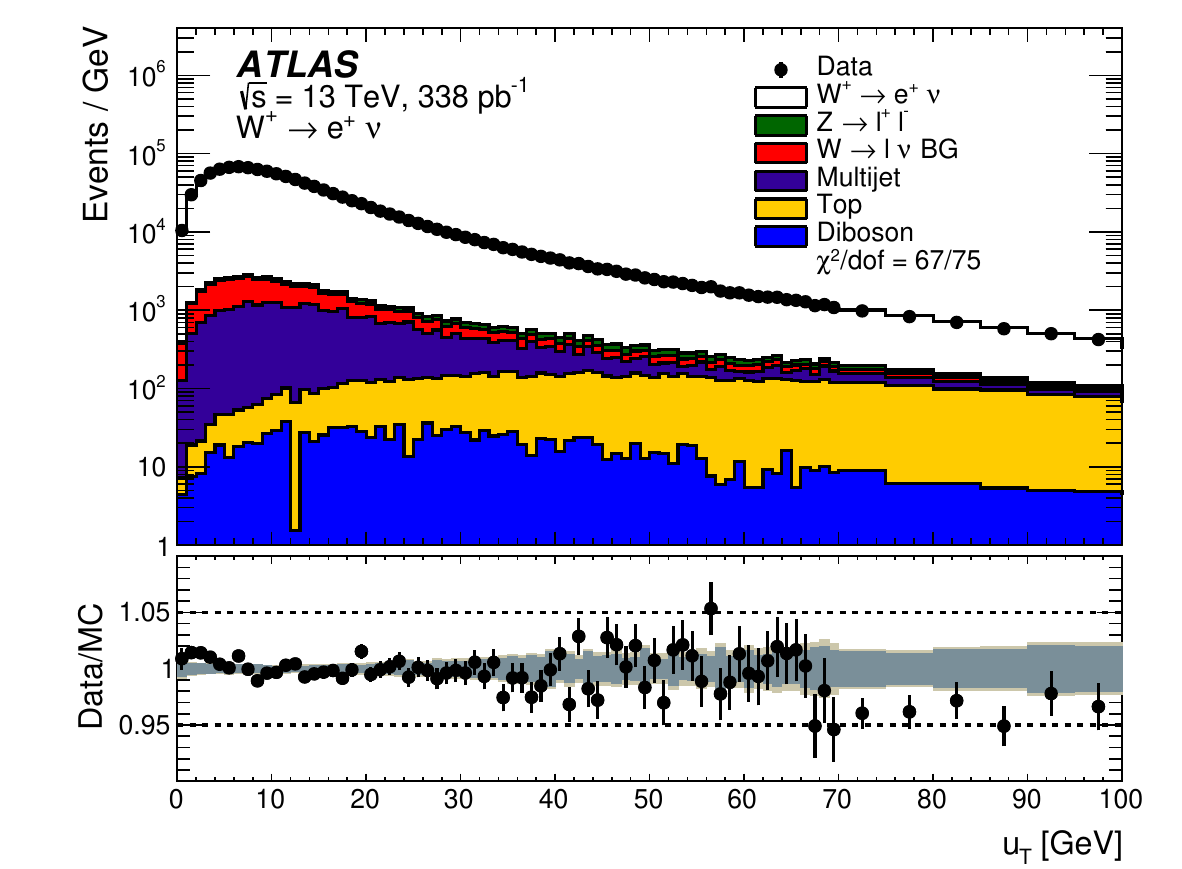}}

\subfloat[]{\includegraphics[width=.490\textwidth]{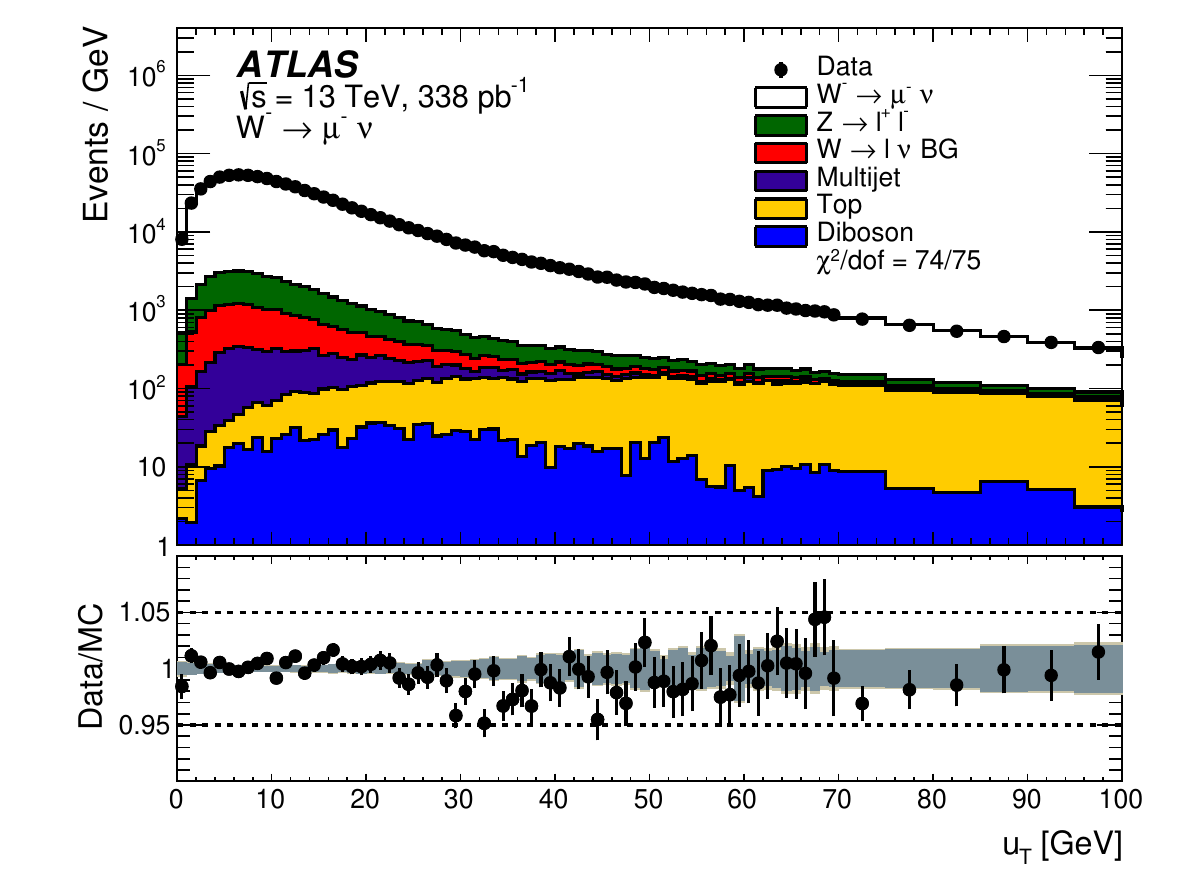}}
\subfloat[]{\includegraphics[width=.490\textwidth]{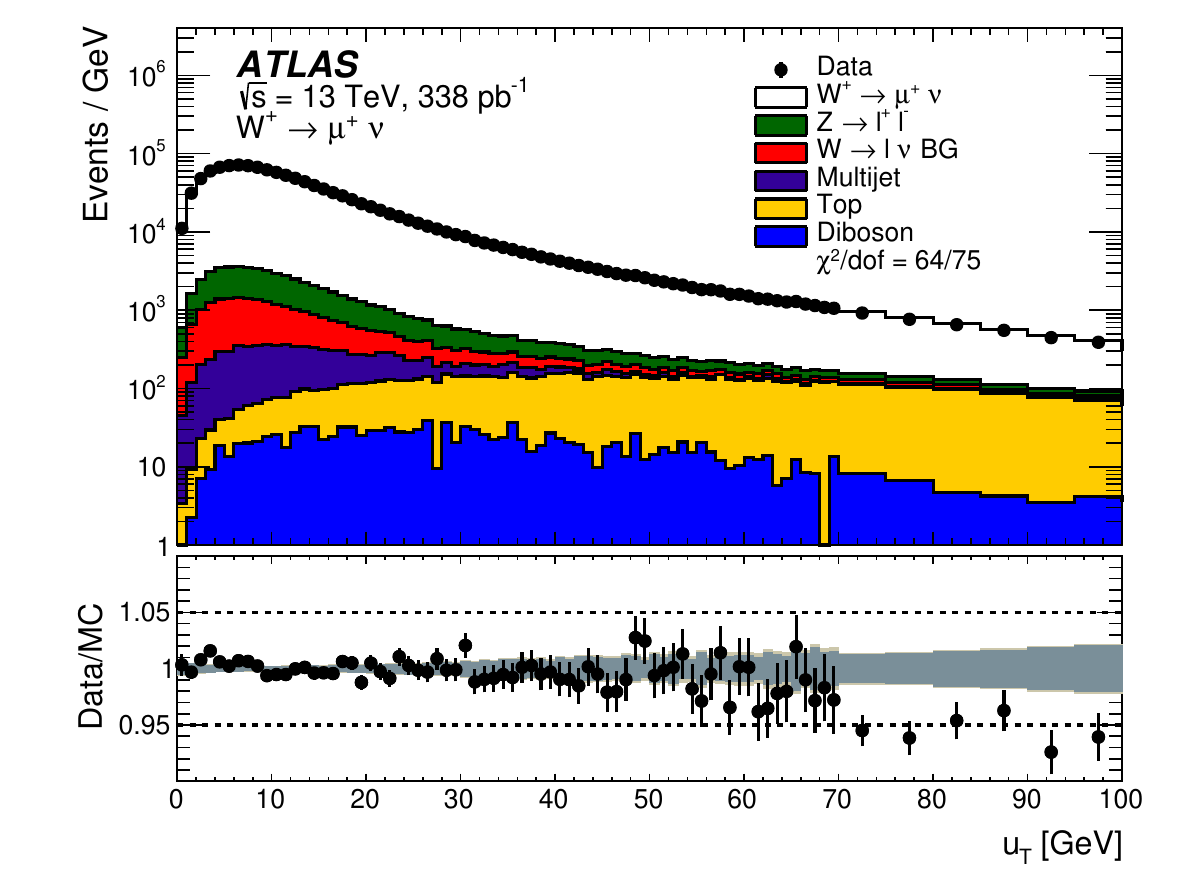}}
\caption{ Hadronic recoil \ut distributions (boson transverse momentum) in the (a) \Wmenu, (b) \Wpenu, (c) \Wmmunu and (d) \Wpmunu channels  for the $\sqrt{s} = 13$~\TeV\ dataset. The data (black points) is compared to the sum
of all expected background and signal contributions, shown as a
solid line, normalised to the total data yield. The lower panel shows the
ratio of the data to the full prediction (black points) as well as the prediction uncertainties around $1$ excluding
(including) the MC simulation statistical component as dark (light) band.
The prediction uncertainties shown exclude those from the luminosity measurement
and the alternative signal modelling, see text. The
$\chi^2/\mathrm{dof}$ is computed using the full covariance matrix
of all shown statistical and systematic uncertainties. The category $\Wln$
BG refers to background from $W\to\tau\nu$ and charge
misidentification in the electron channels.}
\label{fig:ut13_controlplot}
\end{figure}

\begin{figure}[h]
\centering
\subfloat[]{\includegraphics[width=.47\textwidth]{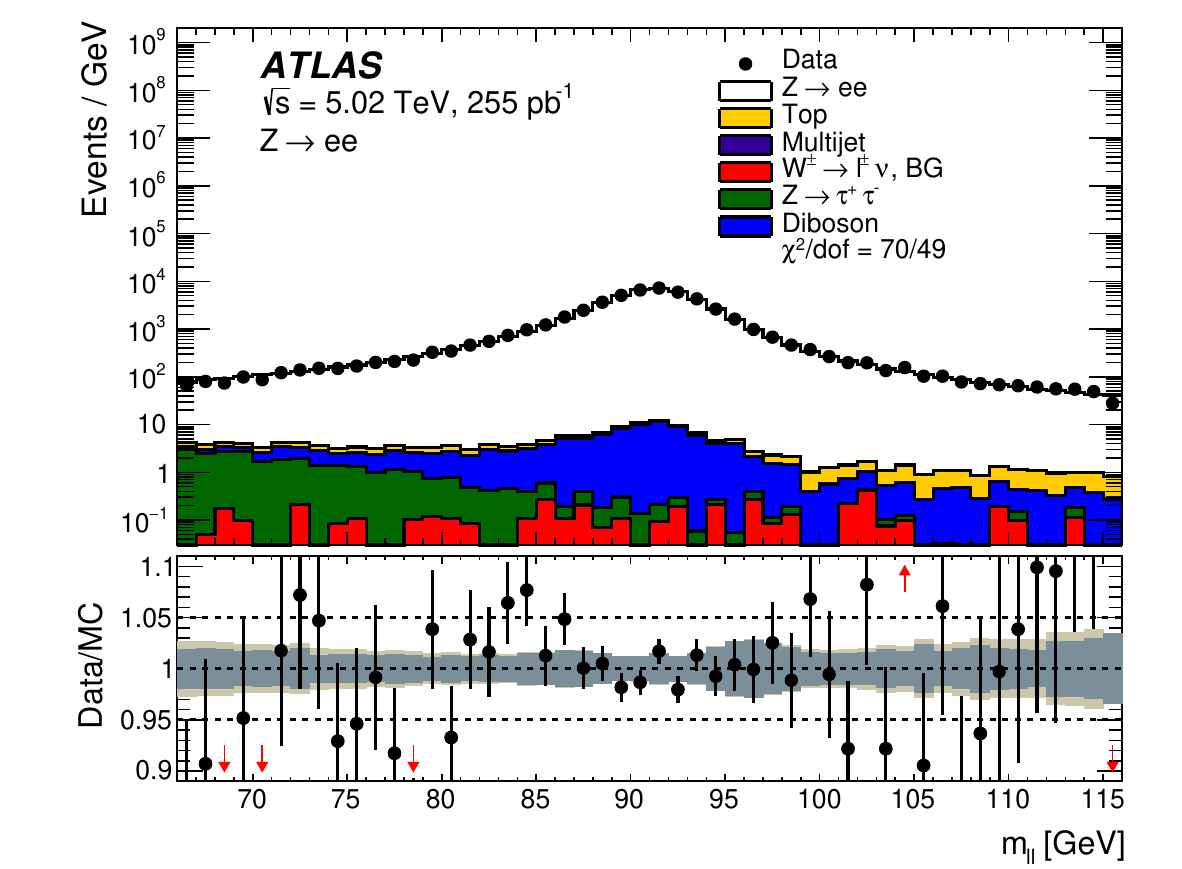}}
\subfloat[]{\includegraphics[width=.47\textwidth]{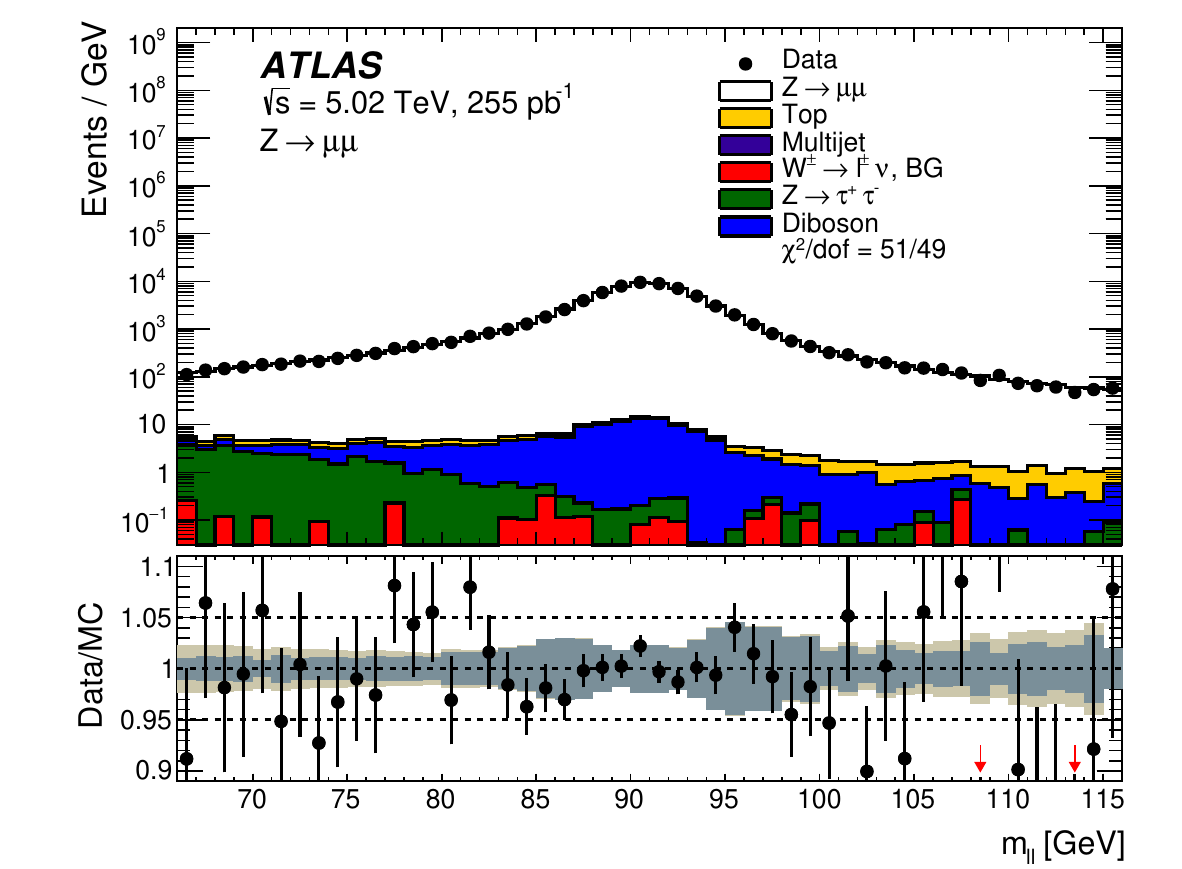}}

\subfloat[]{\includegraphics[width=.47\textwidth]{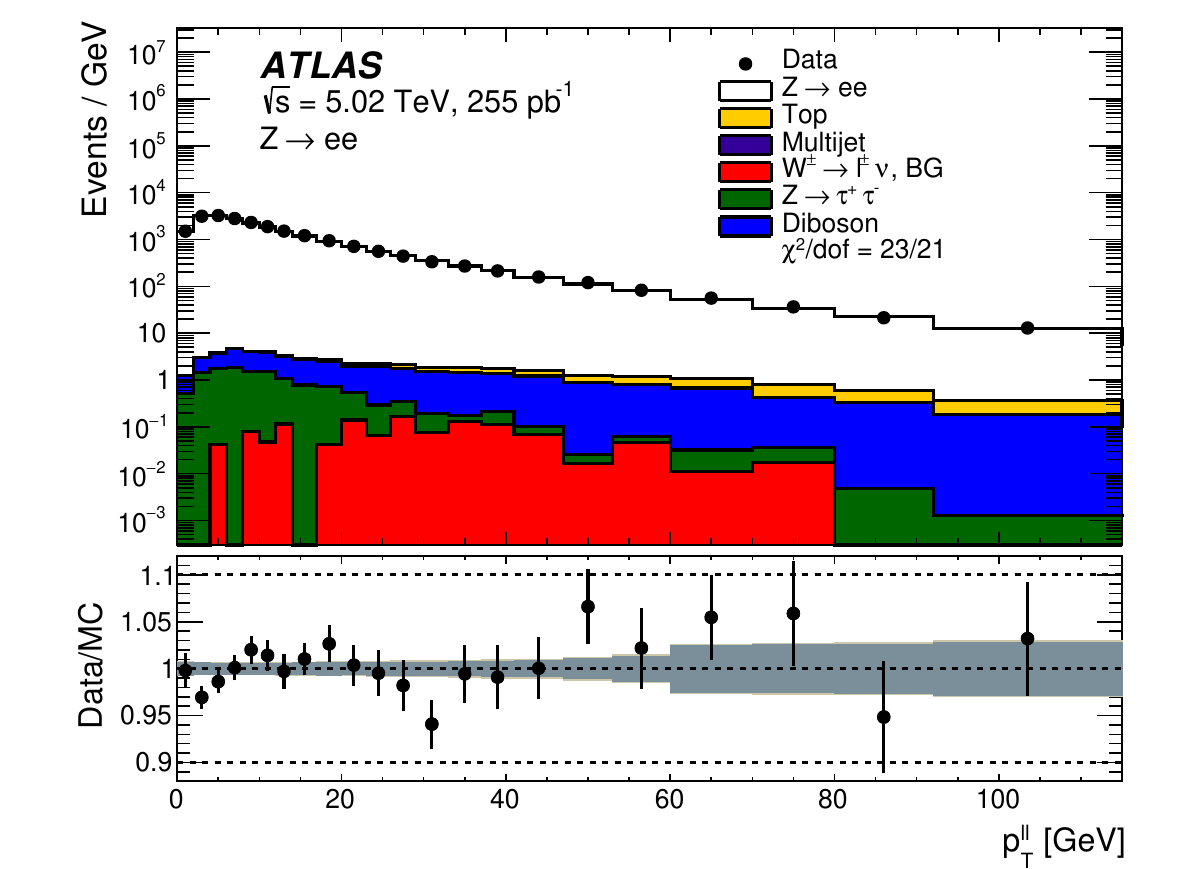}}
\subfloat[]{\includegraphics[width=.47\textwidth]{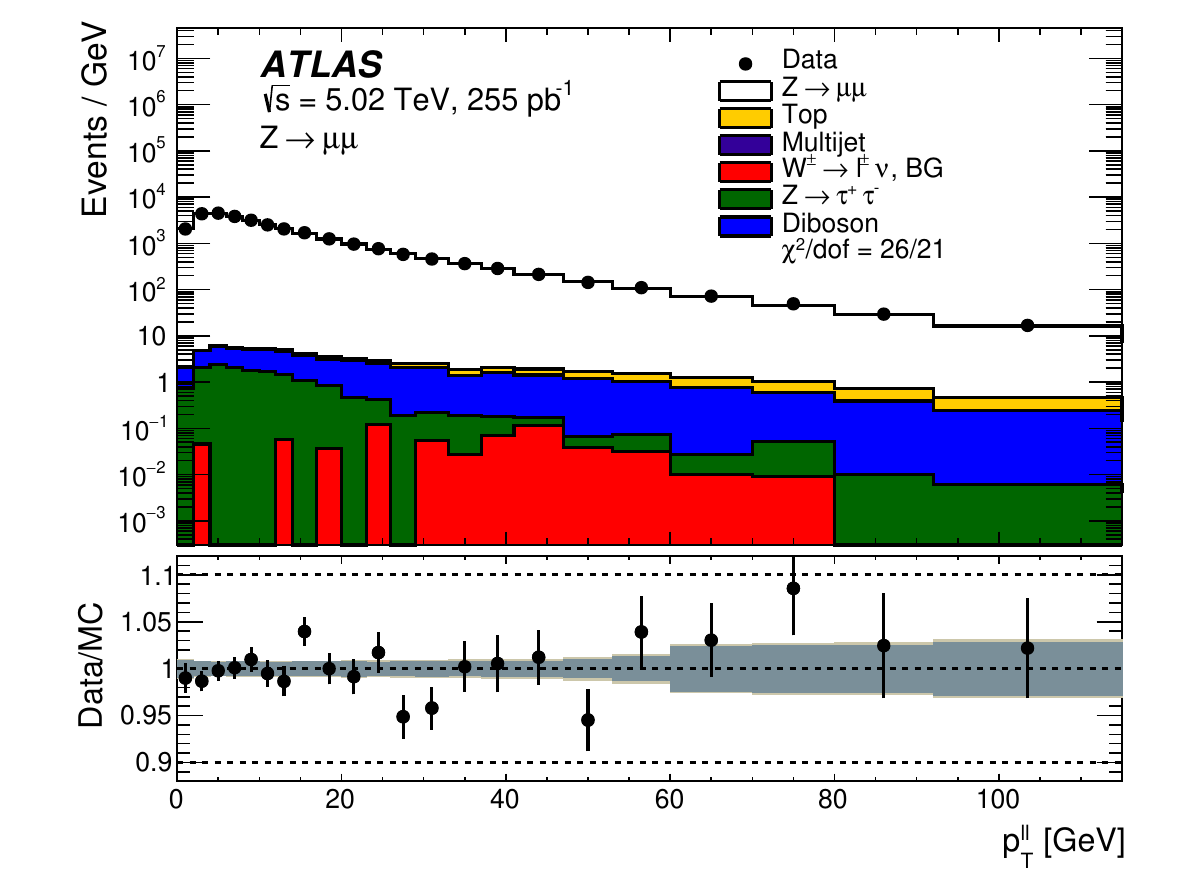}}

\subfloat[]{\includegraphics[width=.47\textwidth]{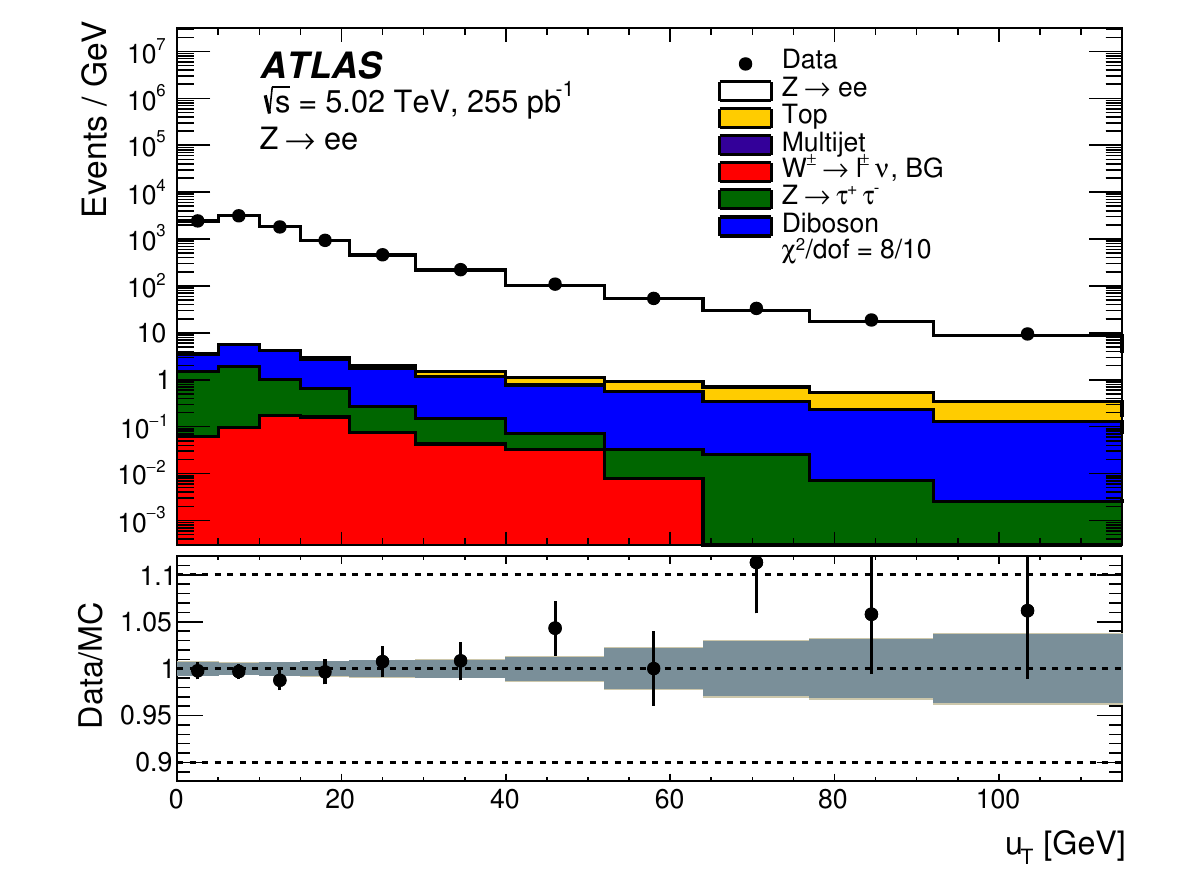}}
\subfloat[]{\includegraphics[width=.47\textwidth]{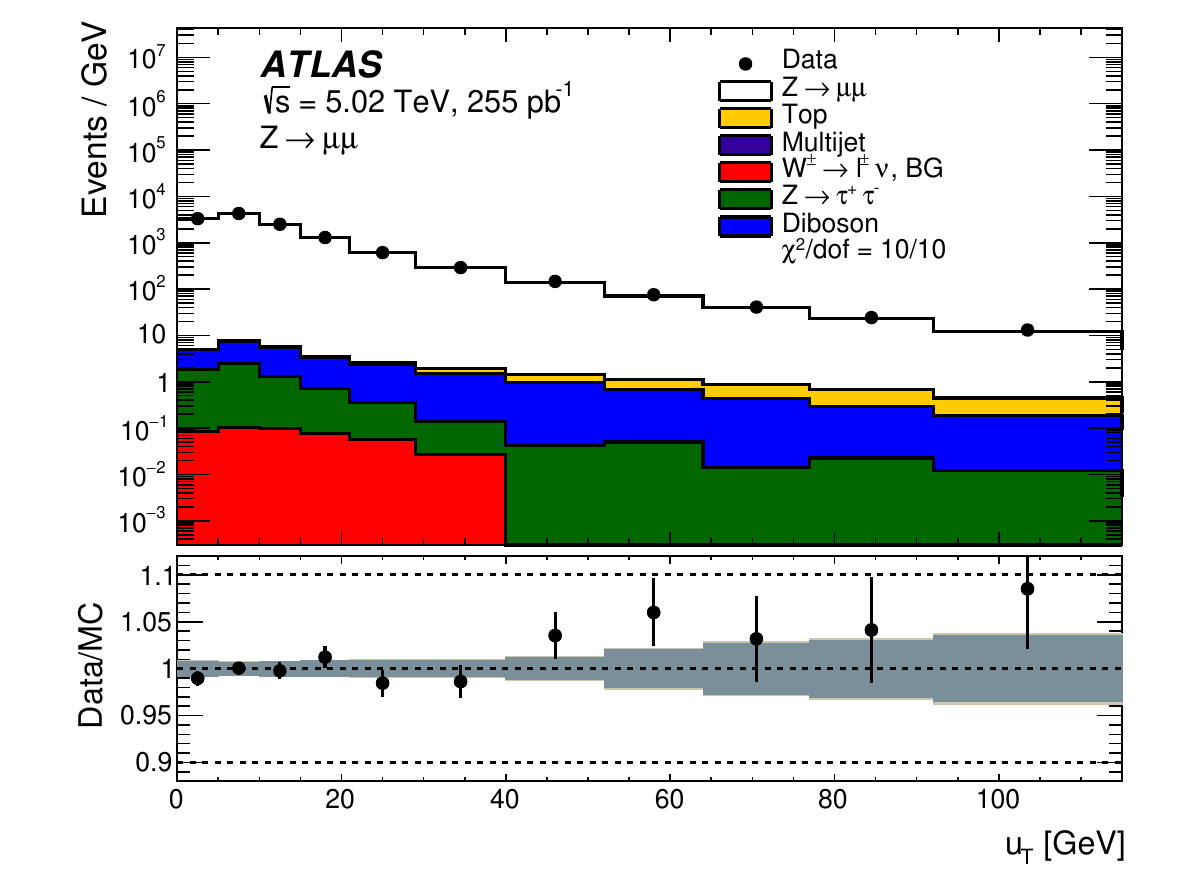}}

\caption{Distributions of dilepton mass \mll in the (a) \Zee and (b) \Zmm channels, of dilepton transverse
momentum \ptll in the (c) \Zee and (d) \Zmm channels, and of hadronic recoil \ut in the (e) \Zee and (f) \Zmm channels in the $\sqrt{s} = \fivetev$
analysis. The data (black points) is compared to the sum
of all expected background and signal contributions, shown as a
solid line, normalised to the total data yield. The lower panel shows the
ratio of the data to the full prediction (black points) and the prediction uncertainties around $1$ excluding
(including) the MC simulation statistical component as dark grey (light brown) band.
The prediction uncertainties shown exclude those from the luminosity measurement
and the alternative signal modelling, see text. The
$\chi^2/\mathrm{dof}$ is computed using the full covariance matrix
of all shown statistical and systematic uncertainties.}\label{f:Z5}
\end{figure}

\begin{figure}[h]
\centering
\subfloat[]{\includegraphics[width=.47\textwidth]{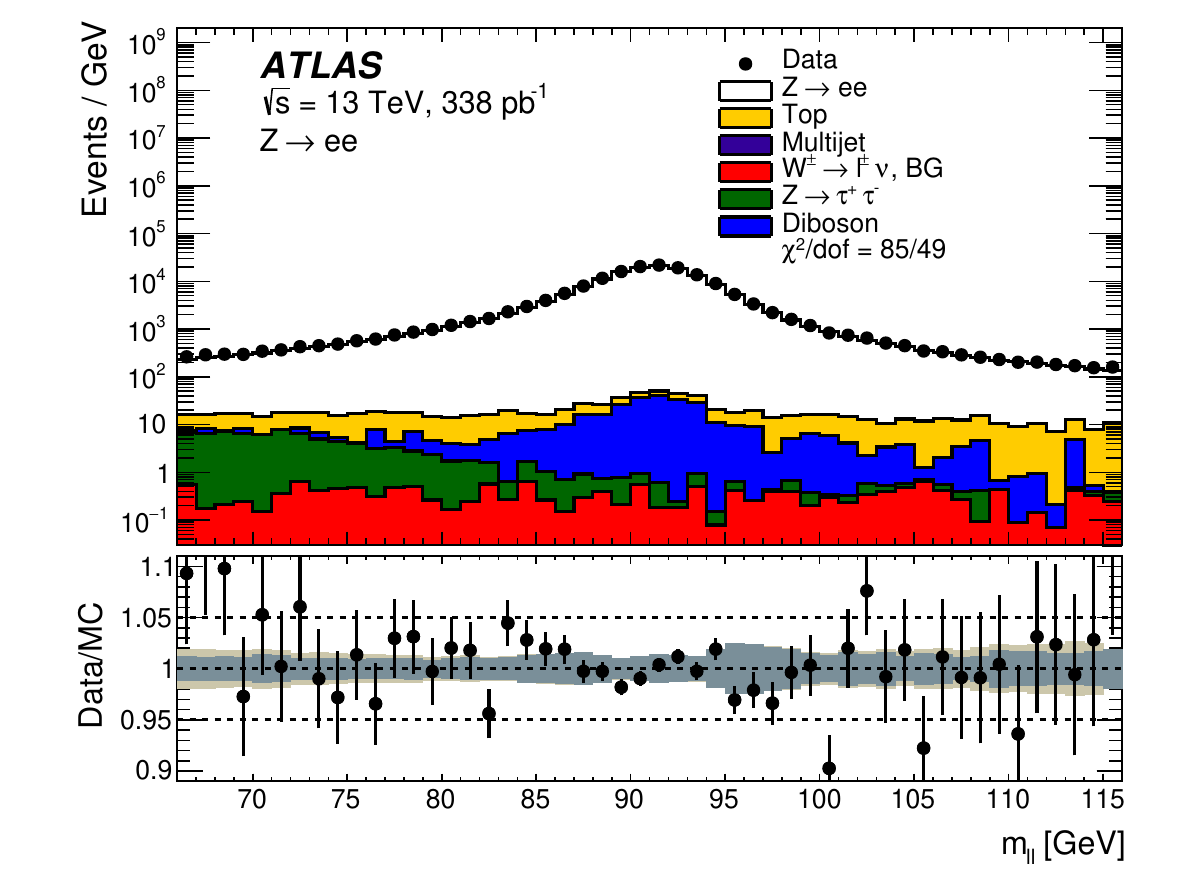}}
\subfloat[]{\includegraphics[width=.47\textwidth]{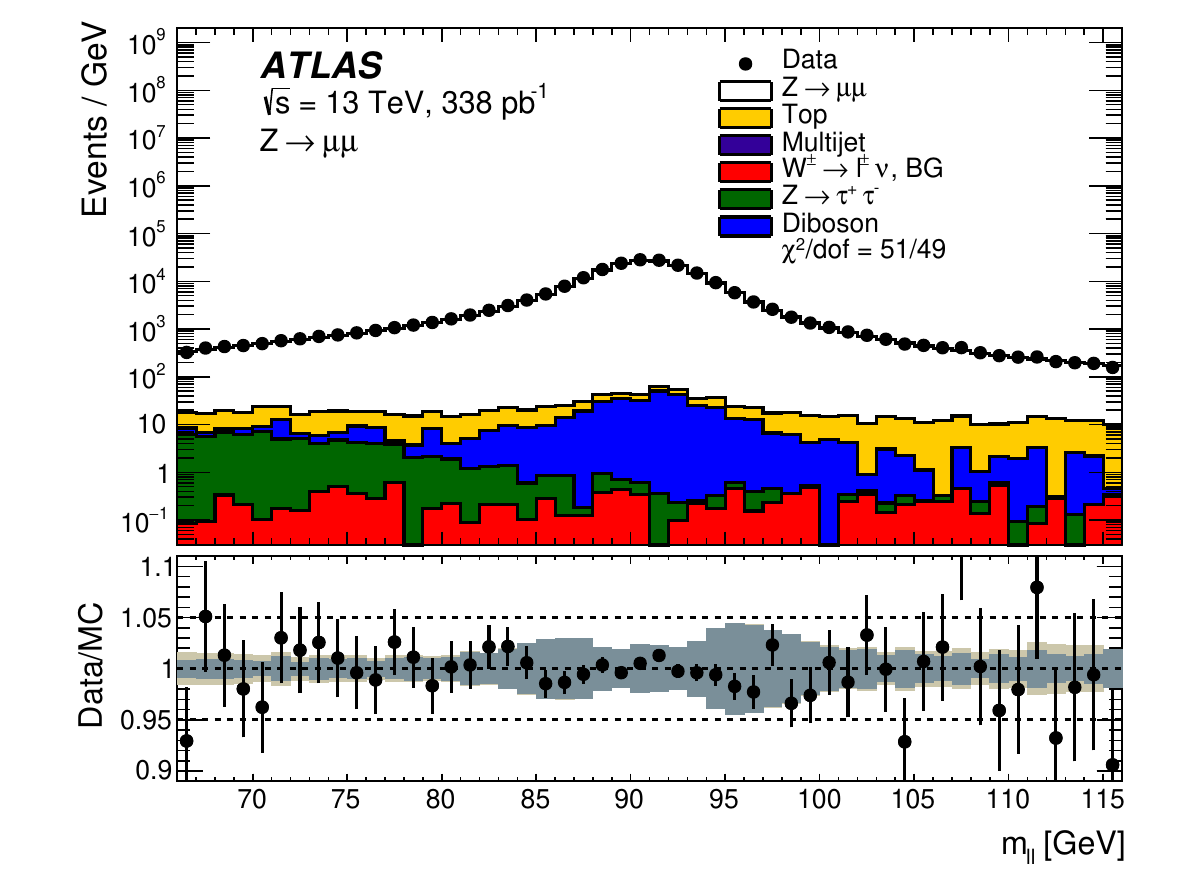}}

\subfloat[]{\includegraphics[width=.47\textwidth]{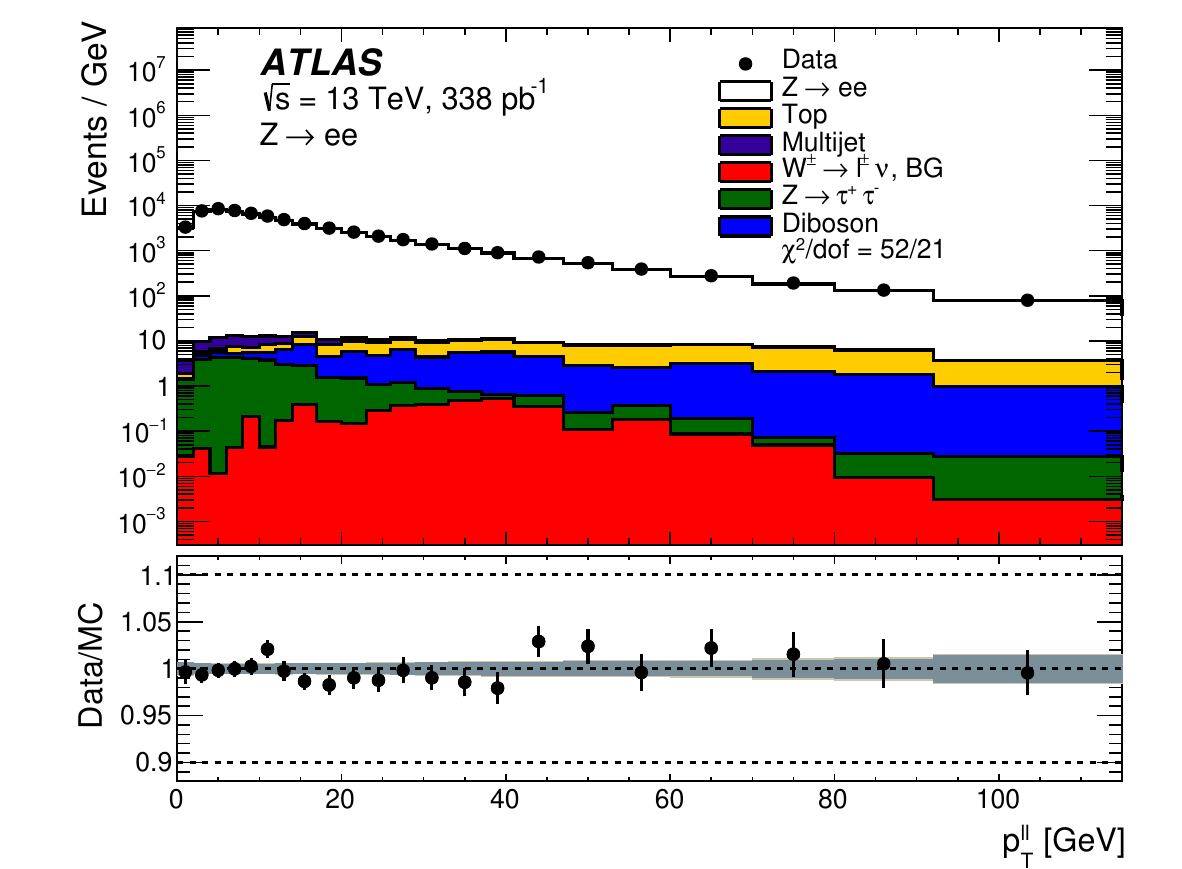}}
\subfloat[]{\includegraphics[width=.47\textwidth]{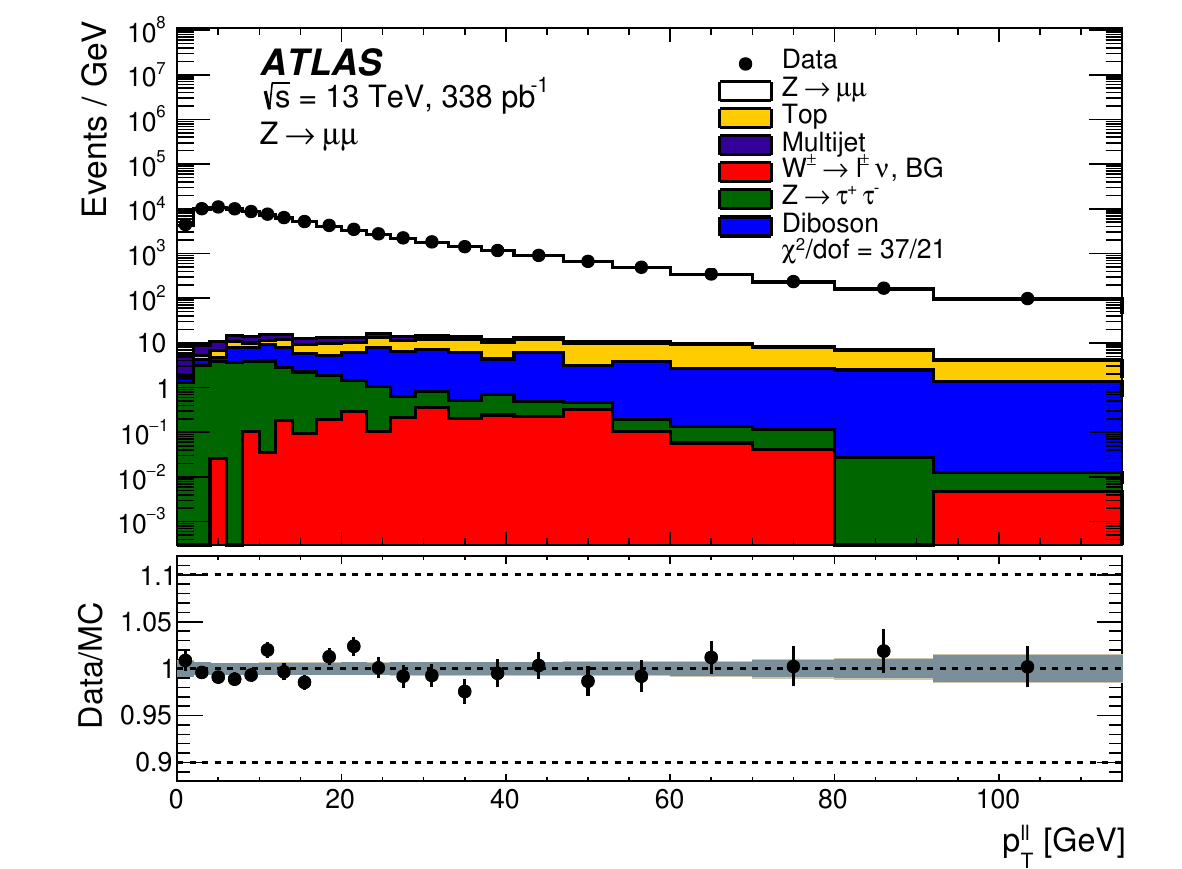}}

\subfloat[]{\includegraphics[width=.47\textwidth]{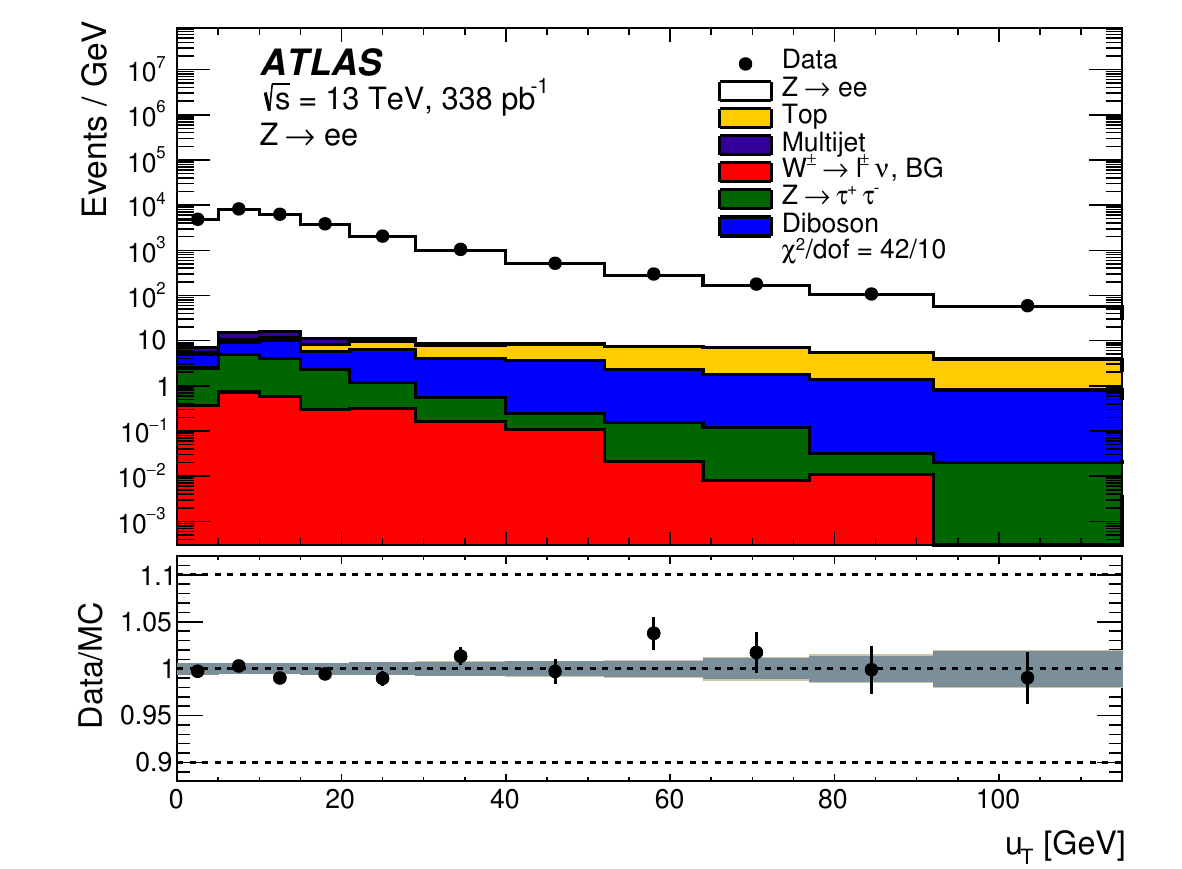}}
\subfloat[]{\includegraphics[width=.47\textwidth]{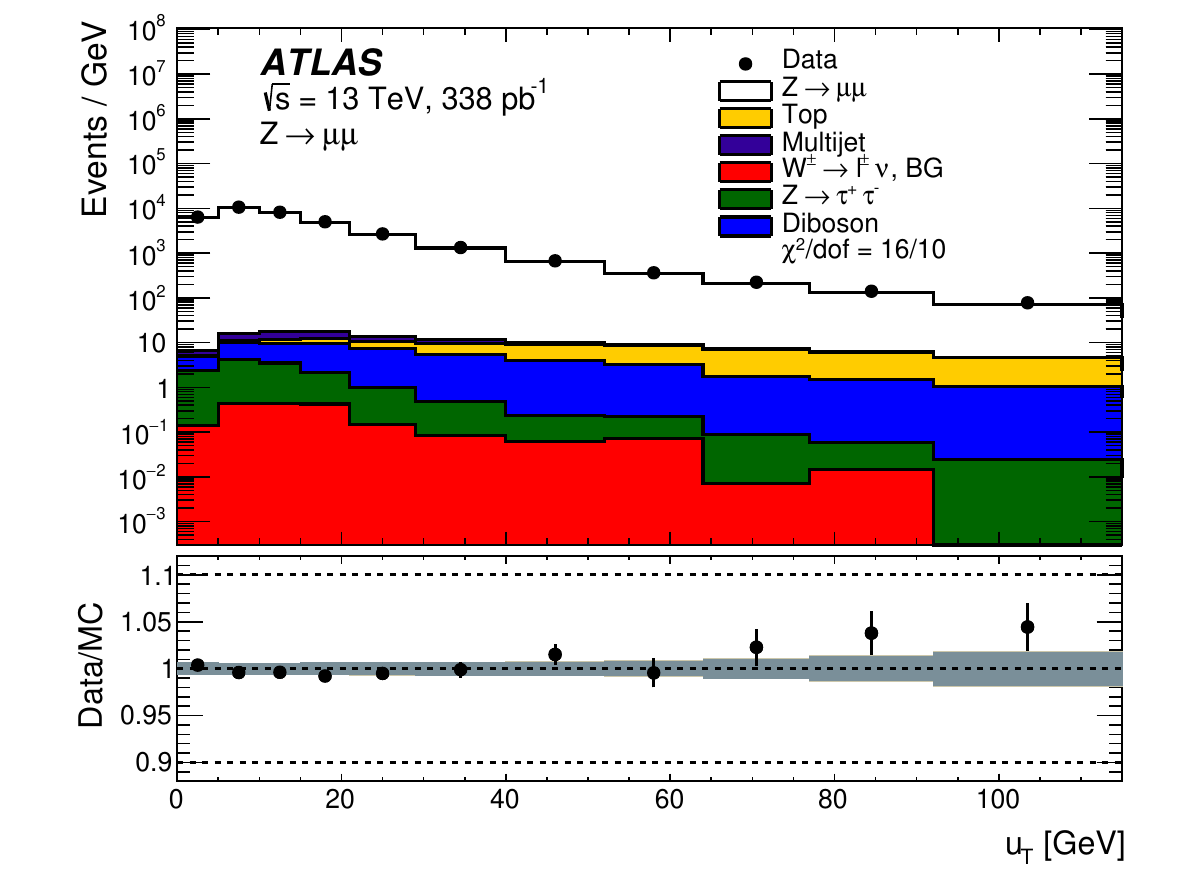}}
\caption{Distributions of dilepton mass \mll in the (a) \Zee and (b) \Zmm channels, of dilepton transverse
momentum \ptll in the (c) \Zee and (d) \Zmm channels, and of hadronic recoil \ut in the (e) \Zee and (f) \Zmm channels
in the $\sqrt{s} = 13\,\TeV$
analysis. The data (black points) is compared to the sum
of all expected background and signal contributions, shown as a
solid line, normalised to the total data yield. The lower panel shows the
ratio of the data to the full prediction (black points) and the prediction uncertainties around $1$ excluding
(including) the MC simulation statistical component as dark grey (light brown) band.
The prediction uncertainties shown exclude those from the luminosity measurement
and the alternative signal modelling, see text. The
$\chi^2/\mathrm{dof}$ is computed using the full covariance matrix
of all shown statistical and systematic uncertainties.}\label{f:Z13}
\end{figure}

Detector-level distributions for the \fivetev dataset and for the $13\,\TeV$
dataset are provided  in
Figures~\ref{fig:met5_controlplot}--\ref{fig:ut5_controlplot}
and Figures~\ref{fig:met13_controlplot}--\ref{fig:ut13_controlplot}
for the \Wboson-boson analyses and in
Figure~\ref{f:Z5} and Figure~\ref{f:Z13}
for the \Zboson-boson analyses. To better
visualise the agreement in the shapes, the MC predictions are
normalised to the integral of the data distribution. The underlying
\pttruthv distributions have been reweighted with a smooth function as
described in detail in Section~\ref{sec:unfold} to match the data as closely
as possible in the \ptv range of $0$~--~$100\,\GeV$.

In the ratio panels, the grey band is the total systematic
uncertainty, while for the brown band the typically small statistical uncertainty in the MC simulation
is added in quadrature. The systematic uncertainty shown
is the experimental uncertainty excluding luminosity, since the MC simulation is
normalised to the data integral, and excluding the uncertainty derived from the alternative
signal simulation using \SHERPA. The $\chi^2/\mathrm{dof}$ shown
includes all shown uncertainties and is calculated with the corresponding
covariance matrix taking into account bin-to-bin correlations of the
systematic uncertainties.

Overall, good agreement is observed between the data and the MC simulation
estimates in all the kinematic distributions.  Several channels at $13\,~\TeV$ show a
modest mismodelling of the high \pt tails
by a few percent, such as the lepton \pt distributions for the \Wboson
boson (Figure~\ref{fig:ptl13_controlplot}). These are due to imperfections in the
\pttruthv\ reweighting. However, the spectra are sufficiently well
modelled to perform the \ptv measurements and corresponding
uncertainties are assessed in the unfolding.

\clearpage

\clearpage


%
\section{Cross-section measurement}
\label{sec:measurement}

The $W^-$, $W^+$ and $Z$-boson transverse momentum distributions are measured separately for the electron and muon decay channels.
Results are reported in a fiducial phase space close to the experimental acceptance, and defined by particle-level kinematics as follows:
\begin{itemize}
\item  $W \rightarrow \ell\nu$ :
$\pT^{\ell}>25~\GeV$, $|\eta^{\ell}|<2.5$, $\pT^{\nu}>25~\GeV$, and $\mt > 50~\GeV$;
\item  $Z \rightarrow \ell\ell$ :
$\pT^{\ell}>25~\GeV$, $|\eta^{\ell}|<2.5$, and $66~\GeV < m_{\ell\ell} <116 ~\GeV$.
\end{itemize}
The lepton kinematics used in the definition of the cross sections corresponds to the Born level for QED final-state
radiation effects. The primary signal model, \Powheg+\PYTHIA{8}, uses \Photos\ to simulate QED FSR. This setup has been
benchmarked extensively in the past, at the sub-permille level on the distributions~\cite{STDM-2012-20}. No explicit
uncertainties are included, although the comparison to \SHERPA\ implicitly uses another QED FSR model. The \Zboson-boson measurements include the small virtual photon contributions, i.e. the full $Z/\gamma^* \to \ell\ell$ process is measured.
The  measured differential cross sections are also used to derive integrated cross sections and ratios.

The measurement bins are chosen by taking into consideration competing demands on the analysis, namely sensitivity to the underlying physics, the statistical precision in each bin, and detector resolution
effects. The latter are particularly important for the hadronic recoil observable at low boson transverse momentum and require a minimum bin size of $7\,\GeV$ at low \ptv to ensure that the unfolding systematic uncertainties are around $1\%$ or lower.
The measurement is performed in \pttruthv bins with bin edges as follows:
\begin{itemize}
\item \ptw at \fivetev : [0, 7, 14, 21, 28, 35, 42, 49, 56, 63, 77, 92, 115, 145, 175, 220]~\GeV;
\item \ptw at $13\,\TeV$ : [0, 7, 14, 21, 28, 35, 42, 49, 56, 63, 77, 92, 115, 145, 175, 220, 310, 600]~\GeV;
\item \ptz at \fivetev : [0, 2, 4, 6, 8, 10, 12, 14, 17, 20, 23, 26, 29, 33, 37, 41, 47, 53, 60, 70, 80, 92, 115, 145, 175, 220]~\GeV;
\item \ptz at $13\,\TeV$ : [0, 2, 4, 6, 8, 10, 12, 14, 17, 20, 23, 26, 29, 33, 37, 41, 47, 53, 60, 70, 80, 92, 115, 145, 175, 220, 310, 600]~\GeV.
\end{itemize}

The \ptz distribution is measured separately using the \ut\ and the \ptdilep\ observables. The former is used to validate the measurement procedures in the same bins as the respective \ptw measurements, as discussed in Section~\ref{sec:recoilcheck}, while the latter is used in the main \ptz\ measurements with the finer binning given above. For the differential $W/Z$ ratio measurements, the \Zboson\-boson distribution is also measured using the \ptdilep\ observable in the same bins as the $W$ distributions.

\subsection{Unfolding}
\label{sec:unfold}

The distributions observed after analysis selections and background subtraction  are corrected for detector effects using an iterative Bayesian unfolding method~\cite{DAgostini:1994fjx,DAgostini:2010hil}. First, the data are corrected for events
that pass the detector-level selection but not the particle-level selection. The iterative Bayesian unfolding technique is then used to correct for the finite detector acceptance, resolution and reconstruction efficiency and estimate the true underlying distribution.

\begin{figure}[th]
\centering
\subfloat[]{\includegraphics[width=.42\textwidth]{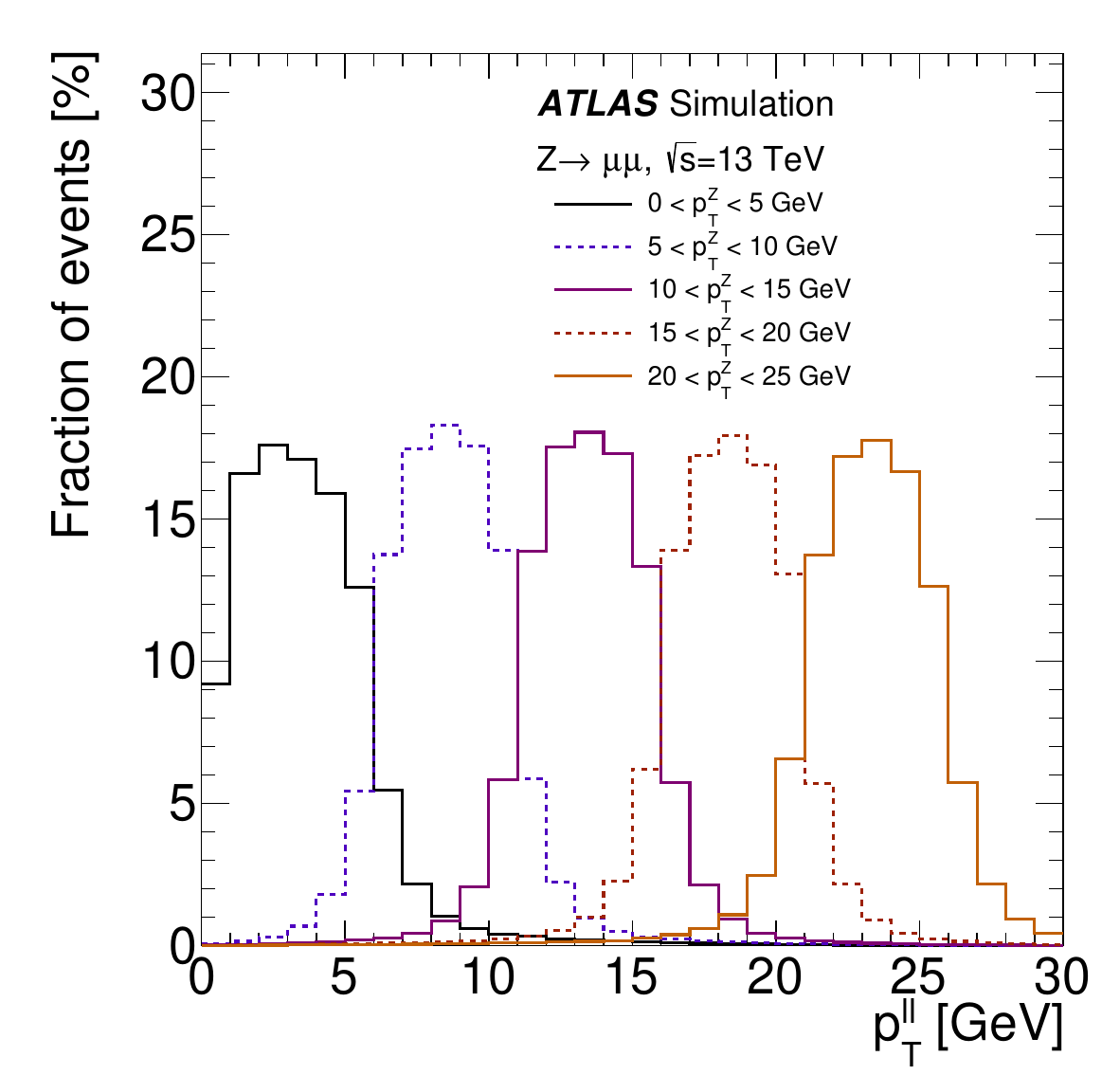}\label{f:purity_pTZmm13}}
\subfloat[]{\includegraphics[width=.42\textwidth]{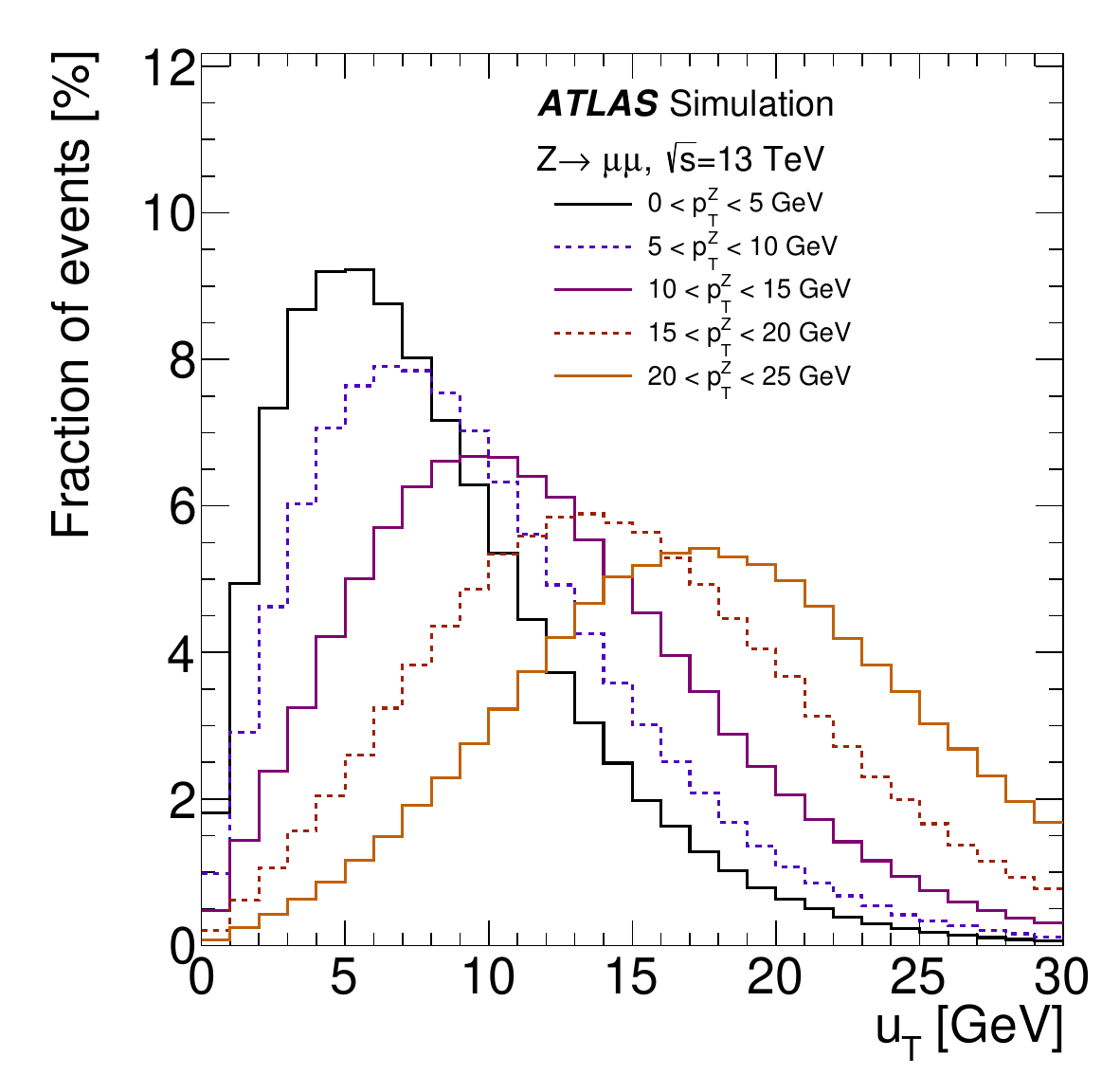}\label{f:purity_pTZmm13UT}}
\caption{Distributions of measured (a) dilepton system \ptdilep and (b) hadronic recoil \ut, shown in five intervals
of the true boson transverse momentum \ptz, for \Zmm events at $\sqrt{s} = 13\,\TeV$.}
\label{fig:slicePurityZ}
\end{figure}

Simulated events are used to determine the response matrices needed to correct for the migration between bins in the detector- and particle-level distributions.
The hadronic recoil resolution leads to significant migrations in \ut. The effect of the detector resolution on the event migrations is illustrated in Figure~\ref{fig:slicePurityZ}, which compares the observed \ptdilep\ and \ut distributions for given intervals in \pttruthv, in simulated \Zmm\ events at 13~\TeV. Significant differences are observed in the corresponding reconstructed distributions.
The particle-level binning is therefore optimised separately for each observable. %
The number of iterations in the unfolding procedure is optimised to minimise the total measurement uncertainty, and
specifically the uncertainty related to possible biases induced by the unfolding prior. Methods to estimate the
unfolding bias and the optimisation of the number of iterations are discussed further in this section.

After calibration corrections, the detector-level transverse momentum
distributions in data and MC are found to disagree, $i.e.$ the difference is found to be significantly larger than the
experimental sensitivity, especially at 13~\TeV. To minimise the related uncertainties, the simulation is reweighted in \pttruthv to optimise the agreement between data
and MC for the reconstructed \ut (or \ptll) distributions for $W$ (and $Z$) analyses.\footnote{To achieve this, the
analysis including some background estimates is iterated. Object calibrations, including the hadronic
recoil are performed such that they are to
first order insensitive to the \ptv modelling in the MC simulation.}
This procedure ensures that the unfolding bias is kept as low as possible.

The agreement between data and simulation is optimised for the \ut distribution by minimising the following $\chi^2$:
\begin{eqnarray}
\chi^2  &=&  \sum_{ij} \, \Delta_i \, (C^{-1})_{ij} \, \Delta_j \label{eqX2} \,,\\
\Delta_i  &=& ( D_i \, - B_i \,) - \, \sum_{k} \, T_{ik} \times w_k \, .
\end{eqnarray}
Here $(C^{-1})_{ij}$ denotes the inverted total covariance between detector-level bins $i$ and $j$, including statistical and systematic uncertainties in the observed and simulated \ut distributions;
$\Delta_i$ is the difference between the background-subtracted data distribution in bin $i$, $D_i-B_i$, and the corresponding reconstruction-level simulated distribution. The latter is obtained from the product between a generator-level weighting coefficient $w_k$ and the transfer matrix $T_{ik}$ that is filled with the number of entries in bins $i$, $k$ at detector and particle level, respectively. The weight, $w_k$, is the average in particle-level bin $k$ of a continuous weighting function $w(\pttruthv)$ that modifies the \ptw
distribution in the MC sample.
The parameters of $w(\pttruthv)$ are the degrees of freedom in the minimisation procedure. A satisfactory agreement is obtained using the empirical function
\begin{equation}
w(\pt)  = N \left[\left(1 + a \, \pttruthv + b \, \left({\pttruthv}\right)^2
\right) \times \left(1-c+c\times r_\mathrm{NNPDF/CT10}(\pttruthv)\right)\right],    \label{eq:Wnom_bin}
\end{equation}
where $N$ is an overall normalisation factor and $r_\mathrm{NNPDF/CT10}(\pttruthv)$ represents the ratio of \textsc{DYTurbo} predictions obtained using NNPDF3.0 and CT10NLO in the full phase space.
The additional parameters ($a,b,c$) provide more flexibility
to minimise the $\chi^2$ of Eq.~\eqref{eqX2}.

\begin{figure}[tbp]
\centering
\subfloat[]{\includegraphics[width=.47\textwidth]{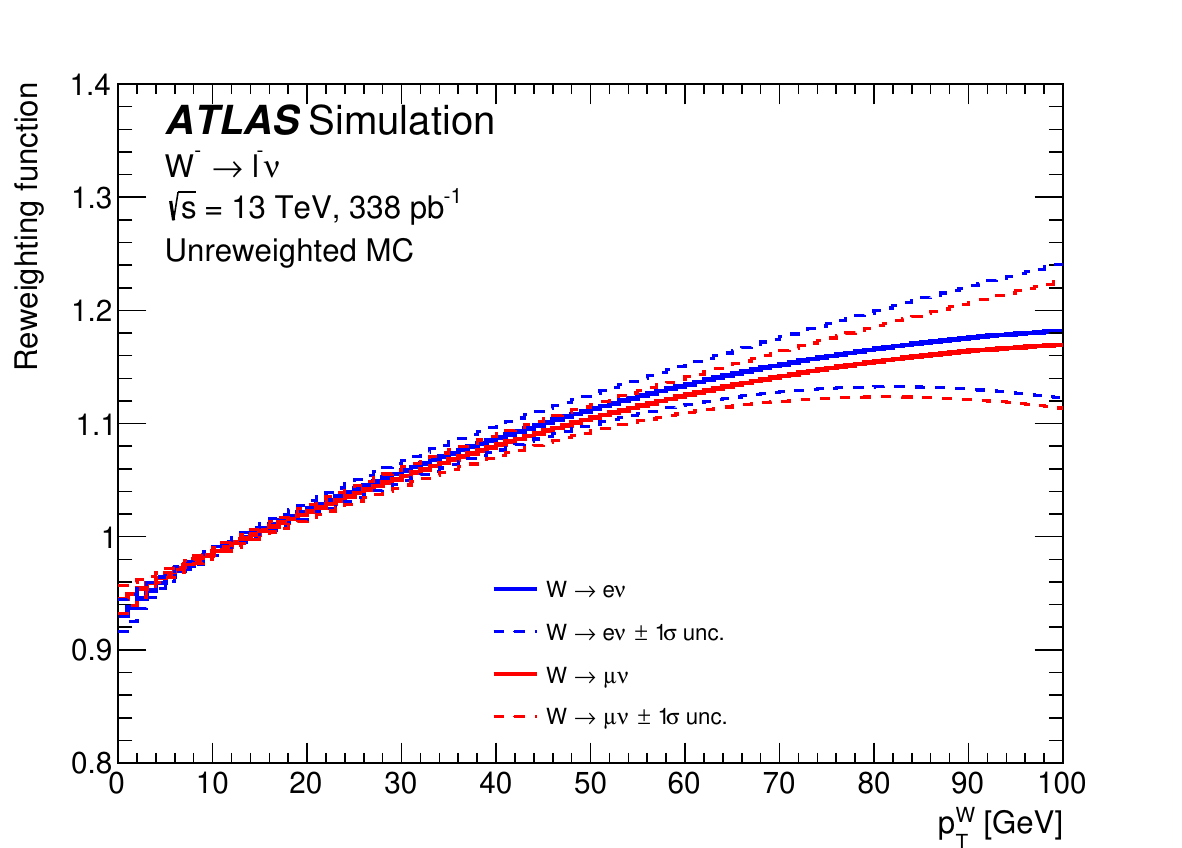}\label{fig:rewFuncBiasCorr_a}}%
~~~~%
\subfloat[]{\includegraphics[width=.47\textwidth]{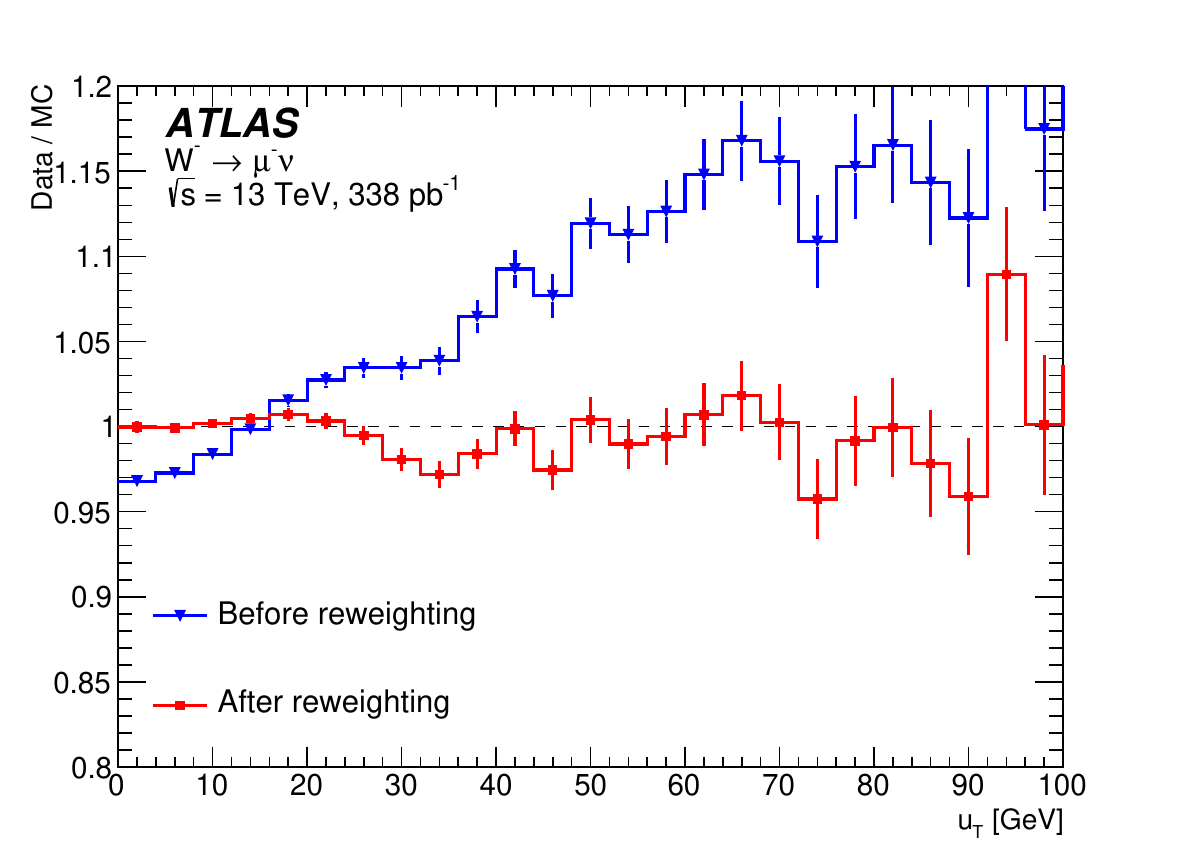}\label{fig:rewFuncBiasCorr_b}}
\caption{(a) Fitted truth reweighting functions $w_{T}$ in the
electron and muon channels (full lines) for \Wmlnu at
$\sqrt{s} = 13\,\TeV$. The dashed lines represent the uncertainty
on the reweighting functions from the fit. (b) The data-to-simulation ratios
of the reconstructed \ut\ distributions before (blue triangles) and
after (red squares) the reweighting. Error bars are statistical only.}
\label{fig:rewFuncBiasCorr}
\end{figure}

The reweighting functions are optimised separately for each process and each centre-of-mass energy. They are derived
over the  $\pttruthv $ range of  $0 - 100\, \gev$ where a good modelling of the spectrum is critical. The value of the correction at $\pttruthv=100\,\GeV$ is used for $\pttruthv > 100\,\gev$, as the impact on the final results
in this region is found to be negligible.
Using the \Wmlnu\ at $13\,\TeV$ as example, the two reweighting functions for the electron and muon channels are shown in Figure~\ref{fig:rewFuncBiasCorr_a}. Since the $e$ and $\mu$ channel are compatible for all final states, the averages of the respective two reweighting functions are later applied to both channels.
The corresponding effect of this \ptv\ reweighting on the reconstructed \ut distributions is shown in Figure~\ref{fig:rewFuncBiasCorr_b}.

The reweighed simulation is used to determine the response matrix used in the unfolding. Both the bin size at low \ptw and
the number of unfolding iterations are optimised given the
considerations discussed above. More unfolding iterations increase the statistical uncertainty while reducing the
unfolding bias. Since the statistical uncertainty is relatively small compared with the unfolding bias, a larger number
of iterations is favoured and allows to obtain a small unfolding bias at the price of a little increase in the
statistical uncertainties. The choice of number of iterations is made by achieving the lowest quadratic sum of
statistical uncertainty, systematic uncertainty and unfolding bias.
A number of 25 iterations is found to be optimal for the $13\,\TeV$
analysis to minimise the total uncertainties. In the smaller \fivetev
samples the optimal number of iterations is found to be nine.

For the \Zboson\ process the same approach is used, with the addition
that separate reweightings are derived using the \ptdilep\ and \ut\
distributions for the respective measurements.\footnote{The
measurement of \ptz using the \ut distribution is only used in the
validation of the measurement procedure presented in
Section~\ref{sec:recoilcheck}.}
Because of the far better resolution in \ptll, only two unfolding iterations are used.
Uncertainties related to the assumed prior for the \ptv\ spectrum in the
unfolding are discussed in the next section.

\subsection{Systematic uncertainties}
\label{sec:sysunc}

The present measurement is affected by statistical uncertainties as well as systematic uncertainties related to detector calibration, the background estimate, the definition of the response matrix and to the unfolding procedure. For each uncertainty, the corresponding source of uncertainty is varied to estimate the effect on the final result. A graphical representation of the uncertainties is given in Appendix~\ref{sec:uncbreakdown}.

The effect on the measurement from the finite size of the data and MC simulated samples is estimated by generating pseudo-experiment variations of the respective samples~\cite{ATL-PHYS-PUB-2021-011}. The resulting statistical uncertainties are uncorrelated between channels, but partially correlated between bins due to the unfolding. In the \ptw{} measurement at \fivetev, where the event samples are
smaller than at $13\,\TeV$, the statistical uncertainty of the data is the co-dominant uncertainty beyond $\ptw=20~\GeV$, going from about $0.5\%$ below this region to
$\approx 2.5\%$ at $\ptw=60~\GeV$. At $13\,\TeV$, it becomes dominant around $\ptw=60~\GeV$. For the \ptz{} measurements, the statistical uncertainty of the data is the dominant source of uncertainty.

Uncertainties in the scale and resolution of the hadronic recoil are among the dominant uncertainties in the \ptw{} measurement. They are estimated by propagating the statistical uncertainties in the calibration procedure and systematic effects as discussed in Section~\ref{sec:recoil}.
In the first \ptw bin these are usually the largest or second-largest systematic uncertainties reaching typically $0.6\%$. They grow
moderately with \ptw depending on the boson charge and on the centre
of mass energy. In the  \ptw  range of $30$--$50$~\GeV, the uncertainty in the measurement for
\Wminus\ events is slightly higher than for \Wplus events, due to observed
differences between the extrapolation uncertainties from the \Zboson-boson. In
the \fivetev data, it reaches $1.5\%$ at $50$~\GeV, and goes up to
$\approx 3\%$ at 100~\GeV. At $13\,\TeV$, since more \Zboson\ events are
available for the calibration, the uncertainty is smaller, and
increases slowly to reach $1.5\%$ at 100~\GeV.

Lepton-related systematic uncertainties due to energy scale and resolution as well as reconstruction and selection efficiency corrections are propagated considering the statistical and systematic uncertainties in the procedures described in Sections~\ref{sec:electrons} and~\ref{sec:muons}. They have only a small effect on both the \ptw{} and the \ptz{} measurements.

Uncertainties in the background estimation are only significant in the \Wboson analysis. Uncertainties induced by the simulated backgrounds are obtained by independently varying the corresponding cross sections. The uncertainty in the multijet background is dominant in the electron channel. The estimated multijet background yields are treated as correlated between the channels since the same techniques are used.

The uncertainty assigned to the unfolding procedure itself is estimated from the dependence of the results on the assumed priors for the \ptv distributions. The nominal particle-level spectrum, obtained from the reweighting procedure described in Section~\ref{sec:unfold}, is varied within the fit uncertainties, taking correlations between the fit parameters into account. Alternative functional forms are used to test the influence of
the \ptv distribution at low values.
The baseline particle-level distribution at a given rapidity
is modified using alternative predictions. A binning in rapidity prevents any modeling bias, since the \ptv\ spectra depend on rapidity. Four  alternative predictions are
chosen: \textsc{DYTurbo} with the NNPDF3.1 and CT10 PDF sets; \Powheg+\HERWIGV{7}, and \PYTHIAV{8}, described in more detail in Section~\ref{sec:predictions}. The \SHERPA{} samples introduced in Section~\ref{sec:mc} are considered but discarded due to the poor description of the data below $20 \,\GeV$.

The closure of the unfolding procedure is further tested using the
alternative \SHERPA{} model for the signal MC simulation, as discussed at the end of
Section~\ref{sec:recoil}. The \SHERPA samples represent a
completely independent implementation of the physics of vector boson
production and decay as compared to \Powheg+\PYTHIAV{8}. In many
respects, such as the vertex efficiency (Section~\ref{sec:pv}) and the
modelling of quantities relevant for the hadronic recoil
(Section~\ref{sec:recoilcalib}), the predictions of the two generators bracket the data.
At \fivetev, the resulting uncertainty
stays subdominant in the region $\ptw<30\,\GeV$, whereas at
$13\,\TeV$, it is the largest or second-to-largest systematic
uncertainty, reaching more than $1\%$ in the second bin.

The uncertainties in the integrated luminosities are $1.0\%$ and
$0.92\%$ for the \fivetev and $13\,\tev$ datasets, following the methodology
discussed in Ref.~\cite{DAPR-2021-01}, using the LUCID-2
detector~\cite{LUCID2} for the primary luminosity measurements,
complemented by measurements using the inner detector and
calorimeters.

\subsection{Validation of recoil-based measurement with $Z$ dilepton}
\label{sec:recoilcheck}

In the \Zboson samples, the transverse momentum spectra can be inferred either from the \ptdilep\ or from the \ut distributions. As shown in Figure~\ref{fig:slicePurityZ}, a sizeable difference is observed between the resolutions of the two observables. Therefore a comparison of the corresponding unfolded distributions is a powerful tool to validate the recoil calibration and the measurement of \ptw.

\begin{figure}[htbp]
\centering
\subfloat[]{\includegraphics[width=.50\textwidth]{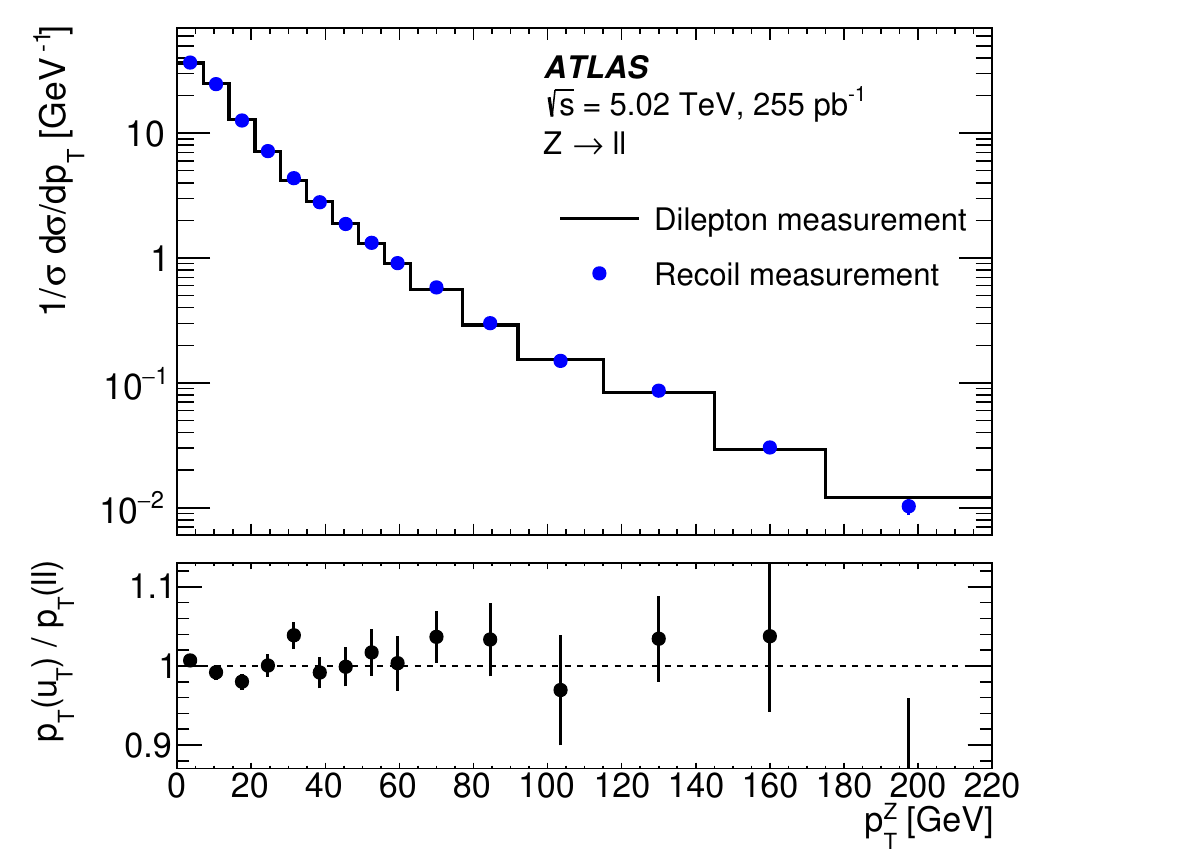}}
\subfloat[]{\includegraphics[width=.50\textwidth]{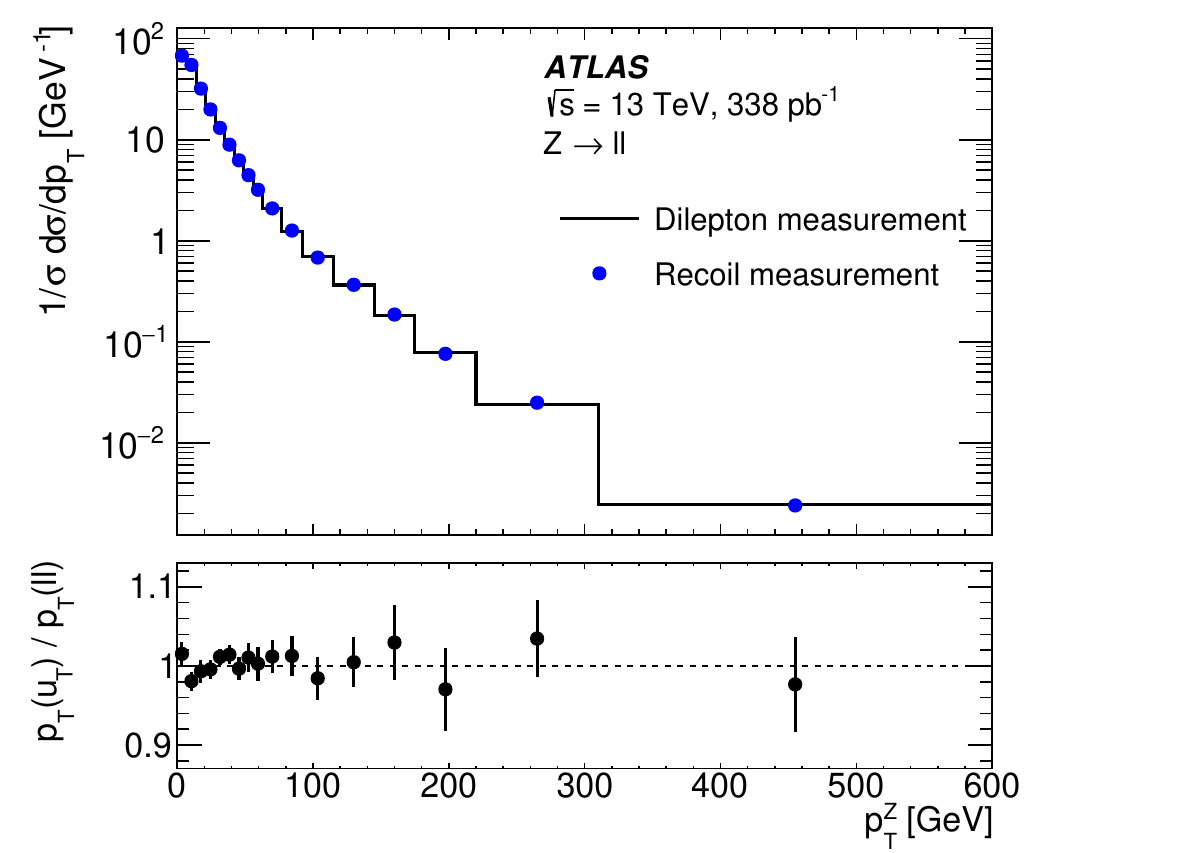}}
\caption{Spectra of \ptz\ measured using either \ptll\ (black line) or \ut\ (blue points),
at (a) $\sqrt{s} = \fivetev$ and (b) $\sqrt{s} = 13\,\TeV$. The
respective measurements, shown in the top panel, largely overlap. The lower panel shows the
the ratio of the  \ut\ over the \ptll\ measurements with the uncertainties
estimated from the square root of the diagonal
elements of the total covariance matrix used to compute the
compatibility between them.
}
\label{fig:ptllutcompatibility}
\end{figure}

For the validation, the $\ptz$ spectra are unfolded from the \ptll and \ut distributions using
the same binning as used for the \ptw measurement. The uncertainty in the difference can be represented in
covariance matrix form, accounting for correlations where
relevant. Since the two measurements are extracted from the same
events, the statistical correlation is estimated by using the bootstrap
method.  The compatibility is estimated
evaluating the $\chi^2$ computed for the differences of the unfolded
distributions. The electron and muon-channel measurements in a given bin are added.
The leading uncertainties arise from data statistical uncertainties, hadronic
recoil calibration and components of the unfolding uncertainty. Lepton calibrations are only significant at $13\,\TeV$.
The number of degrees of freedom is one less than the number of
bins due to the additional constraint of the same total cross section
extracted from the two observables.
Figure~\ref{fig:ptllutcompatibility} compares both measured spectra at each centre-of-mass energy.
The $\chi^2/ \mathrm{dof}$ values computed with the \fivetev and $13\,\TeV$ datasets are
$14.9 / 14$ and $8.7 / 16$, respectively,
which represents an excellent compatibility.

%
\section{Results}
\label{sec:result}

\subsection{Combination of electron and muon channels}
\label{sec:combi}

The electron and muon channel cross sections are combined assuming Gaussian uncertainties, following the best linear unbiased estimator
prescription (BLUE)~\cite{Lyons:1988rp, Valassi:2003mu}.  The $\chi^2$ to be minimised is:
\begin{equation}
\chi^2 = \sum_{i,j=1}^{2N} \left( m_i - \bar{m}_i \right) (C^{-1})_{ij} \left(m_j - \bar{m}_j\right),
\end{equation}
where $m_i$ are the measurements in the $i=1\ldots N$ bins of the distributions in the electron and muon channels, i.e
the $2N$-sized vector $\bm{m} = \{m_1^e, ..., m_N^e ; m_1^\mu , ..., m_N^\mu\}$, $\bm{\bar{m}} = \{\bar{m}_1 , ..., \bar{m}_N ; \bar{m}_1 , ..., \bar{m}_N\}$ is the vector of averages to be
determined, and $C^{-1}$ is the inverse of the $2N\times 2N$ measurement covariance matrix.
The covariance matrix $C$  is constructed as
\begin{equation}
C_{ij} = C^\mathrm{stat}_{ij} + \sum_k C^{\mathrm{syst},k}_{ij},
\end{equation}
where $C^\mathrm{stat}$ is the statistical covariance and can be written as
\begin{equation}
C^\mathrm{stat} =
\begin{pmatrix}
C^\mathrm{stat}_e &  0  \\
0         & C^\mathrm{stat}_\mu
\end{pmatrix}\,,
\end{equation}
and $C^\mathrm{stat}_{e(\mu)}$ are non-diagonal, $N\times N$ covariance matrices resulting from the unfolding of the electron- and muon-channel distributions.
For each systematic source of uncertainty $k$,
\begin{equation}
C^{\mathrm{syst},k}_{ij} = \Gamma_i^k \Gamma_j^k
\end{equation}
is the product of the effects of varying source $k$ in bins $i$ and $j$.

The vector $\bm{\bar{m}}$ that minimises the $\chi^2$ is:
\begin{equation}
\bm{\bar{m}} = (H^T C^{-1} H)^{-1} H^T C^{-1} \bm{m},
\end{equation}
where $H$ is a $2N \times N$ matrix such that $H_{ij}=1$ when measurement bin $i = 1\cdots 2N$ contributes to the combined value $j=1\cdots N$, and $H_{ij}= 0$ otherwise. The covariance matrix of the combined measurement is given by:
\begin{equation}
\bar{C} = (H^T C^{-1} H)^{-1}.
\end{equation}

This procedure is correct in the Gaussian limit and when the uncertainties do not depend on the value of the measured quantity. Many sources of uncertainty are, however, defined relative to the measurement; for example, luminosity or efficiency uncertainties are fixed fractions of the measured cross section. Such uncertainties are then smaller for lower measured values, biasing the average towards the lowest measured cross section. The combination procedure is therefore iterated and the covariance matrix elements are rescaled according to the combined uncertainties of the previous iteration~\cite{Dembinski:2018ihc}. Stable results are obtained after four iterations.

The uncertainty model is detailed, with about 1800 systematic sources. Lepton-related uncertainties are treated as uncorrelated between measurements at different $\sqrt{s}$,
but fully correlated within one $\sqrt{s}$ sample.
Recoil uncertainties are similarly
correlated between the $W^-$, $W^+$ and $Z$ measurements at a given $\sqrt{s}$, with the
exception of the $\setue$ modelling that is only correlated for a
type of vector boson. Background uncertainties are treated as fully correlated across all channels, except for the
QCD multijet between \Wboson and \Zboson samples. Unfolding uncertainties are treated as correlated across a given spectrum
as well as between the electron and muons channels, but uncorrelated otherwise. The
generator uncertainties are separately correlated between all \Wboson and all \Zboson
analyses across both $\sqrt{s}$ datasets, but uncorrelated between \Wboson and \Zboson.
Finally, the luminosity systematic uncertainty is correlated among all analyses using a given $\sqrt{s}$
data sample, but treated as uncorrelated between the \fivetev and
$13\,\TeV$ datasets.

\begin{figure}[p]
\centering
\subfloat[]{\includegraphics[width=.47\textwidth]{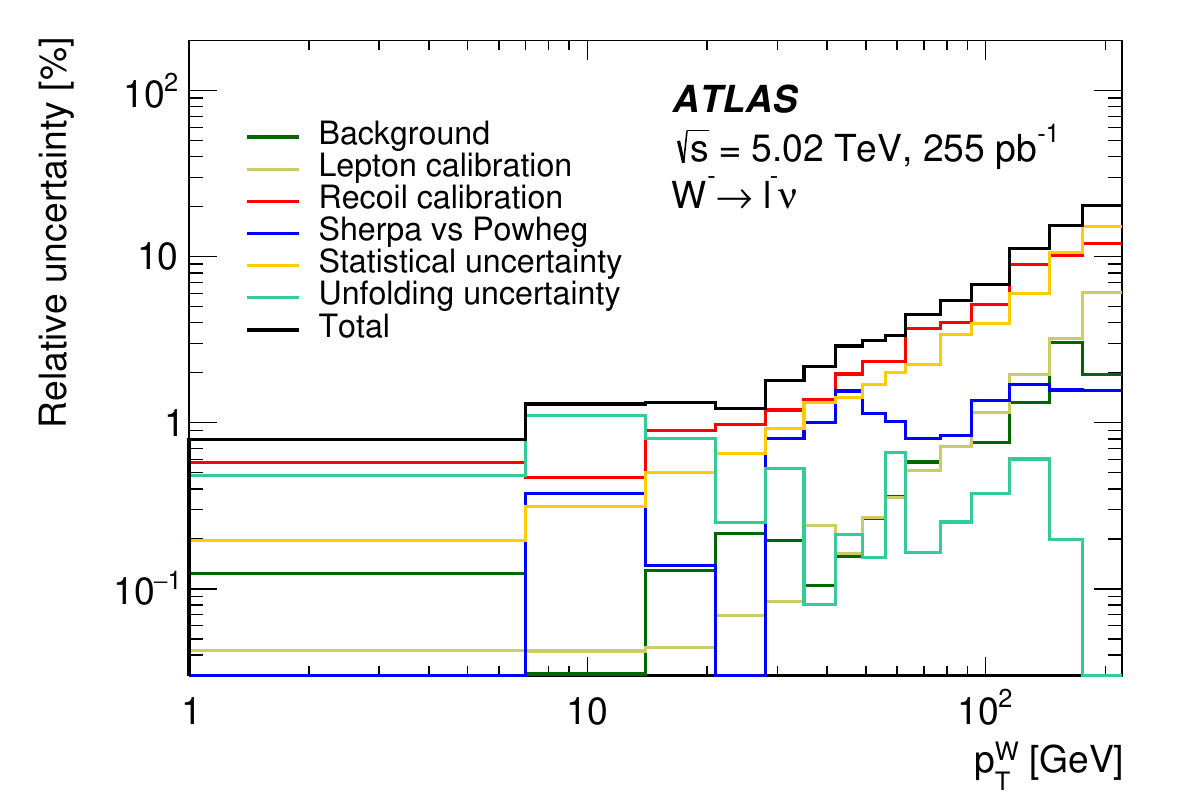}}%
~~~~%
\subfloat[]{\includegraphics[width=.47\textwidth]{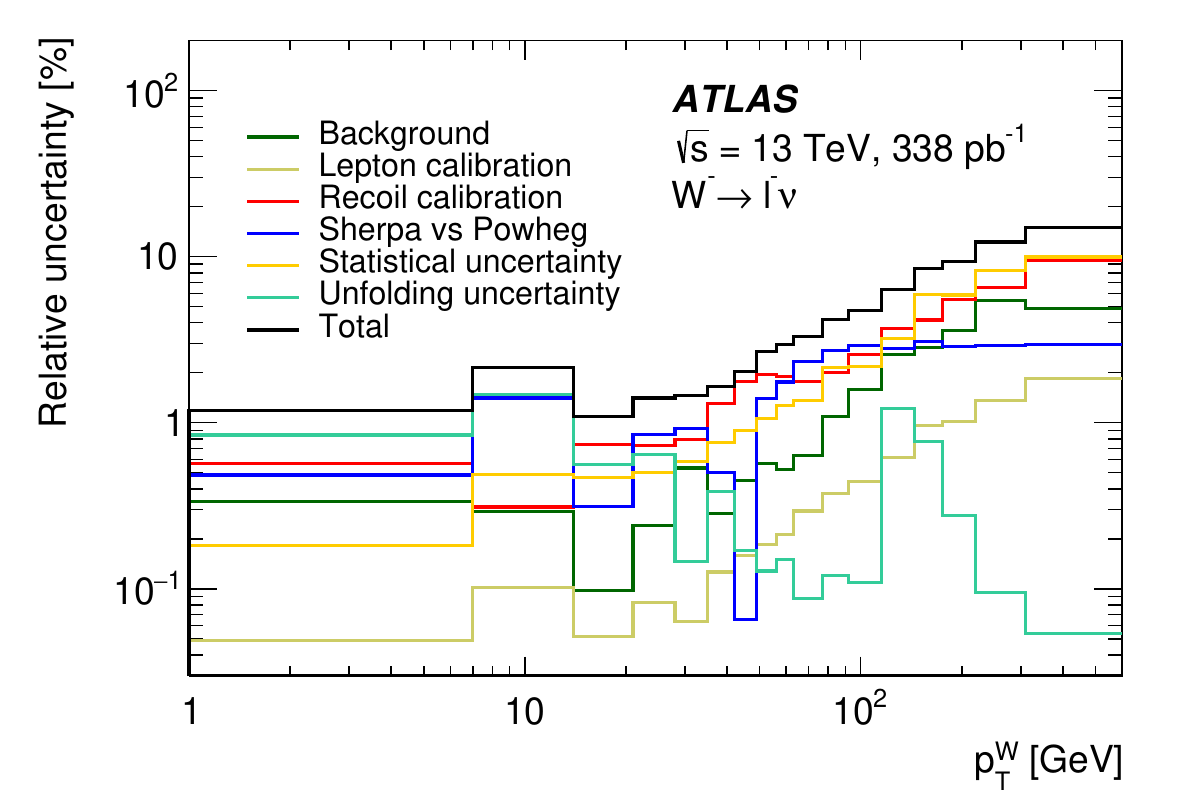}}

\subfloat[]{\includegraphics[width=.47\textwidth]{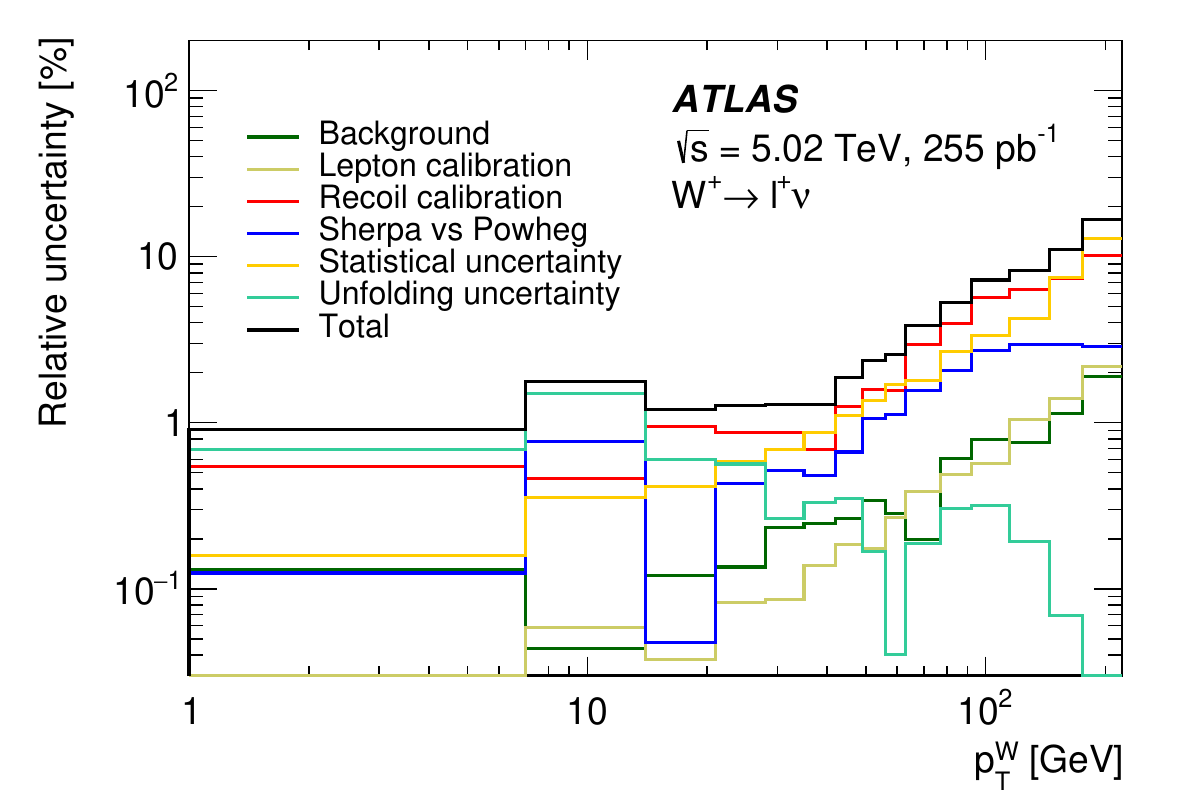}}%
~~~~%
\subfloat[]{\includegraphics[width=.47\textwidth]{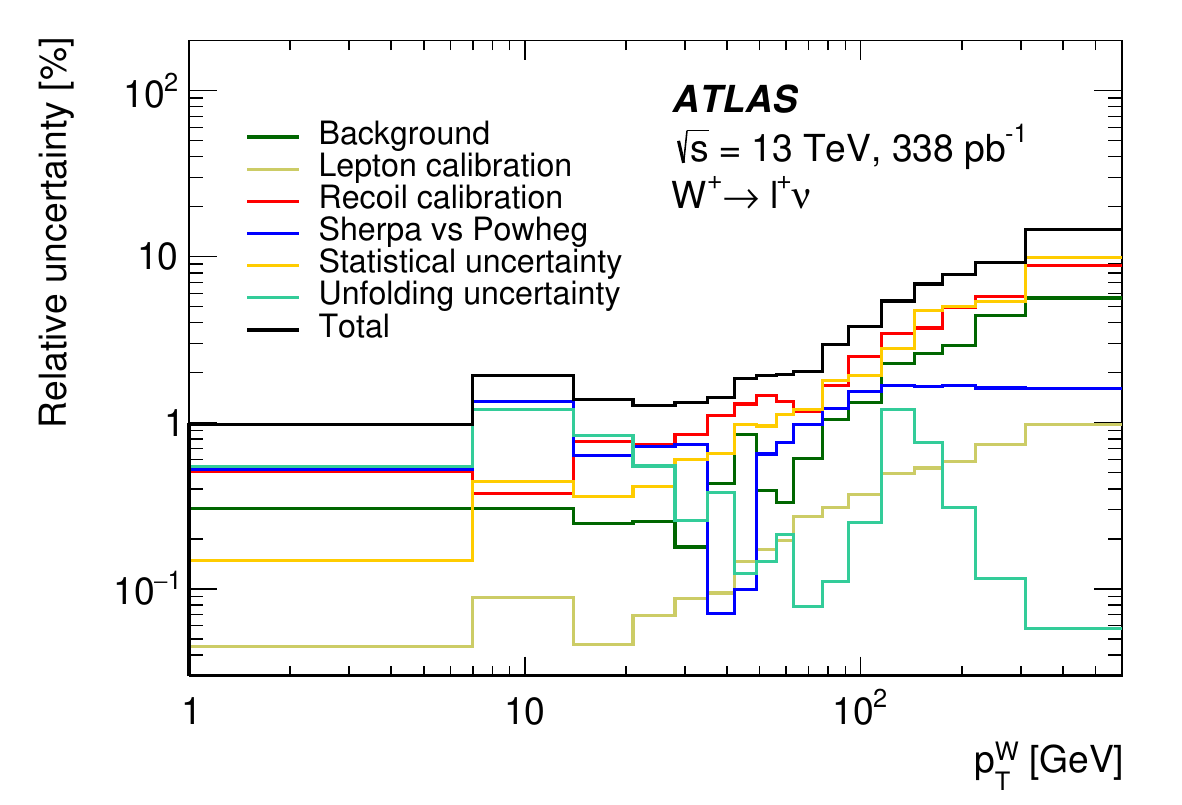}}

\subfloat[]{\includegraphics[width=.47\textwidth]{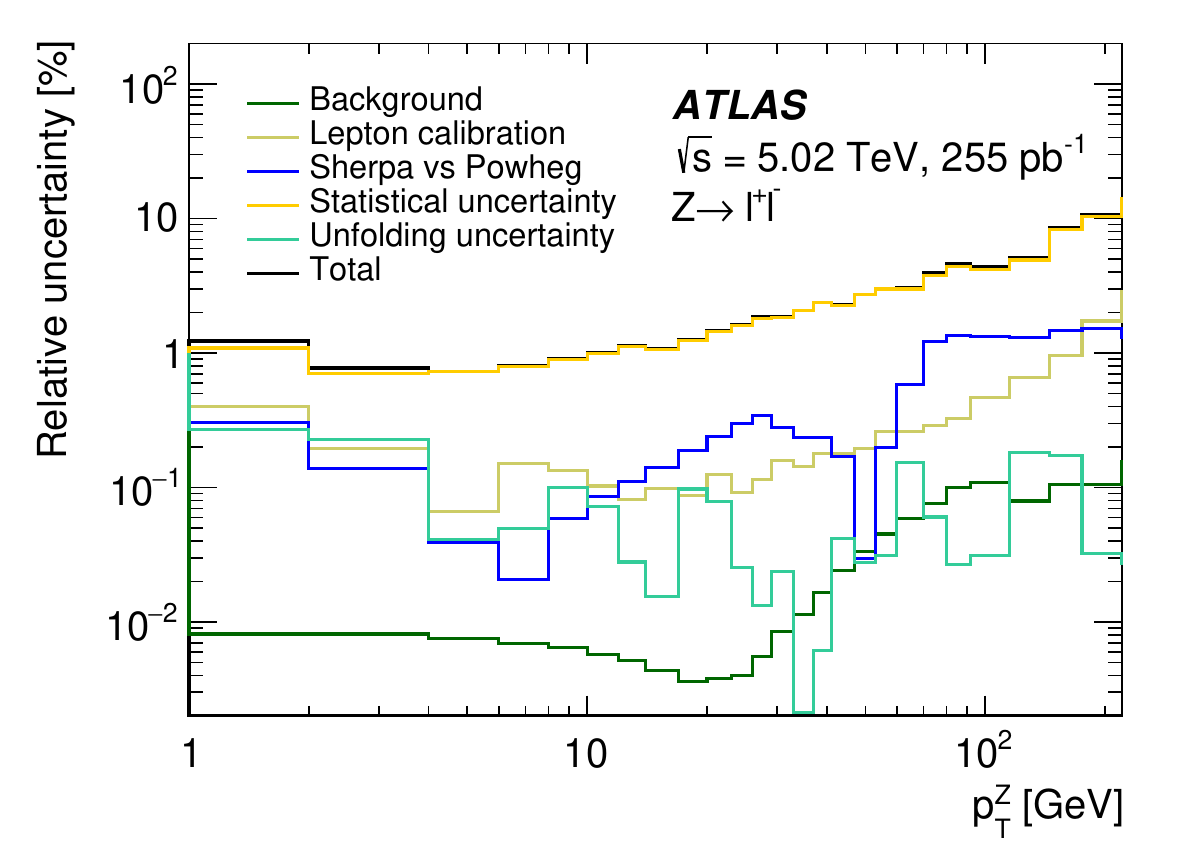}}%
~~~~%
\subfloat[]{\includegraphics[width=.47\textwidth]{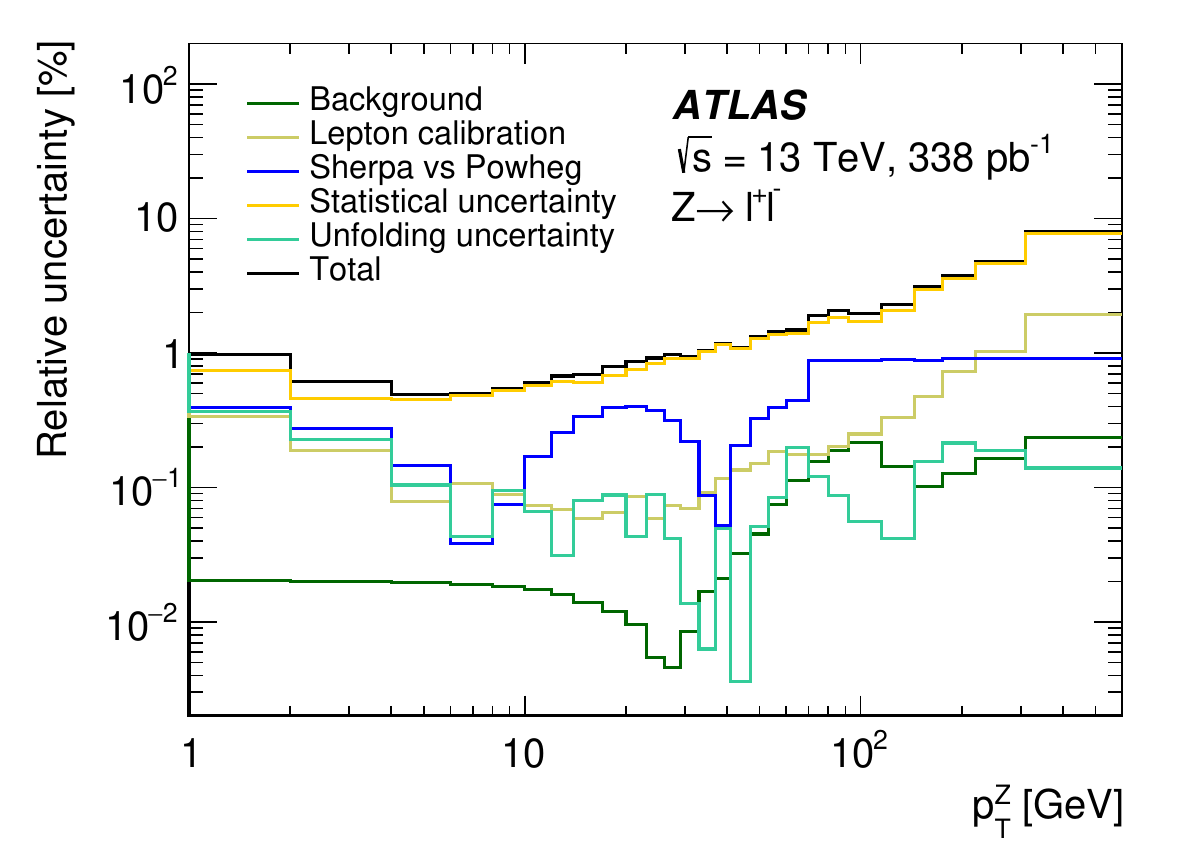}}
\caption{Uncertainty contributions in the combined normalised \pt distributions for $W^-$ at (a) $\sqrt{s} = \fivetev$ and (b) $\sqrt{s} = 13\,\TeV$, $W^+$ at (c) $\sqrt{s} = \fivetev$ and
(d) $\sqrt{s}=13\,\TeV$, and $Z$ at (e) $\sqrt{s} = \fivetev$ and (f) $\sqrt{s} = 13\,\TeV$.}
\label{fig:combiUnc}
\end{figure}

Combinations are performed for absolute and normalised differential cross sections. The
compatibility of the electron and muon channel measurements is generally good, as evidenced by combination $\chi^2$
values close to the number of degrees of freedom.
The combined uncertainties on the normalised spectra are shown as a function of \ptw and \ptz in Figure~\ref{fig:combiUnc}, with an overall precision of about 1\% at low \pt, reaching about 10\% towards the end of the spectrum. The \Wboson-boson uncertainties are dominated by unfolding systematic uncertainties at low \ptw, and statistical uncertainties at high \ptw. The generator dependence of the recoil calibration procedure is sizeable at $13\,\TeV$. The \Zboson-boson uncertainties are fully dominated by statistical uncertainties.

\subsection{Integrated cross sections and distribution ratios}
\label{sec:derived}

Integrated fiducial cross sections are obtained by summing the measured distributions over a given spectrum, and the corresponding uncertainty is given by summing the corresponding covariance matrix elements.
The definition of the fiducial volume is given in Section~\ref{sec:measurement}, and the results are summarised in
Table~\ref{tab:fidxsec}. Experimental systematic uncertainties excluding the luminosity are at the level of 0.4\% and 0.5\% for $W$-boson production at
\fivetev and $13\,\TeV$, respectively. The corresponding numbers for the $Z$-boson cross sections are 0.7\% and
0.9\%. The luminosity uncertainty is 1.0\% at \fivetev, and 0.9\% at $13\,\TeV$.

Similar measurements at \fivetev previously published by ATLAS in
Ref.~\cite{HION-2018-02} are not directly comparable to the new
results due to small differences in the fiducial volume
definition. However, the relative uncertainties are improved by about
a factor of two due to the larger dataset and improved experimental
calibrations, including the luminosity measurement. The measurements
at $13\,\TeV$ are compatible with those performed by ATLAS with the
very early Run 2 data with a luminosity of just
$81\,\ipb$~\cite{STDM-2015-03} and improve the precision by a
factor of 3.5 for the \Wboson measurement and a factor of 1.7 for the \Zboson measurement.
The $13\,\TeV$ \Zboson-boson measurement is also
compatible with the improved ATLAS measurement of
Ref.~\cite{STDM-2016-02} that has a slightly lower precision because of
the preliminary luminosity calibration used then.

The measured ratios of integrated fiducial cross sections are given in
Tables~\ref{tab:fidxsec_Eratio} and \ref{tab:fidratios}. As expected,
all cross sections rise with $\sqrt{s}$, as lower parton momentum fractions $x$ are probed and the densities of sea quarks and antiquarks rise towards low $x$.
The increase in the $W^+$ cross section is the smallest because it has the
largest contribution from higher $x$ valence quarks that contribute less to the phase space measured at higher $\sqrt{s}$.

\begin{table}
\caption{Integrated fiducial cross sections for $W^-$, $W^+$ and $Z$ production at $\sqrt{s} = \fivetev$ and $\sqrt{s} = 13\,\tev$.}
\label{tab:fidxsec}
\centering
\begin{tabular}{ccc}
\toprule
\toprule
Process & $\sigma_\mathrm{fid}(\sqrt{s}=\fivetev)$ [pb] & $\sigma_\mathrm{fid}(\sqrt{s}=13\,\TeV)$ [pb]  \\
\midrule
\Wmlnu    &  1384 $\pm$  ~ 2 (stat.) $\pm$ ~ 5 (syst.) $\pm$ 15 (lumi.) &  3486 $\pm$ ~ 3 (stat.) $\pm$ 18 (syst.) $\pm$ 34 (lumi.) \\
\Wplnu    &   2228 $\pm$ ~ 3 (stat.) $\pm$ ~ 8 (syst.) $\pm$ 23 (lumi.) &  4571 $\pm$ ~ 3 (stat.) $\pm$ 21 (syst.) $\pm$ 44 (lumi.)\\
$Z\to \ell\ell$    &  333.0 $\pm$  1.2 (stat.) $\pm$ 2.2 (syst.) $\pm$ 3.3 (lumi.)   &  780.3 $\pm$  2.6 (stat.) $\pm$ 7.1 (syst.) $\pm$ 7.1 (lumi.)\\
\bottomrule
\bottomrule
\end{tabular}
\end{table}

\begin{table}
\caption{Ratio of integrated fiducial cross sections at $\sqrt{s} = 13\,\tev$ and $\sqrt{s} = \fivetev$.}
\label{tab:fidxsec_Eratio}
\centering
\begin{tabular}{cc}
\toprule
\toprule
Process &  $\sigma_\mathrm{fid}(\sqrt{s}=13\,\TeV) / \sigma_\mathrm{fid}(\sqrt{s}=\fivetev)$\\
\midrule
\Wmlnu   &   2.516 $\pm$  0.005 (stat.) $\pm$ 0.010 (syst.) $\pm$ 0.036 (lumi.)\\
\Wplnu   &   2.050 $\pm$  0.003 (stat.) $\pm$ 0.008 (syst.) $\pm$ 0.029 (lumi.)\\
$Z\to \ell\ell$  &  2.344 $\pm$  0.011 (stat.) $\pm$ 0.011 (syst.) $\pm$ 0.032 (lumi.) \\
\bottomrule
\bottomrule
\end{tabular}
\end{table}

\begin{table}
\caption{Integrated fiducial cross sections ratios at $\sqrt{s} = \fivetev$ and $\sqrt{s} = 13\,\tev$.}
\label{tab:fidratios}
\centering
\begin{tabular}{ccc}
\toprule
\toprule
Processes & Cross-section ratio at $\sqrt{s} = \fivetev$ & Cross-section ratio at $\sqrt{s}=13\,\TeV$  \\
\midrule
$W^+/W^-$ &  1.609 $\pm$  0.003 (stat.) $\pm$ 0.004 (syst.) & 1.308 $\pm$  0.003 (stat.) $\pm$ 0.004 (syst.) \\
$W^-/Z$  & \, 4.16 $\pm$ ~ 0.02 (stat.) $\pm$ ~ 0.03 (syst.)  & ~ 4.46 $\pm$ \, 0.02 (stat.) $\pm$ ~ 0.05 (syst.) \\
$W^+/Z$ & \, 6.69 $\pm$  ~ 0.02 (stat.) $\pm$ ~ 0.05 (syst.) & ~ 5.84 $\pm$ \, 0.02 (stat.) $\pm$ ~ 0.06 (syst.) \\
$W^\pm/Z$  & 10.85 $\pm$ ~ 0.04 (stat.) $\pm$ ~ 0.07 (syst.) &10.30 $\pm$ \, 0.04 (stat.) $\pm$ \; 0.10 (syst.)   \\
\bottomrule
\bottomrule
\end{tabular}
\end{table}

\begin{figure}[p]
\centering
\subfloat[]{\includegraphics[width=.47\textwidth]{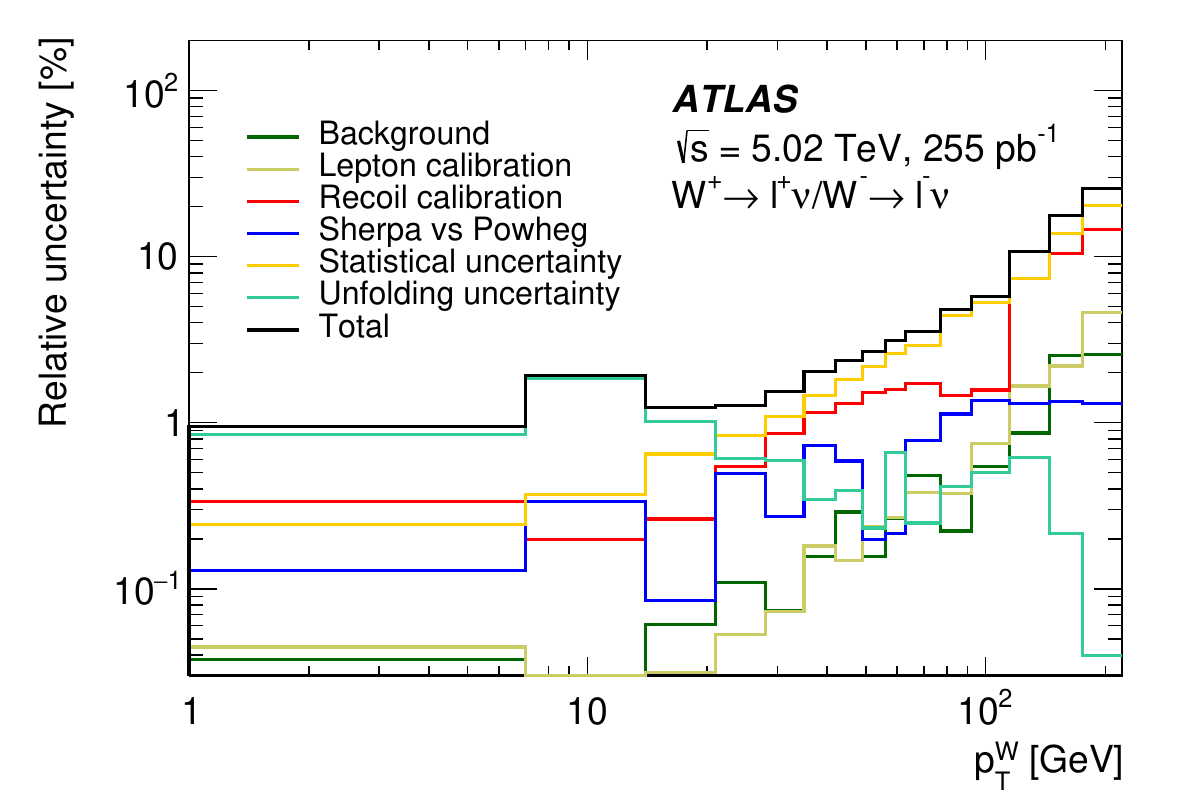}\label{fig:combiUnc_ratio_a}}%
~~~~%
\subfloat[]{\includegraphics[width=.47\textwidth]{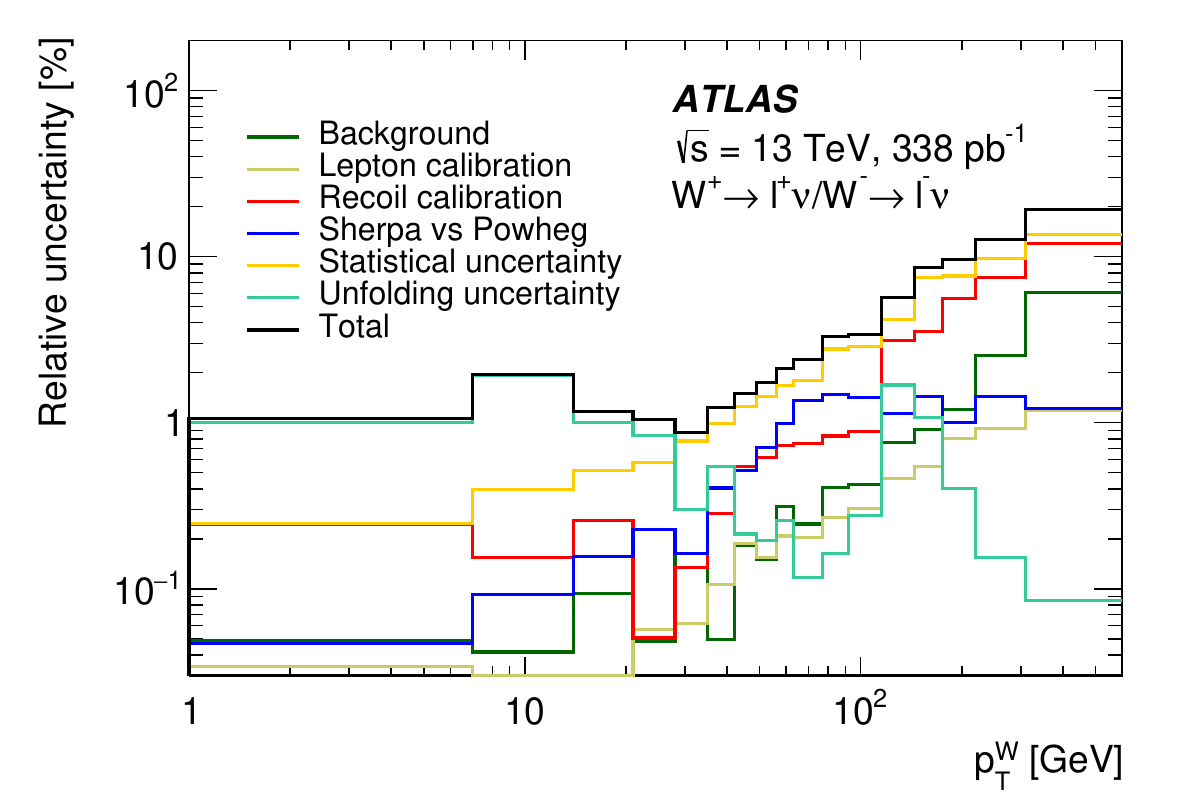}\label{fig:combiUnc_ratio_b}}

\subfloat[]{\includegraphics[width=.47\textwidth]{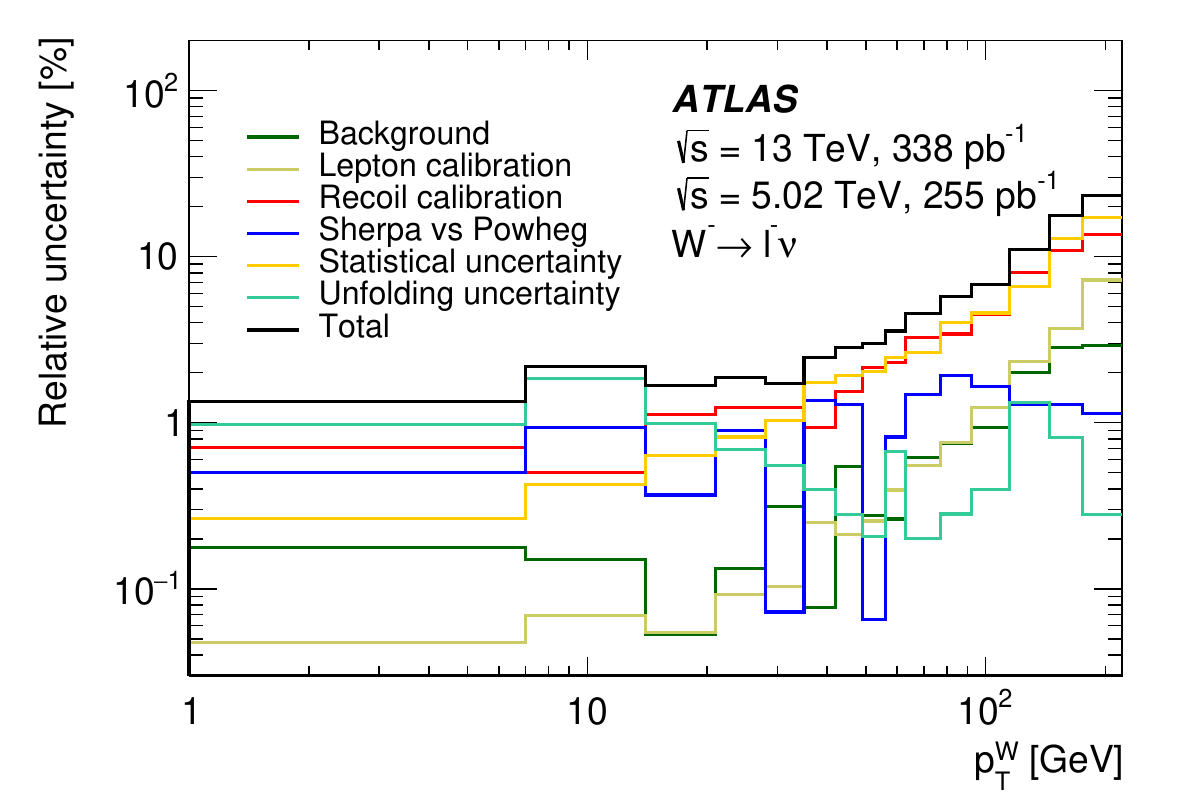}\label{fig:combiUnc_ratio_c}}%
~~~~%
\subfloat[]{\includegraphics[width=.47\textwidth]{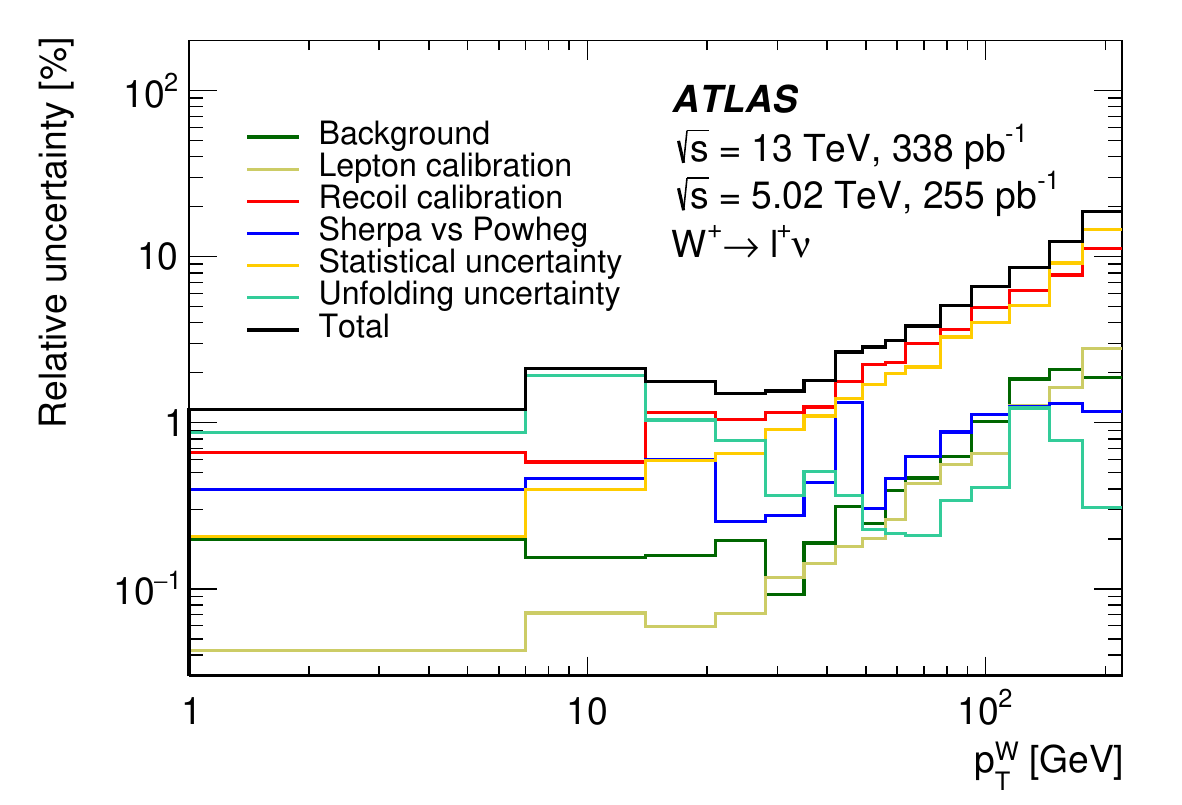}\label{fig:combiUnc_ratio_d}}
\caption{Uncertainty contributions in the ratios of combined
normalised \pt distributions for $W^+/W^-$ at (a) $\sqrt{s} = \fivetev$ and
(b) $\sqrt{s} = 13\,\TeV$ as well as the ratios of measurements at different
centre-of-mass energies $\sqrt{s} = 13\,\TeV$ to $\sqrt{s} = \fivetev$ for (c) $W^-$ and
(d) $W^+$.
}
\label{fig:combiUnc_ratio}
\end{figure}

Cross-section ratios are also formed of the absolute and normalised
differential distributions, both between different processes at a
given $\sqrt{s}$ and between the same processes at different
$\sqrt{s}$.  Theoretically, ratios probe specific aspects of the
modelling of $W$ and $Z$ production. The $W^+/W^-$ and $W^\pm/Z$
ratios are expected to be relatively insensitive to universal
resummation effects, but the low-\pt range is sensitive to the
different initial quark flavours and specifically the contribution of
heavy quarks.
Experimentally, some of these ratios benefit from a partial
cancellation of systematic uncertainties. Most notably, the recoil
calibration at a given $\sqrt{s}$ is strongly correlated between the
\Wplus\ and \Wminus\ measurements as it stems from the same
calibration, mostly using \Zboson\
events. Figure~\ref{fig:combiUnc_ratio_a} and~\ref{fig:combiUnc_ratio_b} illustrates how
the associated uncertainties largely cancel in the $\Wplus/\Wminus$
ratios, resulting in a measurement uncertainty of around $1\%$  at low
\pt. By contrast, since this calibration is obtained independently at
$\sqrt{s} = \fivetev$ and $13~\TeV$, the related uncertainty increases
in a ratio, as shown in Figures~\ref{fig:combiUnc_ratio_c} and~\ref{fig:combiUnc_ratio_d}
and the typical precision is around $2\%$ at lower \pt, increasing
significantly towards higher \pt. The electron--muon combination of
these ratios have generally very good $\chi^2/\mathrm{dof}$.

\subsection{Comparison to predictions}
\label{sec:predictions}

The final measurements, obtained from the combination of electron and
muon channels, are compared to an extensive list of
predictions from higher-order resummed predictions and MC event generators.
As a representative example of resummation predictions, calculations
from \textsc{DYTurbo} v1.2.3 are presented. These
predictions are computed by performing the resummation of
logarithmically-enhanced contributions in the small $q_T$ region of the
lepton pairs at  NNLL accuracy
and also include the corresponding finite-order QCD contributions at
NNLO (corrections up to $\alpha_S^2$). Three different NNLO PDF sets are used:
CT18~\cite{Hou:2019efy}, MSHT20~\cite{Bailey:2020ooq} and
NNPDF3.1~\cite{NNPDF:2017mvq}. Theoretical uncertainties
are estimated through variations of the nominal factorisation,
renormalisation and resummation scales by a factor of two up and down, excluding
variations in opposite directions, and taking the envelope of all
variations. The PDF uncertainties for each set are estimated by calculating
the variations for all eigenvector or replicas and combining
the results following the prescriptions relevant for each PDF set.
When computing the uncertainties in ratios or across differential spectra,
the effect of a certain PDF eigenvector or replica variation is taken as correlated.
Similarly, the effect of each scale variation is treated as fully
correlated between all vector bosons, $\sqrt{s}$ and across spectra.
Additional comparisons at $\sqrt{s} = 13\,\TeV$ to predictions by the \textsc{RadISH} program are presented
in Appendix~\ref{sec:radishcomp}.

The integrated cross-section measurements and ratios presented in
Section~\ref{sec:derived} are compared to the integrated \textsc{DYTurbo}
predictions in Table~\ref{tab:fidxsec_pred}. Good agreement with the
central values of the NNPDF3.1 already within the experimental uncertainties
is seen, while the CT18 and MSHT20 predictions tend to underestimate
the cross sections, similar to previous observations~\cite{HION-2018-02,
STDM-2016-02}. When considering the dominant PDF uncertainties,
the agreement with all PDF sets is satisfactory. Table~\ref{tab:fidxsec_Eratio_pred} compares the
measured ratios of cross sections at $\sqrt{s} = 13\,\TeV$ and
$\sqrt{s} = \fivetev$ to the predictions. The increases in cross
section are well predicted by all PDF sets. Table~\ref{tab:fidratios_pred} then presents ratios at
fixed $\sqrt{s}$, which are again generally compatible with the
predictions. The $W^\pm/Z$ ratios predicted by CT18 are
$1$--$2$ standard deviations above the data, which is similar to an observation at
$\sqrt{s} = 7\,\TeV$ attributed to the strange-quark
density~\cite{STDM-2012-20}.

\begin{table}
\caption{Measured integrated fiducial cross sections for
$W^-$, $W^+$ and $Z$ production at $\sqrt{s} = \fivetev$ and $\sqrt{s} = 13\,\tev$ with
total uncertainties (`Data') are compared to \textsc{DYTurbo} predictions at
NNLO+NNLL using three different NNLO PDF sets: CT18, MSHT20 and NNPDF3.1.
Uncertainties in the theoretical predictions correspond to those obtained by variations of the factorisation and
renormalisation scales (`scale') and those computed from the variations in each PDF given by eigenvector or replica variations (`PDF').}
\label{tab:fidxsec_pred}
\centering

\clearpage

\begin{tabular}{lccc}
\toprule
\toprule
PDF set & $W^-\to \ell^-\nu$ & $W^+\to \ell^+\nu$ & $Z\to \ell\ell$ \\

\midrule
\multicolumn{4}{c}{$\sigma_\mathrm{fid}(\sqrt{s}=\fivetev)$ [pb] } \\
\midrule
Data 	&	1384 $\pm$ 16	&	2228 $\pm$ 25	&	333.0 $\pm$ 4.1 \\
\midrule
CT18	&	1360 $\pm$ 10 (scale)  $^{+30}_{-40}$ (PDF) 	&	2200 $\pm$ 10 (scale)  $^{+40}_{-70}$ (PDF) 	&	320 $\pm$ 1 (scale)  $^{+5}_{-9}$ (PDF) \\[1ex]
MSHT20	&	1351 $^{+5}_{-6}$ (scale)  $^{+22}_{-23}$ (PDF) 	&	2180 $\pm$ 10 (scale)  $^{+30}_{-40}$ (PDF) 	&	324 $\pm$ 1 (scale)  $^{+4}_{-5}$ (PDF) \\[1ex]
NNPDF31	&	1381 $\pm$ 6 (scale)  $\pm$ 16 (PDF) 	&	2232 $^{+8}_{-9}$ (scale)  $\pm$ 25 (PDF) 	&	329 $\pm$ 1 (scale)  $\pm$ 4 (PDF) \\

\midrule
\multicolumn{4}{c}{$\sigma_\mathrm{fid}(\sqrt{s}=13\,\TeV)$ [pb] } \\
\midrule
Data 	&	3486 $\pm$ 38	&	4571 $\pm$ 49	&	780.3 $\pm$ 10.4 \\
\midrule
CT18	&	3410 $^{+40}_{-20}$ (scale)  $^{+60}_{-100}$ (PDF) 	&	4460 $^{+40}_{-30}$ (scale)  $^{+80}_{-130}$ (PDF) 	&	748 $^{+5}_{-4}$ (scale)  $^{+18}_{-25}$ (PDF) \\[1ex]
MSHT20	&	3400 $^{+40}_{-20}$ (scale)  $^{+40}_{-60}$ (PDF) 	&	4460 $^{+40}_{-30}$ (scale)  $^{+60}_{-70}$ (PDF) 	&	763 $^{+6}_{-4}$ (scale)  $^{+9}_{-12}$ (PDF) \\[1ex]
NNPDF31	&	3450 $^{+40}_{-20}$ (scale)  $\pm$ 30 (PDF) 	&	4510 $^{+40}_{-30}$ (scale)  $\pm$ 40 (PDF) 	&	769 $^{+6}_{-4}$ (scale)  $\pm$ 7 (PDF) \\

\bottomrule
\bottomrule
\end{tabular}

\clearpage

\end{table}

\begin{table}
\caption{Ratios of measured integrated fiducial cross sections at $\sqrt{s} = 13\,\tev$
and $\sqrt{s} = \fivetev$ for $W^-$, $W^+$ and $Z$ production with
total uncertainties (`Data') are compared to \textsc{DYTurbo} predictions at
NNLO+NNLL using three different NNLO PDF sets: CT18, MSHT20 and NNPDF3.1.}
\label{tab:fidxsec_Eratio_pred}
\centering
\footnotesize

\clearpage

\begin{tabular}{lccc}
\toprule
\toprule
PDF set & $W^-\to \ell^-\nu$ & $W^+\to \ell^+\nu$ & $Z\to \ell\ell$ \\
\midrule
\multicolumn{4}{c}{ $\sigma_\mathrm{fid}(\sqrt{s}=13\,\TeV) / \sigma_\mathrm{fid}(\sqrt{s}=\fivetev)$} \\
\midrule
Data 	&	2.516 $\pm$ 0.038	&	2.050 $\pm$ 0.030	&	2.344 $\pm$ 0.036 \\
\midrule
CT18	&	2.499 $^{+0.017}_{-0.005}$ (scale)  $^{+0.043}_{-0.046}$ (PDF) 	&	2.029 $^{+0.010}_{-0.007}$ (scale)  $^{+0.029}_{-0.023}$ (PDF) 	&	2.335 $^{+0.012}_{-0.009}$ (scale)  $^{+0.049}_{-0.038}$ (PDF) \\[1ex]
MSHT20	&	2.515 $^{+0.017}_{-0.005}$ (scale)  $^{+0.018}_{-0.020}$ (PDF) 	&	2.040 $^{+0.010}_{-0.007}$ (scale)  $^{+0.014}_{-0.015}$ (PDF) 	&	2.358 $^{+0.012}_{-0.009}$ (scale)  $^{+0.018}_{-0.022}$ (PDF) \\[1ex]
NNPDF31	&	2.500 $^{+0.017}_{-0.005}$ (scale)  $\pm$ 0.031 (PDF) 	&	2.022 $^{+0.010}_{-0.007}$ (scale)  $\pm$ 0.029 (PDF) 	&	2.335 $^{+0.012}_{-0.009}$ (scale)  $\pm$ 0.026 (PDF) \\
\bottomrule
\bottomrule
\end{tabular}

\clearpage

\end{table}

\begin{table}
\caption{Ratios of  measured integrated fiducial cross sections of $W^+/W^-$ and
$W^\pm/Z$ for the $\sqrt{s} = \fivetev$ and $\sqrt{s} = 13\,\tev$ data with
total  uncertainties  (`Data') are compared to \textsc{DYTurbo} predictions at
NNLO+NNLL using three different  NNLO PDF sets: CT18, MSHT20 and NNPDF3.1.}
\label{tab:fidratios_pred}
\centering

\clearpage

\begin{tabular}{lcc}
\toprule
\toprule
PDF set & $W^+/W^-$ & $W^\pm/Z$\\

\midrule
\multicolumn{3}{c}{Cross-section ratio at $\sqrt{s}=\fivetev$} \\
\midrule
Data 	&	1.609 $\pm$ 0.005	&	10.85 $\pm$ 0.08 \\
\midrule
CT18	&	1.612 $^{+0.001}_{-0.003}$ (scale)  $^{+0.018}_{-0.028}$ (PDF) 	&	11.13 $^{+0.04}_{-0.06}$ (scale)  $\pm$ 0.11 (PDF) \\[1ex]
MSHT20	&	1.618 $^{+0.001}_{-0.003}$ (scale)  $\pm$ 0.008 (PDF) 	&	10.92 $^{+0.04}_{-0.06}$ (scale)  $^{+0.07}_{-0.05}$ (PDF) \\[1ex]
NNPDF31	&	1.616 $^{+0.001}_{-0.003}$ (scale)  $\pm$ 0.018 (PDF) 	&	10.98 $^{+0.04}_{-0.06}$ (scale)  $\pm$ 0.15 (PDF) \\

\midrule
\multicolumn{3}{c}{Cross-section ratio at $\sqrt{s}=13\,\TeV$} \\
\midrule
Data 	&	1.308 $\pm$ 0.005	&	10.30 $\pm$ 0.11 \\
\midrule
CT18	&	1.309 $^{+0.000}_{-0.003}$ (scale)  $^{+0.012}_{-0.007}$ (PDF) 	&	10.53 $\pm$ 0.03 (scale)  $^{+0.13}_{-0.15}$ (PDF) \\[1ex]
MSHT20	&	1.312 $^{+0.000}_{-0.003}$ (scale)  $^{+0.005}_{-0.004}$ (PDF) 	&	10.29 $\pm$ 0.03 (scale)  $^{+0.07}_{-0.05}$ (PDF) \\[1ex]
NNPDF31	&	1.307 $^{+0.000}_{-0.003}$ (scale)  $\pm$ 0.003 (PDF) 	&	10.37 $\pm$ 0.03 (scale)  $\pm$ 0.06 (PDF) \\

\bottomrule
\bottomrule
\end{tabular}

\clearpage

\end{table}

Differential predictions are also obtained from MC event generators
that combine LO or NLO
matrix elements and match them to parton showers and models for
underlying event, hadronisation and particle decay. The MC samples
already described in Section~\ref{sec:mc}, namely \Powheg+\PYTHIAV{8} with the
AZNLO tune and \SHERPA[2.2.1/2.2.5], are included,
however without the reweighting to match the reconstruction-level data described in Section~\ref{sec:unfold}.
A \Powheg+\HERWIGV{7}
sample is simulated using a similar set-up as \Powheg+\PYTHIAV{8}, but using
\HERWIGV{7}~\cite{Bellm:2015jjp,Bellm:2017jjp} for the parton
shower, hadronisation and underlying event using the default tune. A prediction with \PYTHIAV{8} alone
matches LO matrix elements to a parton shower with the \CTEQ[6L1] PDF set and the
AZ parameter set tuned to describe $7\,\TeV$ $\ptll$ and $\phi^*$
measurements~\cite{STDM-2012-23}. Further predictions at $13\,\TeV$ with
multijet-merged predictions at NLO as described in
Ref.~\cite{PMGR-2021-01} are prepared with \SHERPA[2.2.11] and
\textsc{MadGraph\_aMC@NLO FxFx}+\PYTHIAV{8}~\cite{Alwall:2014hca, Frederix:2012ps} and
presented in Appendix~\ref{sec:mjcomp}. The fiducial predictions are extracted with
the \textsc{Rivet} toolkit~\cite{Bierlich:2019rhm}.

\begin{figure}[tbph]
\centering
\subfloat[]{\includegraphics[width=.42\textwidth]{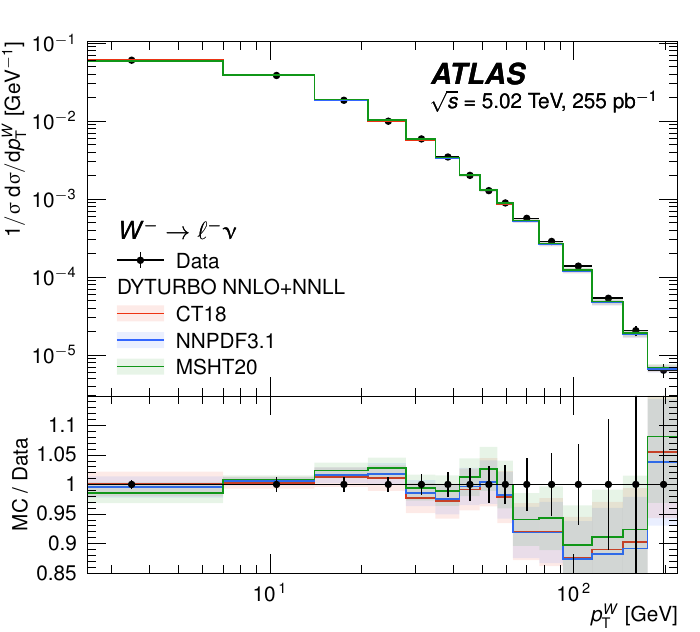}}%
~~~~%
\subfloat[]{\includegraphics[width=.42\textwidth]{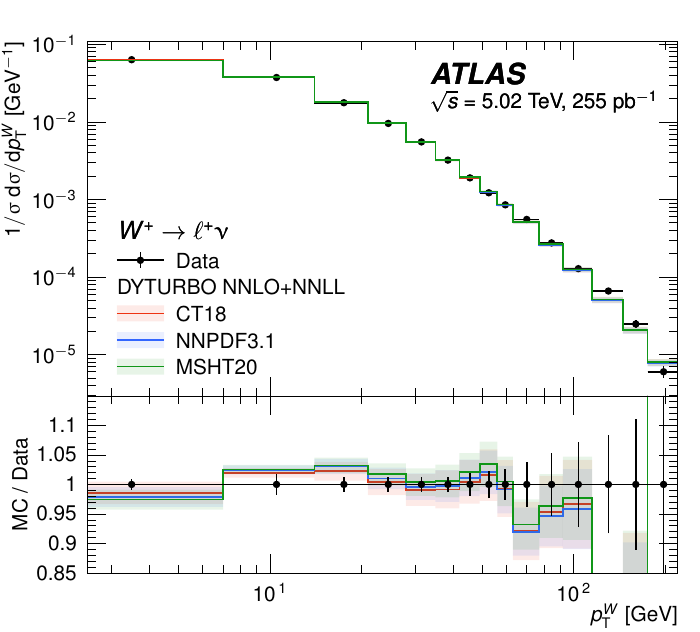}}

\subfloat[]{\includegraphics[width=.42\textwidth]{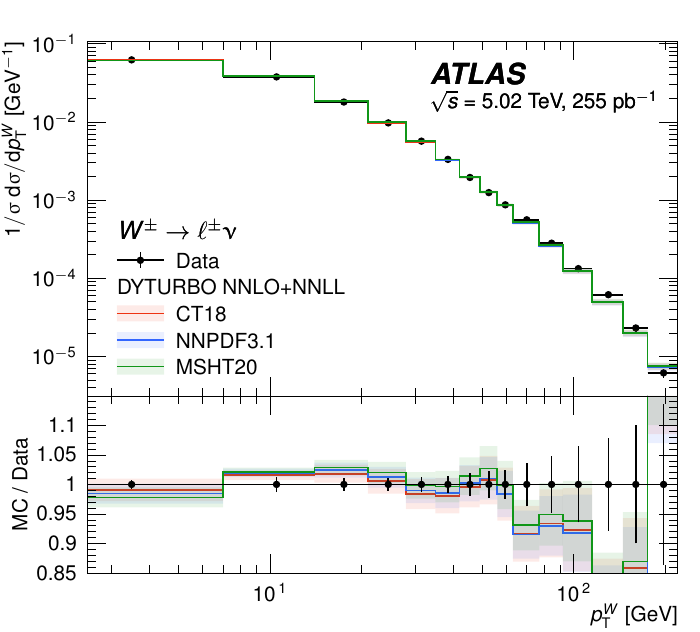}}%
~~~~%
\subfloat[]{\includegraphics[width=.42\textwidth]{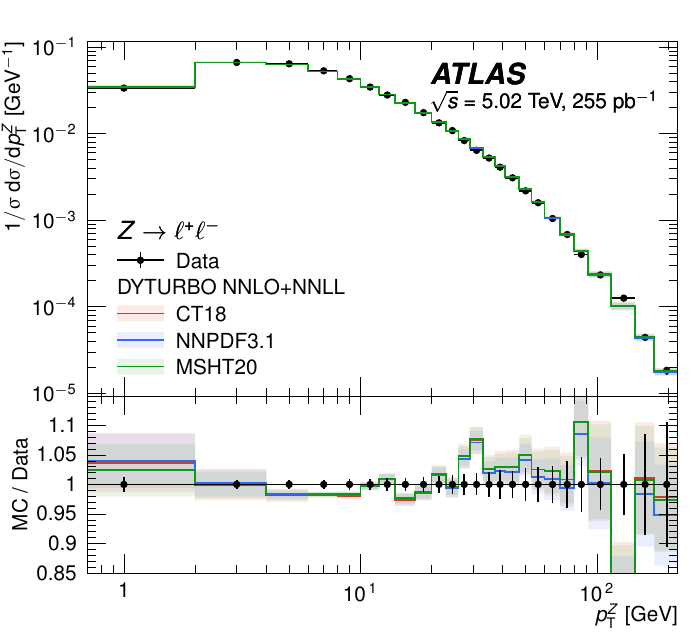}}

\subfloat[]{\includegraphics[width=.42\textwidth]{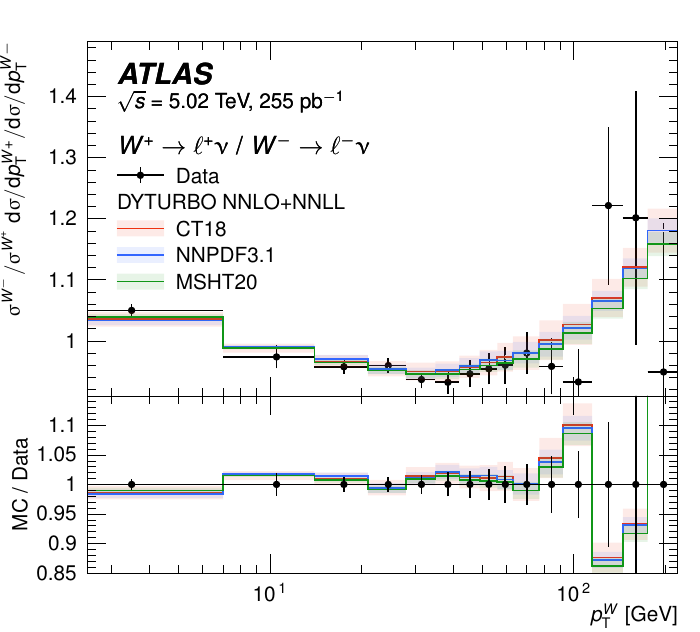}}%
~~~~%
\subfloat[]{\includegraphics[width=.42\textwidth]{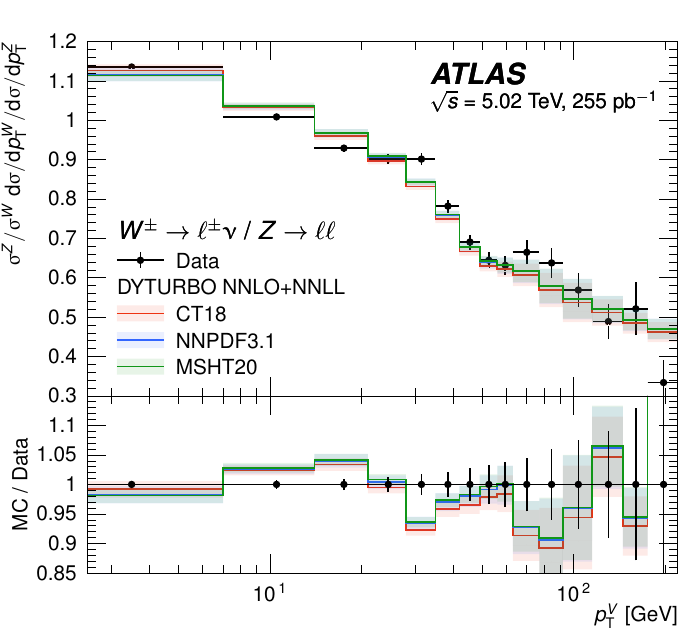}}
\caption{Measurements of normalised differential distributions at
$\sqrt{s} = \fivetev$ (black points) for (a) $W^-$, (b) $W^+$, (c) the sum $W^\pm$, (d) $Z$ as
well as the ratios (e) $W^+/W^-$ and (f) $W^\pm/Z$ compared to \textsc{DYTurbo}
predictions with different PDF sets (coloured lines and PDF and scale uncertainties as shaded band) as described in the
text. The lower panels show the ratio of prediction to data with
data markers centred at one and error bars giving the size of the total measurement uncertainties.}
\label{fig:finalpred_5TeV_norm_dyturbo}
\end{figure}

\begin{figure}[tbph]
\centering
\subfloat[]{\includegraphics[width=.42\textwidth]{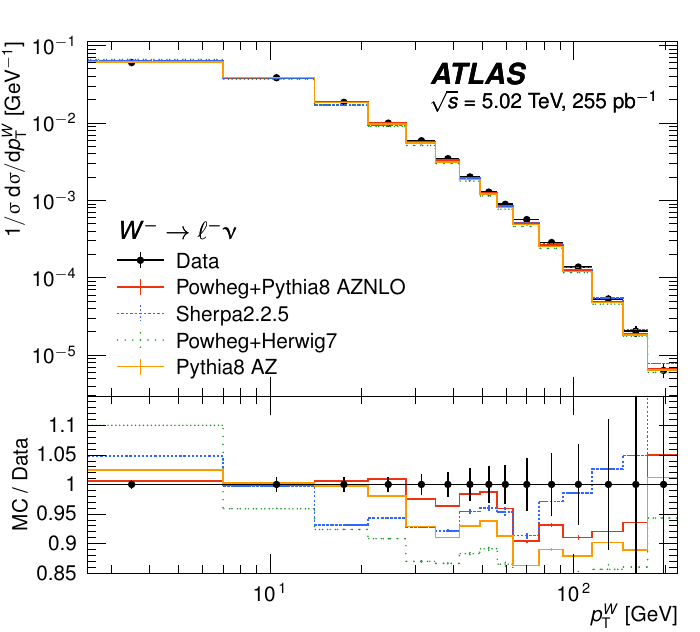}}%
~~~~%
\subfloat[]{\includegraphics[width=.42\textwidth]{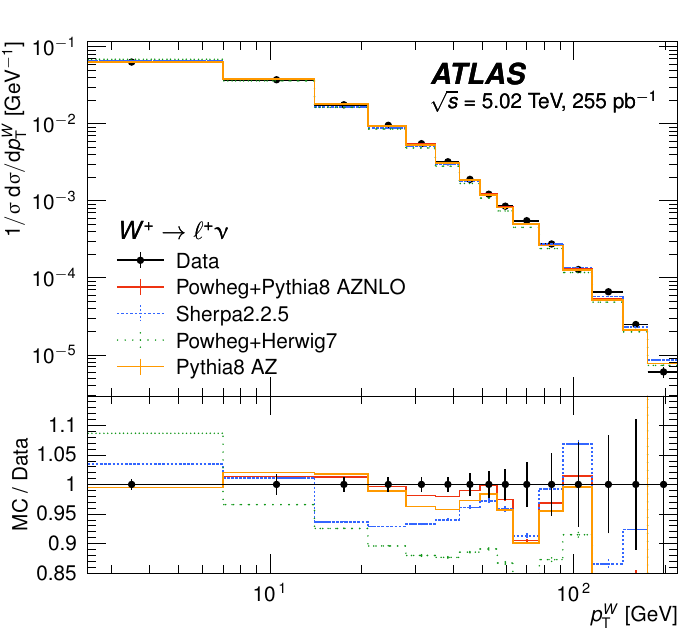}}

\subfloat[]{\includegraphics[width=.42\textwidth]{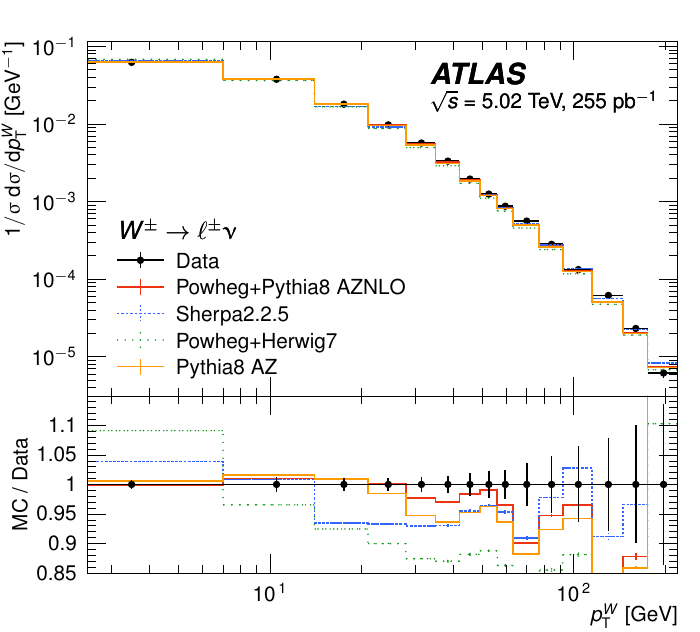}}%
~~~~%
\subfloat[]{\includegraphics[width=.42\textwidth]{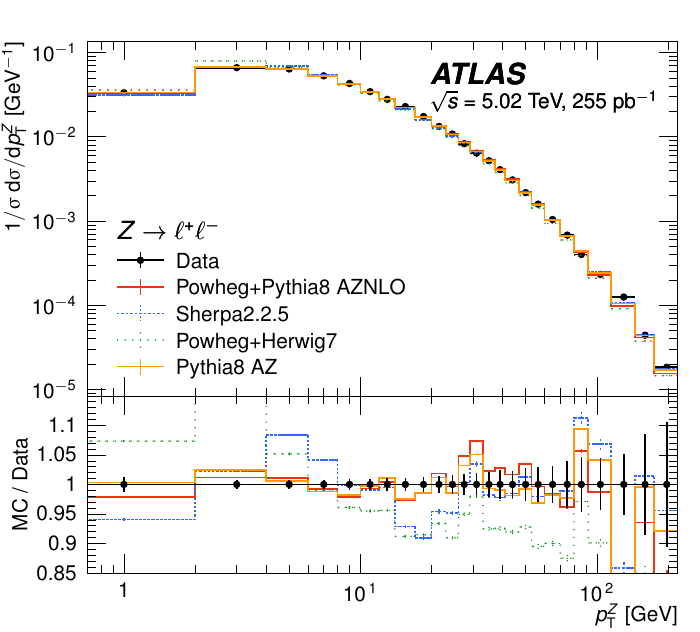}}

\subfloat[]{\includegraphics[width=.42\textwidth]{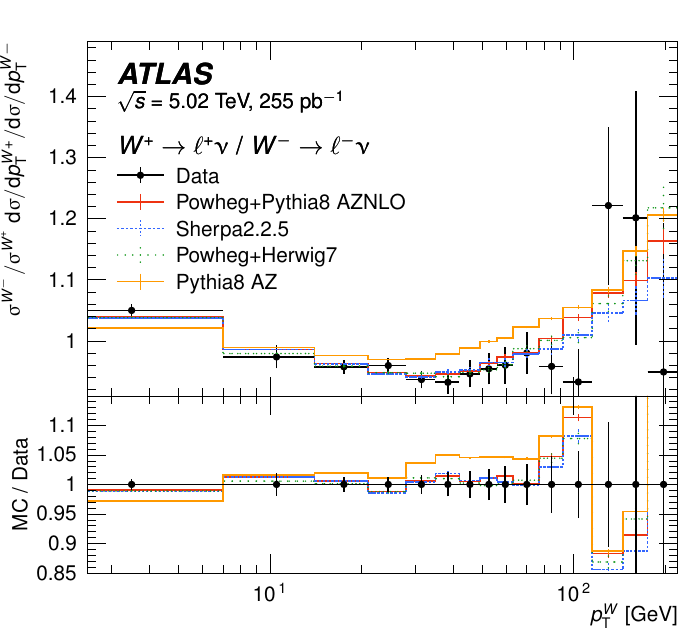}}%
~~~~%
\subfloat[]{\includegraphics[width=.42\textwidth]{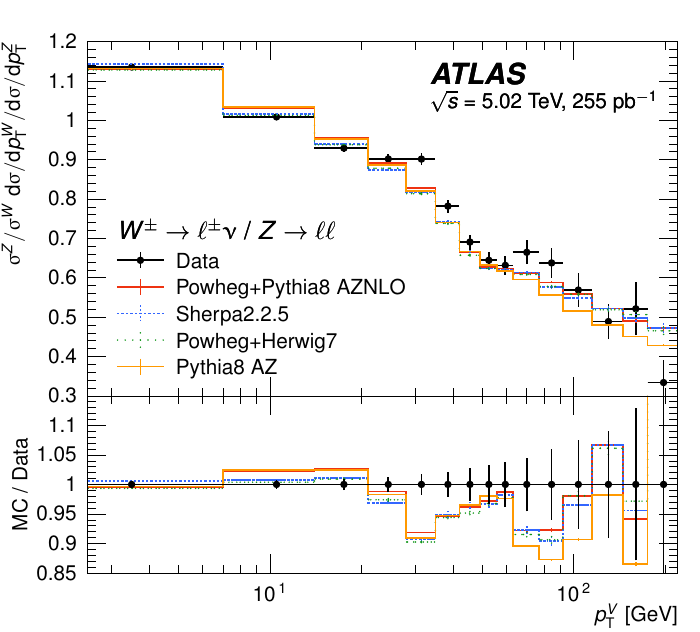}}%
\caption{Measurements of normalised differential distributions at
$\sqrt{s} = \fivetev$ (black points) for (a) $W^-$, (b) $W^+$, (c) the sum $W^\pm$, (d) $Z$ as
well as the ratios (e) $W^+/W^-$ and (f) $W^\pm/Z$ compared to a
variety of MC predictions (coloured lines) as described in the
text. The lower panels show the ratio of prediction to data with
data markers centred at one and error bars giving the size of the total measurement uncertainties.}
\label{fig:finalpred_5TeV_norm}
\end{figure}

\begin{figure}[tbph]
\centering
\subfloat[]{\includegraphics[width=.42\textwidth]{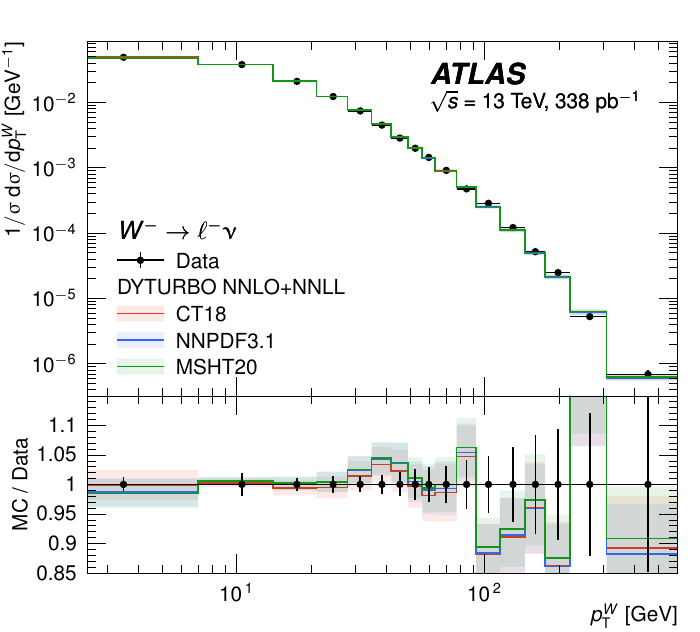}}%
~~~~%
\subfloat[]{\includegraphics[width=.42\textwidth]{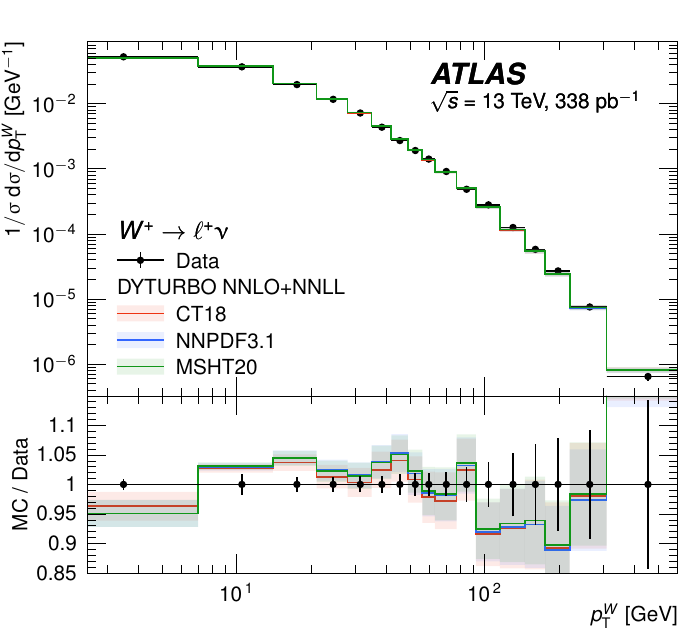}}

\subfloat[]{\includegraphics[width=.42\textwidth]{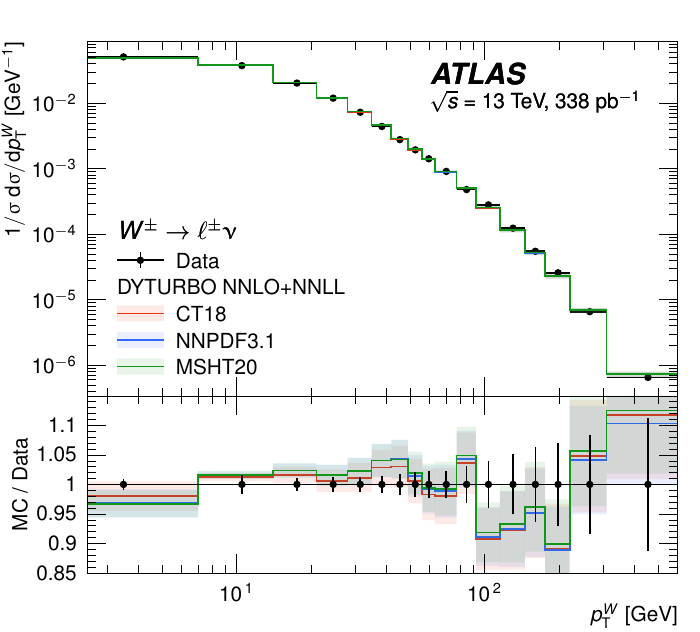}}%
~~~~%
\subfloat[]{\includegraphics[width=.42\textwidth]{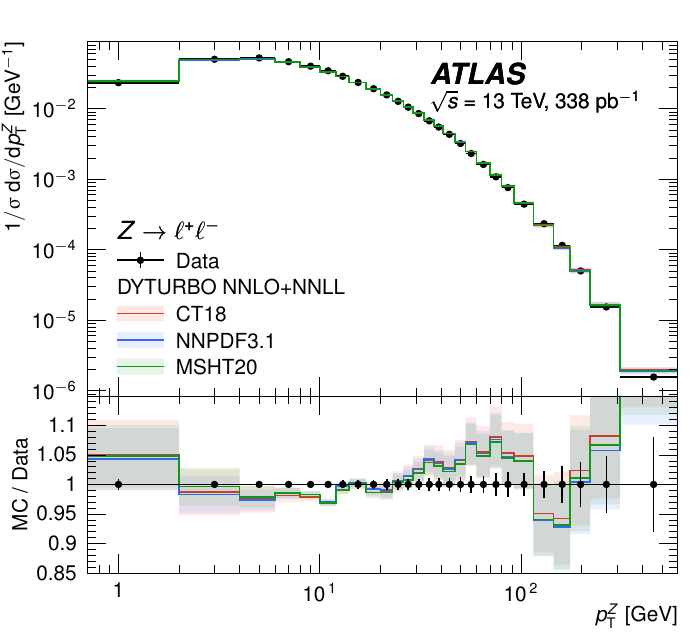}}

\subfloat[]{\includegraphics[width=.42\textwidth]{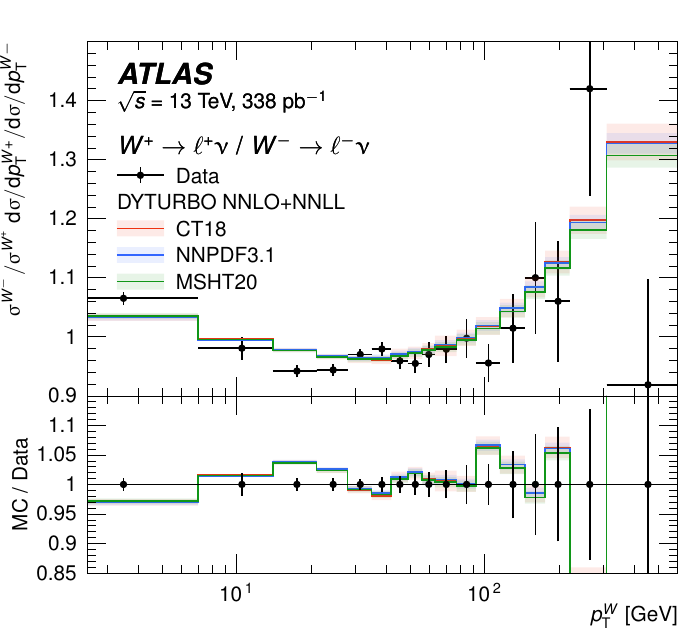}}%
~~~~%
\subfloat[]{\includegraphics[width=.42\textwidth]{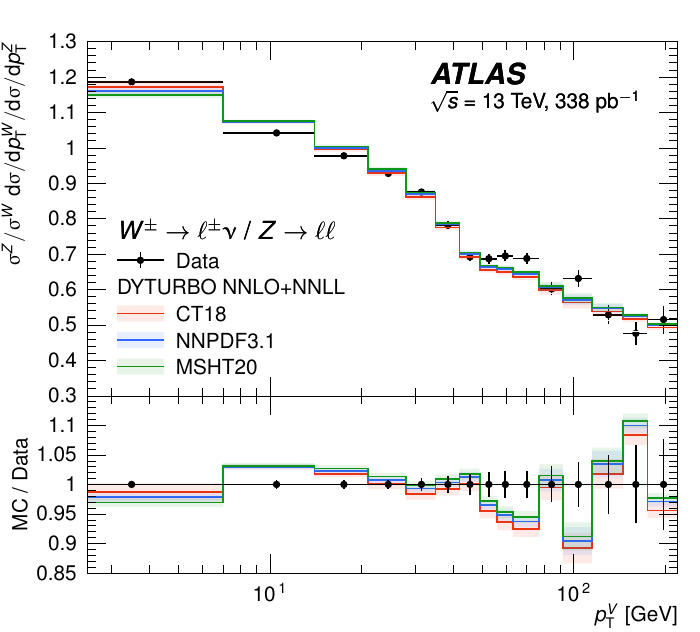}}
\caption{Measurements of normalised differential distributions at
$\sqrt{s} = 13\,\TeV$ (black points) for(a) $W^-$, (b) $W^+$, (c) the sum $W^\pm$, (d) $Z$ as
well as the ratios (e) $W^+/W^-$ and (f) $W^\pm/Z$ compared to \textsc{DYTurbo}
predictions with different PDF sets (coloured lines and PDF and scale uncertainties as shaded band) as described in the
text. The lower panels show the ratio of prediction to data with
data markers centred at one and error bars giving the size of the total measurement uncertainties.}
\label{fig:finalpred_13TeV_norm_dyturbo}
\end{figure}

\begin{figure}[tbph]
\centering
\subfloat[]{\includegraphics[width=.42\textwidth]{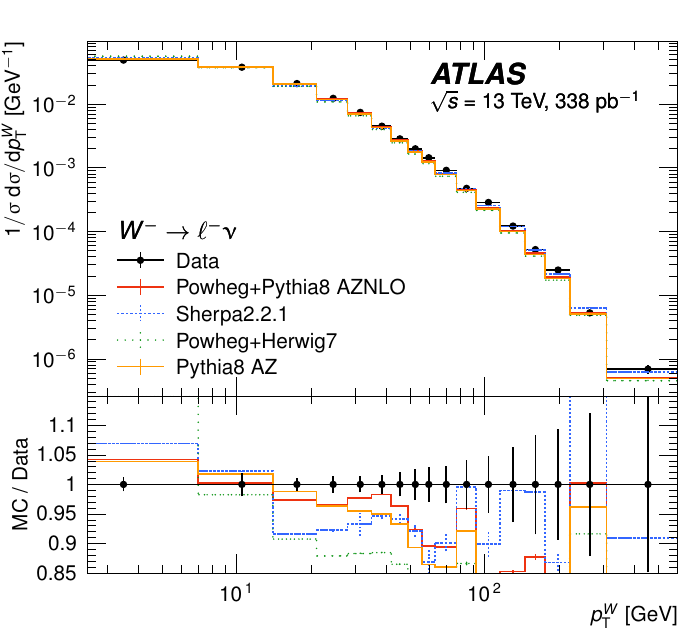}}%
~~~~%
\subfloat[]{\includegraphics[width=.42\textwidth]{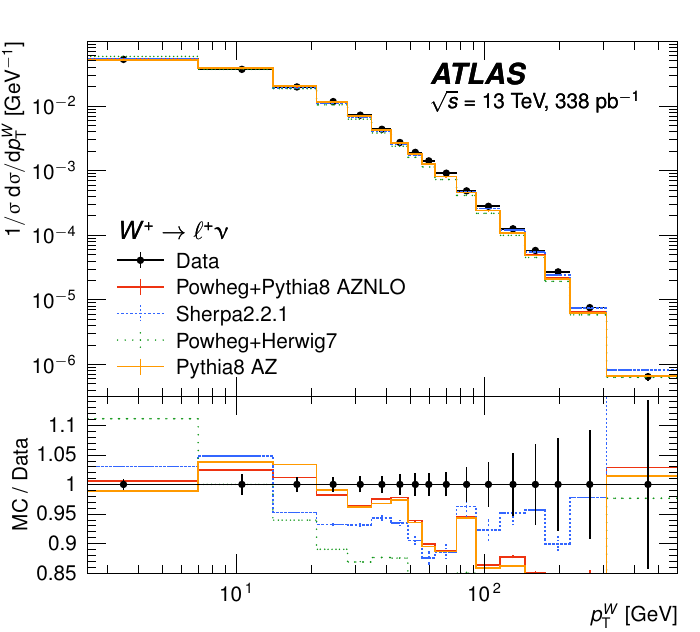}}

\subfloat[]{\includegraphics[width=.42\textwidth]{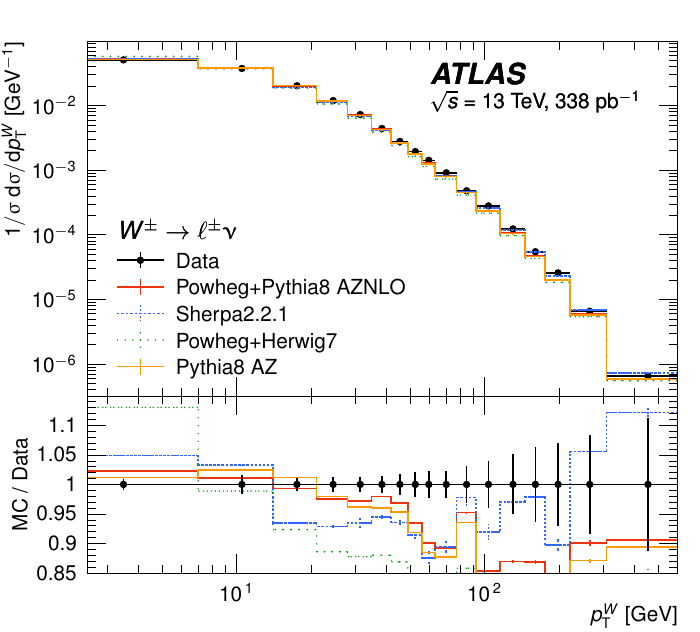}}%
~~~~%
\subfloat[]{\includegraphics[width=.42\textwidth]{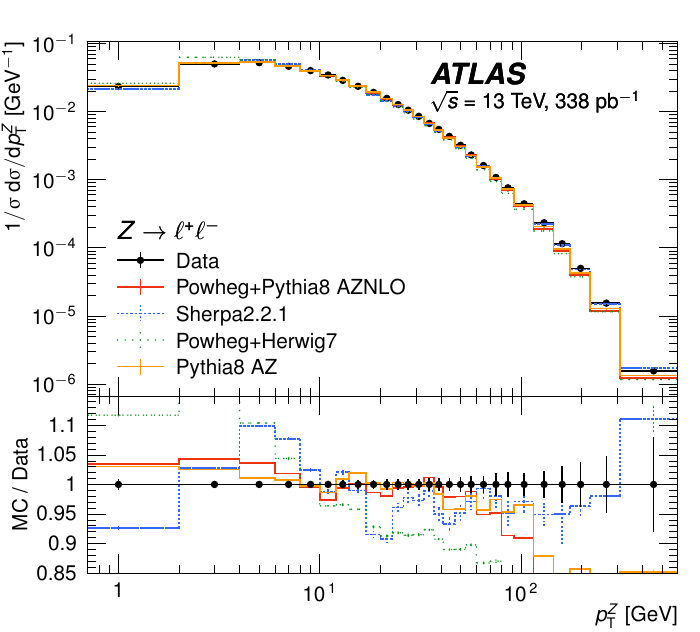}}

\subfloat[]{\includegraphics[width=.42\textwidth]{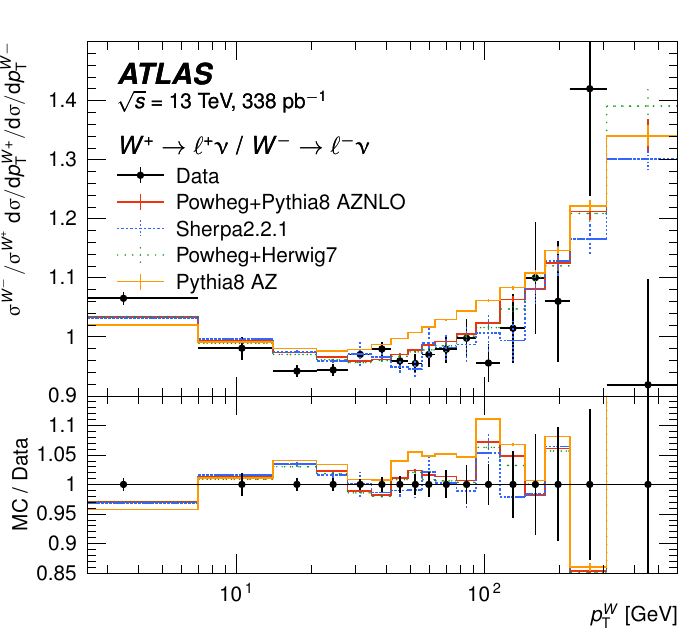}}%
~~~~%
\subfloat[]{\includegraphics[width=.42\textwidth]{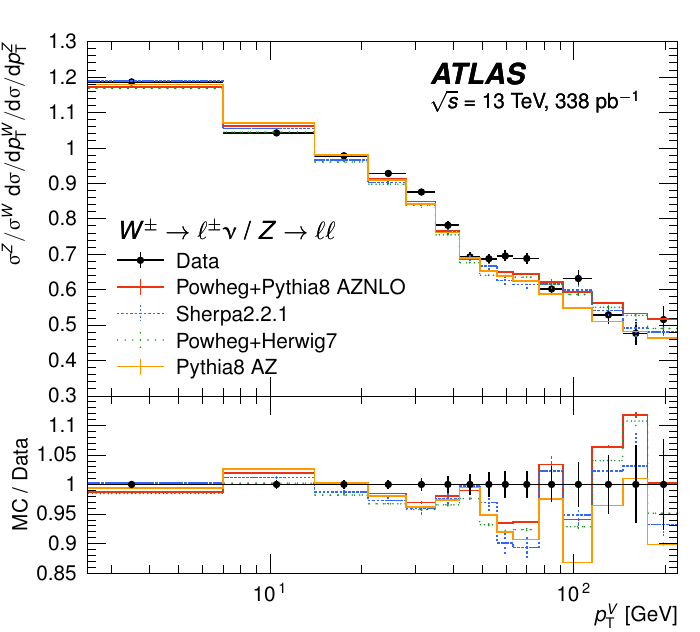}}
\caption{Measurements of normalised differential distributions at
$\sqrt{s} = 13\,\TeV$ (black points) for (a) $W^-$, (b) $W^+$, (c) the sum $W^\pm$, (d) $Z$ as
well as the ratios (e) $W^+/W^-$ and (f) $W^\pm/Z$ compared to a
variety of MC predictions (coloured lines) as described in the
text. The lower panels show the ratio of prediction to data with
data markers centred at one and error bars giving the size of the total measurement uncertainties.}
\label{fig:finalpred_13TeV_norm}
\end{figure}

To separate overall normalisation and shape effects, all
differential comparisons are shown using normalised spectra.
Figures~\ref{fig:finalpred_5TeV_norm_dyturbo} and
\ref{fig:finalpred_5TeV_norm} compare the measurements and some ratios
at $\sqrt{s} = \fivetev$ to \textsc{DYTurbo} and the MC predictions,
respectively. The equivalent comparisons at $\sqrt{s} = 13\,\tev$
are given in Figures~\ref{fig:finalpred_13TeV_norm_dyturbo} and
\ref{fig:finalpred_13TeV_norm}.

\begin{figure}[tbph]
\centering
\subfloat[]{\includegraphics[width=.42\textwidth]{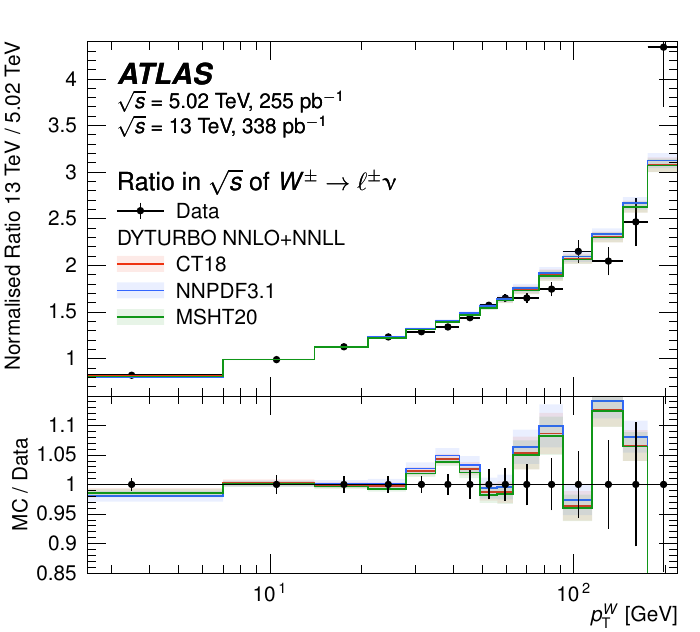}}%
~~~~%
\subfloat[]{\includegraphics[width=.42\textwidth]{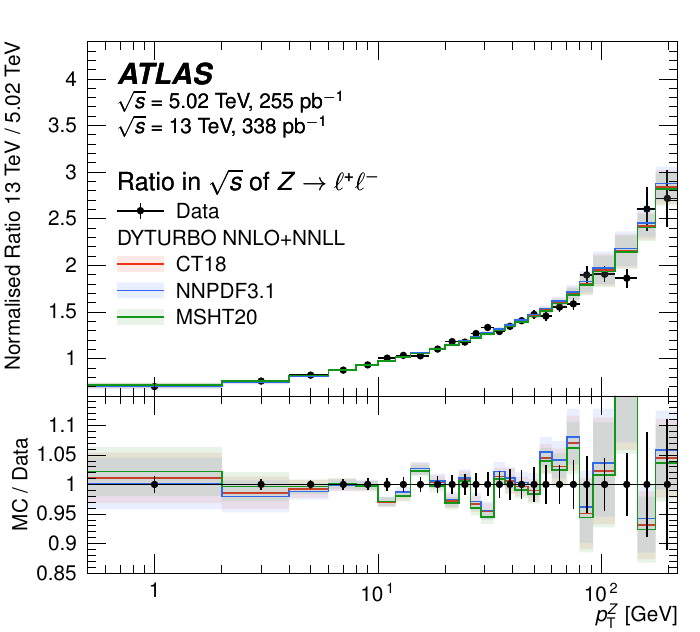}}

\subfloat[]{\includegraphics[width=.42\textwidth]{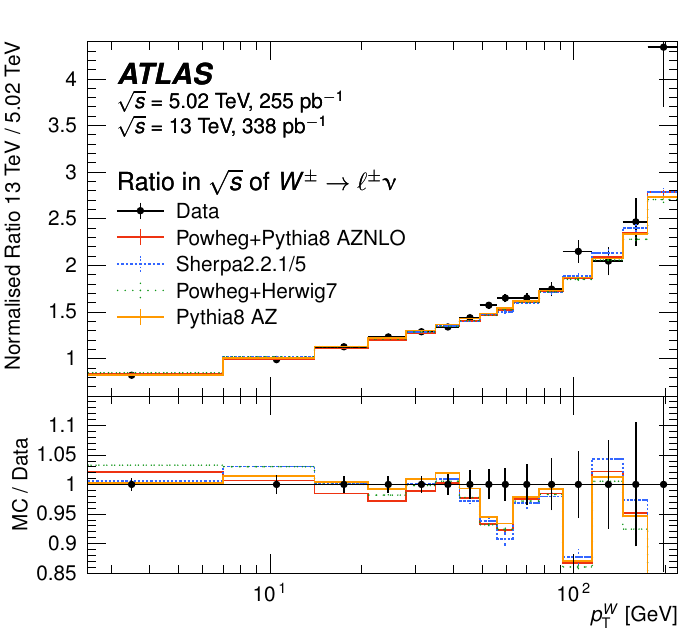}}%
~~~~%
\subfloat[]{\includegraphics[width=.42\textwidth]{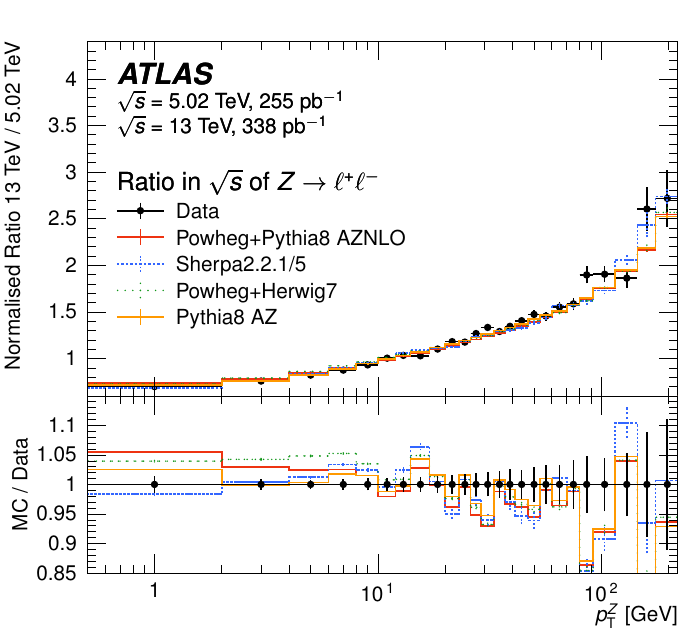}}

\caption{Ratios of normalised differential distributions at
$\sqrt{s} = 13\,\TeV$ to $\sqrt{s} = \fivetev$ (black points) for
(a,c) $W^\pm$ and (b,d) $Z$ compared to (a,b) \textsc{DYTurbo} predictions
with different PDF sets or (b,d) a variety of MC predictions
(coloured lines, PDF and scale uncertainties as shaded band for \textsc{DYTurbo}) as described in the text. The lower panels show
the ratio of prediction to data with data markers centred at one
and error bars giving the size of the total measurement
uncertainties.}
\label{fig:finalpred_513TeV_norm}
\end{figure}

As expected, many features are broadly observed across all final states
and in both $\sqrt{s}$ datasets.
First, the data is precise and discriminating up to about
$\ptv=100\,\GeV$ at \fivetev and $\ptv=200\,\GeV$ at $13\,\TeV$.
The \textsc{DYTurbo} resummed predictions show the best agreement across the
spectrum and generally match the data at the percent level.
When focusing on the lower $\ptv < 30\,\gev$ region, the $W^+$ measurements show overall the largest
deviation from the \textsc{DYTurbo} predictions.
The three investigated PDF sets produce only small differences in the shapes of the distributions,
but theoretical uncertainties at higher \ptv are significant, dominated by the scale variations.

The MC predictions that were tuned to $\sqrt{s} = 7\,\tev$ data,
\Powheg+\PYTHIAV{8} AZNLO and \PYTHIAV{8} AZ, describe reasonably well
the low-\ptv region that is relevant for the measurement of the
\Wboson-boson mass. However, they show worse agreement for $\ptv > 40\,\GeV$.  The
\Powheg+\HERWIGV{7} predictions does not perform well across the whole
spectrum. The \SHERPA[2.2.1/2.2.5] predictions match the data best at
higher \ptv, but they deviate significantly in the region
$\ptv \lesssim 20\,\GeV$, a behaviour that was improved in
\SHERPA[2.2.11] by optimising the matching
conditions~\cite{PMGR-2021-01}, see also Appendix~\ref{sec:mjcomp}.

The ratios $W^+ / W^-$ and $W^\pm/Z$ are measured to percent-level
precision and show some interesting features at both $\sqrt{s}$ values.
The differences between the predictions and uncertainties shrink in these ratios.
The \PYTHIAV{8} AZ
prediction is clearly ruled out by the data in the $W^+/W^-$ ratio.
However, this is an effect of mismodelled angular decay distributions
and thus the fiducial acceptance. The ratios at $\sqrt{s} = 13\,\gev$ show the largest
differences between data and predictions at lower $\ptv < 30\,\gev$.

The ratios of normalised differential distributions at
$\sqrt{s} = 13\,\TeV$ to $\sqrt{s} = \fivetev$ are shown in
Figure~\ref{fig:finalpred_513TeV_norm}. Here \textsc{DYTurbo} is in excellent
agreement with the data, while several MC predictions fail to follow
the energy dependence in the low \ptv\ region correctly.



%
%
\FloatBarrier

%
\section{Conclusion}
\label{sec:conclusion}

This paper
presented an extensive set of measurements of $W$- and $Z$-boson
\pT{} distributions at two centre-of-mass energies:
$\sqrt{s} = \fivetev$ and $\sqrt{s} = 13\,\tev$. The measurements
were performed with $pp$ collision data delivered by the LHC and
recorded by the ATLAS detector in special low pile-up conditions in
2017 and 2018.

While the \ptz distribution had been measured in great
detail previously using the dilepton system produced in $Z$-boson decays, the direct measurement of
the \ptw distribution relies on the full reconstruction of
the hadronic recoil in $W$-boson events. The dedicated datasets, reconstruction and
calibration techniques allowed an unprecedented granularity of
about $7\,\gev$ in \ptw and a final precision at the
level of $1-2\%$. All measurements have been performed in the electron and
muon decays of the bosons and the results are combined.

A unique feature of the analysis presented is the wide coverage of six
high-precision absolute and normalised cross section distributions for
the three vector boson final states $W^-, W^+$ and $Z$ at two
well separated centre-of-mass energies. In addition, integrated fiducial cross sections and a
variety of ratios are presented. The integrated results in particular
profit from the very precise luminosity measurement with an
uncertainty of $0.9-1.0\%$, leading to improvements in the precision
compared to existing results by a factor of two or more.
The integrated results and ratios are generally compatible with
\textsc{DYTurbo} predictions at order NNLO+NNLL.

The comparison of differential spectra to different MC simulations shows a variety of deficiencies in
the description of the transverse momentum distribution; many
are common for all vector bosons. The agreement is better at
$\sqrt{s} = \fivetev$, especially for those predictions that were
tuned to  $Z$-boson measurements at $\sqrt{s} = 7\,\tev$. The ratios show
generally a better agreement as expected due to the similarity in the
processes. The higher-order resummed predictions from
\textsc{DYTurbo} match the data best across the spectrum. All predictions show
similar disagreements with the data in the $W^+/W^-$ and $W^\pm/Z$ ratios, especially at
$\sqrt{s} = 13\,\tev$.

These detailed measurements are of great importance for future
measurements of the $W$-boson mass that rely on a good modelling of
the \ptw spectrum. The results can be used to test and constrain
various theoretical uncertainties affecting the predictions. Different contributions
may be understood from the different ranges of parton momentum fractions and initial parton
flavours that are probed by $W^-, W^+$ and $Z$ production at $\sqrt{s} = \fivetev$ and $\sqrt{s} = 13\,\tev$.


%
\section*{Acknowledgements}
%

%
%

This paper is dedicated to the memory of Max Klein, a leading physicist, humanist, and advocate of international collaboration.

We thank CERN for the very successful operation of the LHC and its injectors, as well as the support staff at
CERN and at our institutions worldwide without whom ATLAS could not be operated efficiently.

The crucial computing support from all WLCG partners is acknowledged gratefully, in particular from CERN, the ATLAS Tier-1 facilities at TRIUMF/SFU (Canada), NDGF (Denmark, Norway, Sweden), CC-IN2P3 (France), KIT/GridKA (Germany), INFN-CNAF (Italy), NL-T1 (Netherlands), PIC (Spain), RAL (UK) and BNL (USA), the Tier-2 facilities worldwide and large non-WLCG resource providers. Major contributors of computing resources are listed in Ref.~\cite{ATL-SOFT-PUB-2023-001}.

We gratefully acknowledge the support of ANPCyT, Argentina; YerPhI, Armenia; ARC, Australia; BMWFW and FWF, Austria; ANAS, Azerbaijan; CNPq and FAPESP, Brazil; NSERC, NRC and CFI, Canada; CERN; ANID, Chile; CAS, MOST and NSFC, China; Minciencias, Colombia; MEYS CR, Czech Republic; DNRF and DNSRC, Denmark; IN2P3-CNRS and CEA-DRF/IRFU, France; SRNSFG, Georgia; BMBF, HGF and MPG, Germany; GSRI, Greece; RGC and Hong Kong SAR, China; ISF and Benoziyo Center, Israel; INFN, Italy; MEXT and JSPS, Japan; CNRST, Morocco; NWO, Netherlands; RCN, Norway; MEiN, Poland; FCT, Portugal; MNE/IFA, Romania; MESTD, Serbia; MSSR, Slovakia; ARIS and MVZI, Slovenia; DSI/NRF, South Africa; MICINN, Spain; SRC and Wallenberg Foundation, Sweden; SERI, SNSF and Cantons of Bern and Geneva, Switzerland; NSTC, Taipei; TENMAK, T\"urkiye; STFC/UKRI, United Kingdom; DOE and NSF, United States of America.

Individual groups and members have received support from BCKDF, CANARIE, CRC and DRAC, Canada; CERN-CZ, PRIMUS 21/SCI/017 and UNCE SCI/013, Czech Republic; COST, ERC, ERDF, Horizon 2020, ICSC-NextGenerationEU and Marie Sk{\l}odowska-Curie Actions, European Union; Investissements d'Avenir Labex, Investissements d'Avenir Idex and ANR, France; DFG and AvH Foundation, Germany; Herakleitos, Thales and Aristeia programmes co-financed by EU-ESF and the Greek NSRF, Greece; BSF-NSF and MINERVA, Israel; Norwegian Financial Mechanism 2014-2021, Norway; NCN and NAWA, Poland; La Caixa Banking Foundation, CERCA Programme Generalitat de Catalunya and PROMETEO and GenT Programmes Generalitat Valenciana, Spain; G\"{o}ran Gustafssons Stiftelse, Sweden; The Royal Society and Leverhulme Trust, United Kingdom.

In addition, individual members wish to acknowledge support from Chile: Agencia Nacional de Investigaci\'on y Desarrollo (FONDECYT 1190886, FONDECYT 1210400, FONDECYT 1230812, FONDECYT 1230987); China: National Natural Science Foundation of China (NSFC - 12175119, NSFC 12275265, NSFC-12075060); Czech Republic: PRIMUS Research Programme (PRIMUS/21/SCI/017); EU: H2020 European Research Council (ERC - 101002463); European Union: European Research Council (ERC - 948254), Horizon 2020 Framework Programme (MUCCA - CHIST-ERA-19-XAI-00), European Union, Future Artificial Intelligence Research (FAIR-NextGenerationEU PE00000013), Italian Center for High Performance Computing, Big Data and Quantum Computing (ICSC, NextGenerationEU), Marie Sklodowska-Curie Actions (EU H2020 MSC IF GRANT NO 101033496); France: Agence Nationale de la Recherche (ANR-20-CE31-0013, ANR-21-CE31-0013, ANR-21-CE31-0022, ANR-22-EDIR-0002), Investissements d'Avenir Idex (ANR-11-LABX-0012), Investissements d'Avenir Labex (ANR-11-LABX-0012); Germany: Baden-Württemberg Stiftung (BW Stiftung-Postdoc Eliteprogramme), Deutsche Forschungsgemeinschaft (DFG - CR 312/5-1); Italy: Istituto Nazionale di Fisica Nucleare (FELLINI G.A. n. 754496, ICSC, NextGenerationEU); Japan: Japan Society for the Promotion of Science (JSPS KAKENHI JP21H05085, JSPS KAKENHI JP22H01227, JSPS KAKENHI JP22H04944); Netherlands: Netherlands Organisation for Scientific Research (NWO Veni 2020 - VI.Veni.202.179); Norway: Research Council of Norway (RCN-314472); Poland: Polish National Agency for Academic Exchange (PPN/PPO/2020/1/00002/U/00001), Polish National Science Centre (NCN 2021/42/E/ST2/00350, NCN UMO-2019/34/E/ST2/00393, UMO-2020/37/B/ST2/01043, UMO-2021/40/C/ST2/00187); Slovenia: Slovenian Research Agency (ARIS grant J1-3010); Spain: BBVA Foundation (LEO22-1-603), Generalitat Valenciana (Artemisa, FEDER, IDIFEDER/2018/048), La Caixa Banking Foundation (LCF/BQ/PI20/11760025), Ministry of Science and Innovation (MCIN \& NextGenEU PCI2022-135018-2, MICIN \& FEDER PID2021-125273NB, RYC2019-028510-I, RYC2020-030254-I, RYC2021-031273-I, RYC2022-038164-I), PROMETEO and GenT Programmes Generalitat Valenciana (CIDEGENT/2019/023, CIDEGENT/2019/027); Sweden: Swedish Research Council (VR 2018-00482, VR 2022-03845, VR 2022-04683, VR grant 2021-03651), Knut and Alice Wallenberg Foundation (KAW 2017.0100, KAW 2018.0157, KAW 2018.0458, KAW 2019.0447); Switzerland: Swiss National Science Foundation (SNSF - PCEFP2\_194658); United Kingdom: Leverhulme Trust (Leverhulme Trust RPG-2020-004); United States of America: U.S. Department of Energy (ECA DE-AC02-76SF00515), Neubauer Family Foundation.

%
%


%
\clearpage
\appendix
\part*{Appendix}
\addcontentsline{toc}{part}{Appendix}

\section{Measurement uncertainty breakdown for all channels}
\label{sec:uncbreakdown}

\begin{figure}[tbph]
\centering
\subfloat[]{\includegraphics[width=.47\textwidth]{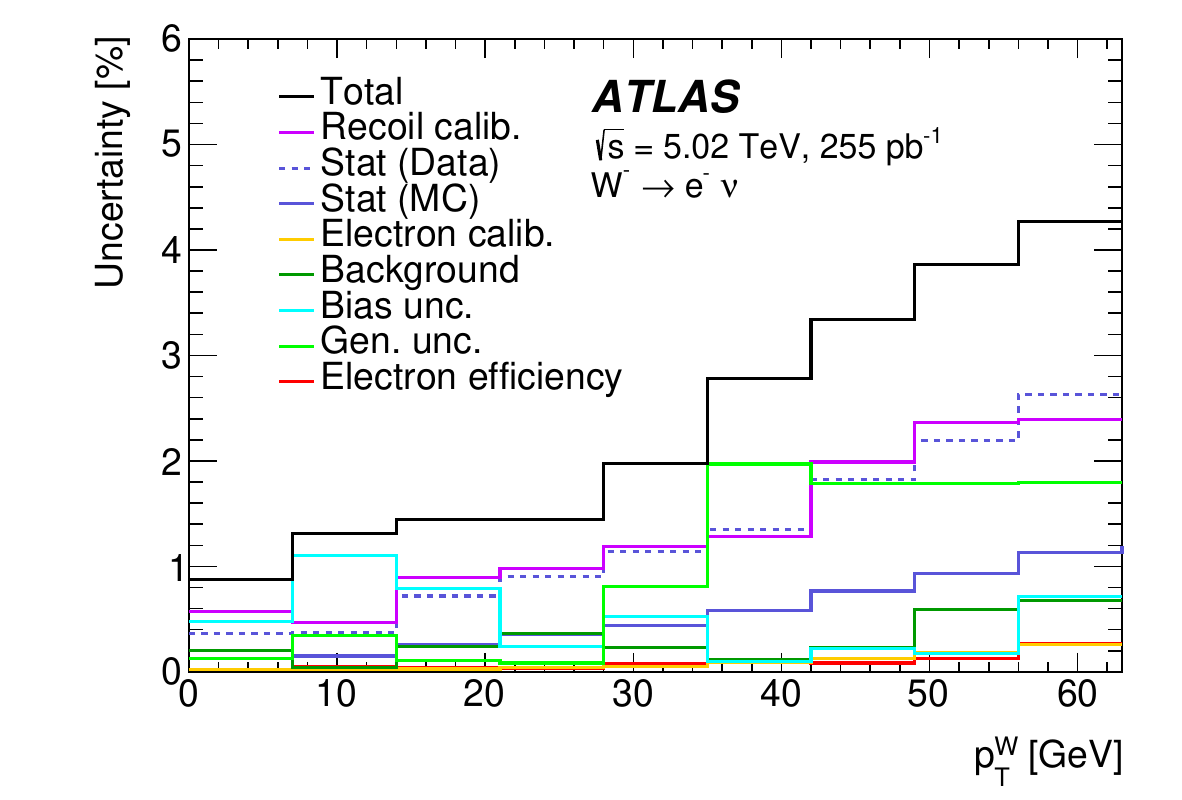}}
\subfloat[]{\includegraphics[width=.47\textwidth]{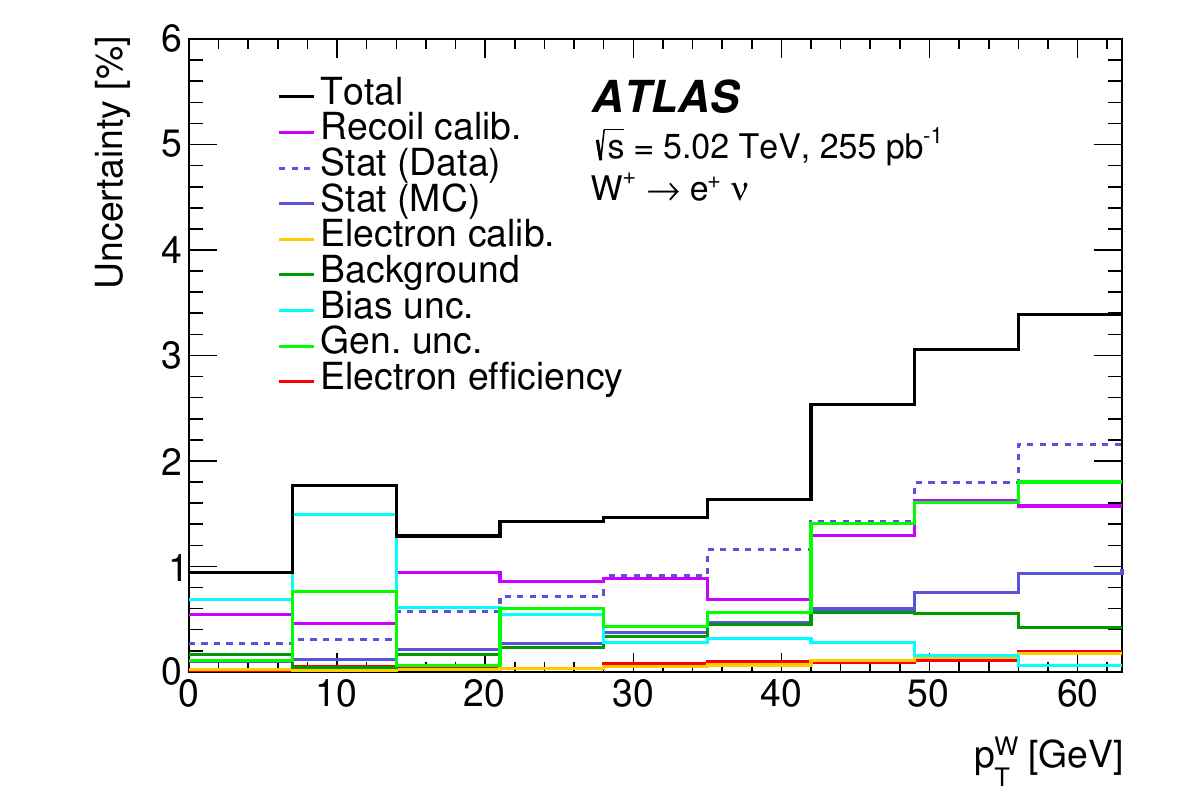}}

\subfloat[]{\includegraphics[width=.47\textwidth]{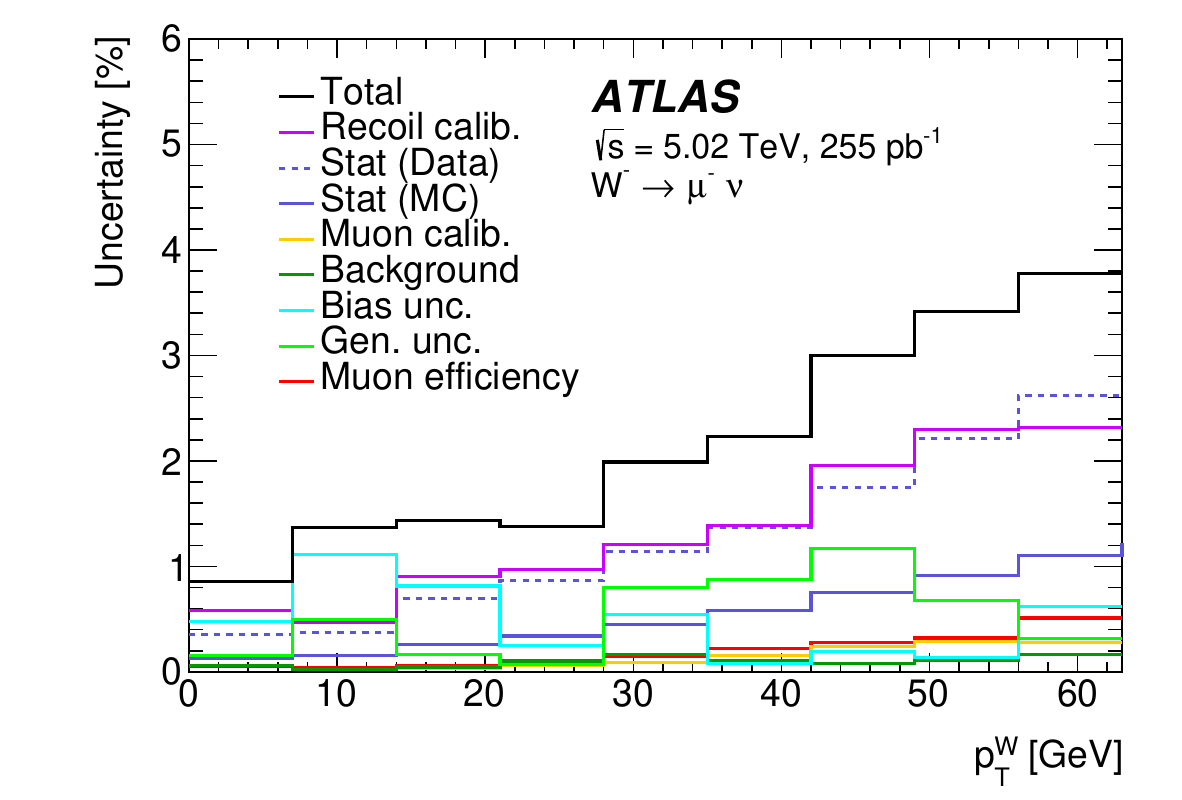}}
\subfloat[]{\includegraphics[width=.47\textwidth]{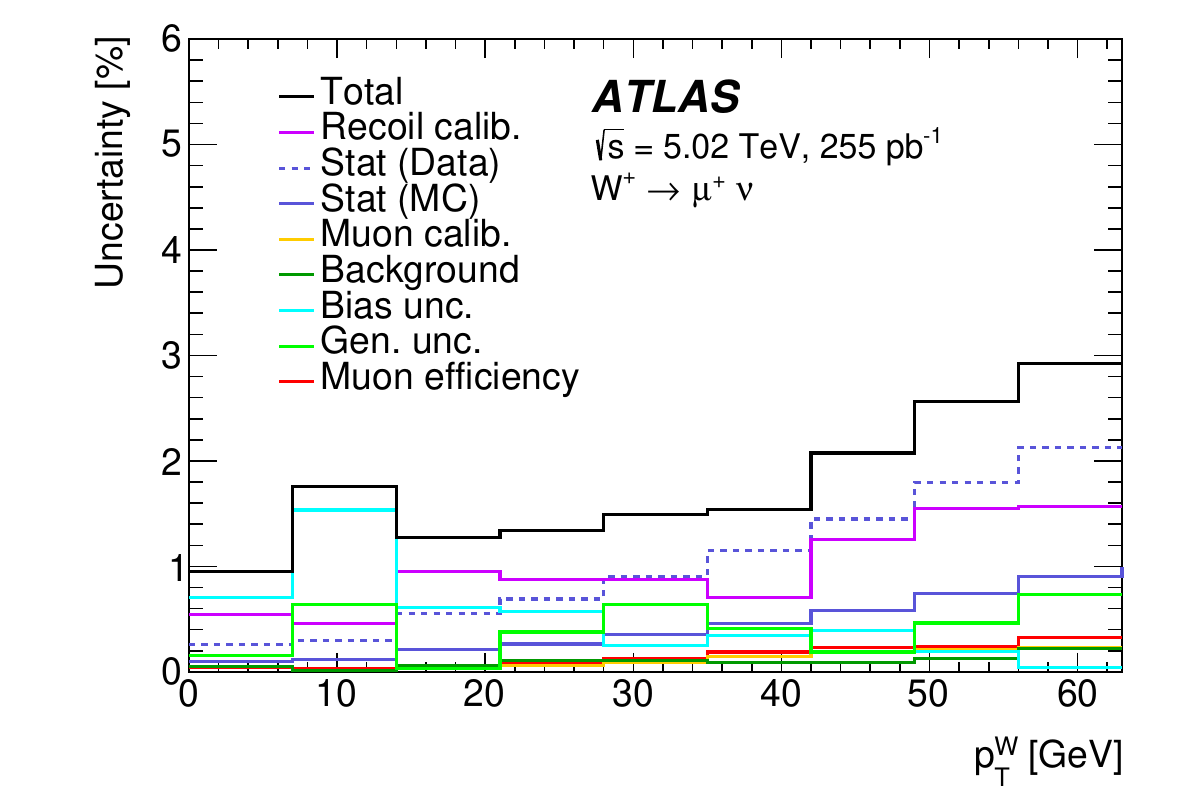}}
\caption{Uncertainty contributions in the normalised \ptw\ distribution in the (a) \Wmenu, (b) \Wpenu, (c) \Wmmunu and (d) \Wpmunu channels for the $\sqrt{s} = \fivetev$ dataset, focusing on the low $\ptw <63\,\GeV$ region. The
‘Background’ component refers to the sum of uncertainties in the normalisation of the simulated background samples and
the uncertainties in the yield and shape of the multijet background estimate. The `Electron efficiency' and `Muon efficiency' components refer to the total uncertainty in the lepton efficiency scale factors : electron trigger, reconstruction, identification, isolation (electron channel) and muon trigger, identification, track-to-vertex association, isolation (muon channel).
The uncertainties are shown as a percentage of the unfolded data per \pT bin.}
\label{fig:unf_sys_ShapeUnc_5}
\end{figure}

\begin{figure}[!tbph]
\centering
\subfloat[]{\includegraphics[width=.47\textwidth]{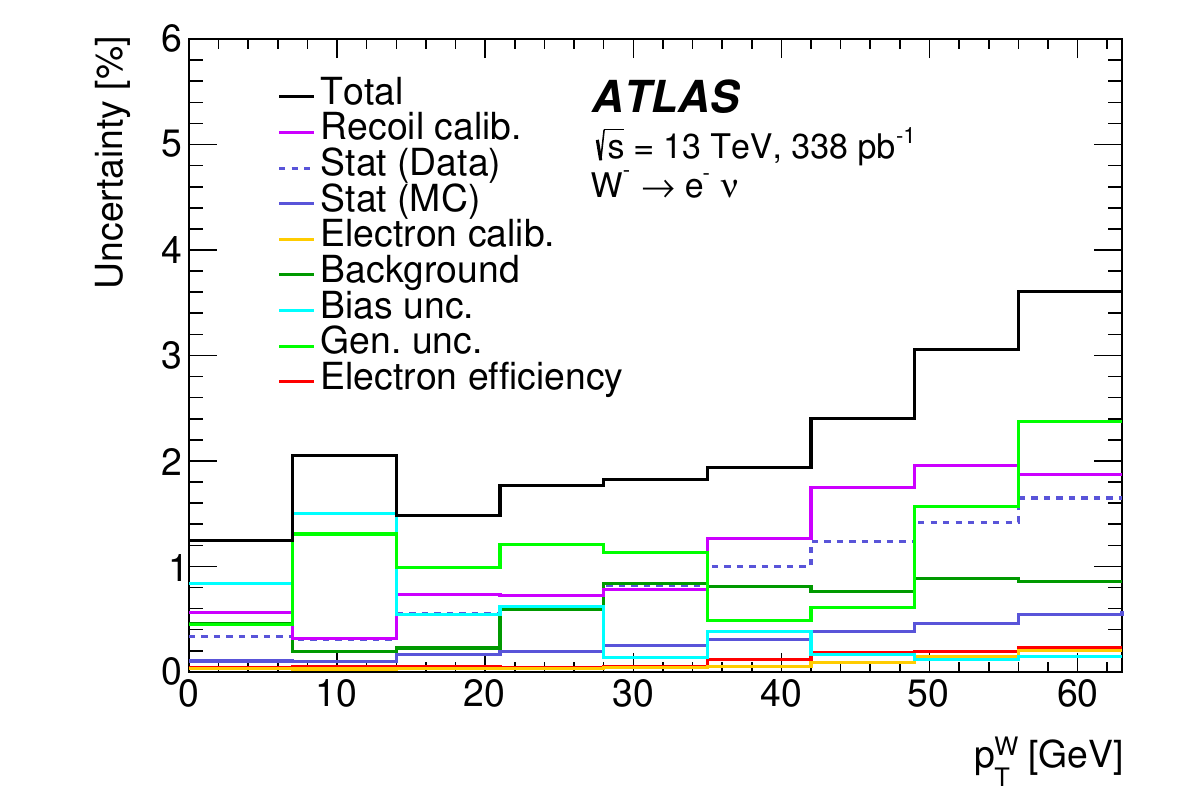}}
\subfloat[]{\includegraphics[width=.47\textwidth]{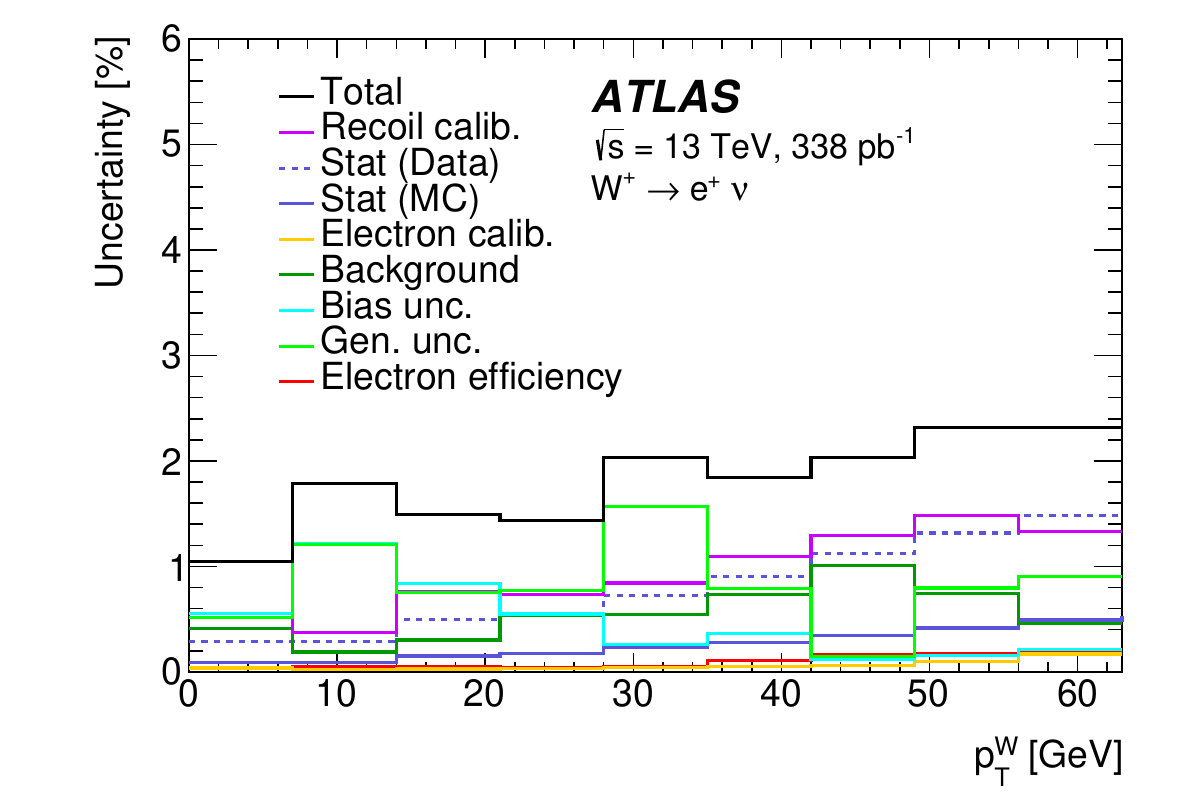}}

\subfloat[]{\includegraphics[width=.47\textwidth]{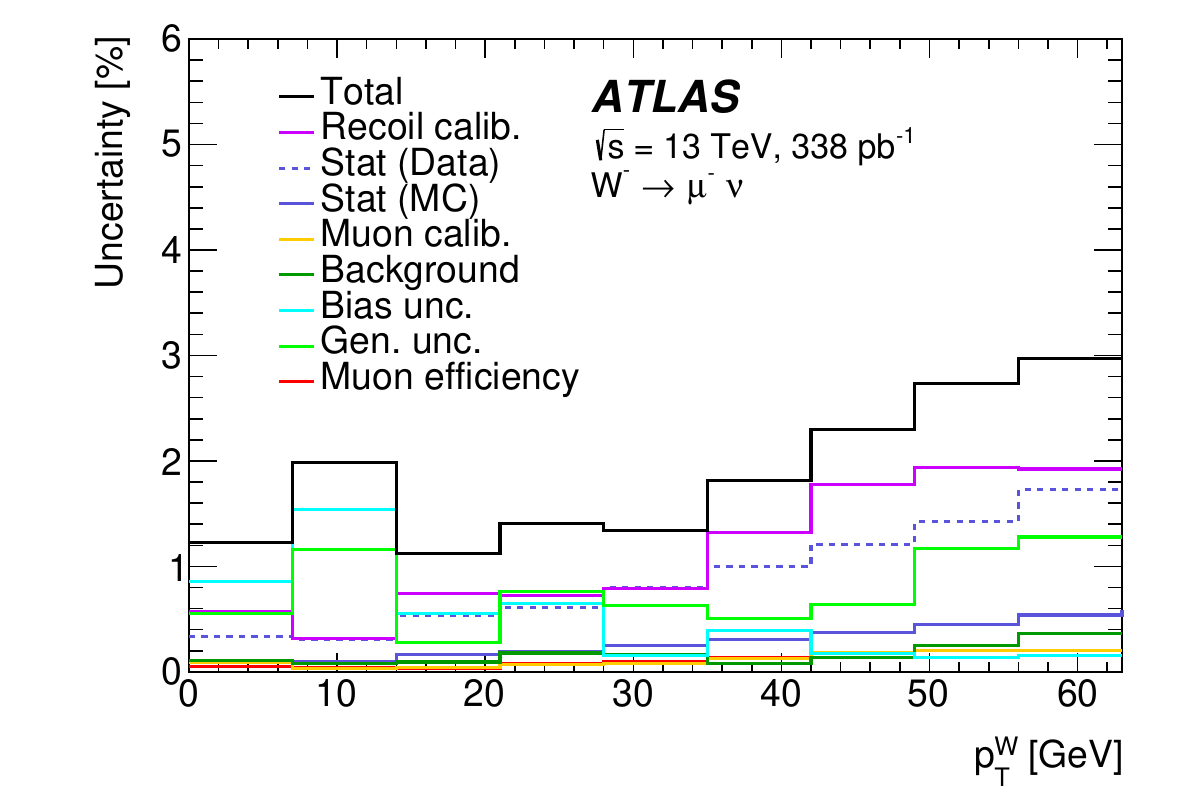}}
\subfloat[]{\includegraphics[width=.47\textwidth]{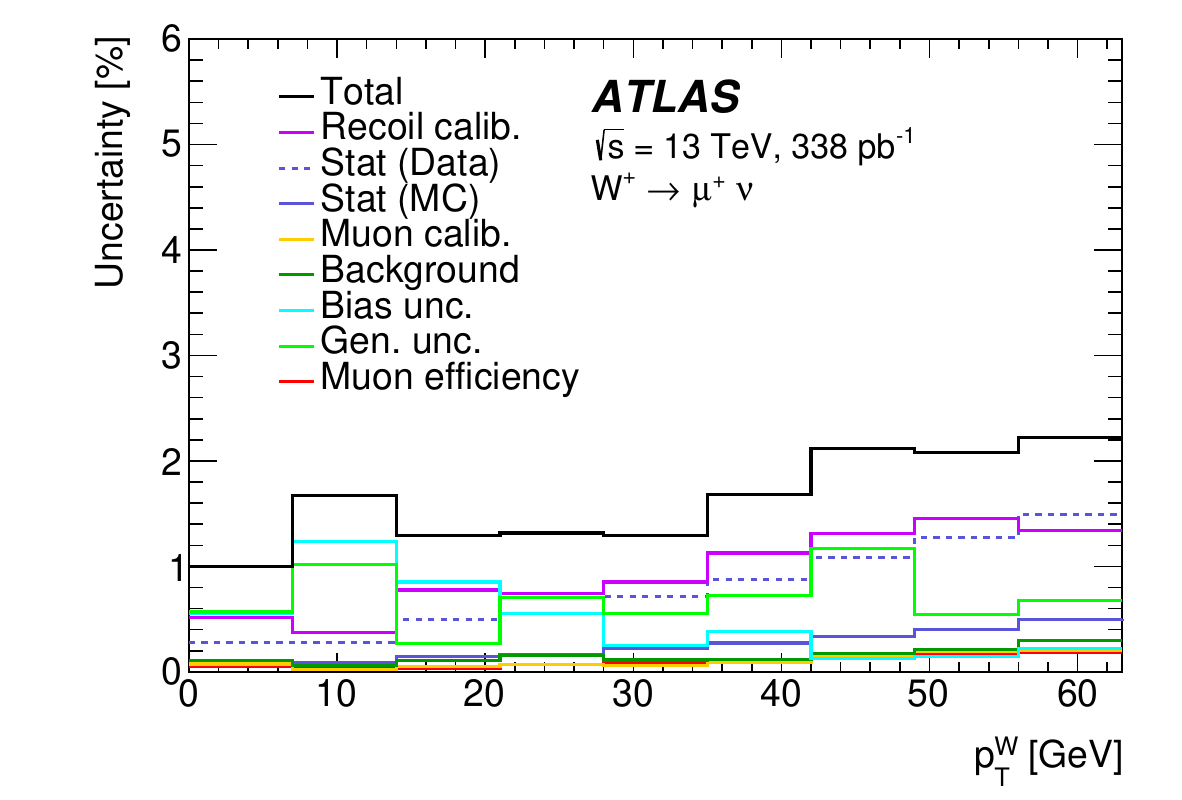}}
\caption{Uncertainty contributions in the normalised \ptw\ distribution in the (a) \Wmenu, (b) \Wpenu, (c) \Wmmunu and (d) \Wpmunu channels for the $\sqrt{s} = 13\,\TeV$ dataset, focusing on the low $\ptw <63\,\GeV$ region. The
‘Background’ component refers to the sum of uncertainties in the normalisation of the simulated background samples and
the uncertainties in the yield and shape of the multijet background estimate. The `Electron efficiency' and `Muon efficiency' components refer to the total uncertainty in the lepton efficiency scale factors : electron trigger, reconstruction, identification, isolation (electron channel) and muon trigger, identification, track-to-vertex association, isolation (muon channel). The uncertainties are shown as a percentage of the unfolded data per \pT bin.}
\label{fig:unf_sys_ShapeUnc_13}
\end{figure}

\begin{figure}[h]
\centering
\subfloat[]{\includegraphics[width=.47\textwidth]{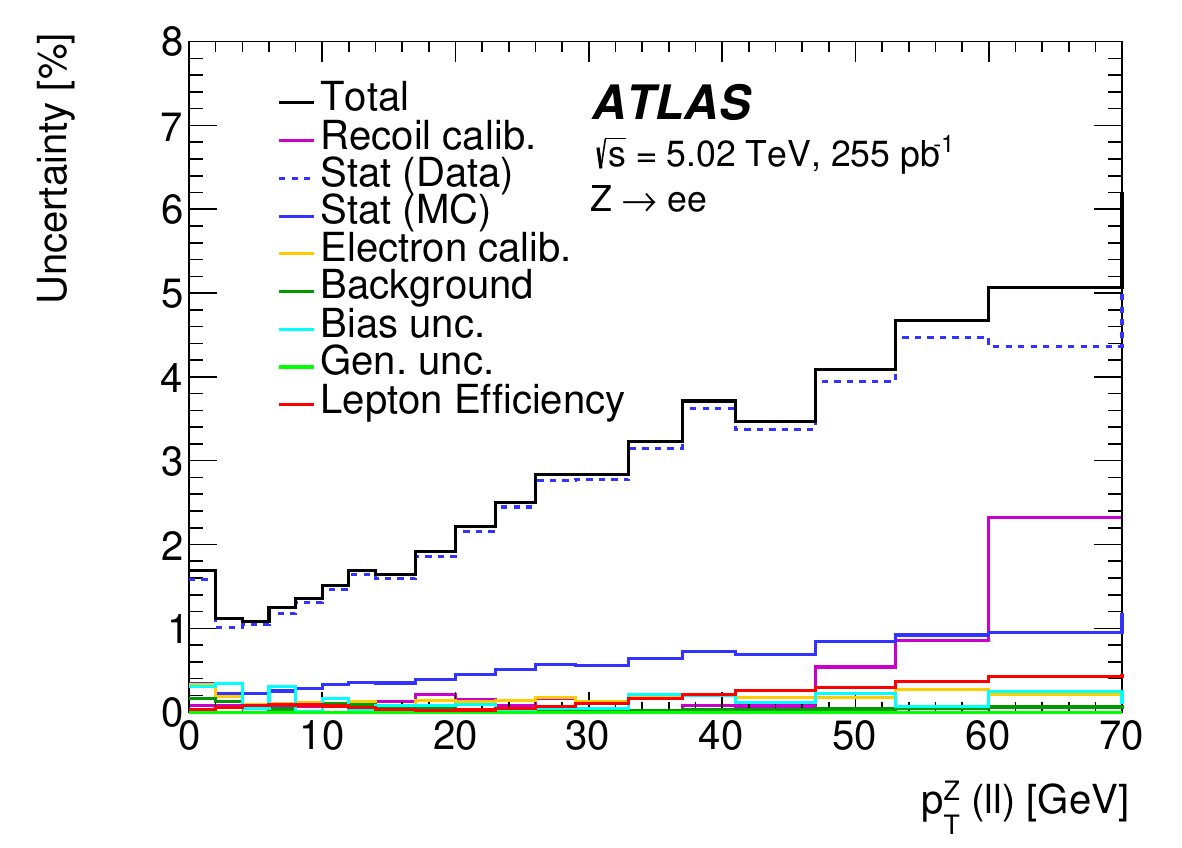}\label{f:AlluncZee_pT5TeV_shape}}
\subfloat[]{\includegraphics[width=.47\textwidth]{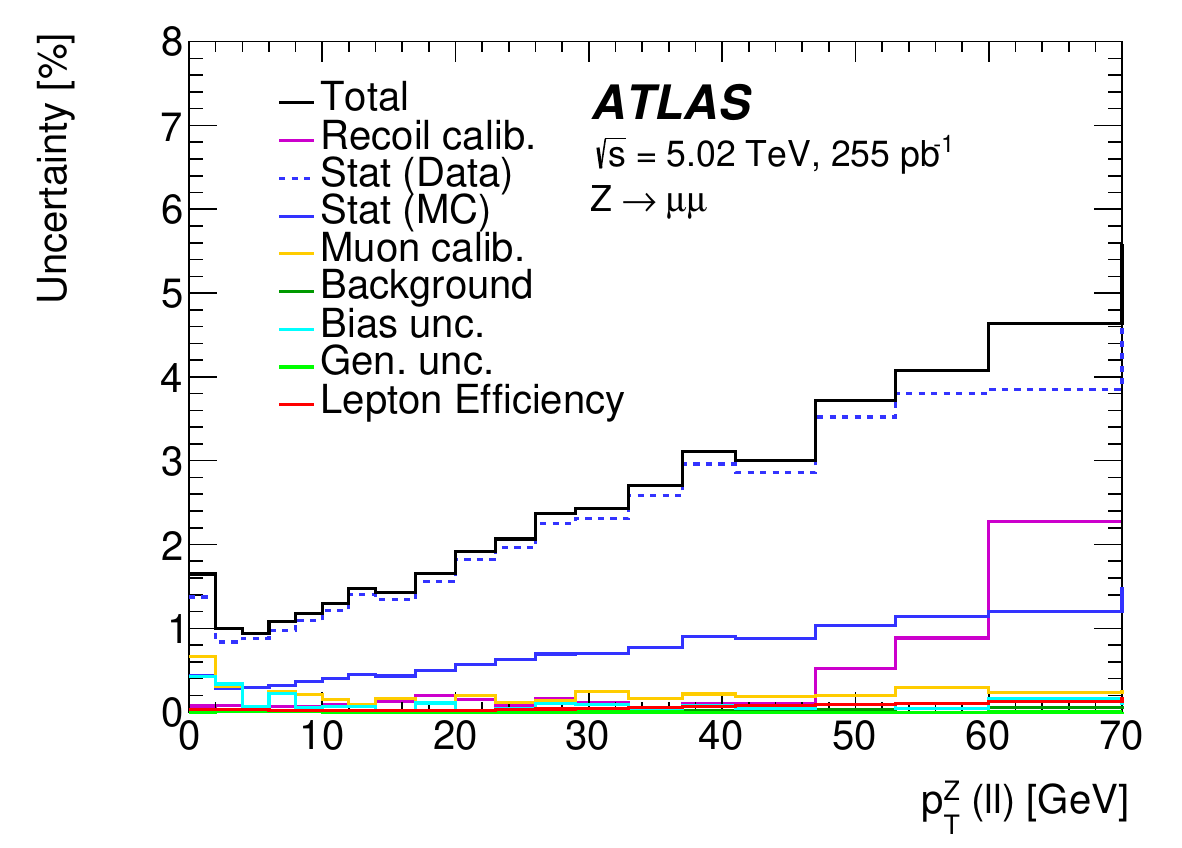}\label{f:AlluncZmm_pT5TeV_shape}}

\caption{Uncertainty contributions in the normalised \ptz\ distribution in the (a) \Zee and (b) \Zmm decay channels for the $\sqrt{s} = \fivetev$ dataset. The
‘Background’ component refers to the sum of uncertainties in the normalisation of the simulated background samples and
the uncertainties in the yield and shape of the multijet background estimate. The `SF tot.' component refers to the total uncertainty in the lepton efficiency scale factors: electron trigger, reconstruction, identification, isolation (electron channel) and muon trigger, identification, track-to-vertex association, isolation (muon channel). The uncertainties are shown as a percentage of the unfolded data per \pT bin.}
\label{f:AlluncZ_5TeV_shape}
\end{figure}

\begin{figure}[h]
\centering
\subfloat[]{\includegraphics[width=.47\textwidth]{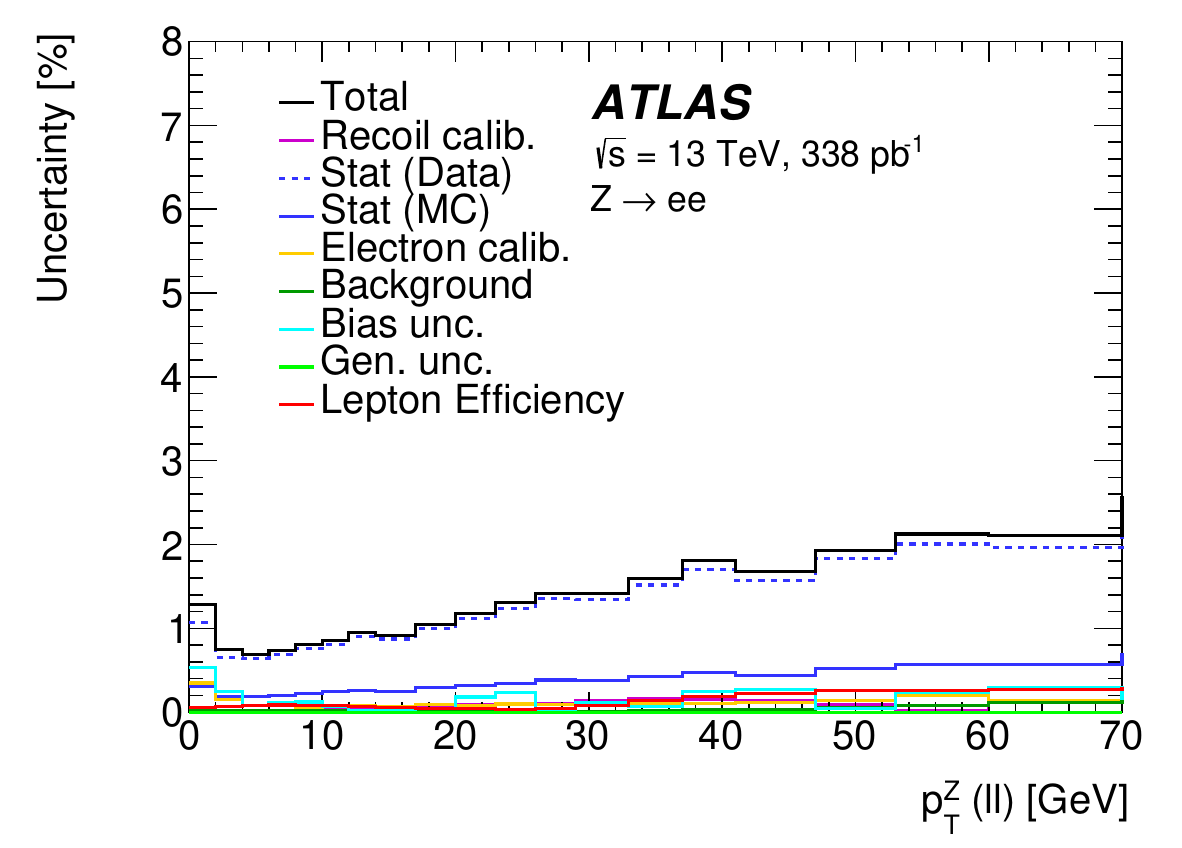}\label{f:AlluncZee_pT13TeV_shape}}
\subfloat[]{\includegraphics[width=.47\textwidth]{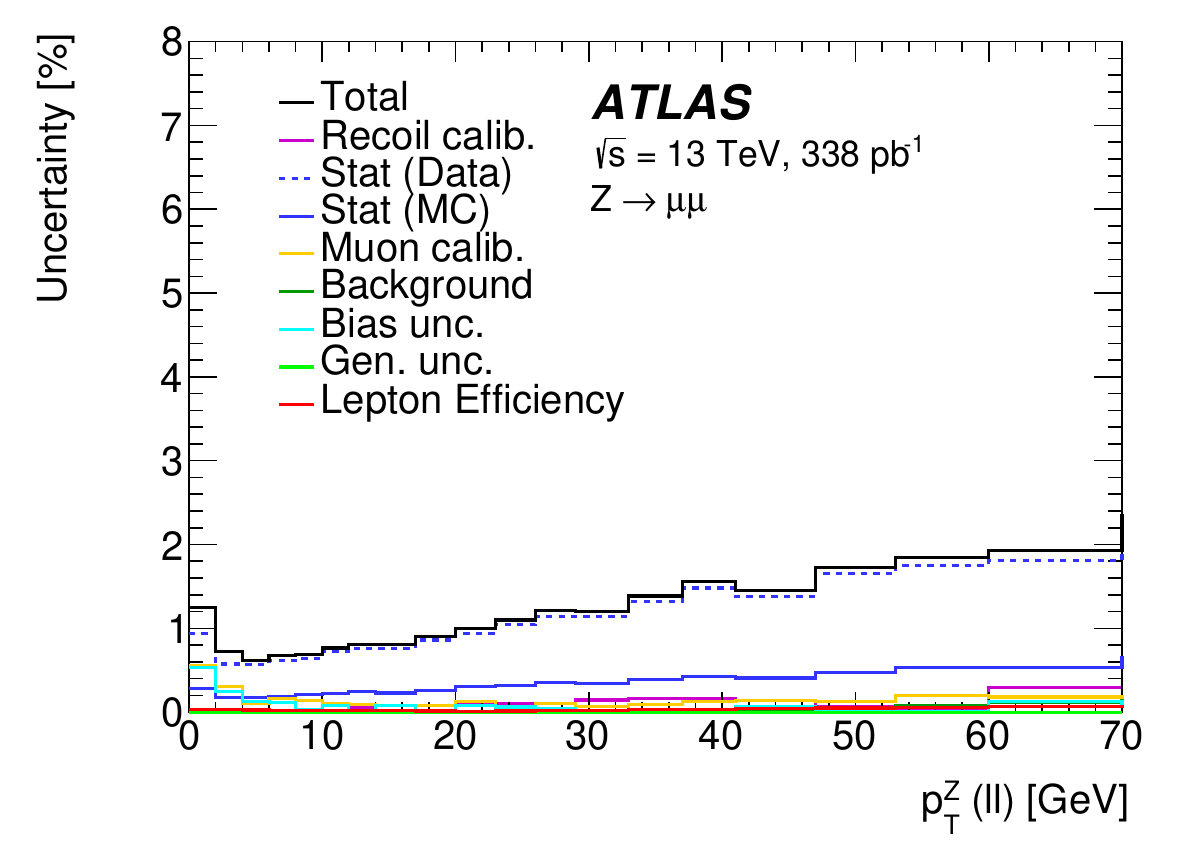}\label{f:AlluncZmm_pT13TeV_shape}}

\caption{Uncertainty contributions in the normalised \ptz\ distribution in the (a) \Zee and (b) \Zmm decay channels for the $\sqrt{s} = 13\,\TeV$ dataset. The
‘Background’ component refers to the sum of uncertainties in the normalisation of the simulated background samples and
the uncertainties in the yield and shape of the multijet background estimate. The `SF tot.' component refers to the total uncertainty in the lepton efficiency scale factors: electron trigger, reconstruction, identification, isolation (electron channel) and muon trigger, identification, track-to-vertex association, isolation (muon channel). The uncertainties are shown as a percentage of the unfolded data per \pT bin.}\label{f:AlluncZ_13TeV_shape}
\end{figure}

The uncertainty contributions in the unfolded, normalised \ptw{}
and \ptz{} spectra for each decay channel are
given in
Figures~\ref{fig:unf_sys_ShapeUnc_5}~(\ref{fig:unf_sys_ShapeUnc_13})
and \ref{f:AlluncZ_5TeV_shape}~(\ref{f:AlluncZ_13TeV_shape}) for
the \Wboson-boson and \Zboson-boson analyses on the \fivetev
($13\,\TeV$) datasets.

\clearpage

\section{Comparisons to \textsc{RadISH+NNLOjet} predictions and \textsc{DYTurbo} uncertainty composition}
\label{sec:radishcomp}

\begin{figure}[tbph]
\centering
\subfloat[]{\includegraphics[width=.42\textwidth]{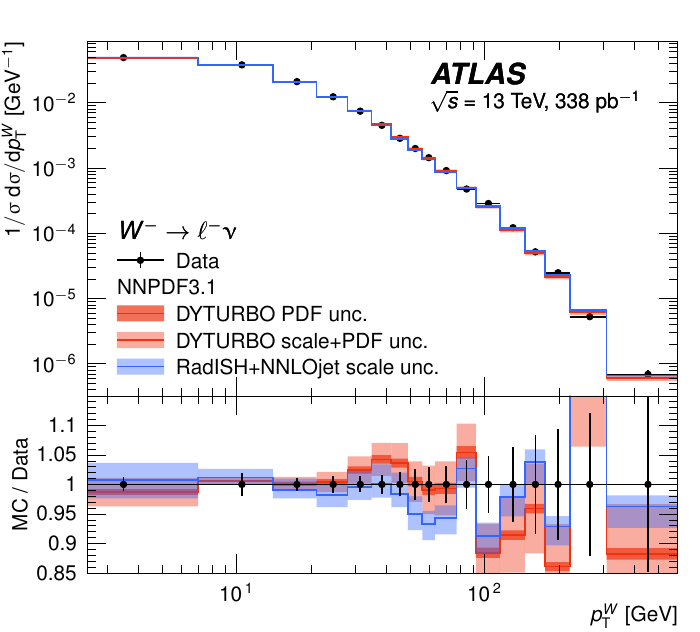}}%
~~~~%
\subfloat[]{\includegraphics[width=.42\textwidth]{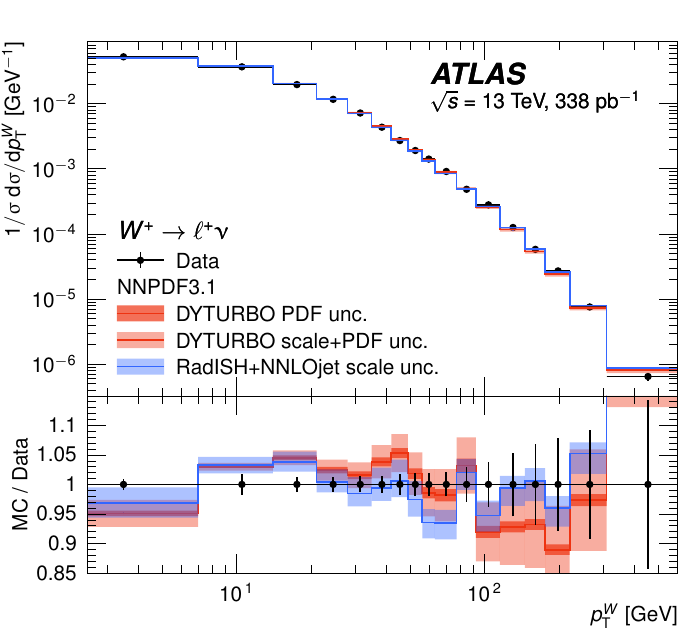}}

\subfloat[]{\includegraphics[width=.42\textwidth]{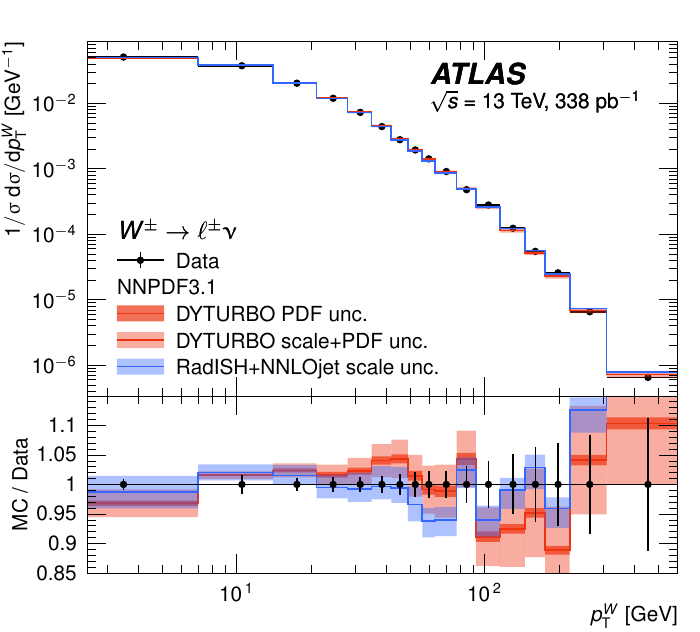}}%
~~~~%
\subfloat[]{\includegraphics[width=.42\textwidth]{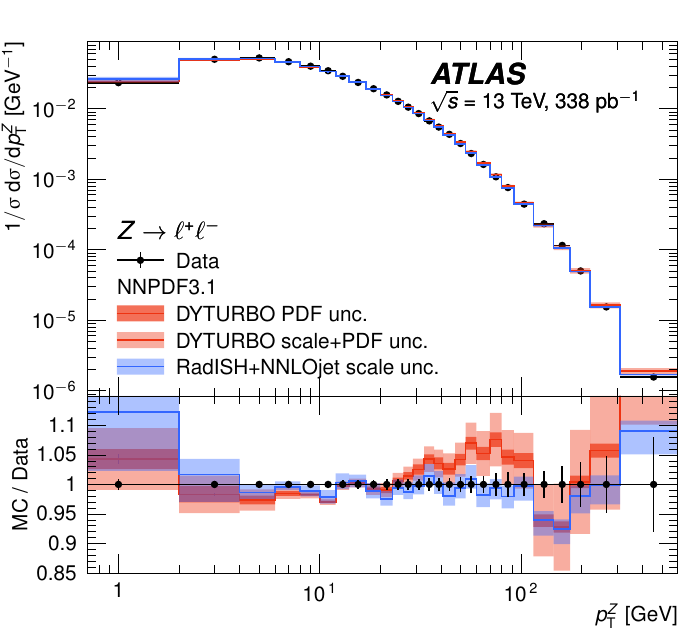}}

\subfloat[]{\includegraphics[width=.42\textwidth]{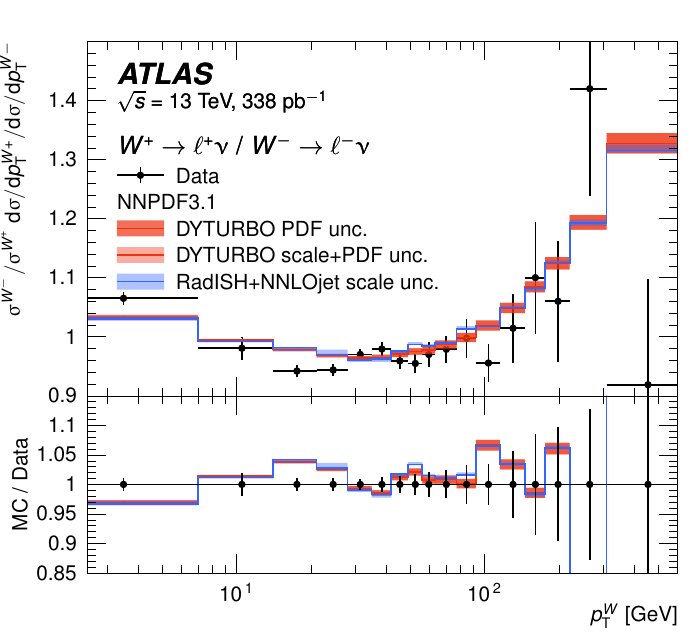}}%
~~~~%
\subfloat[]{\includegraphics[width=.42\textwidth]{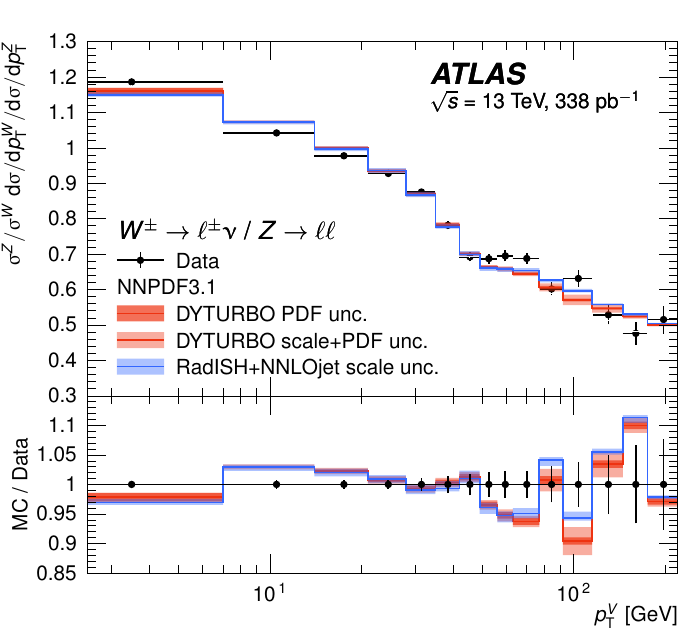}}
\caption{Measurements of normalised differential distributions at
$\sqrt{s} = 13\,\TeV$ (black points) for (a) $W^-$, (b) $W^+$, (c) the
sum $W^\pm$, (d) $Z$ as well as (e) the ratios $W^+/W^-$ and
(f) $W^\pm/Z$ compared to \textsc{RadISH+NNLOjet} and \textsc{DYTurbo} predictions with the
NNPDF3.1 PDF set as described in the text. The shaded areas show
the theoretical uncertainties derived from variations of the QCD
scales and PDFs, the latter only for the \textsc{DYTurbo} predictions. The
lower panels show the ratio of prediction to data with data
markers centred at one and error bars giving the total
measurement uncertainties.}
\label{fig:finalpred_13TeV_norm_radish}
\end{figure}

Figure~\ref{fig:finalpred_13TeV_norm_radish} compares the
measurements at $\sqrt{s} = 13\,\TeV$  to predictions obtained by the
\textsc{RadISH} program~\cite{Bizon:2019zgf, Bizon:2018foh}. Compared to the
\textsc{DYTurbo} predictions shown in the main body, these were computed at a
higher perturbative accuracy by matching to a NNLO prediction of
$Z$+jet production ($O(\alphas^3)$) from
\textsc{NNLOjet}~\cite{Gehrmann-DeRidder:2016jns} with resummation of
$\log(\mll/\pt)$ terms at next-to-next-to-next-to-leading-logarithm
(N$^3$LL) accuracy~\cite{Bizon:2017rah}. The NNPDF3.1 set of
PDFs is used with QCD scales set to
$\muR=\muF=\sqrt{(\mll)^2+(\pt^V)^2}$ and the resummation scale set to
$Q = \mll/2$. Uncertainties in this prediction are derived from
variations of $\muR$ and $\muF$ and two variations of $Q$ by a factor
of two up and down. As comparison, the \textsc{DYTurbo} predictions are shown
as computed with the same PDF set and the uncertainty decomposed into
those originating from the PDF variations and the scale variations.

\begin{figure}[tbph]
\centering
\subfloat[]{\includegraphics[width=.42\textwidth]{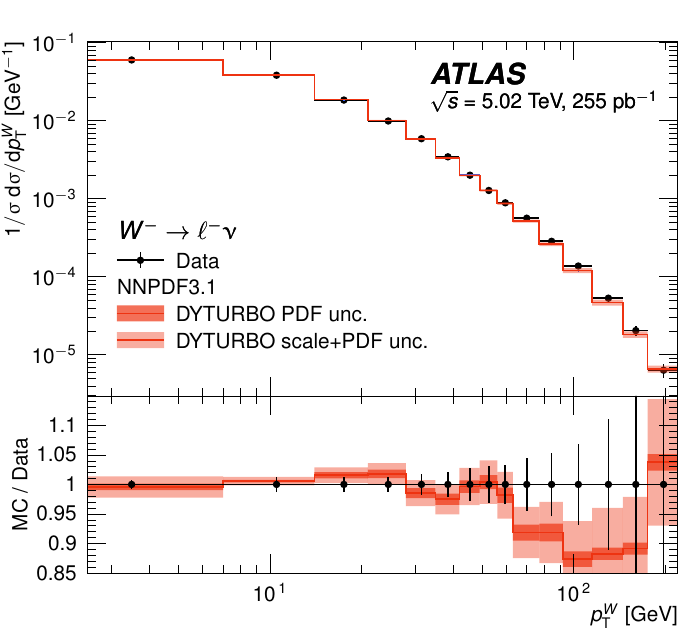}}%
~~~~%
\subfloat[]{\includegraphics[width=.42\textwidth]{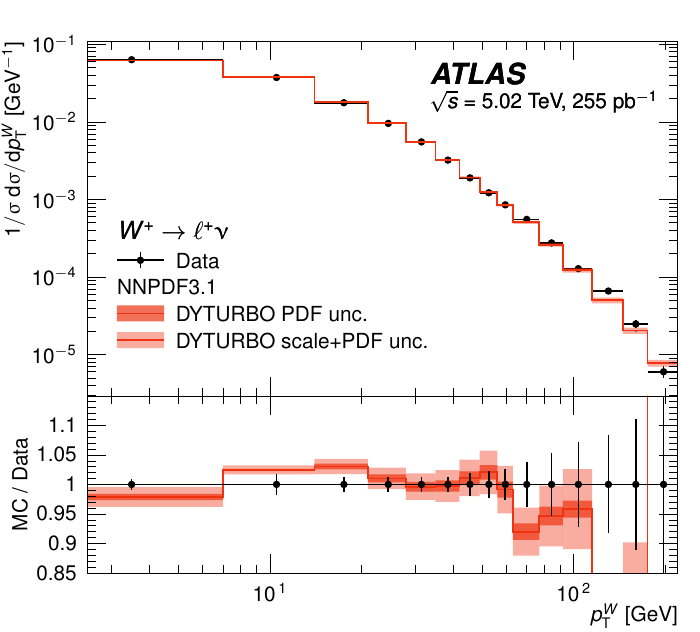}}

\subfloat[]{\includegraphics[width=.42\textwidth]{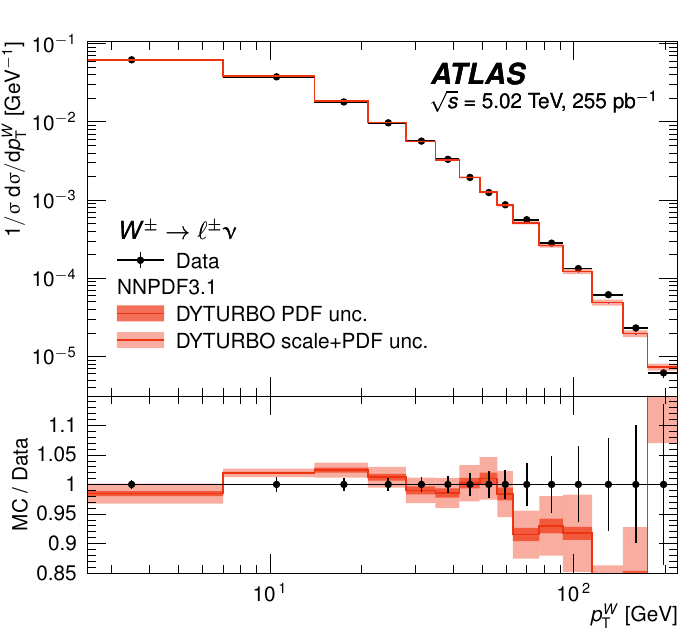}}%
~~~~%
\subfloat[]{\includegraphics[width=.42\textwidth]{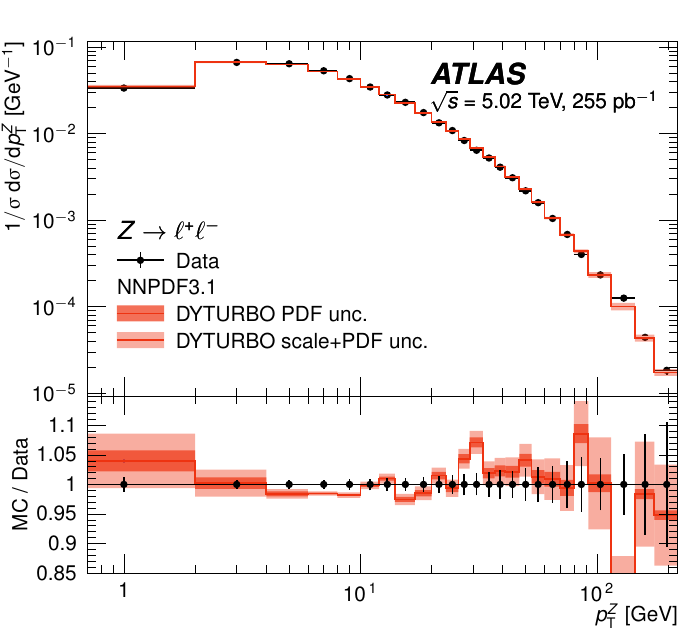}}

\subfloat[]{\includegraphics[width=.42\textwidth]{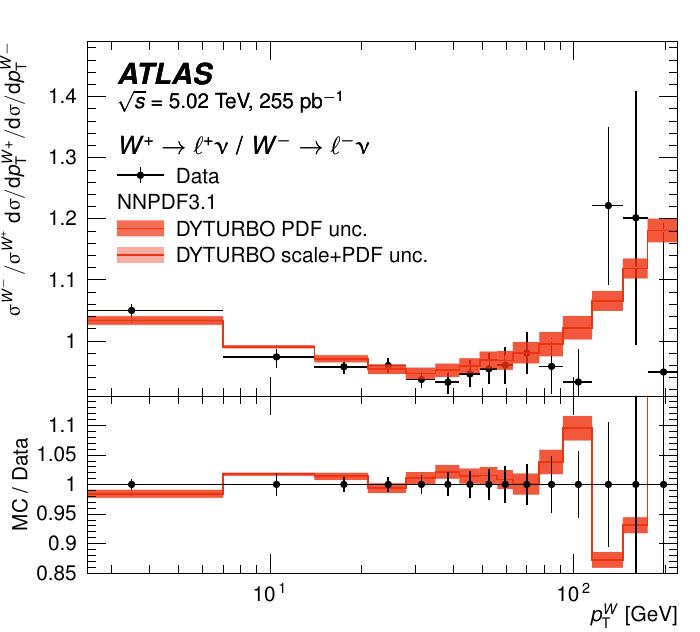}}%
~~~~%
\subfloat[]{\includegraphics[width=.42\textwidth]{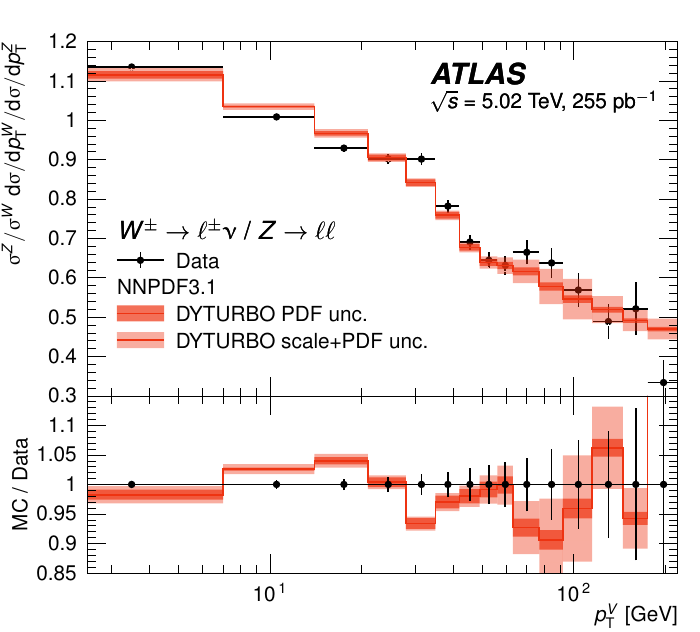}}
\caption{Measurements of normalised differential distributions at
$\sqrt{s} = \fivetev$ (black points) for (a) $W^-$, (b) $W^+$, (c) the
sum $W^\pm$, (d) $Z$ as well as (e) the ratios $W^+/W^-$ and
(f) $W^\pm/Z$ compared to \textsc{DYTurbo}
predictions with different PDF sets (coloured lines) as described in the
text. The lower panels show the ratio of prediction to data with
data markers centred at one and error bars giving the total measurement uncertainties.}
\label{fig:finalpred_5TeV_norm_radish}
\end{figure}

\begin{figure}[tbph]
\centering
\subfloat[]{\includegraphics[width=.42\textwidth]{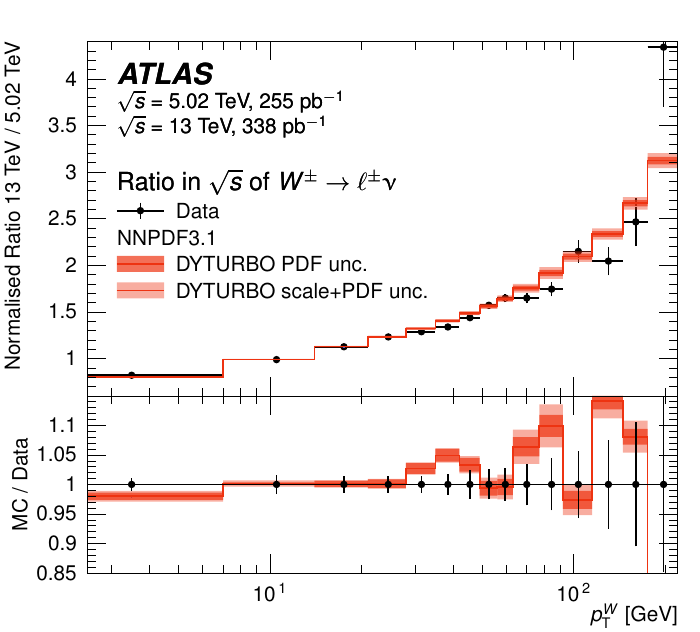}}%
~~~~%
\subfloat[]{\includegraphics[width=.42\textwidth]{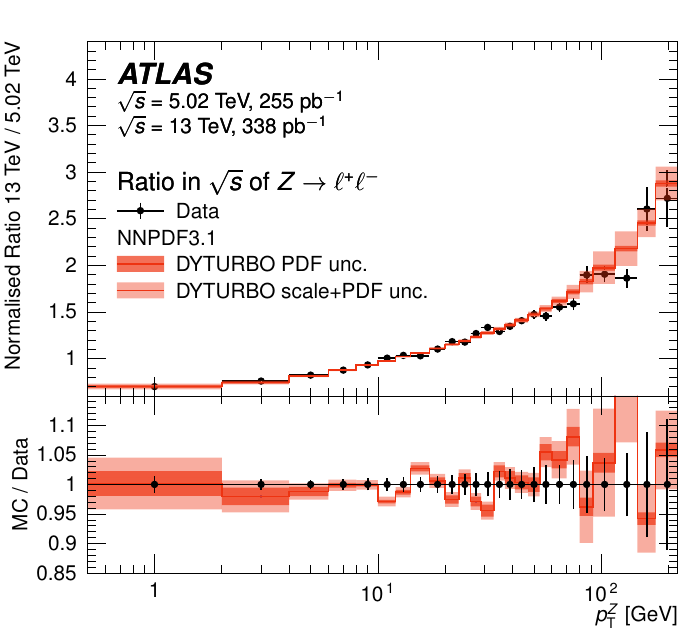}}

\caption{Ratios of normalised differential distributions at
$\sqrt{s} = 13\,\TeV$ to $\sqrt{s} = \fivetev$ (black points) for
(a,c) $W^\pm$ and (b,d) $Z$ compared to (a,b) \textsc{DYTurbo} predictions
with different PDF sets or (b,d) a variety of MC predictions
(coloured lines) as described in the text. The lower panels show
the ratio of prediction to data with data markers centred at one
and error bars giving the size of the total measurement
uncertainties.}
\label{fig:finalpred_513TeV_norm_radish}
\end{figure}

At $\sqrt{s} = \fivetev$ the \textsc{RadISH+NNLOjet} predictions were not available,
therefore Figures~\ref{fig:finalpred_5TeV_norm_radish} and
\ref{fig:finalpred_513TeV_norm_radish} repeat the \textsc{DYTurbo} predictions
as shown in the main body, but provide the same decomposition of the uncertainties
into those estimated by QCD scale variations and those from PDFs.

\clearpage

\section{Comparisons to multijet-merged predictions}
\label{sec:mjcomp}

\begin{figure}[tbph]
\centering
\subfloat[]{\includegraphics[width=.49\textwidth]{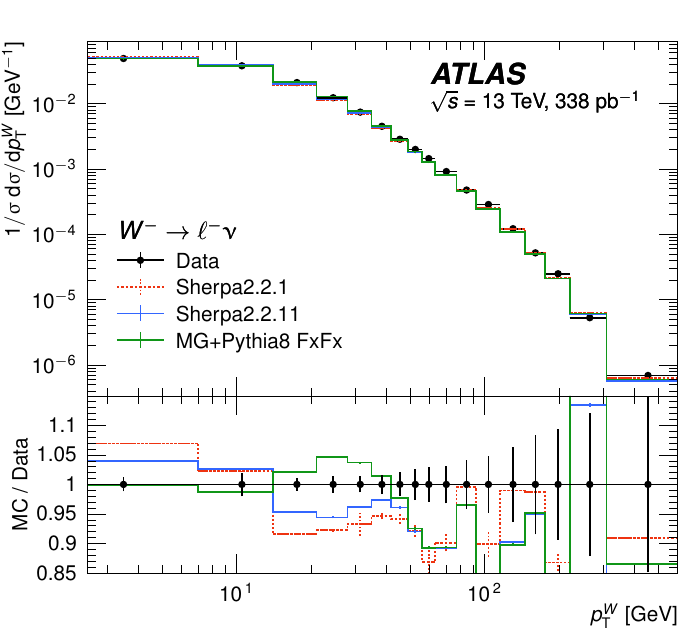}}%
~~~~%
\subfloat[]{\includegraphics[width=.49\textwidth]{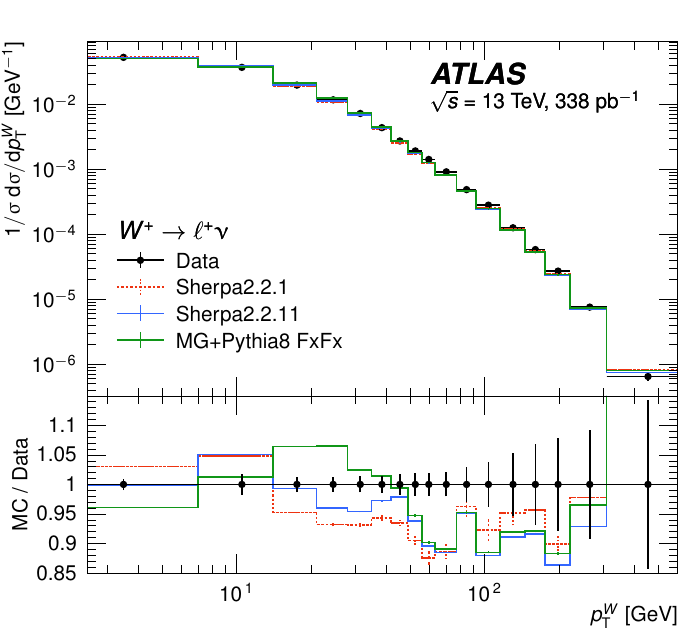}}

\subfloat[]{\includegraphics[width=.49\textwidth]{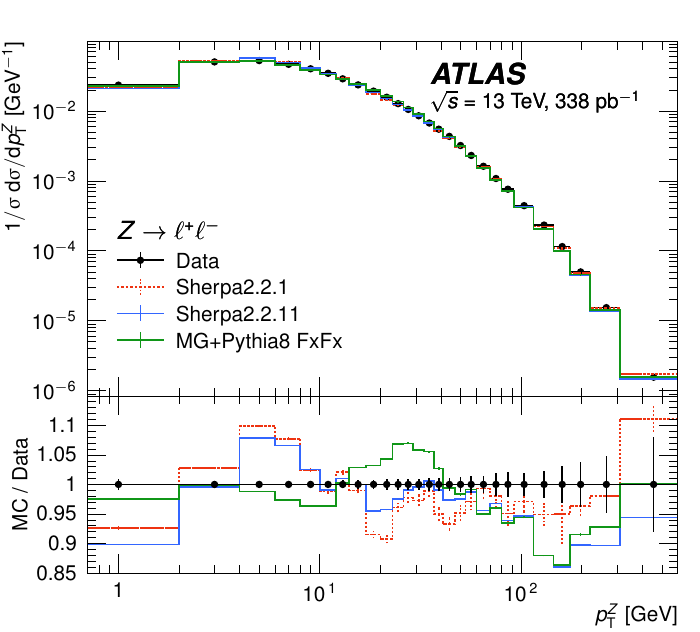}}%
~~~~%
\subfloat[]{\includegraphics[width=.49\textwidth]{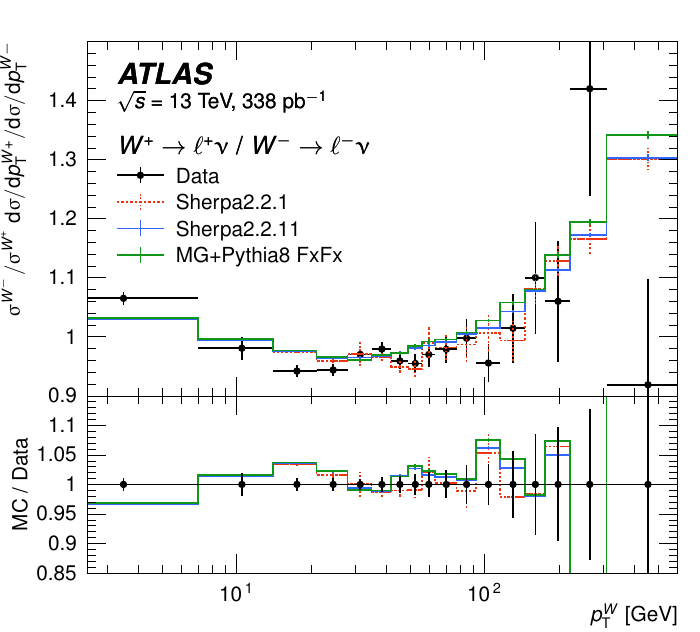}}
\caption{Measurements of normalised differential distributions at
$\sqrt{s} = 13\,\tev$ (black points) for (a) $W^-$, (b) $W^+$, (c) $Z$ as well as (d) the
$W^+/W^-$ ratio compared to a variety MC predictions of multijet-merged
NLO \Sherpa and \textsc{MadGraph\_aMC@NLO FxFx}+\PYTHIAV{8} samples (coloured
lines) as described in the text. The lower panels show the ratio of prediction to data with
data markers centred at one and error bars giving the size of the total measurement uncertainties.}
\label{fig:finalpred_W2_13TeV_norm}
\end{figure}

Figure~\ref{fig:finalpred_W2_13TeV_norm} compares the
measurements at $\sqrt{s} = 13\,\TeV$  to three multijet-merged
predictions that were discussed in detail in
Ref.~\cite{PMGR-2021-01}. All samples show a similar behaviour across
the $W^-, W^+, Z$ measurements and are very close to the data for the
$W^+/W^-$ ratio. The \SHERPA[2.2.11] sample improves the behaviour
of the \SHERPA[2.2.1] sample in the region $\pt = 20\,\GeV$, where
matrix-element and parton shower emissions are matched. Both \SHERPA
samples have some room for improvement in the region of lowest
$\pt \lesssim 10\,\GeV$. The \textsc{MadGraph\_aMC@NLO FxFx}+\PYTHIAV{8} is
very close to \SHERPA[2.2.11] at higher $\pt > 40\,\GeV$ and appears
to behave better at low $\pt < 10\,\GeV$. It shows, however, some excess over data for
$\pt = 20$--$30\,\GeV$.

\clearpage

\section{Total integrated $W$ and $Z$ cross sections and ratios}
\label{sec:totalxsec}

The total production cross section for $pp \to W \to \ell\nu$ and
$pp \to Z/\gamma^* \to \ell\ell$ may be obtained from the fiducial
measurements presented in this paper in Table~\ref{tab:fidxsec}
through theoretically calculated acceptance factors $A$ as
$\sigma_\mathrm{tot} = \sigma_\mathrm{fid}/A$. For the $Z/\gamma^*$
case, the total phase space retains the invariant mass range of
$66 < \mll <116 \,\GeV$.  Table~\ref{tab:totxsec1} summarises the $A$
values as computed from the nominal \Powheg+\PYTHIAV{8} samples using
the CT18 PDF set. The uncertainties are estimated from variations of
QCD scales, PDFs and the difference to the alternative \SHERPA samples. The
corresponding total cross sections are given as well. The typical
uncertainty on the acceptance correction is $2\%$.
The equivalent computations for cross-section ratios are presented in Table~\ref{tab:totxsec2}.

\begin{table}[p]
\caption{Acceptance factors $A$ to derive total production cross
sections for $W^-$, $W^+$ and $Z$ production at \fivetev and
$13\,\tev$ with total theory uncertainties. The measured fiducial cross sections and ratios
are repeated from Table~\ref{tab:fidxsec} and the resulting total cross sections are given.}
\label{tab:totxsec1}
\centering

\clearpage

\begin{tabular}{lcccc}
\toprule
\toprule
& $W^-\to \ell\nu$ & $W^+\to \ell\nu$ & $W^\pm \to \ell\nu$ & $Z\to \ell\ell$ \\

\midrule
\multicolumn{5}{c}{$\sqrt{s} =$ \fivetev} \\
\midrule
$A$	& 0.475 $\pm$ 0.004	& 0.506 $\pm$ 0.008	& 0.494 $\pm$ 0.006	& 0.492 $\pm$ 0.006\\
\midrule
\midrule
$\sigma_\mathrm{fid}$ [pb]	& 1384 $\pm$ 16	& 2228 $\pm$ 25	& 3612 $\pm$ 40	& 333.0 $\pm$ 4.1\\[1ex]
$\sigma_\mathrm{tot}$ [pb]	& 2913 $\pm$ 41	& 4401 $\pm$ 85	& 7316 $\pm$ 124	& 677 $\pm$ 12\\
\midrule

\midrule
\multicolumn{5}{c}{$\sqrt{s} =$ $13\,\TeV$} \\
\midrule
$A$	& 0.399 $\pm$ 0.006	& 0.386 $\pm$ 0.009	& 0.391 $\pm$ 0.008	& 0.393 $\pm$ 0.011\\
\midrule
\midrule
$\sigma_\mathrm{fid}$ [pb]	& 3486 $\pm$ 38	& 4571 $\pm$ 49	& 8057 $\pm$ 88	& 780.3 $\pm$ 10.4\\[1ex]
$\sigma_\mathrm{tot}$ [pb]	& 8740 $\pm$ 160	& 11850 $\pm$ 310	& 20580 $\pm$ 460	& 1986 $\pm$ 59\\
\midrule

\bottomrule
\bottomrule
\end{tabular}

\clearpage

\end{table}

\begin{table}[p]
\caption{Acceptance factors $A$ to derive total production cross-section ratios at \fivetev and
$13\,\tev$ with total theory uncertainties. The measured fiducial ratios
are repeated from Table~\ref{tab:fidratios} and the resulting total cross-section ratios are given.}
\label{tab:totxsec2}
\centering

\clearpage

\begin{tabular}{lcccc}
\toprule
\toprule
& $W^+/W^-$ & $W^\pm/Z$ \\

\midrule
\multicolumn{3}{c}{$\sqrt{s} =$ \fivetev} \\
\midrule
$A^V/A^{V'}$	& 1.065 $\pm$ 0.012	& 1.003 $\pm$ 0.008\\
\midrule
\midrule
$\sigma^V_\mathrm{fid}/\sigma^{V'}_\mathrm{fid}$	& 1.609 $\pm$ 0.005	& 10.85 $\pm$ 0.08\\[1ex]
$\sigma^V_\mathrm{tot}/\sigma^{V'}_\mathrm{tot}$	& 1.511 $\pm$ 0.017	& 10.82 $\pm$ 0.12\\
\midrule

\midrule
\multicolumn{3}{c}{$\sqrt{s} =$ $13\,\TeV$} \\
\midrule
$A^V/A^{V'}$	& 0.968 $\pm$ 0.011	& 0.996 $\pm$ 0.010\\
\midrule
\midrule
$\sigma^V_\mathrm{fid}/\sigma^{V'}_\mathrm{fid}$	& 1.308 $\pm$ 0.005	& 10.30 $\pm$ 0.11\\[1ex]
$\sigma^V_\mathrm{tot}/\sigma^{V'}_\mathrm{tot}$	& 1.351 $\pm$ 0.017	& 10.34 $\pm$ 0.15\\
\midrule

\bottomrule
\bottomrule
\end{tabular}

\clearpage

\end{table}

Tables~\ref{tab:atlascms1} and \ref{tab:atlascms2} display extrapolation factors $E$ that
may be used to translate the measured fiducial cross sections and ratios, respectively, to the fiducial phase
space of the CMS analysis~\cite{CMS:2024myi}:
\begin{itemize}
\item  $W \rightarrow \ell\nu$ :
$\pT^{\ell}>25~\GeV$, $|\eta^{\ell}|<2.4$, $\pT^{\nu}>0$, and $\mt > 40~\GeV$;
\item  $Z \rightarrow \ell\ell$ :
$\pT^{\ell}>25~\GeV$, $|\eta^{\ell}|<2.4$, and $60~\GeV < m_{\ell\ell} <120 ~\GeV$.
\end{itemize}
As the phase spaces are similar, these factors are generally close to unity (within about
10\%) and the theoretical uncertainties, computed in the same way as for $A$ factors, are small.

\begin{table}[p]
\caption{Extrapolation factors $E$ to translate the fiducial
integrated cross sections to the phase space of the analysis
performed by the CMS Collaboration~\cite{CMS:2024myi}. The
measured fiducial cross sections are repeated from
Table~\ref{tab:fidxsec} and the resulting
extrapolated cross sections in the CMS fiducial volume are given.}
\label{tab:atlascms1}
\centering

\clearpage

\begin{tabular}{lccc}
\toprule
\toprule
Experiment & $W^-\to \ell\nu$ & $W^+\to \ell\nu$ & $Z\to \ell\ell$ \\

\midrule
\multicolumn{4}{c}{$\sqrt{s} =$ \fivetev} \\
\midrule
$E$	& 1.0951 $\pm$ 0.0053	& 1.1009 $\pm$ 0.0042	& 0.9637 $\pm$ 0.0008\\
\midrule
\midrule
ATLAS [pb]	& 1384 $\pm$ 16	& 2228 $\pm$ 25	& 333.0 $\pm$ 4.1\\
ATLAS@CMSfid [pb]	& 1516 $\pm$ 19	& 2453 $\pm$ 29	& 320.9 $\pm$ 4.0\\

\midrule
\multicolumn{4}{c}{$\sqrt{s} =$ $13\,\TeV$} \\
\midrule
$E$	& 1.1156 $\pm$ 0.0031	& 1.1154 $\pm$ 0.0014	& 0.9621 $\pm$ 0.0011\\
\midrule
\midrule
ATLAS [pb]	& 3486 $\pm$ 38	& 4571 $\pm$ 49	& 780.3 $\pm$ 10.4\\
ATLAS@CMSfid [pb]	& 3889 $\pm$ 44	& 5099 $\pm$ 55	& 750.7 $\pm$ 10.0\\

\bottomrule
\bottomrule
\end{tabular}

\clearpage

\end{table}

\begin{table}[p]
\caption{Extrapolation factors $E$ to translate the fiducial
integrated cross-section ratios to the phase space of the analysis
performed by the CMS Collaboration~\cite{CMS:2024myi}. The
measured fiducial cross-section ratios are repeated from
Table~\ref{tab:fidratios} and the resulting
extrapolated cross-section ratios in the CMS fiducial volume are given.}
\label{tab:atlascms2}
\centering

\clearpage

\begin{tabular}{lccccc}
\toprule
\toprule
Experiment & $W^+/W^-$ & $W^\pm/Z$ \\

\midrule
\multicolumn{3}{c}{$\sqrt{s} =$ \fivetev} \\
\midrule
$E^V/E^{V'}$	& 1.0052 $\pm$ 0.0013	& 1.1401 $\pm$ 0.0050\\
\midrule
\midrule
ATLAS	& 1.609 $\pm$ 0.005	& 10.85 $\pm$ 0.08\\
ATLAS@CMSfid	& 1.617 $\pm$ 0.005	& 12.37 $\pm$ 0.11\\

\midrule
\multicolumn{3}{c}{$\sqrt{s} =$ $13\,\TeV$} \\
\midrule
$E^V/E^{V'}$	& 0.9998 $\pm$ 0.0033	& 1.1594 $\pm$ 0.0009\\
\midrule
\midrule
ATLAS	& 1.308 $\pm$ 0.005	& 10.30 $\pm$ 0.11\\
ATLAS@CMSfid	& 1.308 $\pm$ 0.007	& 11.94 $\pm$ 0.13\\

\bottomrule
\bottomrule
\end{tabular}

\clearpage

\end{table}

\FloatBarrier


%
%
\printbibliography

\clearpage
 
\begin{flushleft}
\hypersetup{urlcolor=black}
{\Large The ATLAS Collaboration}

\bigskip

\AtlasOrcid[0000-0002-6665-4934]{G.~Aad}$^\textrm{\scriptsize 102}$,
\AtlasOrcid[0000-0002-5888-2734]{B.~Abbott}$^\textrm{\scriptsize 120}$,
\AtlasOrcid[0000-0002-1002-1652]{K.~Abeling}$^\textrm{\scriptsize 55}$,
\AtlasOrcid[0000-0001-5763-2760]{N.J.~Abicht}$^\textrm{\scriptsize 49}$,
\AtlasOrcid[0000-0002-8496-9294]{S.H.~Abidi}$^\textrm{\scriptsize 29}$,
\AtlasOrcid[0000-0002-9987-2292]{A.~Aboulhorma}$^\textrm{\scriptsize 35e}$,
\AtlasOrcid[0000-0001-5329-6640]{H.~Abramowicz}$^\textrm{\scriptsize 151}$,
\AtlasOrcid[0000-0002-1599-2896]{H.~Abreu}$^\textrm{\scriptsize 150}$,
\AtlasOrcid[0000-0003-0403-3697]{Y.~Abulaiti}$^\textrm{\scriptsize 117}$,
\AtlasOrcid[0000-0002-8588-9157]{B.S.~Acharya}$^\textrm{\scriptsize 69a,69b,m}$
% ...
\bigskip
\\

% ...
$^{29}$Physics Department, Brookhaven National Laboratory, Upton NY; United States of America.\\
% ...
$^{35}$$^{(a)}$Facult\'e des Sciences Ain Chock, Universit\'e Hassan II de Casablanca;$^{(b)}$Facult\'{e} des Sciences, Universit\'{e} Ibn-Tofail, K\'{e}nitra;$^{(c)}$Facult\'e des Sciences Semlalia, Universit\'e Cadi Ayyad, LPHEA-Marrakech;$^{(d)}$LPMR, Facult\'e des Sciences, Universit\'e Mohamed Premier, Oujda;$^{(e)}$Facult\'e des sciences, Universit\'e Mohammed V, Rabat;$^{(f)}$Institute of Applied Physics, Mohammed VI Polytechnic University, Ben Guerir; Morocco.\\
% ...
$^{49}$Fakult\"{a}t Physik , Technische Universit{\"a}t Dortmund, Dortmund; Germany.\\
% ...
$^{55}$II. Physikalisches Institut, Georg-August-Universit\"{a}t G\"ottingen, G\"ottingen; Germany.\\
% ...
$^{69}$$^{(a)}$INFN Gruppo Collegato di Udine, Sezione di Trieste, Udine;$^{(b)}$ICTP, Trieste;$^{(c)}$Dipartimento Politecnico di Ingegneria e Architettura, Universit\`a di Udine, Udine; Italy.\\
% ...
$^{102}$CPPM, Aix-Marseille Universit\'e, CNRS/IN2P3, Marseille; France.\\
% ...
$^{117}$Department of Physics, New York University, New York NY; United States of America.\\
% ...
$^{120}$Homer L. Dodge Department of Physics and Astronomy, University of Oklahoma, Norman OK; United States of America.\\
% ...
$^{150}$Department of Physics, Technion, Israel Institute of Technology, Haifa; Israel.\\
$^{151}$Raymond and Beverly Sackler School of Physics and Astronomy, Tel Aviv University, Tel Aviv; Israel.\\
% ...
$^{a}$ Also Affiliated with an institute covered by a cooperation agreement with CERN.\\
$^{b}$ Also at An-Najah National University, Nablus; Palestine.\\
$^{c}$ Also at Borough of Manhattan Community College, City University of New York, New York NY; United States of America.\\
$^{d}$ Also at Center for High Energy Physics, Peking University; China.\\
$^{e}$ Also at Center for Interdisciplinary Research and Innovation (CIRI-AUTH), Thessaloniki; Greece.\\
$^{f}$ Also at Centro Studi e Ricerche Enrico Fermi; Italy.\\
% ...
$^{m}$ Also at Department of Physics, King's College London, London; United Kingdom.\\
% ...

\end{flushleft}


%
%

%

\end{document}